# NUCLEAR REACTIONS
## MECHANISM AND SPECTROSCOPY
## VOLUME I

Prof. Ron W. Nielsen
(aka Jan Nurzynski)

Griffith University

2016

# Nuclear Reactions

## Mechanism and Spectroscopy
## Volume I

## Prof. Ron W. Nielsen
### (aka Jan Nurzynski)
### Griffith University, Gold Coast, Qld, 4222, Australia
### ronwnielsen@gmail.com

**About the Author:** 

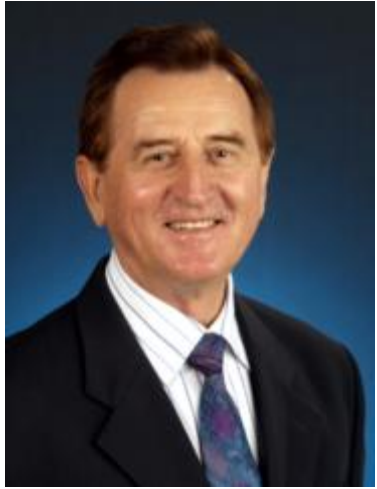

**Abstract:** Volume I of two. This document could be of interest to anyone who wants to have a comprehensive inside information about the research in nuclear physics from its early beginnings to later years. It describes highlights of my research work from the late 1950s to the late 1980s, during the best years of nuclear research, when this field of study was wide opened for its exploration. It presents a panorama of experimental and theoretical methods used in the study of nuclear reactions (their mechanism and their application to the study of nuclear structure), the panorama ranging from simple detection techniques used in the early research to more complicated in later years, from simple theoretical interpretations to more complicated descriptions. This document describes my research work in Poland, Australia, Switzerland and Germany using various particle accelerators and a wide range of experimental and theoretical techniques. It presents a typical cross section of experimental and theoretical work in the early and later stages of nuclear research in the field of nuclear reactions.

# Contents

Volume I

# Preface

This document presents highlights of my research work in nuclear physics carried out over around 30 years. It starts with the description of the first ever research work in Poland in the field of nuclear reactions and continues with my research in Australia, Switzerland and Germany. After moving to Australia, I have introduced a study of direct nuclear reactions. Methods used in the analysis of these reactions were later applied in the study of heavy-ion induced reactions.

### Research objectives

Objectives of my research were to study the mechanism of nuclear reactions and nuclear spectroscopy by using

- direct nuclear reactions,
- polarization phenomena in nuclear reactions, and
- heavy-ion-induced reactions.

### Research institutes

I have carried out my research work in the following research centres:

- Institute of Nuclear Physics, Cracow, Poland
- Department of Nuclear Physics, Institute of Advanced Studies, Australian National University, Canberra, ACT, Australia
- Laboratorium für Kernphysik, ETHZ, Zürich, Switzerland
- Schweizerische Institut für Nuklearforschung, Villigen, Switzerland
- Max-Plank Institut für Kernphysik, Heidelberg, Germany
- Institut für Angewandte Kernphysik and Zyklotron-Laboratorium Kernforschungszentrum Karlsruhe, Karlsruhe, Germany

### Accelerators

My research work was supported by the following particle accelerators:

- U-120 cyclotron in Poland
- EN tandem electrostatic accelerators in Canberra and Zurich
- 14 UD Pelletron accelerator in Canberra
- Cyclograph in Canberra
- Isochronous cyclotron in Karlsruhe
- Injector cyclotron at Schweizerische Institut für Nuklearforschung in Villigen

### Polarized ion sources

Polarized ion sources used in the study of polarization phenomena were:

- The atomic beam polarized ion sources
- Lamb-shift polarized ion sources

### Data acquisition systems

I have used a broad range of particle detection systems. They were:

- Nuclear emulsions
- Proportional counters
- Scintillation counters
- Solid state detectors

- Resistive-wire gas proportional detector
- Detector telescopes
- Magnetic spectrometers

Other data acquisition and processing systems included:

- Hutchinson-Scarrott pulse hight analyser
- 400 channel RIDL analyser
- 512 channel RCL analyser
- PDP computers (PDP-8 and PDP-11)
- IBM 1620 computer
- IBM 1800 computer
- Hewlett-Packard (HP2100A) computers
- VAX online computers (VAX750, VAX1000, VAX2000, VAX3100, VAX3200, and VAX 4000)

### Theoretical frameworks

In addition to the experimental work, I have also carried out theoretical analysis of my data. This work was supported by internationally shared computer codes, which I have adapted to available computers and modified whenever necessary. I have also written many other computer codes, when necessary, to support my research work.

Theoretical frameworks used in my research included:

- Optical model
- Diffraction theory
- Plane wave theory
- Distorted wave theory
- Coupled channels formalism
- Phase-shift analysis
- Resonating group theory
- Faddeev formalism
- R-matrix theory

### Mainframe computers

The mainframe computers used to support the evaluation of data and the theoretical analysis of experimental results included:

- IBM 360/50
- UNIVAC-1108, 1100/42, 1100/82
- VAX-780

The following summary is arranged in approximately chronological order. It starts with my pioneering research work in Poland and ends with experiments in the field of heavy-ion-induced reactions.

Ron W. Nielsen
(aka Jan Nurzynski)
December, 2016
Gold Coast, Australia

# 1

## Neutron Polarization in the $^{12}$C(d,n)$^{13}$N Stripping Reaction

***Key features:***

1. A challenging and complex experiment involving a detection of low-yield neutrons and the measurements of neutron polarization without using a polarized ion source but rather by using the "double scattering" method. (Polarized ion source is a complex apparatus and was not available for this experiment.)

2. The first experimental work in the field of nuclear reactions in Poland.

3. The first ever published results on the neutron polarization at medium deuteron energies.[1]

4. The first attempt to compare experimental results at these energies with the early theoretical predictions.

5. Large polarization detected in our experiment suggested that this reaction could serve as a source of polarized fast neutrons.

***Abstract***: The polarization of neutrons from the stripping reaction $^{12}$C(d,n)$^{13}$N induced by 12.9 MeV deuterons was measured using a "double scattering" method. Measurements were carried out in the angular range of $15^0 - 60^0$ (lab). Results of measurements are compared with the early theoretical predictions. They demonstrate the violation of the theoretical sign rule and thus challenge the early interpretations of the mechanism of the nucleon polarization.

## In the beginning…

Research in nuclear physics in Poland commenced in 1955 in the School of Physics of the Jagiellonian University, in Cracow. Established in 1364, it is one of the oldest universities in Europe. Its well-known alumni include Nicolaus Copernicus and Pope John Paul II.

The University is located in the centre of Crocow, and thus in the old town area. Most of the walls of the original Cracow were destroyed and were replaced by a public park, called *Planty*. The School of Physics is located next to the park but it is also close to the original Collegium Maius and next to the new Collegium Novum. My office was on the first floor of the School of Physics, facing a courtyard. In the middle of the office, incongruously for such a historical surrounding, was an ion source my colleague and I were constructing. A few other groups and individuals were also carrying out their work in various rooms and in the basement of this old building. Meanwhile, a new Institute of Nuclear Physics was being constructed on the outskirts of Cracow and eventually we have all moved to new offices.

Research work at that time was carried out in three major fields: Nuclear Physics; Physics of Structure of Solids, Liquids and Gases; and Applied Nuclear Physics. The layout of the Institute in 1965 is shown in Figure 1.1. By that time, the Institute established close links with various research centres around the world. The link with Australia, shown in Figure 1.2, was my presence in the Department of Nuclear Physics in the Australian National University.

---

[1] Our results (at $E_d$ = 12.9 MeV) were published in 1959 (Budzanowski *et al.* 1959). Two years later, Haeberli (1961) published results of measurements for low energy deuterons ($E_d$ < 3.6 MeV). However, the same results were mentioned earlier in a form of a short abstract at a meeting of the American Physical Society (Haeberli and Rolland 1957).

The first director of the Institute was Professor Henryk Niewodniczanski,[2] who was amiably called *Papa* by those who knew him well. After his premature death of heart attack, the name of the Institute was changed to The Henryk Niewodniczanski Institute of Nuclear Physics.

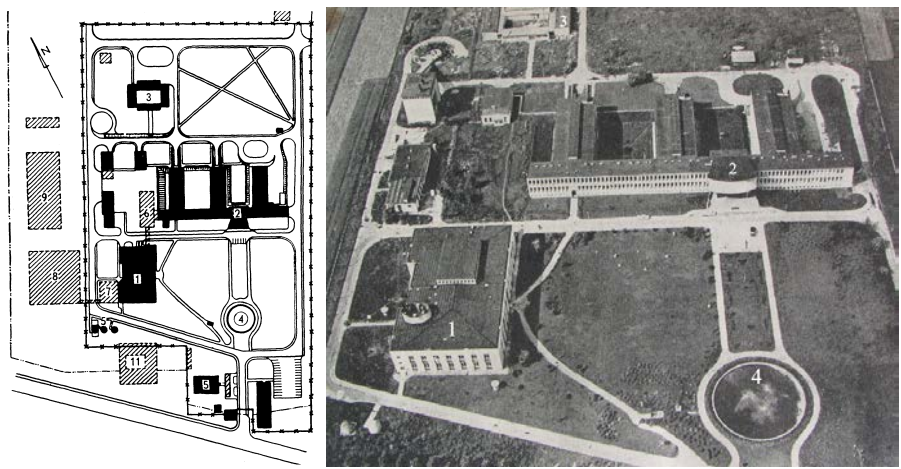

Figure 1.1. The map of the Institute of Nuclear Physics in 1965, on the 10th anniversary of the commencement of nuclear research in Poland. 1 – The location of the U -120 cyclotron where the first experimental work in nuclear physics in Poland was done and where the first in the world measurements of neutron polarization at medium deuteron energy were successfully carried out; 2 – The main building containing other laboratories, library, lecture theatre, workshops and administration offices; 3 – The Low Temperature Laboratory; 4 – A fountain made of the water from the cyclotron cooling system; 5 – A building for storing radioactive materials. Shaded areas mark future (at that time) developments: 6 – A Computer Centre and Library; 7 – Extension for a new target hall of the U -120 cyclotron; 8 – Proposed heavy ion cyclotron; 9 Workshops. (Reproduced from IFJ 1965.)

The major research facility in the Institute was the U-120 cyclotron, which was purchased from Russia. However, at the time when I commenced my research work there, two experimental halls belonging to the cyclotron were empty and the necessary equipment had to be either constructed or purchased.

In 1959 we have successfully completed the first ever experiment in nuclear reactions in Poland. This was also the first experimental work in the world on the neutron polarization in deuteron-induced stripping reactions at medium deuteron energies.

Research in nuclear physics was everywhere at that time still in its early stages. Only a few years earlier, it was discovered that nuclear reactions do not proceed necessarily via a compound nucleus but can also involve direct transitions. The new concept of direct nuclear reactions has thus been introduced and we were interested in studying their mechanism and their application in nuclear spectroscopy. The distinction between the compound nucleus reactions and the direct nuclear reactions is illustrated schematically in Figure 1.4.

---

[2] Called after his death "the Rutherford of Poland" for introducing nuclear research in Poland (Hodgson, P. E. http://www.zwoje-scrolls.com/zwoje36/text04.htm).

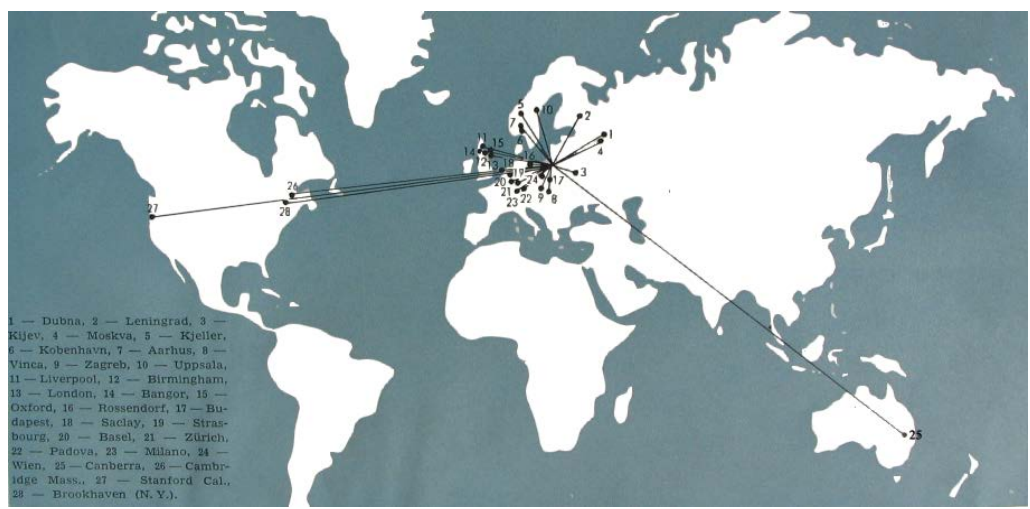

Figure 1.2. International connections of the Institute of Nuclear Physics in Cracow in 1965 in the form of exchange of scientists for longer periods. The link with Australia marks my presence in the Department of Nuclear Physics of The Australian National University in Canberra. (Reproduced from IFJ 1965.)

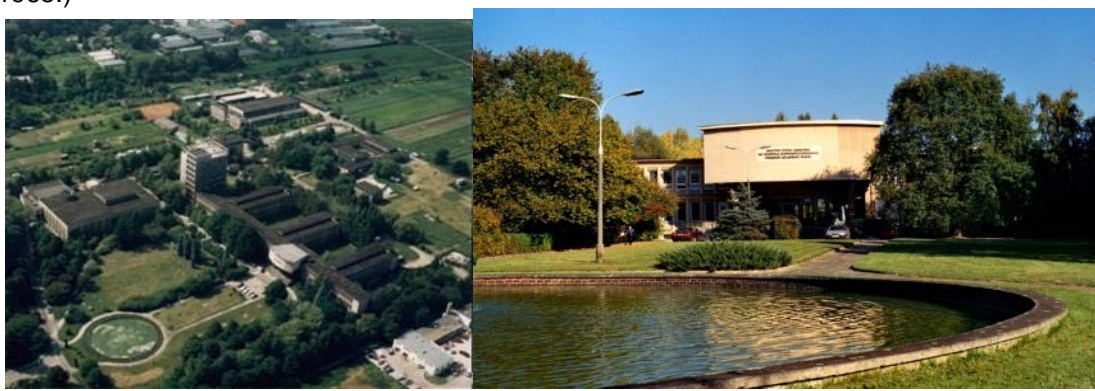

Figure 1.3. Left-hand side: The aerial view of The Henryk Niewodniczanski Institute of Nuclear Physics, in 2006 (EPPOG 2006). Right-hand side: The view of the main building in 2006. In the foreground is the pool for the cyclotron cooling system and behind it the lecture theatre located in the main building (building 2 in Figure 1.1). (Photo credit: P. Zielinski, IFJ 2005.)

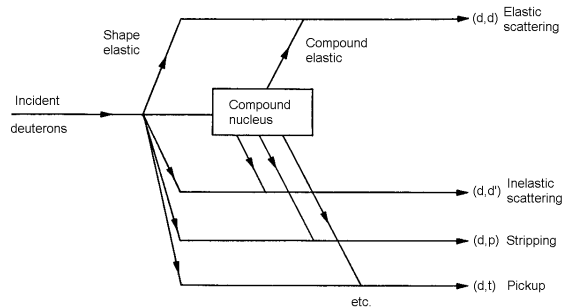

Figure 1.4. Schematic illustration showing two types of reactions, direct and compound nucleus (modified from Hodgson 1971). Reactions can proceed directly, via compound nucleus, or both.

The first experimental evidence for the direct nuclear reactions was provided by Burrows, Gibson and Rotblat (1950) and by Holt and Young (1950). These authors observed forward peaking in reactions induced by 8 MeV deuterons, which could not be explained by the compound nucleus mechanism.

The first theories describing the direct reaction mechanism were proposed by Butler (1950, 1951), Hubby (1950), Hubby and Newns (1951), and Bhatia *et al.* (1952). The first theoretical descriptions of direct nuclear reactions were based on a simple plane-wave approximation. Later, distorted wave approximation and coupled channels formalisms were introduced.

## Introduction

Early predictions of nucleon polarization in deuteron stripping reactions were presented by Newns (1953), Hittmair (1956); Horowitz and Messiah (1953), Cheston (1954), Sawicki (1957), Newns and Refai (1958), and Sachtler (1959, 1960). Experimentally, the polarization was observed for the first time by Hillman (1956) for protons in the reaction $^{12}$C(d,p)$^{13}$C leading to the ground state in $^{13}$C. Later proton polarization from the $(d,p)$ reaction has been also reported by other authors who used $^{9}$Be, $^{10}$B, $^{12}$C, $^{28}$Si and $^{40}$Ca as target nuclei (Bokhari *et al.* 1958; Hensel and Parkinson 1958; Hird *et al.* 1959; Jurić and Čirilov 1959; Juveland and Jentschke 1958).

When we started our experiment, there were no published results on neutron polarization in the $(d,n)$ reaction. However, in the course of our measurements a brief abstract appeared about the measurements of Haeberli and Rolland (1957). Their measurements were for low-energy deuterons of around 2.4-3.6 MeV. Apart from this brief abstract, no additional information was available about their results until about four years later (Haeberli 1961). There were no published results for higher deuteron energies at that time.

Simple theories were used to explain the polarization generated in deuteron stripping reactions. For reactions at small angles, the sign of the polarization was linked with the total angular momentum of the transferred nucleon (Newns 1953; Newns and Rafai 1958; Satchler 1959). Our aim was to study neutron polarization for the $^{12}$C(d,n)$^{13}$N reaction induced by 12.9 MeV deuterons and compare the experimental results with the early theoretical predictions.

## Experimental arrangement

Measurements of neutron polarization were carried out using the so-called "double scattering" method (see Figure 1.5). The method consists of two steps, which involves two targets. The first target is used to produce the polarization and the second to measure it. In our case, neutron polarization was produced by the $^{12}$C(d,n)$^{13}$C reaction and analysed using the n-$\alpha$ scattering.

This method of measurements has been described by Wolfenstein (1956; see also Ohlsen 1972). The differential cross section for the second target (the analyser) is given by the following relation:

$$\sigma(\theta', \phi) = \sigma_0(\theta')(1 + P_1 P_2 \cos \phi)$$

where $\sigma_0(\theta', \phi)$ is the cross section for the scattering of unpolarized neutrons, $P_1$ is the polarization produced in the $^{12}$C(d,n)$^{13}$N reaction, $P_2$ is the analyzing power of the n-$\alpha$ scattering, and $\phi$ is the azimuthal angle of the polarization analyser. The polarization $P_1$ is positive if it is in the direction of $\vec{k}_d \times \vec{k}_n$ (see Figure 1.5). The azimuthal angle $\phi$ is defined by the product $\vec{P}_1 \cdot \vec{n}$, where $\vec{n}$ is the unit vector in the direction of $\vec{k}_n \times \vec{k}_{n'}$.

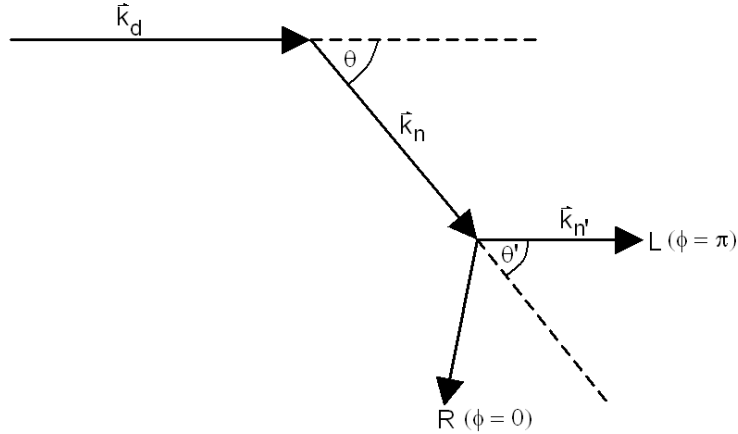

Figure 1.5. The "double scattering" (or the two-step) method of polarization measurements used in the study of neutron polarization for the $^{12}C(d,n)^{13}N$ reaction induced by 12.9 MeV deuterons. Neutron polarization is produced in the first reaction and measured in the second reaction.

We can see that

$$L \equiv \sigma(\theta', \phi = \pi) = \sigma_0(\theta')(1 - P_1 P_2)$$

$$R \equiv \sigma(\theta', \phi = 0) = \sigma_0(\theta')(1 + P_1 P_2)$$

and therefore

$$P_1 P_2 = \frac{R - L}{R + L}$$

Consequently, by measuring the right-left asymmetry in the second step and by using its known analyzing power $P_2$, one can determine the polarization $P_1$ produced in the first step.

We can also notice that if the detectors are positioned in the up ($U$) or down ($D$) direction with respect to the plane defined by vectors $\vec{k}_d$ and $\vec{k}_n$ we have:

$$U \equiv \sigma(\theta', \phi = \pi / 2) = \sigma_0(\theta')$$

$$D \equiv \sigma(\theta', \phi = 3\pi / 2) = \sigma_0(\theta')$$

And consequently

$$\frac{U - D}{U + D} = 0$$

Thus, by measuring the up and down asymmetries one can check the degree of the spurious asymmetries created by imperfections in the experimental setup.

In our measurements, we have used a beam of 12.9 MeV deuterons from the 120cm Cracow cyclotron. The beam passed through quadrupole magnetic lenses, was deflected by a deflecting magnet, and focussed on a target situated in the target hall about 12 m from the cyclotron. The experimental setup is shown in Figure 1.6.

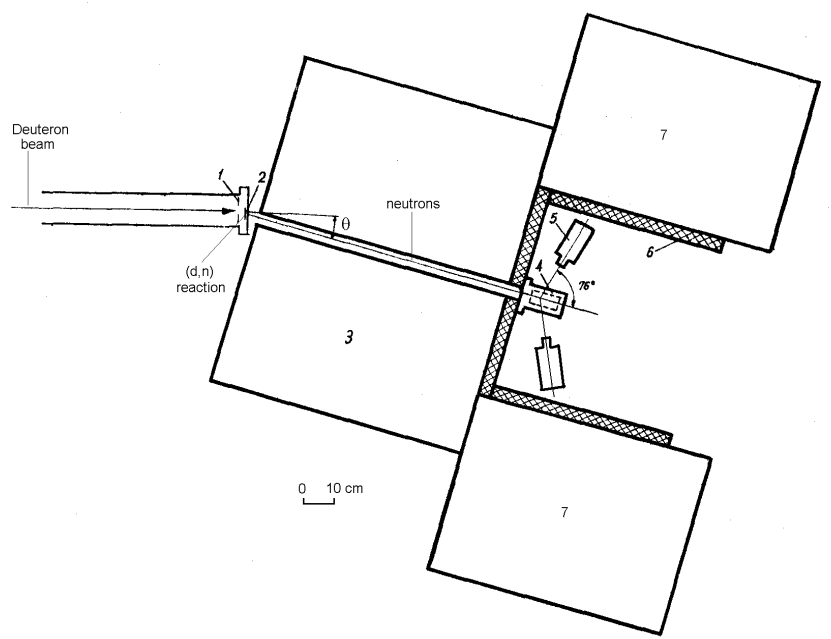

Figure 1.6. The experimental setup used in the measurements of neutron polarization from the reaction $^{12}C(d,n)^{13}N$. 1 – Tungsten ring; 2 – Carbon target; 3 – Paraffin collimator; 4 – Proportional counter filled with helium; 5 – Hornyak-type scintillation counter; 6 – Lead shilling; 7 – Water shielding; $\theta$ – reaction angle for the $^{12}C(d,n)^{13}N$ stripping reaction.

The intensity of the beam current on the target was 3-4μA. The target was made of a water-cooled 0.3 mm thick tungsten foil covered with a layer of about 7mg/cm$^2$ thick carbon. The first angle of the reaction, $\theta_{lab}$ = 15° (lab) corresponding to $\theta_{c.m.}$ = 17° was chosen because it corresponded to a relatively large product of the expected differential cross section and polarization (Middleton, el-Bedawi and Tai 1953; Satchler 1959).

The elastic n-α scattering was used as the polarization analyser. The degree of the polarization of neutrons scattered elastically from helium was determined by Seagrave (1953) and Levintov, Miller and Shamshev (1957).

To reduce the background level of fast neutrons, coincidence method was used during the measurements. Neutrons scattered elastically on helium nuclei in the proportional counter were detected using a Hornyak-type fast neutron scintillation counter (Hornyak 1952).[3] Coincidences between the recoil helium nuclei and scattered neutrons were registered for both the right and left detectors.

The proportional counter was filled with the spectroscopically pure helium at the pressure of 11 atm. The scattering angle of neutrons was 90$^0$ in the centre-of-mass (c.m.) system. The measurements were controlled by a fixed number of counts in the proportional counter. The background produced by accidental coincidences was determined by periodically switching on and off a 25-microsecond delay circuit between the Hornyak counter and the coincidence system.

---

[3] A fast neutron detector based on silver activated zinc sulphite, ZnS(Ag).

Supplementary measurements carried out using a pure tungsten target (i.e. without carbon) had shown that the background of neutrons from the stripping reaction on tungsten was lower than 10% of the yield produced by a target containing a layer of carbon.

We have also carried out measurements of asymmetries in the up and down direction to check for the possible instrumental contributions to the measured asymmetries. They were found to be negligible. For instance, for the measurements at $\theta_{lab}$ = 15⁰

$$\frac{U-D}{U+D} = 0.03 \pm 0.04$$

## Results and discussion

Results of our measurements are shown in Table 1.1. Neutron polarization from the $^{12}$C(d,n)$^{13}$N reaction at $E_d$ = 12.9 MeV increases from around -0.39 at 15⁰ (lab) to around +0.60 at 60⁰. Of particular interest is the polarization at 15⁰ because it can be readily compared with the early theoretical predictions.

Due to the unsatisfactory energy resolution of our polarization analyser we were not able to separate precisely the neutron groups corresponding to different energy levels in $^{13}$N. However, the neutron spectrum was dominated by the transition to the 3.50/3.55 MeV doublet in $^{13}$N. The contribution from the transitions to the ground state and first excited states were small and could be regarded as negligible.

Similar strong transition to the 3.50/3.55 MeV doublet was observed over a wide range of angles in measurements of angular distributions of the differential cross sections for the $^{12}$C(d,n)$^{13}$C induced by 8.1 MeV deuterons (Middleton, el-Bedawi and Tai 1953). Likewise, measurements of the differential cross sections of protons for the mirror reaction $^{12}$C(d,p)$^{13}$C also showed a clearly dominant transition to the 3.7/3.9 MeV doublet in $^{13}$C

Table 1.1
Neutron polarization for the reaction $^{12}$C(d,n)$^{13}$C induced by 12.9 MeV deuterons.

| Reaction angle (lab.) | Neutron polarization (%) | Experimental error (%) |
|---|---|---|
| 15 | -39 | 11 |
| 30 | +3 | 8 |
| 45 | +25 | 8 |
| 60 | +55 | 20 |

The doubled at 3.50/3.55 MeV is made of states with spins $^3/_2{}^-$ and $^5/_2{}^+$. Both states are excited by the $j = l + 1/2$ neutron transfer. If the classical and early quantum-mechanical models of nucleon polarization are correct and if the reaction is dominated by the deuteron-nucleus interaction, then the observed polarization should be *positive* at $15^0$ (see the Appendix A). Our results show that the measured polarization at this angle is *negative*. This would mean that either the reaction mechanism is associated with a strong proton-nucleus interaction or that the observed polarization cannot be described using the proposed models.

The deuteron is a weakly bound particle. The domination of the proton absorption over deuteron absorption is both physically and theoretically unlikely (see the Appendix A). Consequently, we can conclude that our results demonstrated a violation of the sign rule and thus showed that the early theories provided inadequate description of the polarization mechanism.

A similar demonstration of the violation of the sign rule was shown experimentally for the proton polarization 5 years after the publication of our results (Boschitz and Vincent 1964). These authors measured angular distributions of both the differential cross sections and proton polarization for the $^{12}C(d,p)^{13}C$ induced by 21 MeV deuterons and leading to the ground state in $^{13}C$. This reaction is associated with the $l = 1$, $j = l - 1/2$ transfer and in compliance with the theoretically claimed sign rule, the proton polarization within the range of the first stripping maximum should be negative. In contrast, Boschitz and Vincent (1964) observed positive polarization.

Classical and early quantum-mechanical theories give useful but oversimplified description of the polarization in deuteron stripping reactions. The sign of the polarization at forward angles is an unreliable test of the reaction mechanism. The sign may depend on the deuteron energy and on the reaction angle within the range of the first stripping maximum of the differential cross sections.

Our measurements at $60^0$ suggest a violation of another simple classical and quantum-mechanical rule about the maximum value of the polarization (Appendix A). For the transition to the observed doublet, the theoretical maximum of the absolute value of the polarization should be 39%. This value agrees with the measured value at $15^0$ but is only just within the experimental error for $60^0$.

From the application point of view, the large absolute polarization values observed at $15^0$ and $60^0$ suggest that the $^{12}C(d,n)^{13}N$ could serve as a good source of the polarized fast neutrons.

In conclusion, we have demonstrated for the first time the violation of the semiclassical sign rule and thus pointed out the limitations of the early theoretical descriptions polarization phenomena.

Our results have also a historical value. They were the first experimental results in nuclear physics in Poland, and the first in the world for the neutron polarization at the medium deuteron energy.

### References

Bhatia, A. B., Huang, K., Huby, R., and Newns, H. C. 1952, *Phil. Mag.* **53**:485.

Budzanowski, A., Grotowski, K., Niewodniczanski, H., and Nurzynski, J. 1959, *Bulletin de l'Acedemie Plolonaise des Sciences*, Série des sci. math., astr. at phys. **7**:583.

Burrows, H. B., Gibson, W. H. and Rotblat, J. 1950, *Phys. Rev.* **80**:1095.

Bokhari, M. S. Cookson, J. A., Hird, B. and Weesakul, B. 1958, *Proc. Phys. Soc.* **72**:88.

Boschitz, E. T. and Vincent, J. S. 1964, *NASA TR R-218*, National Aeronautics and Space Administration report, Washington, D. C., December 1964.

Cheston, W. B. 1954, *Phys. Rev.* **96**:1590.

Butler, S. T. 1950, *Phys. Rev.* **80**:1095.

Butler, S. T. 1951, *Proc. Phys.* Soc. A208:559.

EPPOG 2006, 'Hands on Physics', http://wyp.teilchenphysik.org/inst_pl_krakow_en.htm

Haeberli, W. 1961, *Helv. Phys. Acta*, Suppl. VI, 157.

Haeberli, W. and Rolland, W. W. *Bull, Am. Phys.* Soc. **2**:234.

Hansel, J. C. and Parkinson, W. C. 1958, *Phys. Rev.* **110**:128.

Hird, B., Cookson, J. A. and Bokhari, M. S. 1959, *Comptes Reudus du Con-gres International de Physique Nucléaire*, Paris 7-12 Juillet 1958, p. 470, Dunod, Paris.

Hittmair, O. 1956, *Z. Physik* **144**:449.

Hodgson, P. E. 1971, *Nuclear Reactions and Nuclear Structure*, Clarendon Press, Oxford.

Holt, J. R. and Young, C. T. 1950, *Proc. Phys. Soc.* (London) **78**:833.

Hornyak, W. F. 1952, *Rev. Sci. Instr.* **23**:264.

Horowitz, J. and Messiah, A. M. L. 1953, *J. Phys. Radium* **14**:731.

Hubby, R. 1950, *Nature* **166**:552.

Hubby, R. and Newns, H. C. 1951, *Phil. Mag.* **42**:1442.

IFJ 1965, *Institute of Nuclear Physics, Cracow*, Wojskowe Zaklady Kartograficzne, Krakow, Poland

IFJ 2005, 'The Henryk Niewodniczanski Institute of Nuclear Physics', http://www.ifj.edu.pl/intro/?lang=en

Jurić, M. K. and Čirilov, S. D. 1969, *Comptes Rendus du Congres International de. Physique Nucléaire*, Paris 7-12 Juillet, 1958, p. 473, Dunod, Paris.

Juveland, A. C. and Jentschke, W. 1958, *Phys. Rev.* **110**:456.

Levintov, I. I., Miller, A. V. and Shamshev, V. N. 1957, *Nucl. Phys.* **3**:221.

McGruer, I. N., Warburton, E. K. and Bender, S. R. 1955, *Phys. Rev.* 100:235.

Middleton, R., el-Bedewi, P. A. and Tai, C. T. 1953, *Proc. Phys. Soc.* **A66**:95.

Newns, H. C. 1953, *Proc. Phys. Soc.* **66**:477.

Newns, H. C. and Refai, M. Y. 1958, *Proc. Phys. Soc.* **71**:627.

Ohlsen, G. G. 1972, *Rep. Prog. Phys.* **35**:717.

Sawicki, J. 1957, *Phys. Rev.* **106**:172.

Satchler, G. R. 1959, *Comptes Rendus du Congres International de Physique Nucléaire*, Paris 7-12 Juillet 1958, p. 101, Dunod, Paris.

Satchler, G. R. 1960, *Nucl. Phys.* **18**:110.

Seagrave, J. D. 1953, *Phys. Rev.* **92**:1222.

Tobocman, W. 1956, *Tech. Rep.* No 29, Case Institute of Technology

Wolfenstein, L. 1956, *Ann. Rev. Nucl. Sci.* **6**:43.

---

# A Systematic Discontinuity in the Diffraction Structure

***Key features:***

1. We have measured angular distributions of the differential cross sections for the elastic scattering of 12.8 MeV deuterons on Be, C, Mg, and Ca nuclei. We have observed an irregularity in the diffraction structure for C and Ca.

2. The subsequent compilation of available data in the vicinity of the 12.8 MeV bombarding energy revealed a systematic discontinuity in the diffraction pattern in the form of a vanishing maximum as the mass number of the target nucleus increases beyond a certain value.

3. In order to reconstruct the observed pattern by using the diffraction theory it is necessary to assume a contribution of two different geometries for scattering in the forward direction and backward directions.

4. Optical model can reproduce the observed irregularities in the diffraction structure. They are interpreted as a 'geometric' effect.

**Abstract**: Angular distributions for the elastic scattering of 12.8 MeV deuterons on Be, C, Mg, and Ca nuclei have been measured at $5^0$ intervals for angles $15^0 - 145^0$ (lab) using a counter telescope. In all cases, the measured distributions show a pronounced diffraction pattern. A systematic breakdown in the diffraction structure is demonstrated.

## Introduction

Measurements of the elastic scattering of 11.8 MeV (Igo, Lorenz, and Schmidt-Rohr 1961) and 13.6 MeV deuterons on carbon (Gofman and Nemec 1961) revealed a curious and unexpected feature (see Figure 2.1). The angular distribution of the differential cross sections measured at 11.8 MeV has only one maximum in the angular range of around $40^0 - 110^0$, but the distribution at 13.6 MeV has two. The two measurements are separated by only about 2 MeV and one would expect a smooth transition between the two energies without a change in the number of oscillations. The observed curious discontinuity was puzzling and merited further investigation.

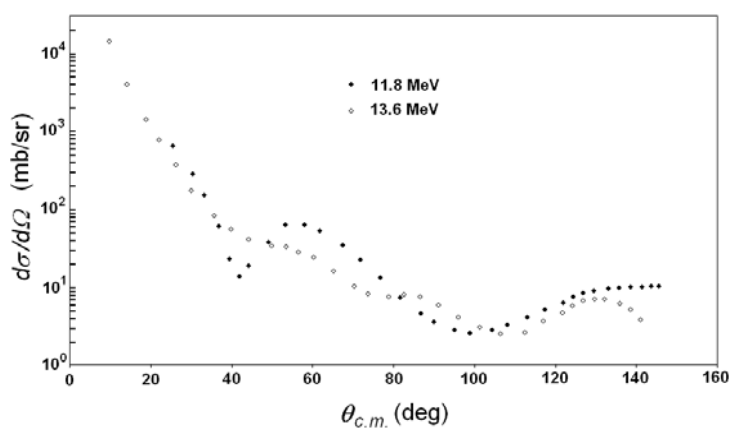

Figure 2.1. The angular distributions of the differential cross sections for the elastic scattering of 11.8 MeV (Igo, Lorenz, and Schmidt-Rohr 1961) and 13.6 MeV (Gofman and Nemets 1961) deuterons. The measurements suggested a discontinuity in the diffraction pattern. (One oscillation maximum in the middle range of scattering angles for the 11.8 MeV deuterons is replaced by two when the deuteron energy is increased to 13.6 MeV.)

The scattering of nucleons (protons or neutrons) can be relatively easily described theoretically. However, it has not been immediately obvious whether simular treatment could be also applied to deuterons. The deuteron is a weakly bound particle. Its bounding energy is only 2.225 MeV, which is about four times lower than the bounding energy of tritons (8.482 MeV) or $^3$He (7.718 MeV). Such a weakly bound particle can be expected to lead readily to a stripping reaction, where one component moves away relatively undisturbed while the other is captured by the target nucleus. Another likely process is a deuteron break-up.

A study of deuteron scattering was giving a convenient insight into the stability of this weakly bound system. Consequently, the observed anomaly presented itself as a phenomenon that should be further investigated.

The energy available in our laboratory was 12.8 MeV, which was between the two energies for which previous measurements were carried out. We have therefore decided to study the observed anomaly by carrying the measurements of elastic scattering not only for carbon but also for a few other nuclei, Be, Mg, and Ca.

## Experimental procedure

Angular distributions of the differential cross sections for the elastic scattering of 12.8 MeV deuterons were measured using deuterons accelerated in the 120-cm cyclotron of the Institute of Nuclear Physics in Cracow (Poland). The beam energy of the 120cm cyclotron was estimated to be (12.8 ±0.3) MeV.

The accelerated deuteron beam was directed to the target area by a system of magnetic quadrupole lenses and by a horizontal deflection magnet. The target was located about 12 meters from the cyclotron. The beam of deuterons was cut by a collimating entrance aperture to a spot of 4 mm diameter on the target. The target chamber with the detecting system is shown in Figure 2.2.

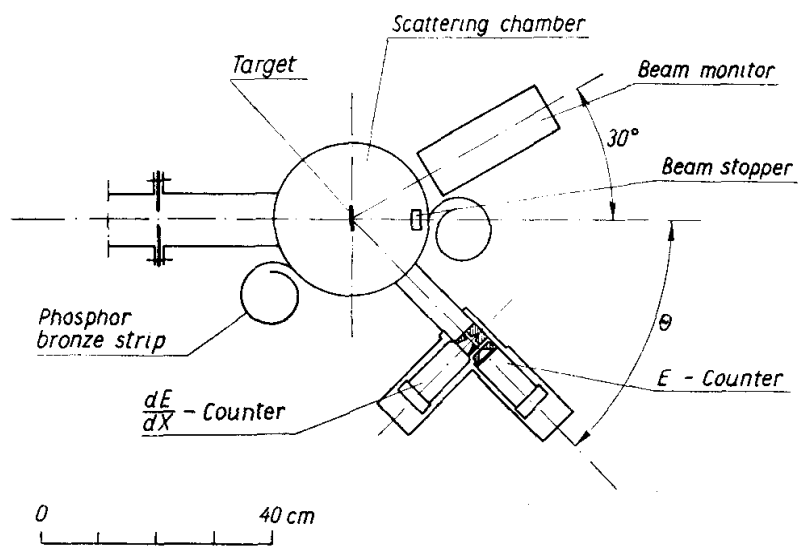

Figure 2.2. The experimental arrangement.

In order to minimize the interference of protons from the (d,p) reaction it was necessary to have a particle identification system. Our detector system consisted of a counter telescope, which was made of two scintillation counters. The first counter with a thin CsI(Tl) or plastic scintillator acted as a *dE/dx* detector. With the deuteron energy loss of about 1.5 MeV in the scintillator the energy resolution of the counter was about 11%.

The energy loss per unit path, *dE/dx*, can be expressed as (Bethe and Livingstone 1937):

$$-\frac{dE}{dx} = \frac{4\pi e^4 Z^2}{m\upsilon^2} nz \left\{ log\, \frac{2mV^2}{I} - log\left(1 - \beta\right)^2 - \beta^2 \right\}$$

where *e* is the electron charge, *m* − the mass of the electron, *Z* − the atomic number of the detected particle, *υ* − the velocity of the particle, *z* − the atomic number of the stopping material, *n* − the number of atoms per cubic centimetre in the stopping material, *β* − the ratio of the particle velocity to the velocity of light, and *I* − the average ionisation energy of the stopping material.

For *β* << 1 this expression can be significantly simplified:

$$-\frac{dE}{dx} = C_1 \frac{MZ^2}{E} log\, C_2 \frac{E}{M}$$

where *M* is the mass of the detected particle and *$C_1$* and *$C_2$* are constants.

The logarithmic term is a slow varying function of energy and thus the above expression can be further simplified as:

$$\frac{dE}{dx} = C_3 \frac{MZ^2}{E}$$

In general, the discrimination between various types of particles is done by using the product:

$$\frac{dE}{dx} \cdot E = C_3 MZ^2$$

However, in our measurements clean spectra were obtained by setting a gate on the output of the *dE/dx* detector, which was run in coincidence with the *E* − counter made of a CsI(Tl) scintillator sufficiently thick to stop the entering deuterons.

The energy resolution of the detection system was about 5%. The gated pulses from the *E* - counter were fed into a 100-channel Hutchinson-Scarrott amplitude analyser. In all cases, except for measurements at largest angles, the resolution of the detection system ensured a good separation of elastically scattered deuterons from protons. The defining aperture of the counter telescope subtended an angle of 1.2°.

The detecting system was attached to a phosphor-bronze strip making the sidewall of the scattering chamber. This arrangement allowed for selecting a scattering angle in the range of 0° - 145° without disturbing the vacuum in the scattering chamber. The accuracy of the angle settings was about ±0.25°.

The beam current was integrated by means of a current integrator connected to a beam stopper, which was placed behind the target. In addition, a CsI(Tl) scintillation monitor, which detected particles at a fixed angle of 30° was used.

The determination of the absolute values of the differential cross-sections was checked by using the cross-section for the elastic scattering of deuterons from gold. Separate measurements indicated that 12.8 MeV deuterons scattering from gold could be described using Rutherford formula for angles of up to about 35° (see Figure 2.3). The elastic scattering from gold at small angles was measured several times during the experiment and used in the calculations of the absolute values of differential cross-sections.

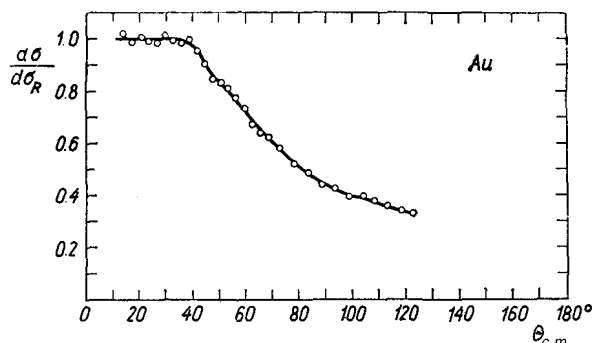

Figure 2.3. The ratio of the measured and Rutherford cross-sections for the elastic scattering of 12.8 deuterons scattered elastically from Au target.

The targets were in the form of thin foils: 0.42 mg/cm$^2$ thick for Au, 7.6 mg/cm$^2$ for Be, 4.1 mg/cm$^2$ for Mg and 4.6 mg/cm$^2$ for Ca. A polystyrene foil of 2.11 mg/cm$^2$ was used as a carbon target.

## Results and discussion

### Results

The differential cross-sections were measured in steps of 5° for angles between 15° and 145° (lab). Results are presented in Figures 2.4 and 2.5. The relative errors take into account the statistical errors, the inaccuracies in beam integration and errors associated with angle inaccuracies. The absolute cross-sections for C, Mg, and Ca were estimated at ±10%, and for Be at ±15%.

A convenient way to see the structure of the angular distributions for the elastic scattering is to plot them as ratios to the Rutherford cross-sections. The expression for the Rutherford scattering cross section has the following form (Rutherford 1911):

$$\left(\frac{d\sigma}{d\Omega}\right)_R = \left(\frac{zZe^2}{2\mu\upsilon}\right)^2 \frac{1}{\sin^4(\theta/2)}$$

where $ze$ and $Ze$ are the charges of the incident particle and of the target nucleus, $\mu$ is the reduced mass of the interacting particles, $\upsilon$ is the initial velocity of the projectile, and $\theta$ is the centre-of-mass angle of the scattered projectile.

The Rutherford scattering cross section can be expressed in terms of more convenient variables and in mb/sr. After substituting the relevant variables one gets:

$$\left(\frac{d\sigma}{d\Omega}\right)_R = 1.2959964 \left(\frac{zZ}{E}\right)^2 \left(1 + \frac{A_p}{A_t}\right)^2 \frac{1}{sin^4(\theta/2)} \quad \text{(in mb/sr)}$$

where, $E$ is the laboratory energy of the projectile in MeV, $A_p$ is the atomic mass number of the projectile and $A_t$ the atomic mass number of the target nucleus.

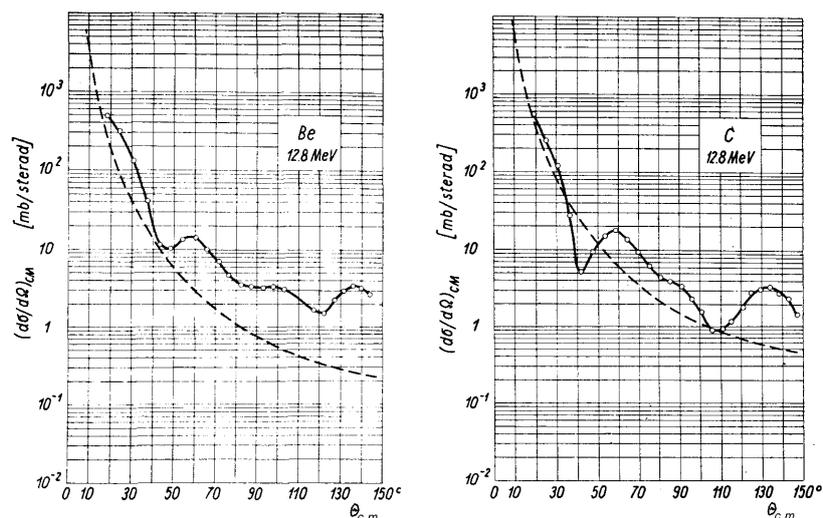

Figure 2.4. The measured differential cross sections (open circles) for the elastic scattering 12.8 MeV deuterons scattered elastically from Be and C nuclei are compared with the Rutherford cross sections (dashed lines). The solid lines through the data points are to guide the eye.

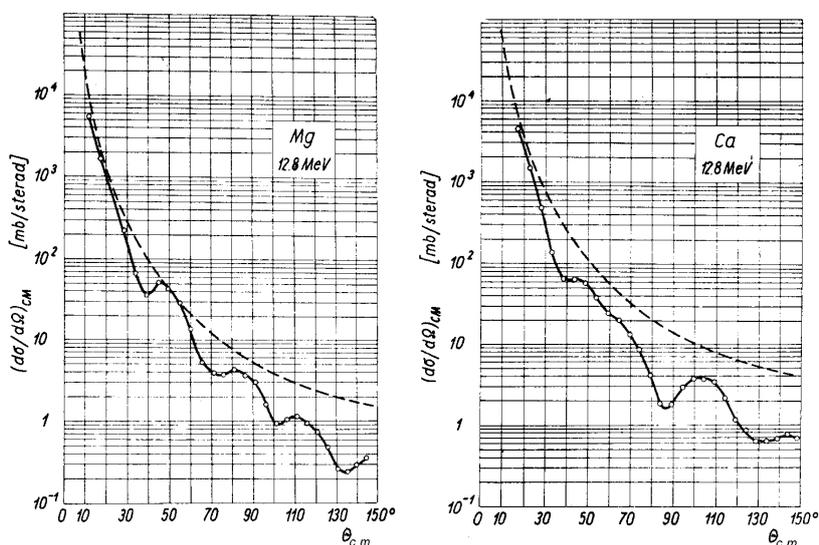

Figure 2.5. The measured differential cross sections (open circles) for the elastic scattering 12.8 MeV deuterons scattered elastically from Mg and Ca nuclei are compared with the Rutherford cross sections (dashed lines). The solid lines through the data points are to guide the eye.

Plots of the measured differential cross-sections divided by the Rutherford cross-sections ($d\sigma/d\sigma_R$) are presented in Figure 2.6. As can be seen, the measured cross-sections show pronounced diffraction structure. For the lightest elements, Be and C, the values of the measured cross-sections exceed considerably the values of the Rutherford cross-sections, especially at backward angles. Both Be and Mg data display a regularly spaced diffraction pattern. However, the data for C and Ca show two nearly merged maxima in the middle range of the reaction angles.

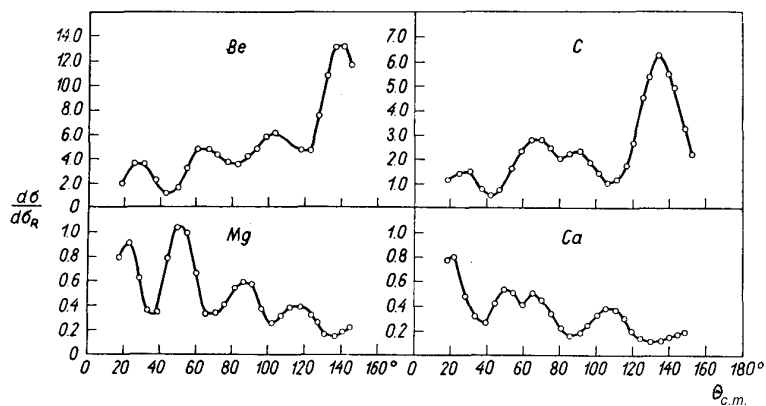

Figure 2.6. Angular dependence of the ratios of the measured and Rutherford scattering cross-sections for Be, C, Mg and Ca.

### Comparing with other data

Our measurements for carbon target can now be compared with the previous measurements (Gofman and Nemets, 1961; Igo, Lorenz, and Schmidt-Rohr, 1961) mentioned in the Introduction. This is done in Figure 2.7. As can be seen, the measured distributions display a clear transition from one maximum in the $40^0$-$100^0$ range at 11 MeV, via two nearly merging maxima at 12.8 MeV to two distinct maxima at 13.6 MeV.

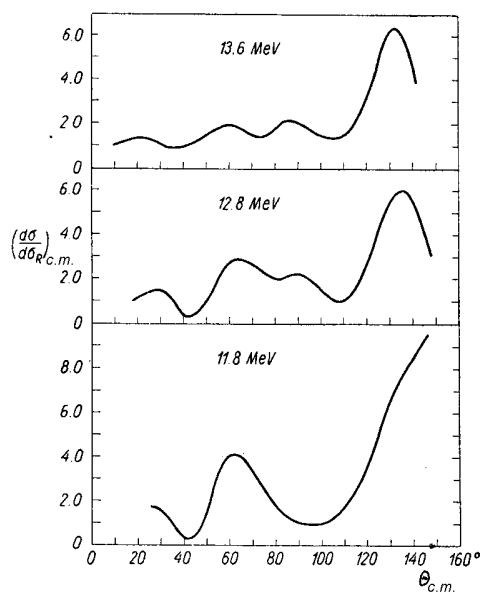

Figure 2.7. Our results at 12.8 MeV are compared with the measurements of Igo, Lorenz, and Schmidt-Rohr (1961) at 11.8 MeV and Gofman and Nemets (1961) at 13.6 MeV. They show a gradually emerging maximum around $90^0$ when the projectile energy is increasing.

In view of this now clearly demonstrated smoothly varying pattern for carbon and the curious irregular pattern for Ca, I have decided to compile and examine all the available data on deuteron scattering in the vicinity of the energy used in our experiment. I have also decided that the easiest way to study the variations in the diffraction pattern is to draw the positions of the maxima and the minima in the measured angular distributions. Results of my study are shown in Figure 2.8.

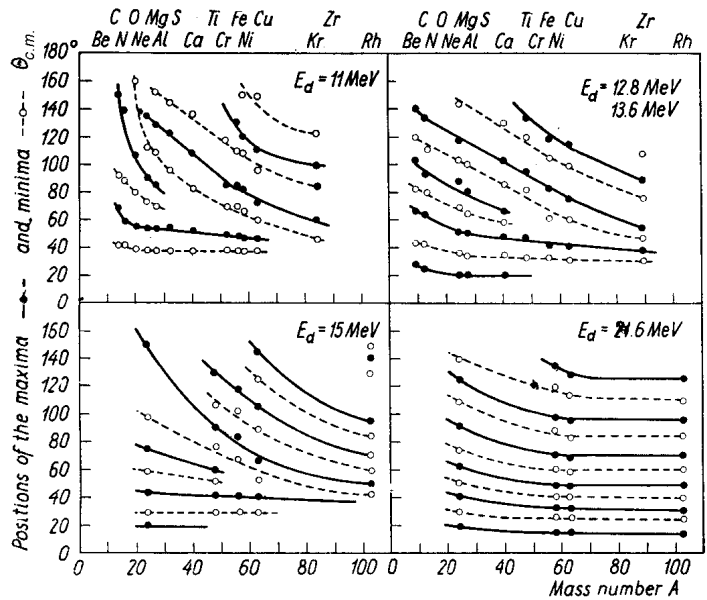

Figure 2.8. The positions of maxima and minima in the differential cross-sections as a function of the mass number and incident deuteron energy: 11 MeV (Takeda 1960); 12. 8 MeV, our results work; 13. 6 MeV for Ti, Fe, Cu, Zr (Gofman and Nemec 1961); 15 MeV (Cindro and Wall 1960); 21. 6 MeV (Yntema 1959).

The figure shows a smooth dependence of the positions of maxima and minima on the mass of the target nucleus and on the incident deuteron energy. The positions are shifted towards smaller angles with the increasing deuteron energy and mass number, which is consistent with the diffraction interpretation of elastic scattering (see below). However, Figure 2.8 shows also a systematic discontinuity in the diffraction pattern: one of the maximum disappears when either the mass number or incident energy is increased beyond a certain value. Insufficient data at 21.6 MeV did not allow for a systematic study of changes in the diffraction pattern at this energy.

Following the publication of our results, Tjin a Djie, Udo, and Koerts (1964) published their results for scattering of 26 MeV deuterons from a range of target nuclei. They also observed a systematic change in the diffraction structure and a disappearance of a maximum. Their results are presented in Figure 2.9, which displays similar pattern as in the compilation presented in Figure 2.8

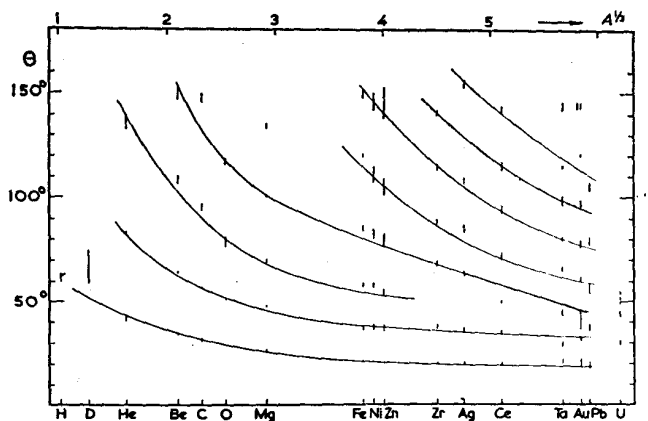

Figure 2.9. The positions of maxima of angular distributions for the elastic scattering of 26 MeV deuterons (Tjin a Djie, Udo, and Koerts 1964).

### *Different types of irregularities*

It is important to distinguish between this observed systematic discontinuity and the apparent random irregularities observed either for elastic scattering or transfer reactions.

For instance, the disappearance or emergence of diffraction maxima was observed in $^{16}O(^{3}He,\alpha_0)^{15}O$ reaction in the energy range of 9.8 – 11.2 MeV for the incident $^{3}He$ particles (Bray, Nurzynski, and Bourke 1966). This feature is displayed in Figure 2.10. The left-hand side of the figure shows the angular distributions and the right-hand side the positions of maxima and minima. The patterns are irregular and they appear to be associated with compound nucleus contributions.

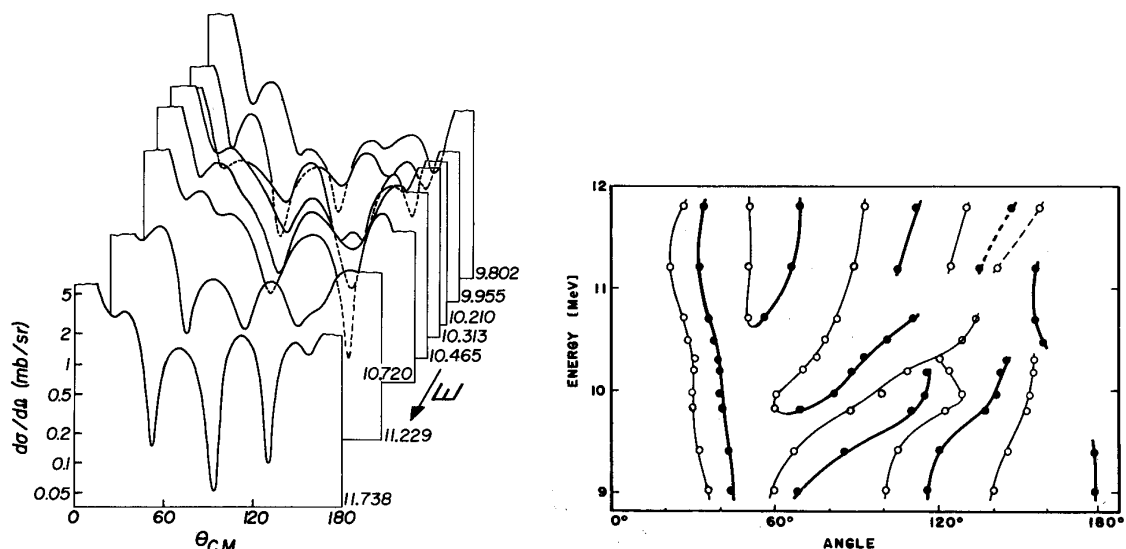

Figure 2.10. Angular distributions (the left-hand side of the figure) and the positions of maxima and minima (closed and open circles, respectively on the right-hand side of the figure) for the $^{16}O(^{3}He,\alpha_0)^{15}O$ reaction (Bray, Nurzynski, and Bourke 1966).

Irregularities in the diffraction structure were also observed for the elastic scattering of 27.0-35.5 MeV $\alpha$ – particles (Mikumo 1961). Resonances in excitation functions were identified and the observed irregularities were also interpreted as arising from the compound nucleus contributions.

Another good example of irregularities associated with compound nucleus mechanism is for the elastic scattering $^{12}C(\alpha,\alpha)^{12}C$ in the energy range of around 11 – 23 MeV (Atneosen *et al.* 1964; Carter, Mitchel, and David 1964). Using their angular distributions, I have prepared a plot of the positions of maxima and minima, which is presented in Figure 2.11.

Excitation functions measured at 165.8[0] for 10-19 MeV $\alpha$ - particles revealed a number of resonances clustered around 11.5, 13, 15, and 18.5 MeV (Carter 1962; Carter, Mitchell, and Davis 1964). Their positions are marked by arrows in Figure 2.11 and as can be seen, they are roughly located where strong irregularities in the angular distributions are observed.

No excitation function was measured for energies 20-23 MeV. However, if we use the data at 180[0] we can reproduce the excitation function and find a strong maximum at around 22 MeV. At this energy, which is also indicated by an arrow in Figure 2.11, a maximum at around $150^0$ is replaced by a maximum at around $105^0$ when the incident energy of $\alpha$ particles is increased.

In contrast, the patterns observed for the elastic scattering of deuterons in the energy range of 11-15 MeV, as shown in Figure 2.8, are of an entirely different nature. They are regular and systematic, and they also involve a wide range of target nuclei. They do not therefore appear to be in any way connected with the formation of a compound nucleus but must be caused by some other mechanism.

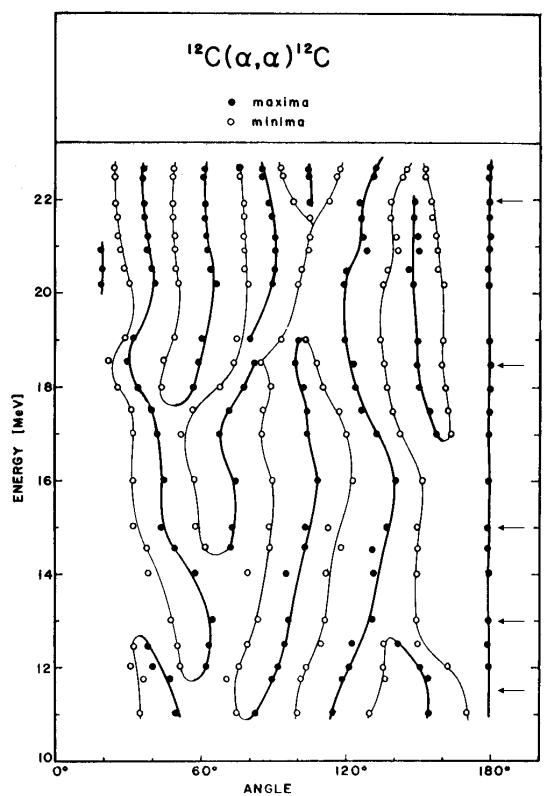

Figure 2.11. Irregularities in the positions of maxima and minima of the elastic scattering angular distributions, which can be identified as being associated with compound nucleus interaction. The horizontal arrows show the positions of resonances in the excitation functions. They roughly coincide with the discontinuities in the diffraction structure of the angular distributions. The plots are based on the measurements of Atneosen *et al.* (1964), Carter 1962, and Carter, Mitchell, and Davis (1964).

### *The impact parameter*

If we examine the Figure 2.8 we shall notice that in all three examples where one maximum disappears, the curves seem to follow two different patterns. The gradient for the lowest curves appear to be distinctly smaller than the gradient for the upper curves. The difference appears to be larger than normally expected for the normal diffraction scattering.

As the mass number for the target nucleus increases, the top curves come quickly into the region of the lower curves and one of the maximum belonging to the higher curves cannot be accommodated. This feature is observed as a disappearing maximum. However, this distinction is less clear for the 26 MeV data, where results for very light target nuclei (H, D, and He) have been included.

The lower curves belong to a large impact parameter, i.e. when the approaching deuteron passes the target nucleus at a relatively large distance. The upper curves belong to deuterons with smaller impact parameters, i.e. when the passing deuteron feels better the influence of nuclear forces.

### *The diffraction model analysis*

In order to study the mechanism of the observed systematic discontinuities in the diffraction structure of the angular distributions I have carried out a conceptual analysis based on the diffraction theory of nuclear interaction (see the Appendix B). Diffraction theories have been used extensively in the early interpretations of experimental data (see for instance Dar, 1963, 1964 and Bassichis and Dar 1965)

In its simplest form, the differential cross section for scattering from a black disc can be expressed as

$$\frac{d\sigma}{d\Omega} = (k_i R_0^2)^2 \left( \frac{J_1(x)}{x} \right)^2$$

where $k_i \equiv 1/\hbar_i$ is the wave number for the incident particles, $R_0$ the nuclear radius, and $J_1(x)$ is the cylindrical Bessel function of the first order.

The variable $x = q R_0$, where $q = \left| \vec{k}_i - \vec{k}_f \right|$ is the momentum transfer.

$$q^2 = k_i^2 + k_f^2 - 2 k_i k_f \cos\theta$$

where $k_f$ is the wave number for the outgoing particles and $\theta$ is the scattering angle. For the elastic scattering $k_i = k_f$ and therefore $x = 2 k_i R_0 \sin(\theta/2)$.

A graph of the universal function $[J_1(x)/x]^2$ is shown in the Appendix B.

In order to calculate theoretical locations of the positions of maxima in the angular distributions we first have to determine the locations of the maxima of the function $[J_1(x)/x]^2$. This can be done by calculating the derivative of this function. The maxima $x_i$ of the $[J_1(x)/x]^2$ satisfy the equation

$$\frac{d}{dx}\left[ \frac{J_1(x)}{x} \right]^2 = 2\frac{d}{dx}\left[ \frac{J_1(x)}{x} \right] = 0$$

Using the following relation for the Bessel functions

$$n J_n(x) - x J_n'(x) = x J_{n+1}(x)$$

where

$$J_n'(x) \equiv \frac{d}{dx} J_n(x)$$

we can find that

$$\frac{d}{dx}\left[ \frac{J_1(x)}{x} \right] = -\frac{J_2(x)}{x}$$

Thus, the location of the maxima $x_i$ of the $[J_1(x)/x]^2$ function is given by the equation

$$\frac{J_2(x)}{x} = 0$$

The first maximum is at $x_0 = 0$ because

$$\lim_{x \to 0}\left[\frac{J_2(x)}{x}\right] = 0$$

Other positions coincide with the $x_i$ values for which $J_2(x) = 0$. They can be calculated by a linear interpolation of the tabulated values near the $J_2(x) = 0$ value.

Alternatively, the position of the maxima of the $[J_1(x)/x]^2$ function can be calculated by fitting polynomials to points around the maxima of this function.

The $x_i$ values, calculated using both methods are listed in Table 2.1. The last number calculated using the derivative

$$\frac{d}{dx}\left[\frac{J_1(x)}{x}\right] = 0$$

represents an approximate value because the table of $J_2(x)$, which I have used ends at $x$ = 17.5 (Abramowitz and Stegun 1964).

Table 2.1

Positions $x_i$ of the maxima of the $[J_1(x)/x]^2$ function

| Method | $x_0$ | $x_1$ | $x_2$ | $x_3$ | $x_4$ | $x_5$ |
|--------|-------|-------|-------|-------|-------|-------|
| Derivative | 0.00 | 5.14 | 8.42 | 11.62 | 14.80 | 18.00 |
| Polynomial | 0.00 | 5.14 | 8.43 | 11.62 | 14.80 | 17.97 |

The determined positions of the maxima of the $[J_1(x)/x]^2$ function can now be translated into the mass and angle dependence of the positions of the maxima in the angular distributions. The relevant conversion formula is

$$\theta = 2\sin^{-1}\left(\frac{x_i}{2kR_0}\right)$$

where $x_i$ is the position of an $i$th maximum of the $[J_1(x)/x]^2$ function.

The convenient expression for $k$ is:

$$k = 0.2187\sqrt{\mu E_{c.m.}} \qquad \text{(fm}^{-1}\text{)}$$

where $\mu$ is the reduced mass

$$\mu = \frac{M_p M_t}{M_p + M_t} \approx \frac{A_p A_t}{A_p + A_t}$$

and

$$E_{c.m.} = \frac{E_{lab}}{1 + M_p / M_t} \approx \frac{E_{lab}}{1 + A_p / A_t} \qquad \text{(MeV)}$$

Alternatively, using $E_{lab}$

$$k = 0.2187 \mu \sqrt{E_{lab} / A_p}$$

Figure 2.12 shows the calculated angle-mass dependence of the positions of maxima in the angular distributions for the elastic scattering of 12.8 MeV deuterons. The three upper curves were calculated using $R_0 = r_0 A^{1/3}$ and the lowest two curves using $R_0 = r_0 A^{1/3} + 4.2$. For both sets, the parameter $r_0 = 1.7$ fm was used.

This conceptual diagram shows that in order to reproduce the patterns observed experimentally using diffraction model, one has to assume contributions from two different geometries. Whether such contributions can be physically justified is not clear.

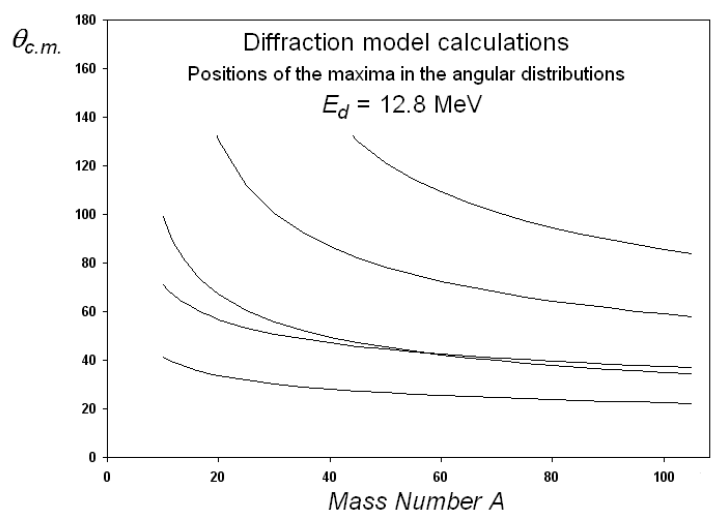

Figure 2.12. The positions of the maxima in the elastic scattering angular distributions calculated using the diffraction model. The two lower curves were calculated using $R_0 = r_0 A^{1/3} + 4.2$ fm and the upper curves using $R_0 = r_0 A^{1/3}$ fm. In both cases, $r_0 = 1.7$ fm has been used.

It is now well known that the optical model formalism can account for the observed systematic irregularities. They are interpreted as reflecting a 'geometrical' effect (Hodgson 1964, 1966; Wilmore and Hodgson 1964). This approach does not explain the physics of the observed phenomenon but only shows that by adjusting optical model parameters one can describe with some degree of success the measured angular distributions.

We have also analysed our Ca data using optical model (see Chapter 4). As can be seen in Figure 2.6, the angular distribution for this nucleus shows an irregularity in the form of two merging maxima. We were able to reproduce the experimental data by introducing a spin-orbit component in the optical model potential.

## Summary and conclusions

Prompted by observed irregularities in the angular distributions for the elastic scattering of 11.8 and 13.6 MeV deuterons from carbon (Gofman and Nemec 1961; Igo, Lorenz, and Schmidt-Rohr 1961) we have carried out measurements of the elastic scattering at 12.8 MeV for Be, C, Mg, and Ca nuclei. Our measurements demonstrated a smooth transition between 11.8 and 13.6 MeV. We have observed clear discontinuities in the diffraction structure for C and Ca in the form of two nearly merging maxima in each case.

My compilation and analysis of existing data in the vicinity of 12.8 MeV revealed a puzzling systematic discontinuity in the diffraction structure in the form of a vanishing maximum as the mass of the target number increases beyond a certain value. This systematic discontinuity cannot be explained by a compound nucleus mechanism and must have a different physical interpretation.

Following the publication of our results, similar systematic discontinuity was also reported for 26 MeV deuterons scattered from a wide-range of target nuclei (Tjin a Djie, Udo, and Koerts 1964).

In an attempt to understand the physics of the observed phenomenon, I have carried out calculations using a simple black-disc diffraction theory. I have found that in order to reproduce the observed pattern I had to assume a contribution from two geometries for the elastic scattering. The two geometries are distinguished by two radii: $R_0 = r_0 A^{1/3} + 4.2$ and $R_0 = r_0 A^{1/3}$ fm with $r_0 = 1.7$ fm in both cases. Whether such two contributing geometries can be physically justified is not clear.

Later optical model calculations indicated that the observed systematic discontinuities can be reproduced by suitably adjusting optical model parameters. The discontinuities are interpreted as a geometrical effect. However, such optical model analyses do not explain the physics of the observed phenomenon.

## References

Abramowitz M. and Stegun, I. A. 1964, Handbook of Mathematical Functions, United States Department of Commerce, National Bureau of Standards, Washington, DC.

Atneosen, R. A., Wilson, H. L., Sampson, M. B., and Miller, D. W. 1964, *Phys. Rev.* **135**:B660.

Bassichis, W. and Dar, A. 1965, *Phys. Rev. Lett.* **14**:648.

Bethe, H. A. and Livingstone, M. S. 1937, *Rev. Modern Phys.* **9**:245.

Bray, K. H., Nurzynski, J., and Bourke, W. P. 1966, *Phys. Lett.* **21**:536.

Carter, E. B. 1962, Ph.D. Thesis, Florida State University (unpublished).

Carter, E. B., Mitchell, G. E., and Davis, R. H. 1964, *Phys. Rev.* **133**:B1421.

Cindro, N., and Wall, N. S. 1960, *Phys. Rev.*, **119**:1340.

Dar, A. 1963, *Phys. Lett.* **7**:339.

Dar, A. 1964, *Nucl. Phys.* **55**:305.

Gofman, Yu. V., and Nemets, O. F. 1961, *Soviet Phys. JETP* **12**:1035; *Soviet Phys. JETP* **13**:333.

Hodgson, P. E. 1964, *Compt. Rend. du Congrès International de Physique Nucléaire*, Paris, Vol. I, p. 257.

Hodgson, P. E. 1966, *Advances in Physics*, **15**:329.

Igo, G., Lorenz, W., and Schmidt-Rohr. U. 1961, *Phys. Rev.* **124**:832.

Mikumo, T. 1961, *J. Phys. Soc. Japan*, **16**:1066.

Rutherford, E. 1911, *Phil. Mag.* **21**:669.

Takeda, M. 1960, *J. Phys. Soc. Japan*, **15**:557.

Tjin a Djie, H. R. E., Udo, F., and Koerts, L. A. 1964, *Nucl. Phys.* **53**:625.

Wilmore, D. and Hodgson, P. E. 1964, *Nucl. Phys.* **55**:673.

Yntema, J. L. 1959, *Phys. Rev.*, **113**:261.

_______________________________________________________

# Core Excitations in $^{27}$Al

***Key features:***

1. Nuclear excitations are usually interpreted using either single-particle or collective models. However, it has been suggested (De Shalit 1961; Lawson and Uretsky 1957) that some excited states in odd-mass nuclei could be described as a coupling of a single nucleon or hole with an excited collective state of an even-mass core. This mode of excitations is now known as core excitations.

2. Originally, I have intended to study the possibility of such excitations by using a pair of $^{24}$Mg and $^{25}$Mg nuclei. However, the prohibitive price of enriched $^{25}$Mg isotope forced me to opt for more readily available $^{27}$Al and $^{28}$Si pair of nuclei. We have measured elastic and inelastic scattering of 12.8 MeV deuterons from these nuclei.

3. I have carried out an analysis of our experimental results using diffraction theories (with a sharp and smooth cut off radii) and a plain wave theory. Our results were also analysed using a strong coupling model.

4. This study resulted in a textbook demonstration[4] of the existence of core-excited states in $^{27}$Al.

5. A broad-range magnetic spectrograph we have designed and constructed was used for the first time in these measurements. Consequently, much of our experimental work was devoted to testing and bringing it into operation.

**Abstract**: Angular distributions for the elastic and inelastic scattering of 12.8 MeV deuterons to the first few excited states of $^{27}$Al and $^{28}$Si were measured using a broad-range magnetic spectrograph and nuclear emulsions. The overall resolution of around 150 keV allowed to distinguish between deuteron groups belonging to the first and the second excited states in $^{27}$Al, which was essential in the study of core excitations. Experimental data have been analysed using diffraction theories, plane wave theory, and strong coupling model. Results demonstrated that low-lying levels in $^{27}$Al could be interpreted as belonging to a quintuplet of core-excited states.

## Introduction

The low-lying nuclear levels are usually interpreted using single-particle or collective models. Single-particle excitations are associated with configurations characterised by their orbital momenta $l$ and total spins $j$. The remaining nucleons are usually treated as passive observers.

For instance, in a single particle stripping reaction a single nucleon is seen as being deposited into a shell-model orbital characterised by $l$ and $j$. Likewise, a single particle pickup is interpreted as involving a pickup of a single nucleon characterised by specific $l$ and $j$ values. Reactions involving more than one nucleon can also be interpreted as a transfer of single nucleons to or from specific shell-model orbitals.

The collective excitations involve the collective response of all the nucleons in the nucleus. The nucleus can be described either as a rigid rotating or a softer vibrating body or by a combination of the two. Collective excitations can be studied conveniently by inelastic scattering.

At around the mid-1950s, it has been suggested that some excited states in odd-mass nuclei can belong to a different type of excitations, in which a single particle or a hole with a specific shell-model $l$ and $j$ configuration is couple to a collective state of the core made of the remaining nucleons (De Shalit 1961; Lawson and Uretsky 1957). These modes of excitations are now known as core-excitations.

_______________________________________________________

[4] See P. E. Hodgson 1971, *Nuclear Reactions and Nuclear Structure*, Clarendon Press, Oxford, pp. 386-392; and in R. R. Roy and B. P. Nigam 1967, *Nuclear Physics: Theory and Experiment*, John Wiley & Sons, Inc., New York, p. 429, 430.

For instance, let us consider a single particle with spin $j = {}^5/_2{}^+$ coupled to a core with spin $J_c = 2^+$ (see Figure 3.1). The coupling results in a quintuplet of states with spins ranging from $^1/_2{}^+$ to $^5/_2{}^+$.

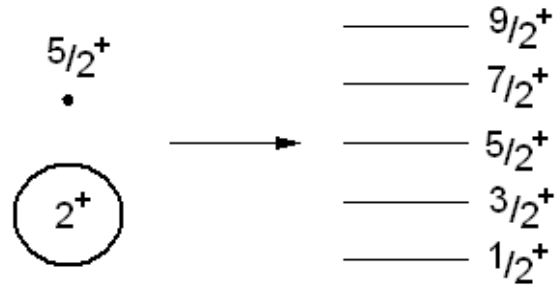

Figure 3.1. The core-excited model

The signatures of core-excited states are as follows:

(1) In an odd-mass nucleus, the coupling of a single particle spin $j$ with an excited even-even mass core spin $J_c$ will results in a multiplet of states with spins $I$:

$$\left| J_c - j \right| \leq I \leq J_c + j$$

(2) The centre of gravity of the multiplet should be approximately equal to the energy of the collective state of the core

$$E_c \approx \frac{\sum_I (2I+1)E_I}{\sum_I (2I+1)}$$

where $E_c$ is the energy of the collective state of the core and $E_I$ are the excitation energies of states in the core-excited multiplet in the odd-mass nucleus.

(3) The shape of the inelastic scattering angular distributions of the differential cross sections for the members of the multiplet and for the excited state of the core should by approximately same. This is because the excitation of members of the multiplet is formed by only one transition: from the ground state to the excited state of the core.

(4) The absolute values of the differential cross sections $\left[ d\sigma / d\Omega \right]_I$ for the inelastic scattering to states $I$ in the odd-mass nucleus and the differential cross section $\left[ d\sigma / d\Omega \right]_{J_c}$ for inelastic scattering to a core state $J_c$ in the even-mass nucleus should satisfy the following relation:

$$\left[ \frac{d\sigma}{d\Omega} \right]_I \approx \frac{(2I+1)}{\sum_I (2I+1)} \left[ \frac{d\sigma}{d\Omega} \right]_{J_c}$$

The sum of the cross sections for the multiplet should be approximately equal to the cross section of the core

$$\left[\frac{d\sigma}{d\Omega}\right]_{J_c} \approx \sum_I \left[\frac{d\sigma}{d\Omega}\right]_I$$

In order to investigate the possibility of the existence of core-excited states I first intended to use the $^{24}$Mg and $^{25}$Mg pair of nuclei. The $^{25}$Mg nucleus has a single neutron in the $1d_{5/2}$ configuration located over a well-known rotational core made of $^{24}$Mg. It also has a compliment of low-lying states with spins ranging from $^{1}/_{2}^{+}$ to $^{9}/_{2}^{+}$. These states presented themselves as good candidates for a core-excitation multiplet formed by coupling the $1d_{5/2}$ single neutron to the first excited $2^+$ state of the $^{24}$Mg core. Unfortunately, in its natural form, Mg contains only 10% of $^{25}$Mg and the price for enriched isotope was too high.

I have therefore decided to try another combination made of the $^{27}$Al and $^{28}$Si pair. The low-lying states in $^{27}$Al form almost a mirror image of the states in $^{25}$Mg. The advantage of using this pair is that natural Al contains 100% of $^{27}$Al and natural Si has 92.2% of $^{28}$Si. Thus, the targets can be made of natural elements, which is both cheap and convenient. The difference between the two systems is that in $^{27}$Al it is a $1d_{5/2}$ proton hole, which would be coupled to the $2^+$ spin of the $^{28}$Si core to produce the familiar core-excited multiplet.

Table 3.1

Similarities between the two pairs of nuclei, $^{24}$Mg – $^{25}$Mg and $^{28}$Si – $^{27}$Al, which could be used to study the core-excited model

| Nucleus | $E_x$ (MeV) | Spin | Nucleus | $E_x$ (MeV) | Spin |
|---|---|---|---|---|---|
| $^{24}$Mg | 1.3687 | $2^+$ | $^{28}$Si | 1.7790 | $2^+$ |
| $^{25}$Mg | g.s. | $^{5}/_{2}^{+}$ | $^{27}$Al | g.s. | $^{5}/_{2}^{+}$ |
| | 0.5850 | $^{1}/_{2}^{+}$ | | 0.8438 | $^{1}/_{2}^{+}$ |
| | 0.9747 | $^{3}/_{2}^{+}$ | | 1.0145 | $^{3}/_{2}^{+}$ |
| | 1.6118 | $^{7}/_{2}^{+}$ | | 2.2111 | $^{7}/_{2}^{+}$ |
| | 1.9646 | $^{5}/_{2}^{+}$ | | 2.7349 | $^{5}/_{2}^{+}$ |
| | 2.5638 | $^{1}/_{2}^{+}$ | | 2.9820 | $^{3}/_{2}^{+}$ |
| | 2.7377 | $^{7}/_{2}^{+}$ | | 3.0042 | $^{9}/_{2}^{+}$ |
| | 2.8015 | $^{3}/_{2}^{+}$ | | 3.6804 | $^{1}/_{2}^{+}$ |
| | 3.4052 | $^{9}/_{2}^{+}$ | | 3.9568 | $^{1}/_{2}^{+}$ |

$E_x$ – Excitation energy.

## Experimental procedure and results

Measurements were carried out using a beam of 12.8 MeV deuterons from the 120-cm cyclotron in the Institute of Physics in Cracow, Poland. The overall diagram of the experimental equipment, including the cyclotron, is shown in Figure 3.2.

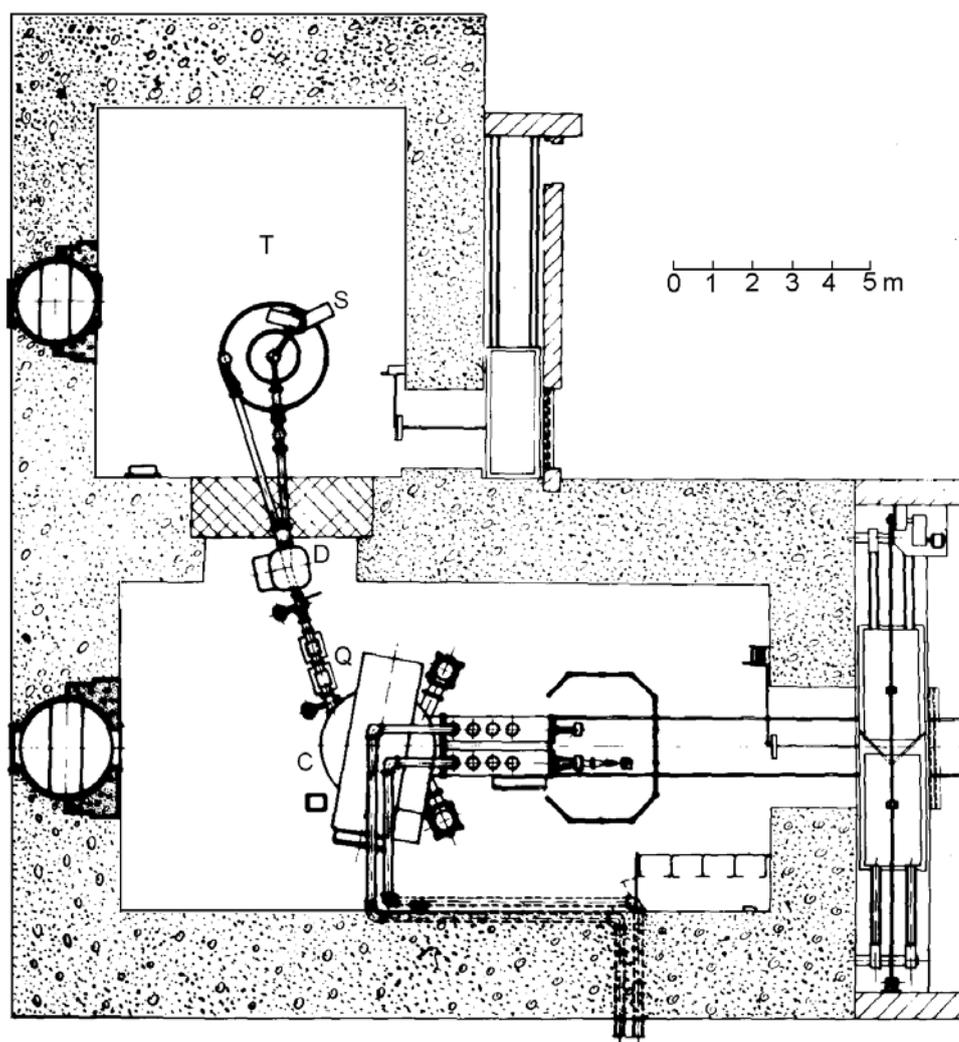

Figure 3.2. The general layout of the experimental equipment: (C) the 120-cm cyclotron, (Q) quadrupole lenses, (D) deflection magnet, (T) target area, (S) the broad-range magnetic spectrometer.

After leaving the cyclotron (C), the accelerated deuterons passed through a system of two quadrupole lenses (Q) and were deflected by $13^0$ by a deflection magnet (D) before entering the target area (T).

The scattering chamber used in these measurements was similar to the one described in Chapter 2. However, certain modifications were introduced, one of them being a system of slits, which was designed to facilitate a convenient control of accelerated deuterons and to improve resolution of detected particles (see Figure 3.3). Other modifications included adding a connection to the broad-range spectrometer. In fact, the whole chamber had to be first removed to install a system of supporting rails for the spectrometer.

### The slit system

An insulated tungsten aperture with the diameter of 6 mm was placed about 800 mm from the target. Current from this aperture was measured and used in focusing the deuteron beam on the target. The tungsten aperture was followed by a collimating system, which contained two 2 mm by 12 mm gold slits and between them three 5 mm diameter antiscattering brass slits. The system collimated the deuteron beam to less than $0.5^0$ and contributed significantly to improving the effective energy resolution of the detected particles. Good resolution was important in our experiment because we aimed at resolving the first two excited states in $^{27}$Al, which were essential in studying the core-excited model. Without taking special precautions, resolution is low for particles accelerated by a cyclotron.

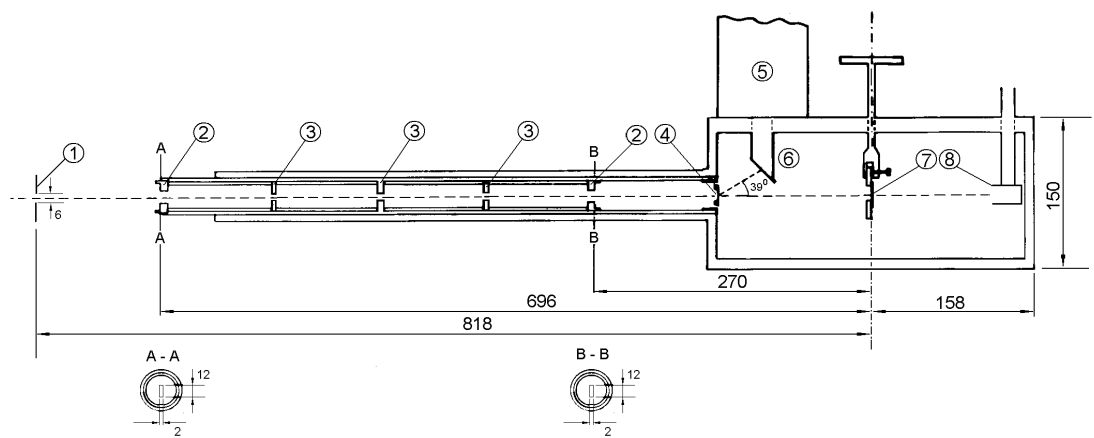

Figure 3.3. Target chamber and the system of entrance slits: 1. An insulated tungsten aperture used to guide the deuteron beam to the target. 2. The Au collimating slits. 3. The antiscattering 5 mm diameter brass slits. 4. The Au beam monitor target. 5. Scintillation counter used as a beam monitor. 6. Light guide. 7. The main target. 8. Faraday cup.

### Beam monitoring

The deuteron beam intensity was monitored using beam integrator connected to a Faraday cup. In addition, deuteron elastic scattering from a thin, 0.4 mg/cm$^2$ Au target at a fixed angle of $39^0$ was used. The scattered deuterons were detected using a thin CsI(Tl) crystal located on a light guide and connected to an EMI 6097 photomultiplier located in an antimagnetic shield.

### Detection of reaction products

Deuterons scattered from the main target ($^{27}$Al or $^{28}$Si) were directed to a broad range magnetic spectrometer, which we have designed and constructed, and were detected using 100 μm thick nuclear emulsions. The chamber was vacuum-sealed using a thin, Melinex polyester film. Nuclear emulsion plates were placed outside the spectrometer chamber and thus could be easily replaced. For the inelastic scattering, two plates 30 mm x 90 mm were used. Elastic scattering was measured using only one plate, 30 mm x 46 mm at each angle. A view of the detection system is presented in Figures 3.4 and 3.5.

### Determination of the focal plane, solid angle and defocusing coefficient

Our spectrometer was used for the first time in this experiment, and a good part of our work was to bring it into operation. One of the tasks was to determine its focal plane. For this purpose, we have used a sufficiently large spectrometer chamber with nuclear plates placed inside. As a target, we used a polystyrene foil and the spectrometer set at $30^0$. With this arrangement, we had 7.6 MeV recoil protons and 12.0 elastically scattered deuterons, which we used to locate the focal plane. Exposures of the plates were taken at various positions and the focal plane was calculated using results of measurements.

With this information, we then constructed a smaller spectrometer chamber with one wall at the position of the focal plane. The focal-plane wall was made of a 6 mg/cm$^2$ Melinex window, to allow for the reaction products to pass through and bombard the nuclear emulsions, which were placed outside the window. This arrangement allowed for a significant reduction in the time of collecting the data because there was no need to open the spectrometer chamber to replace the plates between measurements at various reaction angles. Furthermore, when placed in vacuum, the emulsions were often peeling off the plates. Consequently, by placing them outside the vacuum chamber the danger of such damage was eliminated. The focal plane wall contained an energy scale to assist in correct placing nuclear emulsions plates depending on the reaction angle.

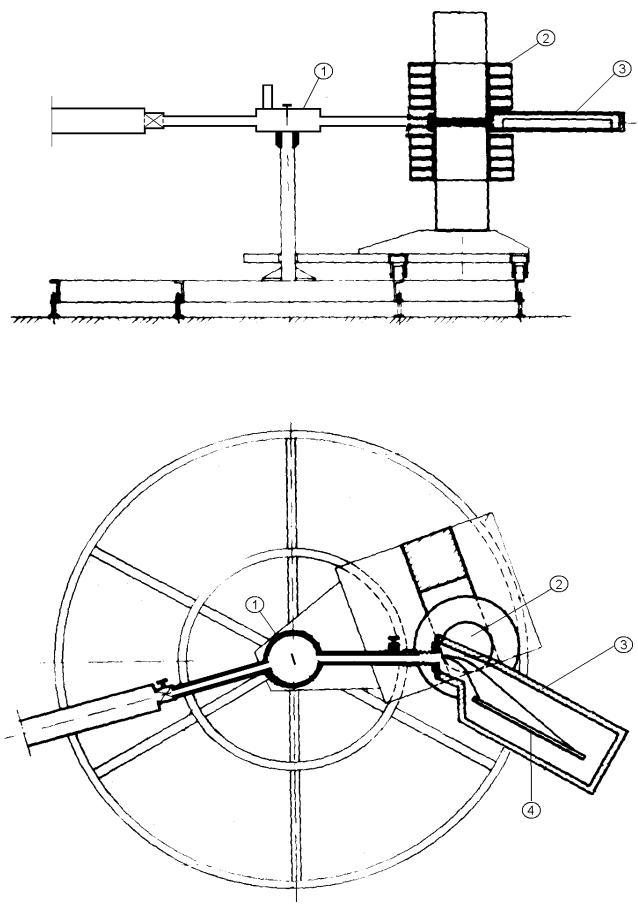

Figure 3.4. Target chamber and the magnetic spectrometer. 1. Target chamber. 2. Magnetic spectrometer. 3. Large spectrometer chamber used to determine the focal plane. After determining the focal plane, the chamber was replaced by a smaller version, which allowed for the nuclear emulsions to be placed outside the vacuum. 4. Nuclear emulsions holder.

Before using the spectrometer in the measurements of angular distributions we also had to determine its solid angle and the defocusing coefficient as a function of the energy of reaction products. For these measurements, we have also used a polystyrene target.

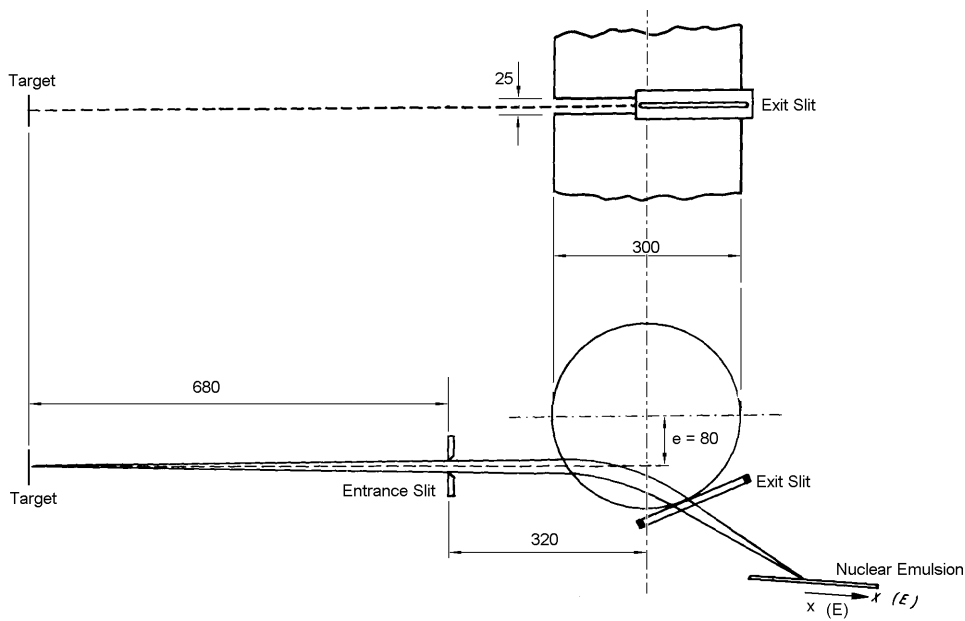

Figure 3.5. Schematic diagram of the detection system.

### Beam alignment and angular scale

Other important introductory work included beam alignment of the accelerated deuterons to ensure that they were delivered at the centre of the main target, and the determination of the zero position for the angular scale of the spectrometer. To assist in the beam alignment, we have installed a flexible section in the beam line between the target chamber and the cyclotron. To determine the zero of the angular scale we have carried out measurements of Rutherford scattering from a thin Au foil for angles between $-30^0$ and $+30^0$.

### Determination of the accelerated beam energy

Finally, we have also carried out measurements to determine the energy of the accelerated deuterons. For this purpose, we have used the $^{27}$Al(d,d)$^{27}$Al and $^{27}$Al(d,p)$^{28}$Al reactions and the reaction angle of $90^0$. This combination was convenient because the $Q$ – value for the (d,p) reaction leading to the ground state had a well-known value of 5.498 MeV. The measurements included corrections for the energy loss in the target. We have found that the energy of accelerated deuterons was $E_d = (12.81 \pm 0.065)$ MeV.

### Targets and results of measurements

The targets used in the measurements of angular distributions were 1.76 mg/cm$^2$ $^{27}$Al self-supporting foil and 2.43 mg/cm$^2$ SiO$_2$ quartz foil. Examples of the energy spectra are presented in Figure 3.6. The energy resolution was about 150 keV. This resolution allowed us to resolve the states at 0.84 and 1.01 MeV excitation energies in $^{27}$Al, which were important in the study of core excited model.

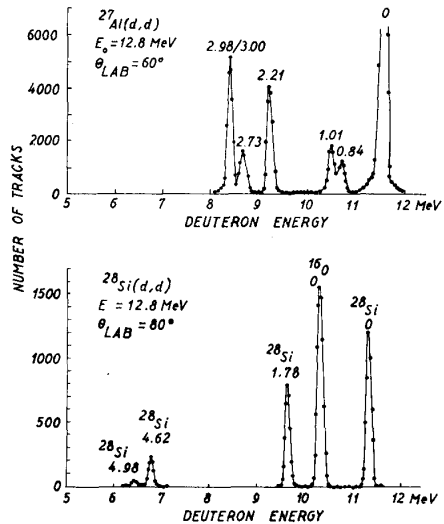

Figure 3.6. Examples of particle spectra for the elastic and inelastic scattering of 12.8 MeV deuterons from $^{27}$Al and SiO$_2$ targets.

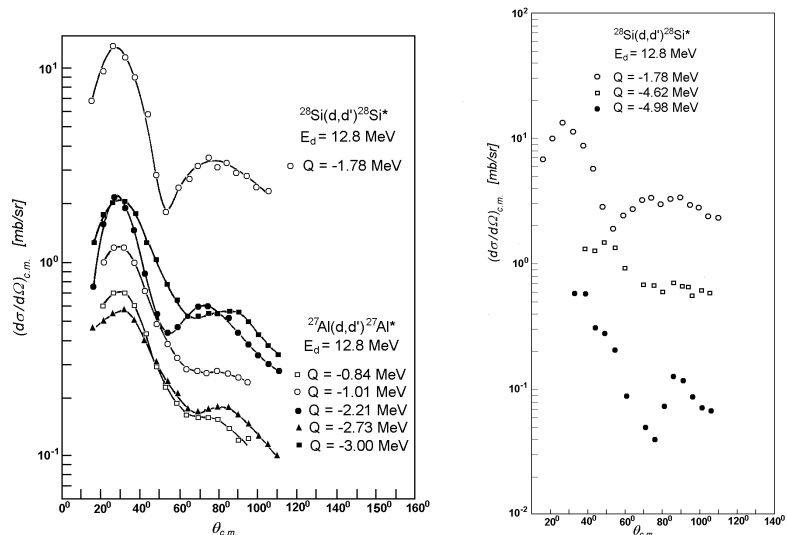

Figure 3.7. Angular distributions of the inelastic scattering of 12.8 MeV deuterons from $^{28}$Si and $^{27}$Al (the left-hand side of the figure). The lines are to guide the eye. Shown also are the angular distributions for all three excited states in $^{28}$Si (the right-hand side of the figure).

Angular distributions of the differential cross sections for the elastic and inelastic scattering were measured in steps of $5^0$. Results are presented in Figures 3.7-3.12.

Relative errors of the cross-section measurements consisted of statistical errors, inaccuracies of scanning of nuclear emulsions (estimated at about 2%) and inaccuracies caused by errors in angle setting. The combined relative errors were between about 5% and 9% depending on the intensity of observed groups and the reaction angle. The absolute values of the differential cross sections are estimated to be accurate to within about 15%.

### Preliminary discussion of data

It may not be completely trivial to observe that all the inelastic scattering angular distributions, relevant to the study of core excitations (Figure 3.7, left-hand side) exhibit features, which are consistent with any of the simpler models for direct interaction, i.e. asymmetry, forward peaking, some oscillatory structure with periodicity consistent with that for the elastic scattering. This is not a trivial point since 12.8 MeV energy is not so large an energy that it is obvious that direct processes will dominate.

The angular distributions are also all very similar, which is consistent with the interpretation that these levels are excited by a direct reaction in a very similar fashion.

Bearing the above observations in mind, one may attach considerable importance to the relative cross sections, as they should be measures of similar nuclear matrix elements between the ground and excited states. It should be noted that while for $^{27}$Al there is a fair difference in relative magnitudes, there is not an order of magnitude difference in their values. Thus, it appears that there are no overpowering selection rules inhibiting any of these transitions. In particular, this bears on the often repeated but never proved tenant that inelastic scattering strongly favours collective excitations. We must either say (a) that some of these states have a single particle character but that single particle excitation is not vastly less than collective excitation or (b) that collective excitation has been spread among all of the states.

It is interesting to observe that there is some correlation between spin $I$ of the excited state and the maximum of the differential cross section, except for the $^5/_2{}^+$ state, which will be discussed later.

**Theoretical analysis**

I have analysed our data using both the diffraction (Blair, 1959, 1961; Blair, Sharp, and Wilets 1962) model of scattering and the plane wave Born approximation (PWBA) theory (Butler 1951, 1957; Butler, Austern, and Pearson 1958). These simple theories give much the same information about nuclear structure and reaction mechanism as their more complex counterparts. However, their advantage is that because of their relative simplicity they can be used without computers. At the time when simple desk calculators or slide rules were as popular as currently used personal computers, the availability of such simple theories was of considerable importance.

Our data were also analysed using strong-coupling model (Buck, 1963; Buck, Stamp, and Hodgson 1963; Chase, Wilets, and Edmonds 1958) and computing facilities in the USA and UK.

***The diffraction model***

The description of the diffraction model is presented in the Appendix B. I have carried out calculations using both a sharp and a smooth cut-off boundaries for the target nuclei. Examples of the results of theoretical analysis are presented in Figure 3.8. Fits to all angular distributions for $^{27}$Al are of the same quality as the fit displayed for the 2.21 MeV level.

Table 3.2 contains the list of the maximum values of the experimental cross sections used in the normalisation of the theoretical cross sections and the resulting matrix elements M (see the Appendix B). In this table, $E_x$ is the excitation energy; $I^\pi$ – spin and parity of the excited state; $\sigma_{max}$ – maximum value of the experimental differential cross section. M is the matrix element related to the deformation distance.

The derived matrix elements were used to calculate model-related parameters $\delta_2$ and $\hbar\omega_2/2C$. The listed values have been calculated using the extreme rotational and vibrational models, respectively. The $\beta_2$ parameters were calculated assuming the radius $R_c = 1.2A^{1/3}$ fm.

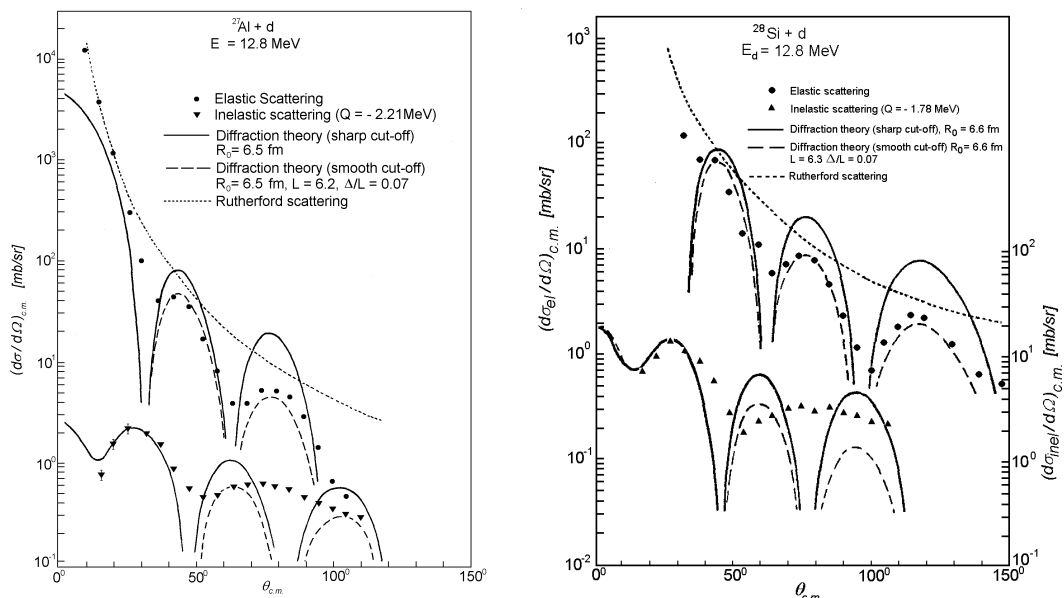

Figure 3.8. The experimental angular distributions (points) of the elastic and inelastic scattering of 12.8 MeV deuterons from $^{27}$Al and $^{28}$Si are compared with the calculations using the diffraction models with a sharp and smooth cut-off radius (full and dashed lines, respectively).

Table 3.2
A summary of the results based on the diffraction model analysis

| Nucleus | $E_x$ | $I^\pi$ | $\sigma_{max}$ | M | $\delta_2$ | $\beta_2$ | $\hbar\omega_2/2C_2$ |
|---|---|---|---|---|---|---|---|
| | (MeV) | | (fm$^2$) | (fm$^2$) | (fm) | | |
| $^{28}$Si | 1.78 | 2$^+$ | 1.35 | 0.1463 | 1.36 | 0.37 | 0.028 |
| $^{27}$Al | 0.84 | $^1\!/_2{}^+$ | 0.072 | 0.0083 | 1.14 | 0.32 | 0.153 |
| | 1.01 | $^3\!/_2{}^+$ | 0.123 | 0.0140 | 1.49 | 0.41 | 0.130 |
| | 2.21 | $^7\!/_2{}^+$ | 0.213 | 0.0237 | 1.94 | 0.54 | 0.110 |
| | 2.73 | $^5\!/_2{}^+$ | 0.058 | 0.0064 | 1.01 | 0.28 | 0.040 |
| | 3.00 | $^9\!/_2{}^+$ | 0.210 | 0.0230 | 1.92 | 0.53 | 0.086 |

## The plane wave (PWBA) theory

The concept of the plane wave theory for the inelastic scattering, as used in my calculations, is outlined in the Appendix C. The theory is also mentioned in the Appendix E. The parameters used in fitting the $^{27}$Al and $^{28}$Si data are listed in Table 3.3.

As explained in the Appendix C, tables of Lubitz (1957) were prepared to assist in the interpretation of experimental results using the PWBA theory. However, for the inelastic scattering of 12.8 MeV deuterons from $^{27}$Al and $^{28}$Si nuclei, the required $R_0$ - independent parameter $y/x$ was well outside the values used by Lubitz. The required values are in the vicinity of 4 whereas the values used by Lubitz are terminated at $y/x = 2$.

Table 3.3
Parameters used in the plane wave analysis of the $^{27}$Al and $^{28}$Si data

| Nucleus | $E_x$ | $\sigma_{max}$ | $B_f$ | $\kappa_f$ | $\tilde{q}$ | $y/x$ | $y$ | $S$ |
|---|---|---|---|---|---|---|---|---|
| | (MeV) | (fm$^2$) | (MeV) | (fm$^{-1}$) | (fm$^{-1}$) | | | |
| $^{27}$Al | 0.78 | 0.072 | 17.05 | 1.23 | 0.509 | 4.88 | 15.65 | 0.0168 |
| | 1.01 | 0.123 | 16.82 | 1.22 | 0.507 | 4.88 | 15.60 | 0.0287 |
| | 2.21 | 0.213 | 15.62 | 1.18 | 0.500 | 4.86 | 15.32 | 0.0495 |
| | 2.73 | 0.058 | 15.10 | 1.16 | 0.500 | 4.83 | 15.19 | 0.0135 |
| | 3.00 | 0.210 | 14.83 | 1.15 | 0.500 | 4.81 | 15.13 | 0.0488 |
| $^{28}$Si | 1.78 | 1.350 | 10.06 | 0.94 | 0.495 | 3.97 | 12.39 | 0.2957 |

$E_x$ – Excitation energy. $\sigma_{max}$ – maximum value of the experimental differential cross-section.

$B_f$ – The projectile binding energy in the outgoing channel. The projectile binding energy in the incoming channel, $B_i$ = 17.83 MeV for the $^{27}$Al + d and 11.84 MeV for $^{28}$Si + d systems.

$B_i$ and $B_f$ were calculated using binding energies compiled by Mattauch, Thiele, and Webstra (1965).

$\kappa_i$ = 1.26 fm$^{-1}$ for $^{27}$Al and 1.02 fm$^{-1}$ for $^{28}$Si. (For the definition of $\kappa_i$, $\kappa_f$ and other parameters used in my calculations see the Appendix C.)

$x_{max}$ = 3.14 for $^{27}$Al and 3.11 for $^{28}$Si, resulting in $R_0 = 6.3$ fm for both nuclei.

$S$ – The normalisation factor.

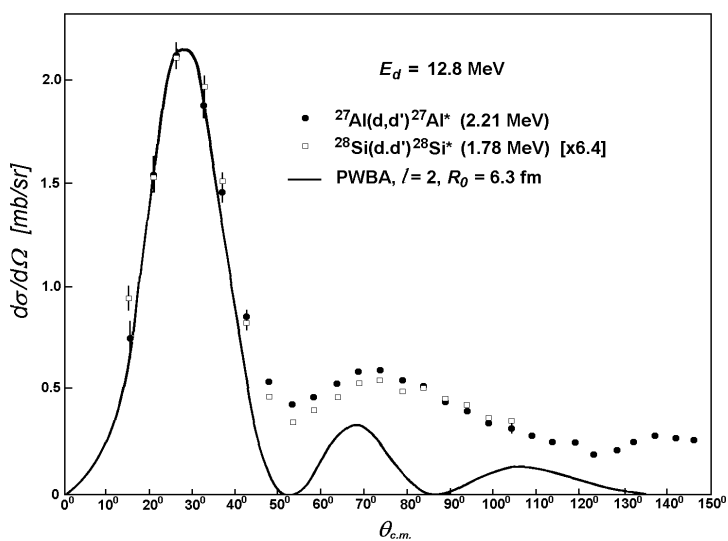

Figure 3.9. The calculated plane wave (PWBA) angular distribution is compared with experimental results for $^{27}$Al and $^{28}$Si. To display the data on the same graph, the experimental differential cross sections for $^{28}$Si have been divided by a factor of 6.4. The calculated curve shows that the observed transitions are associated with the $l = 2$ angular momentum transfer. The figure shows also close similarity between the experimental angular distributions for the two target nuclei as predicted by the core-excited model.

However, for $y/x > 2$, a linear interpolation can be applied and it gives $y \approx 15$ for $^{27}$Al and 12 for $^{28}$Si. Unfortunately, these values are again outside the range used by Lubitz, who tabulates theoretical cross sections for $y \le 6$. Consequently, the tables of Lubitz could not be used and the necessary function $W_2^2(x, y)$ had to be calculated.

I have calculated this function for a series of $y$ values around 15 for $^{27}$Al and 12 for $^{28}$Si and I have found that the position of its prominent maximum $x_{max} = 3.14$ for $^{27}$Al and 3.11 for $^{28}$Si. Using these values and the values of $\tilde{q}$ calculated at the positions of the experimental maxima I have obtained $R_0 = 6.3$ fm for both target nuclei.

The calculated PWBA angular distribution is compared with experimental results in Figure 3.9. For the purpose of displaying two experimental distributions on the same graph, the experimental values for $^{28}$Si have been divided by 6.4. Similar fits have been obtained for the angular distributions to other excited states in $^{27}$Al. The calculations confirmed the $l = 2$ assignment to the observed transitions.

### Strong coupling analysis

The experimental data for the elastic and inelastic scattering ($E_x = 1.78$ MeV) for $^{28}$Si were analysed using the strong coupling optical model (Buck, 1963; Buck, Stamp, and Hodgson 1963; Chase, Wilets, and Edmonds 1958) and computing facilities at Oak Ridge National Laboratory, USA and Oxford University, UK. In this model (see the Appendix D) the elastic and the first inelastic channel are considered explicitly in the calculations. The remaining non-elastic channels are taken into account through an appropriate absorbing potential.

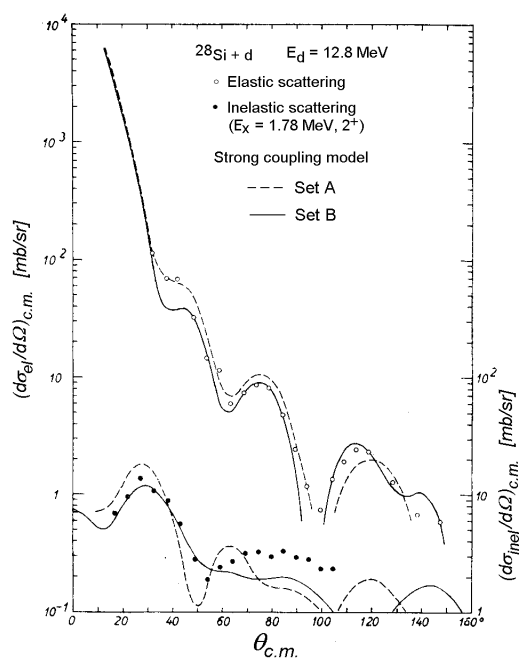

Figure 3.10. The experimental data for the elastic and inelastic scattering of 12.8 MeV deuterons by $^{28}$Si are compared with the strong coupling optical model calculations. The dotted line shows the fit to all the inelastic data and the full line the fit to the inelastic cross-sections at angles less than 50°. The total reaction cross-sections and cross-sections for excitation of the 2⁺ state for these two fits are $\sigma_R$ = 1357 mb, $\sigma_{in}$ = 34.6 mb and $\sigma_R$ =1345 mb, $\sigma_{in}$ = 32.7 mb, respectively. Parameter sets are listed in Table 3.4.

A simple collective model is used to represent the ground and excited states of the target nucleus. The rotational and vibrational models give almost identical results for the $0^+ \rightarrow 2^+$ excitation and the former was used in the present work. The elastic and inelastic angular distributions may then be calculated from the central optical potential together with the nuclear deformation parameter $\beta_2$.

In our analysis, the undeformed optical potential had the form:

$$U(r) = V_C(r) - Vf(r) - iWg(r)$$

where $V_C(r)$ is the Coulomb potential, taken to be that due to a uniformly charged sphere of radius $R_c = r_0 A^{1/3}$, $V$ and $W$ are the real and imaginary central potential depths, $f(r) = \left[1 + \exp\left(r - R/a\right)\right]^{-1}$ is the Woods-Saxon radial form factor, where the nuclear radius $R = r_0 A^{1/3}$ and $a$ is the surface diffuseness parameter, and $g(r)$ is its normalized radial derivative of $f(r)$. Spin-orbit forces were not included in this calculation.

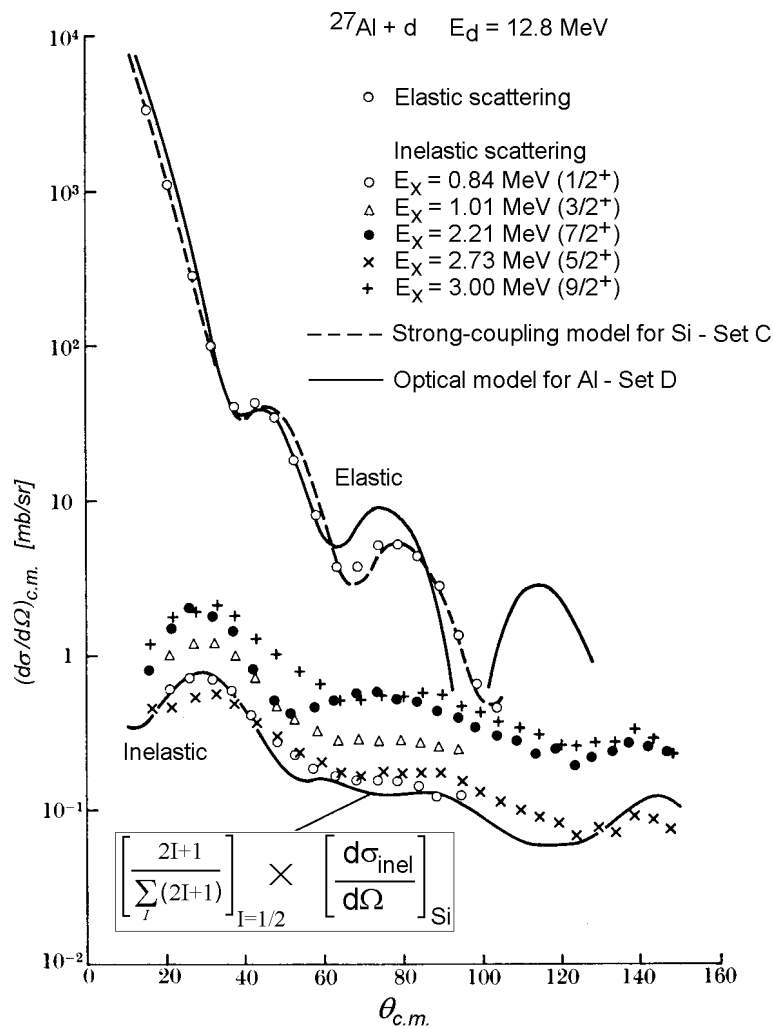

Figure 3.11. Experimental data for the elastic and inelastic scattering of 12.8 MeV deuterons by $^{27}$Al are compared with the distributions generated by the optical model (dashed line) and strong-coupling model (full lines). The corresponding reaction cross-section is 1221 mb. Parameter sets are listed in Table 3.4.

The strong coupling model is in principle applicable to the excitation of any finite number of collective states but at the time of our calculations the only available computer program was for the $0^+ \rightarrow 2^+$ excitation of even nuclei, so only the elastic and first inelastic distributions for silicon were analysed in this way. To make this analysis, the parameters of the above potential were systematically adjusted to give the best fit to the elastic cross-section alone, using a parameter search routine of Maddison (1962). This potential, together with an approximate value of $\beta_2$, was then used in the strong coupling calculation to give the elastic and first inelastic cross-sections.

The elastic distribution differed from that found from the initial fit to the elastic data partly due to the effect of the strong coupling and partly to the different value of the absorbing potential. All the parameters were then iterated to optimise the fit to the elastic and inelastic data, using the strong coupling search routine of Buck (1963).

At first, we have found that the value of $\beta_2$ increased at each iteration up to quite unphysical values. The results of one of these calculations are shown in Figure 3.10. The parameters used in the strong-coupling analysis are listed in Table 3.4.

Table 3.4
Parameters used in the strong-coupling analysis

| Nucleus | Set | $V$ | $W$ | $r_0$ | $a$ | $\beta_2$ | $\delta_2$ |
|---------|-----|-----|-----|-------|-----|-----------|-----------|
|         |     | (MeV) | (MeV) | (fm) | (fm) | | (fm) |
| $^{28}$Si | A | 93.2 | 17.59 | 1.3 | 0.7 | 0.866 | 3.42 |
|         | B | 106.2 | 11.68 | 1.3 | 0.7 | 0.446 | 1.74 |
| $^{27}$Al | C | 106.2 | 11.68 | 1.3 | 0.7 | 0.446 | 1.74 |
|         | D | 61.4 | 23.33 | 1.55 | 0.53 | — | — |

Set C – The parameter $\beta_2 = 0.664$ when normalised to the angular distributions for the 2.21 MeV state. The corresponding $\delta_2 = 2.59$ fm.
Set D – Conventional optical model analysis.

Figure 3.10 shows that the elastic cross-section is quite well fitted, and also the shape of the main forward peak of the inelastic cross-section but the calculated inelastic cross-section falls markedly below the experimental values above 50°. This failure of the model may possibly be due to the neglect of spin-orbit forces. The effect of this discrepancy is that the search program pushes up the inelastic cross-section to give the best overall fit to the data, resulting in an unphysically large value of $\beta_2$. The effect is particularly marked as the saturation region is soon reached, where a large increase of $\beta_2$ gives only a small increase in the inelastic cross-section.

To avoid this difficulty, the calculations were continued omitting the inelastic data above 50°. The iteration then converges to a satisfactory fit to the elastic and inelastic cross-sections, as shown in Figure 3.10, and gives a reasonable value for the nuclear deformation $\beta_2$.

The data for the elastic scattering from $^{27}$Al were analysed using conventional optical model. As the shapes of the experimental angular distributions for the inelastic scattering for $^{27}$Al and $^{28}$Si are virtually the same, the appropriately scaled theoretical distribution for $^{28}$Si has been used to compare it with the distributions for $^{27}$Al. Results are presented in Figure 3.11. It can be seen that the shape of the theoretical angular distribution for the inelastic scattering closely resembles the shapes of the experimental distributions.

## Deformation distance

It should be recognised that it is the deformation distance $\delta_{lm}$, which enters the calculations rather than a dimensionless parameter $\alpha_{lm}$ (or $\beta$) defined by

$$R = R_0(1 + \sum_{lm} \alpha_{lm} Y_{lm}^*)$$

More appropriately, the nuclear radius is written as

$$R = R_0 + \sum_{lm} \delta_{lm} Y_{lm}^*$$

Formally, all this says is that

$$\alpha_{lm} R_0 \equiv \delta_{lm}$$

There is an important consequence of working with $\delta_{lm}$: this parameter is believed to refer to the deformation of the nuclear matter field, while $R_0$ always includes effects related to the experiment in question. For example, in the diffraction model analysis, $R_0$ for scattering of $\alpha$ particles or deuterons on $^{27}$Al or $^{28}$Si is around 6.5 fm. This large size comes in part from the finite extent of $\alpha$ particles or deuterons, and from the fact that the cut-off radius used in the analysis corresponds to a radius well out in the tail of the nuclear potential.[5] On the other hand, for an electron scattering experiment, $R_0$ would be much smaller (less than around 4 fm), since it refers directly to the electric charge distribution. For a proton scattering experiment, $R_0$ would lie between these extremes.

For all these causes, however, we would expect the deformation distance parameters to be essentially the same since they refer to the deformation of the nuclear field alone. There might turn out to be some differences between $\delta_{lm}$ values derived from $\alpha$ or deuteron scattering experiments and $\delta_{lm}$ describing charge density, but it is certainly much more appropriate to compare these quantities rather than $\alpha_{lm}$ parameters derived using different projectiles. These $\alpha_{lm}$ parameters will differ by a factor of around 2 even though the deformation is the same.

## Rotational model for $^{27}$Al

Attempts have been made to apply rotational model to $^{27}$Al (Almqvist *et al* 1960; Bendt and Eidson 1961; Bishop 1960; Lawergren and Ophel 1962a, 1962b; Towle and Gilboy, 1962; Vanhuyse and Vanpraet 1963). In this model the first band with $K = {}^5/_2$ is built on the ground state, the second and third members being the $^7/_2{}^+$ state at 2.21 MeV and $^9/_2{}^+$ at 3.00 MeV. The $^3/_2{}^+$ state at 1.013 MeV and the $^5/_2{}^+$ state at 2.73 MeV are supposed to belong to the $K = {}^1/_2$ rotational band built on the $^1/_2$ 0.84 MeV excited level.

However, the rotational model has some difficulty with $^{27}$Al. This is unexpected because this model gives satisfactory description of the neighbouring nuclei $^{24}$Mg and $^{25}$Mg. An extension of this model to $^{27}$Al would seem a natural one. Some of the difficulties are as follows:

---

[5] In the strong-coupling analysis, a relatively small radius used in the optical model potential was compensated by large deformation parameters giving generally unphysically large values for the deformation distance.

(1) The deformation distance, $\delta_2$, as derived from my diffraction analysis for the $5/2 \rightarrow 7/2$ transition is 1.94 fm if one assumes a permanent deformation.[6] This value is larger than expected in this region, which could mean that rotational model cannot be applied to $^{27}$Al. For instance, similar diffraction model analysis of $^{24}$Mg gives $\delta_2 = 1.43$ fm (Blair 1961).

 (2) The deformation distance for this transition disagrees with the deformation distance calculated from electromagnetic transition.

The spectroscopic quadrupole moment $Q$ is the experimental observable, which in the case of axially deformed nuclei can be related to the intrinsic quadrupole moment $Q_0$ and thus to the nuclear *charge* deformation parameter or the deformation distance.

$$Q = \frac{3K - I_0(I_0 + 1)}{(I_0 + 1)(2I_0 + 3)} Q_0$$

The intrinsic quadrupole moment of a deformed ellipsoidal charge distribution can be expressed as:

$$Q_0 = \frac{3}{\sqrt{5\pi}} eZR_0^2 \beta (1 + 0.36\beta + ...)$$

Thus

$$Q_0 \cong \frac{3}{\sqrt{5\pi}} eZR_0^2 \beta = \frac{3}{\sqrt{5\pi}} eZR_0 \delta_2$$

The measured value for $Q$ is 14 -15 $e$fm$^2$ for $^{27}$Al (Stone 2001). The radius for the nuclear charge distribution can be assumed to be $R_c = 1.2A^{1/3}$ fm, which for $^{27}$Al is 3.6 fm. Using this value and the measured value for $Q$, we can calculate that the deformation distance $\delta_2$ determined by electromagnetic transition is between 1.11 and 1.19 fm, whereas the distance determined by the inelastic scattering using diffraction theory is 1.94 fm.

(3) In the strong coupling model, the only other strongly excited state should be a $^9/_2{}^+$ state located somewhere around 4 MeV or higher. However, what we find is in fact an equally strongly excited level at 3 MeV. It is clear that the rotational band sequence is not being followed.

(4) The intensity patterns are all in disagreement with the rotational prescription. The ratio of $^9/_2$ to $^7/_2$ cross sections should be 1/6 to 10/21, whereas the observed intensities of inelastic scattering are approximately 1 to 1. Thus, the rotational model predicts that only two states (9/2 and 7/2) are strongly excited but that of these two states, the 7/2 MeV level should be most prominent, which is in disagreement with experimental observation. Furthermore, the collective excitations are zero between states belonging to different bands. They may be excited only through changing the single particle orbitals. In contrast, experimental results show significant excitation of states allegedly belonging to the $K = 1/2$ band.

---

[6] As discussed earlier (see also Table 3.4), the analysis based on a more complex strong coupling theory also had a tendency to produce unrealistically large $\beta_2$ and $\delta_2$ parameters.

## Core-excited model of $^{27}$Al

To understand how the core excited model applies to $^{27}$Al we have to look again at its signatures and compare them with experimental results. The concept of core-exited model for this nucleus is shown in Figure 3.12.

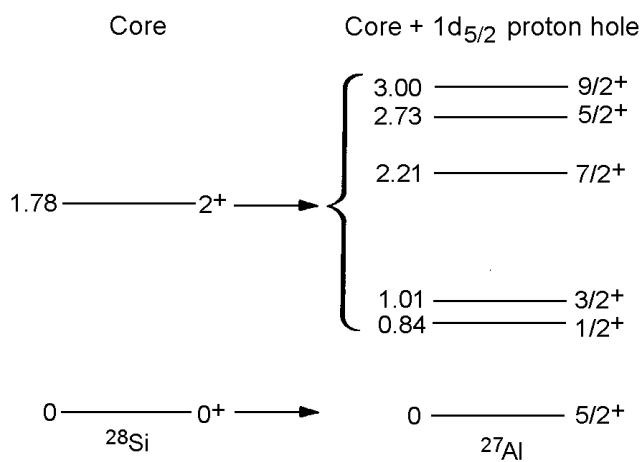

Figure 3.12. The core-excited model of $^{27}$Al.

### The spins of core-excited states

The coupling of a $1d_{5/2}$ single proton hole with the $2^+$ state should produce a quintuplet of states with spins $^1/_2$, $^3/_2$, $^5/_2$, $^7/_2$, and $^9/_2$. As can be seen, $^{27}$Al contains the full complement of these expected states.

### The centre of gravity

The core-excited states should be located around the energy of the $2^+$ state of the $^{28}$Si core. (The centre of gravity energy of the core-excited states should be equal to the excitation energy of the $2^+$ state.)

The centre of gravity of the quintuplet in $^{27}$Al is 2.32 MeV, which is close to the excitation energy 1.78 MeV of the $2^+$ state in $^{28}$Si. The higher than expected centre of gravity energy for the quintuplet can be explained by coupling of its $^5/_2^+$ member with the ground state (see below), which has the same spin and parity. The coupling repels the $^5/_2^+$ excited state and thus pushes it to a higher energy.

### Shapes of the angular distributions

All relevant inelastic scattering angular distributions should have the same shape. (The shapes of the angular distributions for the quintuplet of states and for the $2^+$ state should be the same.) The remarkable feature of our experimental results is that indeed all angular distributions display similar features (see Figure 3.7).

All inelastic scattering angular distributions have a prominent forward peak located approximately at the same scattering angle and all have been fitted using $l = 2$ angular momentum transfer. Examples of theoretical analysis are shown in Figure 3.8 for the diffraction model, in Figure 3.9 for the plane wave theory and in Figure 3.11 for the strong coupling model. Figure 3.11 shows how the scaled down theoretical curve for $^{28}$Si resembles closely the shapes of the inelastic scattering angular distributions for $^{27}$Al.

*The 2I+1 dependence*

The absolute values of the differential cross sections for the core-excited quintuplet should be proportional to $2I+1$, where $I$ is the spin of a member of the quintuplet.

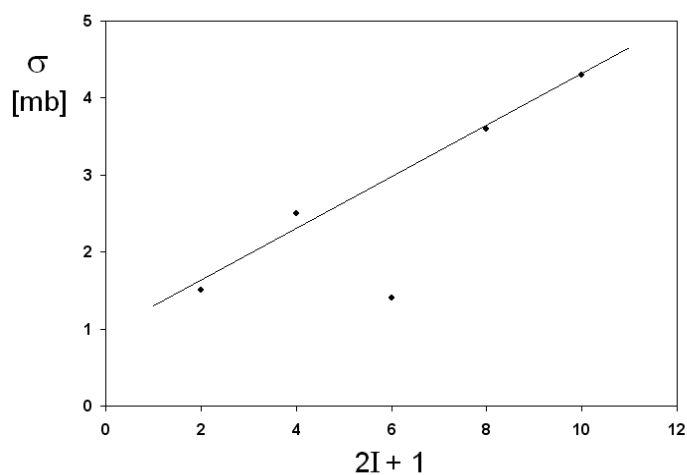

Figure 3.13. The integrated experimental cross sections (points) are compared with the expected linear 2I+1 dependence.

In order to study this signature, the experimental cross sections have been integrated under the first maximum for each member of the quintuplet. Results of the integration are presented as a function of 2I+1 in Figure 3.13. It can be seen that with the exception of one point, which belongs to the ⁵/₂⁺ excited state, the experimental points follow closely the linear 2I+1 dependence as expected for the core-excited model.

The cross section, which is outside the linear 2I+1 dependence, is for the scattering to the ⁵/₂⁺ level at 2.73 MeV. Its non-compliance with the rule may be understood as a consequence of coupling to the ground state, which has the same spin and parity. The coupling reduces the intensity of the inelastic scattering.

The reduction in the intensity of the scattering to the ⁵/₂⁺ state may be used to estimate the degree of mixing between this state and the ground state. The wave functions of the two states can be written as (Vervier 1963):

$$\left|\frac{5}{2}\right\rangle_1 = A\left|0\,\frac{5}{2}\,\frac{5}{2}\right\rangle + \sqrt{1-A^2}\left|2\,\frac{5}{2}\,\frac{5}{2}\right\rangle$$

$$\left|\frac{5}{2}\right\rangle_2 = A\left|2\,\frac{5}{2}\,\frac{5}{2}\right\rangle - \sqrt{1-A^2}\left|0\,\frac{5}{2}\,\frac{5}{2}\right\rangle$$

where $\left|\frac{5}{2}\right\rangle_1$ and $\left|\frac{5}{2}\right\rangle_2$ are the wave functions of the ground state and the excited state, the quantum numbers $\left|J_c\,jI\right\rangle$ refer to the total angular momentum of the $^{28}$Si core, the d$_{5/2}$ hole, and the $^{27}$Al, respectively, and $A$ is the mixing parameter.

The coupling reduces the differential cross section to the ⁵/₂⁺ member of the quintuplet by a factor of $0.38 \pm 0.06$, which corresponds to $A^2 = 0.72 \pm 0.03$.

*Absolute values of the measured cross-sections*

As mentioned in the Introduction, the measured cross sections for a core-excited member of a multiplet should be related to the cross section for the $2^+$ state of the core. The sum of measured cross sections for the multiplet should be equal to the cross section for the excitation of the core.

Table 3.5 lists the measured and calculated cross sections. The measured cross sections were obtained by integrating experimental results over the first maxima of the inelastic scattering angular distributions. The cross sections listed in the last column were calculated using the formula mentioned in the Introduction:

$$\left[\frac{d\sigma}{d\Omega}\right]_I = \frac{(2I+1)}{\sum_J (2I+1)}\left[\frac{d\sigma}{d\Omega}\right]_{J_c}$$

Table 3.5

The measured and calculated cross sections for the inelastic scattering of 12.8 MeV deuterons from $^{27}$Al and $^{28}$Si nuclei

| Nucleus | $E_x$ [MeV] | $I^\pi$ | Measured cross sections [mb] | Calculated cross sections [mb] |
|---------|------|-------|------------------------------|--------------------------------|
| $^{28}$Si | 1·77 | $2^+$ | 18·1 ± 3·0 | — |
| | 0·84 | $\frac{1}{2}^+$ | 1·5 ± 0·2 | 1·2 |
| | 1·01 | $\frac{3}{2}^+$ | 2·5 ± 0·4 | 2·4 |
| $^{27}$Al | 2·21 | $\frac{7}{2}^+$ | 3·6 ± 0·5 | 4·8 |
| | 2·73 | $\frac{5}{2}^+$ | 1·4 ± 0·2 | 3·6 |
| | 3·00 | $\frac{9}{2}^+$ | 4·3 ± 0·7 | 6·0 |

As can be seen, the measured cross sections agree reasonably well with the calculated values. The notable exception is again for the $^5/_2{}^+$ excited state in $^{27}$Al, which as mentioned earlier, can be at least partly explained by coupling to the ground state. When correction is made for the lost intensity, the measured cross section increases from the listed value of 1.4 mb to 2.3 mb. The sum of the cross sections for $^{27}$Al is then 14.2 mb, which is close to the expected 18.1 mb. The difference is caused mainly by the lower than expected excitation of the $^9/_2{}^+$state.

The reason for slightly lower cross section for the $^9/_2{}^+$ state is unclear but it might be due to sharing its intensity with more complex core-coupling schemes, such as coupling of the proton hole with the $4^+$ excited state in $^{28}$Si. Simple models are hardly ever expected to result in a perfect representation of nuclear structure and it is already remarkable that the simple core-coupling model gives such a good description of the observed experimental features.

## Summary and conclusions

Angular distributions of the differential cross sections for the elastic and inelastic scattering of 12.8 MeV deuterons from $^{27}$Al and $^{28}$Si have been measured over a wide range of angles. Experimental results for the inelastic scattering display clear features of direct reaction mechanism.

Results of measurements were analysed using a wide range of theories: the diffraction models, the plane wave Born approximation theory, the optical model and the strong coupling model. As shown by Rost and Austern (1960) the diffraction models are a good approximation of more complex distorted wave descriptions of inelastic scattering. The carried-out calculations confirmed that the observed angular distributions can be described assuming a direct reaction mechanism. They have also resulted in useful structure-related parameters.

A step-by-step comparison of experimental results with the core-excited model shows that the low-lying states in $^{27}$Al can be described by assuming a coupling a $1d_{5/2}$ proton hole with the $2^+$ state of the $^{28}$Si core.

### References

Almqvist, E., Bromley, D. A., Gove, H. A. and Litherland, A. E. 1960, *Nucl. Phys.* **19**:1.

Bent, R. D. and Eidson, W. W. 1961, *Phys. Rev.* **122**:1514.

Bishop, G. R. 1960, *Nucl. Phys.* **14**:376.

Blair, J. S. 1959, *Phys. Rev.* **115**:928.

Blair, J. S. 1961, private communication.

Blair, J. S., Sharp, D., and Wilets, L. 1962, *Phys. Rev.* **125**:1625.

Buck, B. 1963, *Phys. Rev.* **130**:712.

Buck, B., Stamp, A. P. and Hodgson, P. E. 1963, *Phil. Mag.* **8**:1805.

Butler, S. T. 1951, *Proc. Roy. Soc.* (London) **A208**:559.

Butler, S. T. 1957, *Phys. Rev.* **106**:272.

Butler, S. T., Austern, N., and Pearson, C. 1958, *Phys. Rev.* **112**:1227.

Chase, D. M., Wilets, L. and Edmonds, A. R. 1958, *Phys. Rev.* **110**:1080.

De Shalit, A. 1961, *Phys. Rev.* **122**:1530.

Lawergren, B. T. and Ophel, T. R. 1962a, *Phys. Lett.* **2**:265.

Lawergren, B. T. and Ophel, T. R. 1962b, *Proc. Phys. Soc.* **79**:881.

Lawson, R. D., and Uretsky, J. L. 1957, *Phys. Rev.* **108**:1300.

Lubitz, C. R. 1957, *Numerical Table of Butler-Born Approximation Stripping Cross Sections*, Randall Laboratory of Physics, University of Michigan, Ann Arbor.

Maddison, R. N. 1962, *Proc. Phys. Soc.* **79**:264.

Mattauch, J. H. E., Thiele, W., and Webstra, A. H. 1965, *Nucl. Phys.* **67**:1.

Rost, E. and Austern, N. 1960, *Phys. Rev.* **120**:1375.

Stone, N. J. 2001, *Tables of Nuclear Magnetic Dipole and Electric Quadrupole Moments*, Oxford Physics, Clarendon Laboratory, Oxford, UK.

Towle, J. H. and Gilboy, W. B. 1962, *Nucl. Phys.* **39**:300.

Vanhuyse, V. J. and Vanpraet, G. J. 1963, *Nucl. Phys.* **45**:603.

Vervier, J. 1963, *Nuovo Cim.* **28**:1412.

---

<div align="center">

**4**

**$^{40}$Ca(d,d), (d,d'), and (d,p) Reactions with 12.8 MeV Deuterons**

</div>

***Key features:***

1. We have measured angular distributions of the differential cross sections for the elastic scattering of 12.8 MeV deuterons from $^{40}$Ca nuclei using a broad range magnetic spectrometer. We have also measured the distributions for the inelastic scattering to the states 3.35 MeV (0$^+$), 3.74 MeV (3$^-$) and 4.49 MeV (5$^-$) in $^{40}$Ca, and for the deuteron stripping reaction to the ground state ($^7/_2^-$), 1.94 MeV ($^3/_2^-$), and 2.46 MeV ($^3/_2^-$) levels of $^{41}$Ca

2. Optical model analysis of the elastic scattering employed central potential with surface absorption. We have found that by including spin-orbit interaction we can reproduce the diffraction anomaly discussed in Chapter 2.

3. Inelastic scattering angular distributions were analysed using distorted wave Born approximation and deformed optical model potential. We have found that the deformation of both the real and imaginary components was necessary to generate physically meaningful deformation parameters. Our deformation (and deformation distance) parameters are in good agreement with the values based on a study of 55 MeV proton scattering (Yagi *at al.* 1964).

4. Analysis of the stripping reaction was also carried out using the distorted wave Born approximation. We have carried out a detailed study of the local, nonlocal, finite range and zero range approximations. Final results included nonlocal and finite-range effects. The derived spectroscopic factors are lower than theoretically expected and possible reasons for the discrepancy are discussed.

***Abstract***: Differential cross sections were measured for the $^{40}$Ca(d,d), (d,d'), and (d,p) reactions induced by 12.8 MeV deuterons. Inelastic scattering to the 3.35 MeV (0$^+$), 3.74 MeV (3$^-$) and 4.49 MeV (5$^-$) levels, and (d,p) reactions to the ground state ($^7/_2^-$), 1.94 MeV ($^3/_2^-$), and 2.46 MeV ($^3/_2^-$) levels of $^{41}$Ca were studied. Optical model analysis of the elastic scattering was carried out, and potentials obtained were used in the distorted wave analysis of the inelastic scattering and stripping reactions. Spin-orbit interaction was necessary to fit the elastic scattering data. The collective model gave a good fit to the inelastic excitations of the 3$^-$ and 5$^-$ levels. To obtain realistic deformation parameters it was necessary to use a complex form of nuclear interaction. The effects of finite range and nonlocality were included in the deuteron stripping analysis and produced reasonable agreement with the observed cross sections. The spectroscopic factors extracted when the spin-orbit coupling is included in the deuteron optical model potential are significantly smaller than expected.

## Introduction

The direct nuclear reactions have proved to be useful in studies of nuclear structure. While inelastic scattering of particles yield parameters characterizing the collective modes of nuclear excitations, the single-particle aspects of nuclear internal motion can be studied by stripping reactions. Thus, inelastic scattering and transfer reactions give valuable information about collective and single-particle excitations. Theoretical analysis of these reactions gives also useful insights into the details of the mechanism of nuclear interactions.

For our study, we have chosen $^{40}$Ca as a target nucleus because of its double magic properties. Both neutron and proton shells are closed, which means that, in its ground state, $^{40}$Ca should can be described as a spherical nucleus. The excited states of $^{41}$Ca should be then described by a simple single-particle model with a stripped neutron moving around in a spherical potential of a well bound, doubly-closed core.

In the work described here we have studied elastic and inelastic scattering of 12.8 MeV deuterons and single-neutron stripping reaction. Theoretical analysis of the data was carried out using optical model and distorted wave Born approximation.

### Experimental procedure and results

#### *Procedure*

Experimental data were obtained using a beam of 12.8-MeV deuterons from the 120-cm cyclotron of the Institute of Nuclear Physics in Cracow. A broad-range spectrometer (see Chapter 3) with photographic emulsions as detectors has been used for the measurements of angular distributions. The spectra of particles emitted in the reactions were taken in steps of 5° in the angular range of 10° to 110° in the laboratory system.

The integrated beam passing through the $^{40}$Ca target was collected in the Faraday cup placed behind the target and measured by means of a standard current integrator. Beam monitoring was also done by using a thin gold foil placed in the beam at the entrance to the scattering chamber. An additional detector was used to monitor changes in the target thickness.

The $^{40}$Ca targets with a thickness of 2.28 mg/cm$^2$ were prepared by rolling it from metallic samples, which were placed immediately in the vacuum in order to reduce the oxygen content.

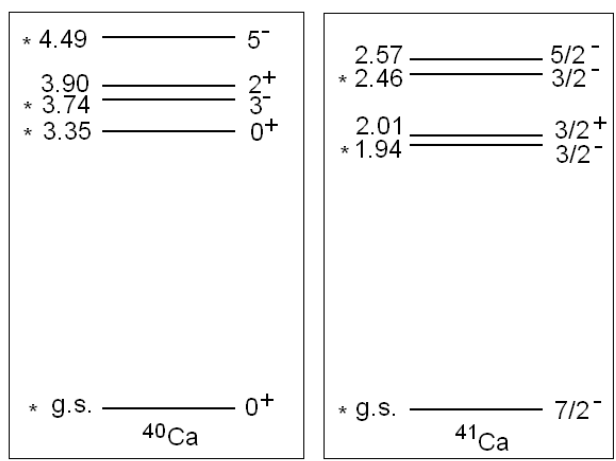

Figure 4.1. The low-lying states in $^{40}$Ca and $^{41}$Ca. States for which angular distributions have been measured are marked with the asterisks.

We have measured inelastic scattering angular distributions for the states at 3.35 MeV (0$^+$), 3.74 MeV (3$^-$), and 4.49 MeV (5$^-$). We have also repeated measurements of elastic scattering (see Chapter 2) using a 5.4 mg/cm$^2$ target. Proton angular distributions for the deuteron stripping reaction $^{40}$Ca(d,p)$^{41}$Ca were measured for transitions to the ground state and two excited states (1.94 and 2.46 MeV) of $^{41}$Ca.

The low-lying energy levels for $^{40}$Ca and $^{41}$Ca are shown in Figure 4.1. States excited with substantial intensity and for which angular distributions have been measured are marked with the asterisks.

### Results

The measured differential cross sections for the elastic and inelastic scattering, together with theoretical curves, are shown in Figures 4.2, 4.4, and 4.6. No significant excitation of the 3.90 MeV level of $^{40}$Ca nucleus by inelastic scattering was observed.

The relative errors included the statistical errors as well as inaccuracies in the angle setting and in scanning of the emulsions. The combined relative errors were around 6%. The error in the absolute values of the cross sections was estimated at 15%. It was associated with inaccuracies in the beam intensity measurements and in the determination of the target thickness.

The energy resolution of the detecting system (150 keV) assured good separation of the detected groups of particles, except for the group of protons leading to the 1.94 MeV state of the $^{41}$Ca nucleus where a small admixture from the very weak level at 2.01 MeV remained unresolved.

The two sets of data for the elastic scattering shown in Figure 4.2 are for two different target thicknesses (4.6 and 5.4 mg/cm$^2$) and two different experimental arrangements. The 4.6 mg/cm$^2$ data were obtained using a counter telescope (see Chapter 2). The 5.4 mg/cm$^2$ data represent the results of repeated measurements with the magnetic spectrometer. The differences between the two sets of measurements are also believed to be associated with slightly different incident energies used in both experiments. As can be seen in Figure 4.2, the two sets of measurements agree well for most angles but diverge for angles larger than around 120$^0$ (c.m.) where differential cross sections are expected to depend strongly on the incident deuteron energies (*cf* Chapter 2).

## Theoretical analysis

### Optical-Model Analysis of the Elastic Scattering

For the purpose of the present analysis, the two available sets of elastic-scattering data were combined and treated as one. The optical potential used assumes a surface-peaked absorption and has the form:

$$U(r) = U_c(r) - Vf(r) - W_D g(r) - V_s h(r) \mathbf{L} \cdot \mathbf{s}$$

where $U_c(r)$ is the Coulomb potential from a uniformly charged sphere with the radius $r_c A^{1/3}$,

$$f(r) = \frac{1}{e^x + 1} \qquad g(r) = 4i\left(\frac{d}{dx'}\right)\frac{1}{e^{x'} + 1} \qquad h(r) = \left(\frac{\hbar}{m_\pi c}\right)^2 \frac{1}{r}\left(\frac{d}{dx}\right)\frac{1}{e^x + 1}$$

$$x = \frac{r - r_0 A^{1/3}}{a} \qquad x' = \frac{r - r_0' A^{1/3}}{a'}$$

We have assumed $r_c$=1.3 fm for the Coulomb potential and we have carried out parameter search for $V$ around 100 MeV. We have carried out three sets of calculations, which resulted in three sets of parameters, which are listed in Table 4.1. The corresponding theoretical calculations are compared with experimental data in Figure 4.2.

Table 4.1

Optical model parameters describing the $^{40}$Ca(d,d)$^{40}$Ca elastic scattering at 12.8 MeV deuteron energy

| Set | $V$ | $r_0$ | $a$ | $W_D$ | $r_0'$ | $a'$ | $V_s$ | $\chi^2$ |
|---|---|---|---|---|---|---|---|---|
| | (MeV) | (fm) | (fm) | (MeV) | (fm) | (fm) | (MeV) | |
| 1 | 111.8 | 1.000 | 0.850 | 15.30 | 1.431 | 0.613 | 0.00 | 7.54 |
| 2 | 131.3 | 0.840 | 1.013 | 16.55 | 1.509 | 0.598 | 0.00 | 6.70 |
| 3 | 91.6 | 1.164 | 0.722 | 8.88 | 1.374 | 0.773 | 9.60 | 2.96 |

The $\chi^2$ value measures the quality of fits to the data. It shows that the best fit is for the set 3, which includes the spin-orbit interaction.

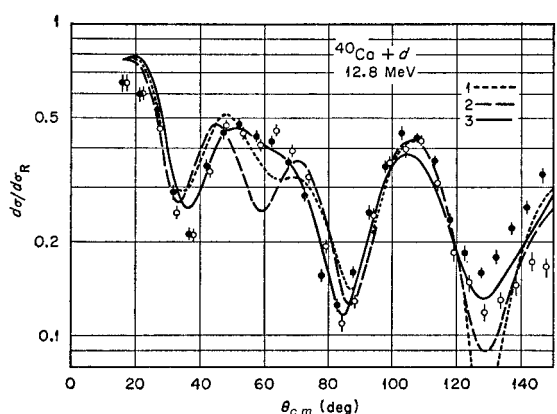

Figure 4.2. Optical model calculations for the $^{40}$Ca(d,d)$^{40}$Ca elastic scattering at 12.8 MeV deuteron energy for three sets of parameters (see Table 4.1) are compared with experimental data. Open circles represent experimental results obtained using a 4.6 mg/cm$^2$ target and a counter telescope as described in Chapter 2. Closed circles are for the current measurements with the 5.4 mg/cm$^2$ target and the magnetic spectrometer. The two sets of measurements were combined and treated as one in the automatic search code employed in our optical model analysis.

There are known to be considerable ambiguities in the choice of optical potentials for deuterons (Halbert 1964; Perey and Perey 1963) and these have been examined in some detail for $^{40}$Ca + d at other energies (Bassel *et al.* 1964). However, there is evidence from analysis of (d,p) reactions that the potential required is one with a real well depth $V$ ~ 100 MeV, and attention was restricted to this potential in the present work.

Previous studies (Bassel *et al.* 1964; Halbert 1964; Perey and Perey 1963) of deuteron scattering from $^{40}$Ca have indicated that the imaginary potential extends to significantly larger radii than the real potential ($r_0' \approx 1.5$ fm while $r_0 \approx 1.0$ fm). These earlier results were used as a starting point for the present searches.

The first calculations were made without spin-orbit coupling, i.e. with $V_s = 0$. It was known from earlier works that the least well determined parameters are $r_0$ and $a$, so initially these parameters were fixed at the value $r_0 = 1.0$ fm and $a = 0.85$ fm previously found for the 12 MeV data (Bassel *et al.* 1964). The optimum values for the other parameters are given in Table 4.1, denoted as set 1. When $r_0$ and $a$ are also allowed to vary, set 2 results; the reduction in $\chi^2$ is only about 10%. The corresponding predicted cross sections are compared to experiment in Figure 4.2.

A considerable improvement was obtained when spin-orbit coupling was introduced, i.e. when $V_s \neq 0$ was assumed. The optimum $\chi^2$ was then reduced by more than a factor of 2. The improvement is particularly evident at the maximum around 60°. The parameters are given in Table 4.1 as set 3, and the comparison of theory with experiment is included in Figure 4.2. The main effect of the introduction of spin-orbit coupling upon the optimum values of the other parameters is to reduce considerably the strength of the absorptive potential.

### *Inelastic Scattering*

Calculations were made for the inelastic scattering to the 3.35 MeV (0⁺, *l* = 0), 3.74 MeV (3⁻, *l* = 3), and 4.49-MeV (5⁻, *l* = 5) states in ⁴⁰Ca, using the collective-model interaction and the distorted-wave approximation. Both the model and the method have been discussed in detail elsewhere (Bassel *et al.* 1962). Suffice it to say that the inelastic angular distributions are determined once the optical-model parameters have been found by fitting the elastic scattering. The one adjustable parameter is then the deformation $\beta_l$ (or equivalently the deformation distance $\delta_l$ defined as $\delta_l = \beta_l r_0 A^{1/3}$), which is chosen to reproduce the *magnitude* of the measured cross section. The predicted cross sections are proportional to $\beta_l^2$.

All three optical model potentials described in the previous section were used. However, since the computer code was unable to include spin-orbit effects for spin-1 particles in both channels, set. 3 was used with $V_s = 0$.

A subsidiary calculation with potential set 3, but treating the deuteron as though it had spin 1/2, showed a slight filling in of the minima in the angular distribution. It is expected that a correct spin-1 calculation would show a similar small effect.

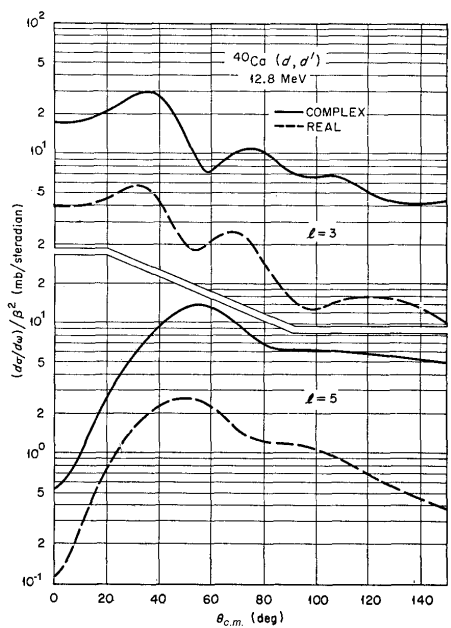

Figure 4.3. The dependence of the calculated angular distributions on whether only real or both real and imaginary components of the optical model were assumed to be deformed (real or complex coupling). Potential set 2 of Table 4.1 was used.

Table 4.2

The deformation $\beta_l$ and deformation distance $\delta_l$ parameters for the low-lying states in $^{40}$Ca

| Potential set | $l = 0$ (3.35 MeV) $\beta_0$ | $l = 3$ (3.74 MeV) $\beta_3$ | $l = 5$ (4.49 MeV) $\beta_5$ | $l = 0$ (3.35 MeV) $\delta_0$ | $l = 3$ (3.74 MeV) $\delta_3$ | $l = 5$ (4.49 MeV) $\delta_5$ |
|---|---|---|---|---|---|---|
| 2 | 0.08 | 0.32 | 0.16 | 0.23 | 0.92 | 0.46 |
| 3 | 0.05 | 0.30 | 0.13 | 0.20 | 1.19 | 0.52 |

The energies of excited states are shown in the parentheses. The deformation distance parameters $\delta_l$ are in fm.

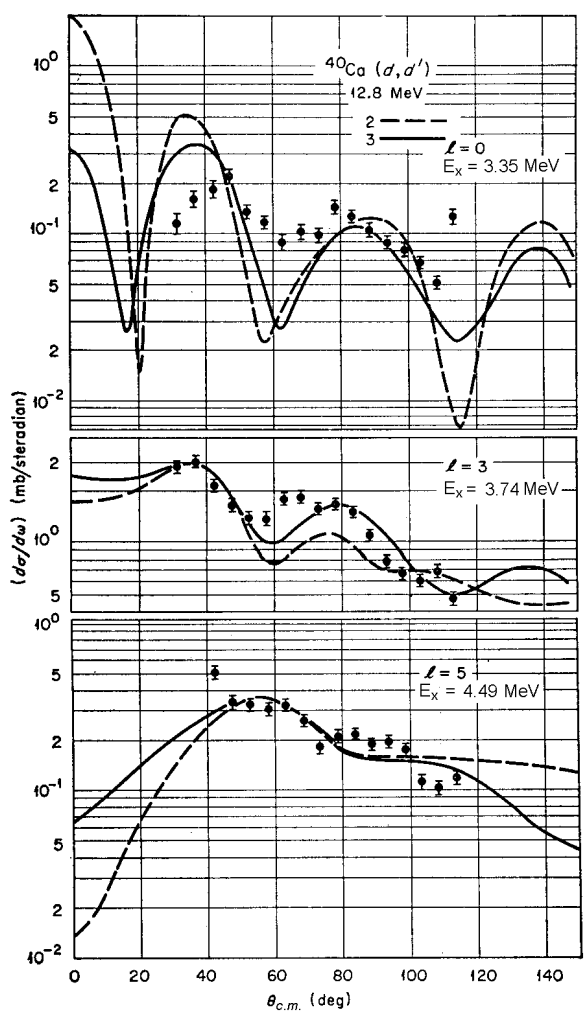

Figure 4.4. Angular distributions for the inelastic scattering of 12.8 MeV deuterons by $^{40}$Ca. Theoretical calculations are for the potential sets 2 and 3 as defined by parameters listed in Table 4.1. The corresponding deformation parameters are listed in Table 4.2.

Two versions of the model were used, one in which only the real part of the optical potential was deformed (the real coupling labelled as "real") and one in which both real and imaginary parts were deformed with the same deformation (the complex coupling labelled as "complex"). Figure 4.3 compares their predictions, for the potential set 2.

It is seen that the complex coupling produces a much larger cross section (for a given $\beta_l$ value) than does real coupling. As a result, when normalising to the experimental distributions, calculations with real coupling would require unacceptably large values of $\beta_l$ (0.704 for $l = 3$ and 0.384 for $l = 5$). This effect has been noticed before for deuteron scattering (Dickens, Perey, and Satchler 1965)

Figure 4.4 compares the theoretical calculations with the experimental data using complex coupling for potential sets 2 and 3. The $\beta_l$ and $\delta_l$ values are listed in Table 4.2. Excitation of the 3⁻ and 5⁻ levels by 55-MeV protons (Yagi *et al.* 1964) yielded deformation parameters (and deformation distances), which are in good agreement with the values derived here. The application of the model to the 0⁺ excited state is perhaps questionable. It might be regarded as a monopole ("breathing mode") vibration, but one would expect the volume integral of the potential well to be conserved approximately during the oscillations, and this would introduce a volume interaction term additional to the surface coupling used here. There is only qualitative agreement between the measurements and the $l = 0$ predictions shown in Figure 4.4. The transition is quite weak, as evidenced by the small values of $\beta_0$ ~ 0.05 or 0.08 required.

Potential sets 1 and 2 give very similar predictions. The angular distributions from potential set 3 are not very different, but the cross-section magnitudes are significantly larger. This result is due to the smaller absorptive strength of this potential (see Table 4.1).

### *Deuteron Stripping*

The application of the distorted-wave method to the $^{40}$Ca(d,p)$^{41}$Ca reaction has been described and discussed in detail by Lee *at al.* (1964). For the analysis of the present data, the deuteron optical potentials described earlier for the elastic scattering were used, but the proton optical potential and the shell-model potential into which the neutron is captured were taken to be the same as used by Lee *at al.* (1964). Spin-orbit coupling was included for both the neutron and proton. Corrections for finite range of the n-p interaction were made in the local energy approximation (Buttle and Goldfarb 1964; Perey and Saxon 1964), which is known to be very accurate for this reaction (Austern1965; Perey 1963). Although the optical potentials used were local, the damping of the wave functions in the nuclear interior, which would arise from using equivalent nonlocal potentials (Hjorth, Saladin, and Satchler 1965) was also calculated in the local energy approximation (Buttle and Goldfarb 1964; Hjorth, Saladin, and Satchler 1965; Perey and Saxon 1964).

The nonlocality range was taken to be $\beta$ =0.85 fm for the nucleons and $\beta$ = 0.54 fm for the deuteron (see the Appendix E). These values assume that all the observed energy dependence of the optical potentials is due to nonlocality rather than any intrinsic energy variation; hence, they give an upper limit to these effects.

Nonlocality reduces the contributions to the reaction amplitude from the nuclear interior by 30-40%, but it also increases the magnitude of the tail of the neutron bound-state wave function by 13% (for the 1f state) or 11% (for the 2p states). Finite range then produces an additional 10-20% reduction in the contributions from the interior.

As a result, the predicted differential cross-section curves with nonlocality and finite range fall between those calculated in zero-range approximation with local potentials with and without a radial cutoff in the nuclear surface, which eliminates all the interior contributions. This is illustrated in Figure 4.5 for the $l = 1$ and 3 transitions.

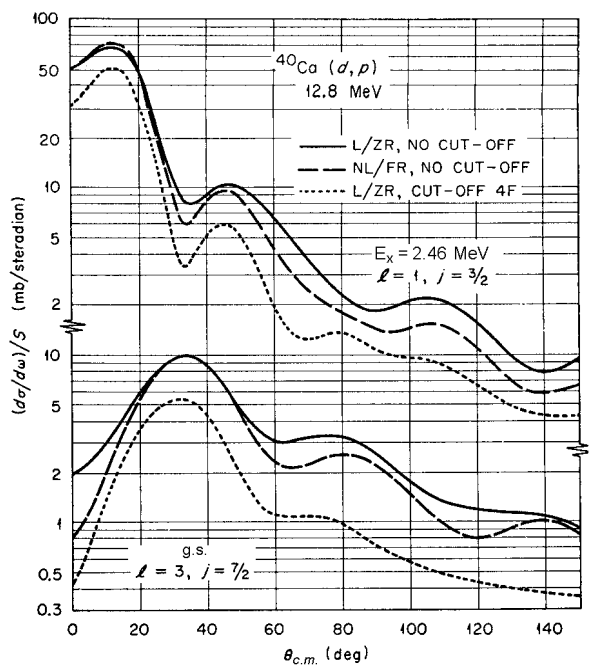

Figure 4.5. Theoretical predictions for the $l = 1$ and $l = 3$ stripping comparing the effects of nonlocality (NL) and finite-range (FR) with those from the use of a radial cutoff in a local (L) or zero-range (ZR) calculation. Potential set 3 was used. The L/ZR calculations with a cutoff radius eliminate completely the contribution from the interior of the nucleus. The remaining assumptions (L/ZR or NL/FR) only reduce these contributions.

Calculations were made for all three deuteron potentials of Table 4.1, including the spin-orbit coupling of set. 3. The results for potential sets 1 and 2 were almost identical. It may be seen that the calculations with a non-zero cutoff radius predict significantly lower differential cross sections then the calculations with no cutoff radius. Thus, the non-zero cutoff radius calculations lead to significantly larger values of the spectroscopic factors.

The experimental data are compared with theoretical calculations in Figure 4.6 for the potential sets 2 and 3, with nonlocal and finite-range effects included. The fits to the angular distributions are as good as those obtained at other energies (Hjorth, Saladin, and Satchler 1965; Lee *et al.* 1964). In particular, the discrepancy remains between theory and experiment for the shape of the second maximum for the $l = 3$ ($1f_{7/2}$) transition.

The spectroscopic factors obtained are given in Table 4.3. These values are not directly comparable to those obtained by Lee *et al.* (1964), because nonlocality corrections were not included in that work. The effect on the tail of the neutron wave function alone would reduce all the spectroscopic factors of Lee *et al.* (1964) by 22% (for 2p transitions) or 26% (for 1f transitions), but these reductions are partly compensated for by the additional damping of the contributions from the nuclear interior.

The spectroscopic factors for the set 3, which resulted in the best fit to the elastic scattering angular distribution, are significantly smaller than the values expected for a closed-shell target. (One would expect the ground-state transition to have the value of unity and the two $p_{3/2}$ transitions to sum to unity.) A similar discrepancy for the $f_{7/2}$ state was noted at deuteron energy of 14.5 MeV (Hjorth, Saladin, and Satchler 1965).

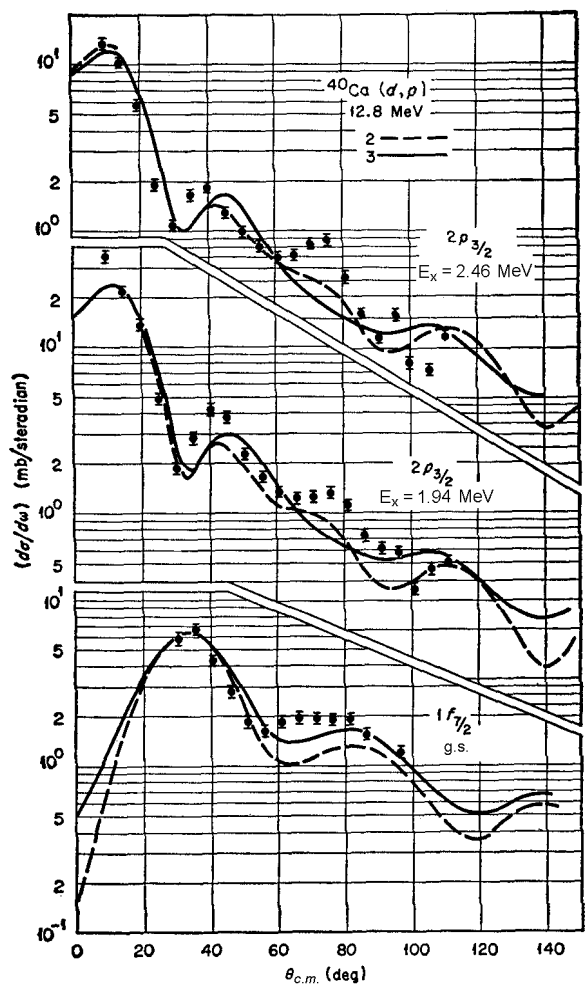

Figure 4.6. Comparison of the measured cross sections for the $^{40}$Ca(d,p)$^{41}$Ca stripping reaction with the distorted-wave predictions using potential sets 2 and 3. Nonlocality and finite-range effects are included, and no cutoff is used.

Table 4.3
Spectroscopic factors for the reaction $^{40}$Ca(d,p)$^{41}$Ca
(Nonlocal and finite range effects are included)

| $E_x$ | Configuration | Spectroscopic factors | |
|---|---|---|---|
| | | Pot. set 2 | Pot. set 3 |
| g.s. | $f_{7/2}$ | 0.95 | 0.65 |
| 1.94 | $p_{3/2}$ | 0.44 | 0.34 |
| 2.46 | $p_{3/2}$ | 0.22 | 0.17 |

## Summary and conclusions

We have measured angular distributions for the elastic and inelastic scattering for the low-lying states of $^{40}$Ca using 12.8 MeV deuterons and a broad range magnetic spectrometer. We have also measured distributions for the deuteron stripping reaction leading to low-lying states in $^{41}$Ca.

We have analysed elastic scattering angular distribution using optical model potential with surface absorption. We have found that by adding spin-orbit interaction we can reproduce the anomaly described in Chapter 2.

To analyse angular distributions for the inelastic scattering we have used deformed optical model potential with the same parameters as used for the elastic scattering. We have found that deformation of both the real and imaginary components was necessary. If the deformation is assumed to be only for the real component, the resulting deformation parameters are unrealistically high. If both components are assumed to be deformed, the deformation parameters determined by our analysis are in good agreement with parameters obtained from a study of 55 MeV proton scattering (Yagi *et al.* 1964).

The analysis of the angular distributions for the deuteron stripping reactions was carried out using distorted wave Born approximation theory. We have studied the effects of local, nonlocal, zero range and final range approximations. We have also carried out calculations using a cutoff radius and compared them with other calculations.

We have found that a cutoff radius lowered significantly the absolute values of the differential cross sections but had little effect on the shape of the angular distributions. In the final calculations, we have included both the nonlocal and finite range effects.

We have found that the derived spectroscopic factors are lower than the theoretically expected values. The spectroscopic factor for the ground state should be 1 but is 0.95 if set 2 of optical model parameters is used or 0.65 for the set 2. The sum of the spectroscopic factors for the $p_{3/2}$ transfers should be also 1 but is 0.66 for the set 2 of the optical model potential or 0.51 for the set 3.

The conclusions to be drawn about the spectroscopic factors are not clear. The results may mean that $^{40}$Ca does not have very pure closed-shell structure, or they may merely reflect a poor choice of parameters for the neutron potential well (Lee *at al.* 1964). Another uncertainty concerns the nonlocality of the neutron potential well; we have no direct evidence on this, and we have seen that using a local potential would increase the spectroscopic factors by approximately 20%.

## References

Austern, N. 1965, *Phys. Rev.* **137**:B752.

Bassel, R. H., Drisko, R. M., Satchler, G. R., Lee, L. L., Schiffer, J. P., and Zeidman, B. 1964, *Phys. Rev.*, **136**:B960.

Bassel, R. H., Satchler, G. R., Drisko, R. M., and Rost, E. 1962, *Phys. Rev.*, **128**:2693.

Buttle, P. J. A. and Goldfarb, L. B. J. 1964, *Proc. Phys. Soc. (London)* **A83**:701.

Dickens, J. K., Perey, F. G., and Satchler, G. R. 1965, *Nucl. Phys.*, **73**:529.

Halbert, E. C. 1964, *Nucl. Phys.* **51**:353.

Hjorth, S. A., Saladin, J. X., and G. R. Satchler 1965, *Phys. Rev.* **138**:B1425.

Lee, L. L., Schiffer, J. P., Zeidman, B., Satchler, G. R., Drisko, R. M., and Bassel, R. H. 1964, *Phys. Rev.*, **136**:B971.

Perey, F. G. 1963, in *Proceedings of the Conference on Direct Interactions and Nuclear Reaction Mechanisms, Padua, 1962*, edited by E. Clement and C. Villi, Gordon and Breach Science Publishers, Inc., New York.

Perey, C. M. and Perey, F. G. 1963, *Phys. Rev.* 132:755.

Perey, F. G. and Saxon, D. 1964, *Phys. Letters* **10**:107.

Yagi, K., Ejiri, H., Furukawa, M., Ishizaki, Y., Koike, M., Matsuda, K., Nakajima, Y., Nonaka, I., Saji, Y., Tanaka, E., and Satchler G. R. 1964, Phys. Letters **10**:186.

<div align="center">

**[5](#)**

# Spectroscopic Applicability of the ($^3$He,$\alpha$) Reactions

</div>

***Key features:***

1. The aim of this work was to study the applicability of ($^3$He,$\alpha$) neutron pickup reactions in nuclear spectroscopy

2. I have measured differential cross sections for the pickup reaction $^{26}$Mg($^3$He,$\alpha$)$^{25}$Mg leading to low-energy states in $^{25}$Mg.

3. I have carried out theoretical analysis of experimental results using direct reaction theory and computer codes, which I have modified and adapted to run at the Australian National University.

4. Nuclear interaction in both incoming and outgoing channels is described using central optical model potential with volume absorption.

5. The study shows a clear preference for deep potentials $V_\tau \approx 150$ MeV ($\tau = {}^3$He) and $V_\alpha \approx 200$ MeV

6. Shapes of the experimental angular distributions can be well reproduced by using fixed sets of optical model parameters with the exception of $V_\alpha$, which should be adjusted around its discrete values. This procedure results also in extracting reliable *relative* ratios of the spectroscopic factors. However, to extract their *absolute* values it is necessary to keep the optical model parameters as closely as possible to the values determined from analysis of elastic data.

7. Spectroscopic factors determined in this study are in excellent agreement with calculations based on the assumption of rotational model for $^{25}$Mg with coupling between rotation and particle motions.

8. In general, when using ($^3$He,$\alpha$) reactions, only relative values of the spectroscopic factors are determined. Finding their absolute values is hindered by significant uncertainties about the zero-range coefficient $D_0^2$ for this reaction. Using theoretical values for the spectroscopic factors, I have found that $D_0^2 = 16.4 \times 10^4$ MeV$^2$·fm$^3$ for these reactions, which is in excellent agreement with the theoretical value of $17.0 \times 10^4$ MeV$^2$·fm$^3$ calculated using Irving-Gunn wave function. This factor can be used to extract absolute values of spectroscopic factors.

9. The single neutron pickup reaction ($^3$He,$\alpha$) can serve as a useful spectroscopic tool. However, my study resulted in formulating certain recommendations, which are summarized in the last section of this Chapter.

**Abstract:** Differential cross sections for the reaction $^{26}$Mg($^3$He,$\alpha$)$^{25}$Mg were measured using 10.2 MeV $^3$He particles. They were then analysed using distorted wave Born approximation (DWBA). The zero-range coefficient $D_0^2$ for ($^3$He,$\alpha$) reactions has been determined and has been found to be in good agreement with the theoretical value calculated using Irving-Gunn wave function. This coefficient can be used to extract absolute values of spectroscopic factors from ($^3$He,$\alpha$) angular distributions for other target nuclei. The derived spectroscopic factors have been compared with the spectroscopic factors calculated using rotational model for $^{25}$Mg with and without the coupling of the rotation and particle motions. Excellent agreement has been obtained between experimentally determined spectroscopic factors and model calculations. My results show that low-energy states in $^{25}$Mg can be described using rotational model with the coupling between the rotation and particle motion.

## Introduction

At the time of this study, my two PhD students were busy with their respective projects. The work described in this Chapter was an extra task I have undertaken to investigate the applicability of ($^3$He,$\alpha$) reactions in nuclear spectroscopy. To this end, it was necessary to understand the dependence of the extracted spectroscopic factors on the optical model parameters employed in fitting the reaction angular distributions. In particular, my aim was to see whether this reaction could be used to determine not only the relative but also absolute values of the spectroscopic factors.

I have chosen $^{26}$Mg because for this target nucleus the ($^3$He,$\alpha$) reaction leads to a well-known deformed nucleus $^{25}$Mg for which spectroscopic factors are known with a high level of confidence. This should allow for a reliable comparison of the *absolute* values of the experimental and theoretical spectroscopic factors, which in turn should help in a reliable determination of the notorious zero-range coefficient $D_0^2$, for which a wide range of values was previously used.

## Experimental arrangement and results

Measurements of ($^3$He,$\alpha$) angular distributions were carried out using 0.2$\mu$A $^3$He$^{++}$ beam from the Australian National University tandem accelerator. The experimental setup is shown in Figure 5.1. The beam from the accelerator entered a 51-cm reaction chamber through a system of four collimator slits and after passing the $^{26}$Mg target was stopped in a Faraday cup placed at the opposite end of the chamber. The target consisted of 103 ± 5 $\mu$g/cm$^2$ enriched magnesium containing 98.22% of $^{26}$Mg evaporated on approximately 10$\mu$g/cm$^2$ carbon backing.[7] The thickness of the target was determined by measuring Coulomb scattering of 4 and 5 MeV $\alpha$ particles at small angles and comparing them with the Rutherford scattering cross sections.

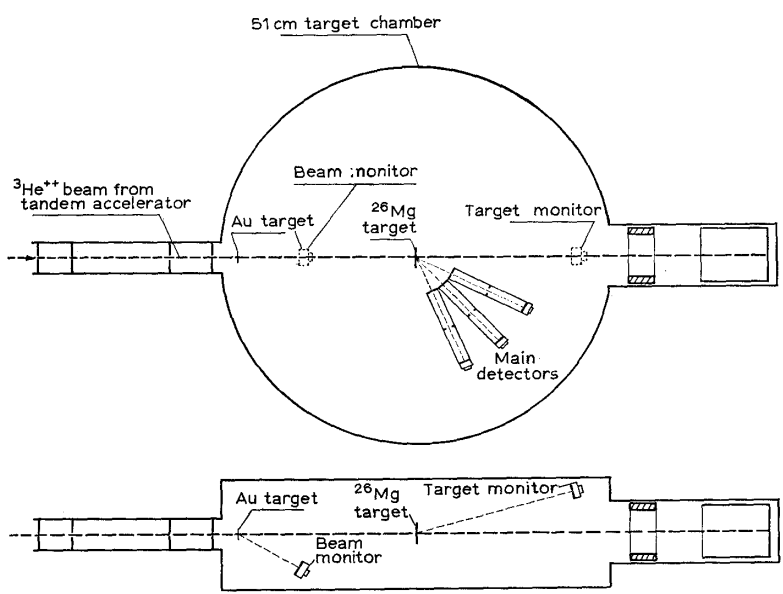

Figure 5.1. Schematic diagram of the experimental setup.

---

[7] The target was supplied by GTJ Arnison, Special Techniques Group, N69 AWRE, Aldermaston, Berks, England. In addition to $^{26}$Mg, it also contained 0.71% of $^{25}$Mg and 0.07% $^{24}$Mg. Other impurities were mainly Ca (0.5%) and Na (0.1%).

The intensity of the incident beam was measured using a standard, Elcor model A309A, current indicator and integrator connected to the Faraday cup, and it was also determined independently by detecting elastically scattered $^3$He particles from a thin Au foil located at the entrance to the target chamber. Variations in the effective target thickness were recorded using a target monitor, which registered $^3$He particles scattered elastically at a fixed angle of approximately $30^0$ from the $^{26}$Mg target.

Reaction products were detected using an array of ORTEC surface barrier semiconductor detectors. ORTEC Model 260 Time Pickoff and Model 262 Inspector units were employed to reduce pileup background for angles below $30^0$. Pulses from movable detectors were fed to the RIDL pulse height analyser and the particle spectra were punched on computer cards. The overall resolution for the particle spectra was around 100keV.

Angular distributions of $\alpha$ particles corresponding to the ground state and the 0.584 MeV ($J^\pi =$$^1$/$_2$$^+$), 0.976 MeV ($^3$/$_2$$^+$), 1.611 MeV ($^7$/$_2$$^+$), and 1.962 MeV ($^5$/$_2$$^+$) excited states in $^{25}$Mg were measured in the range of angles between $10^0$ and $165^0$ (lab). They are shown in Figure 5.4. The relative errors of the measured cross sections include the statistical and current integration inaccuracies. The errors are typically around ±3%. The uncertainties in the absolute values of the cross sections were estimated at ±6%.

**Theoretical analysis**

I have carried out theoretical analysis of the data using the optical model code JIB3 (Perey 1963) and the distorted wave Born approximation (DWBA) code DRC (Gibbs *at al.* 1964), both of which I have modified and adapted to run on the ANU IBM 360/50 mainframe computer. One of the modifications I have introduced to the DRC code was to allow for a six-parameter search of the optical model potential both in the input and output channels. I have also added plotting subroutines to allow for a quick and easy evaluation of theoretical results.

***Optical model potentials for the $^{26}$Mg($^3$He,$\alpha$)$^{25}$Mg reaction***

The parameters of the optical model potentials in the incident ($^3$He+$^{26}$Mg) channel were determined by measuring and analyzing $^3$He elastic scattering at incident energies around 10 MeV.

The energy of $\alpha$ particles produced in the investigated $^{26}$Mg($^3$He,$\alpha$)$^{25}$Mg reaction was between 22 MeV for the transition to the ground state and 19 MeV for the 1.962 MeV excited state. To determine the parameters for the $\alpha$ + $^{25}$Mg channel, I analysed the data of Budzanowski *et al.* (1964) for the elastic scattering of 24.7 MeV $\alpha$ particles from Mg and Al nuclei.

The optical potential had the following form:

$$U(r) = V_C(r) - Vf(x) - iWg(x')$$

where $V_C(r)$ is the Coulomb potential generated by a uniformly charged sphere,

$$f(x) = (1 + e^x)^{-1} \quad g(x') = (1 + e^{x'})^{-1}$$

$$x = \frac{r - r_0 A^{1/3}}{a} \qquad x' = \frac{r - r_0' A^{1/3}}{a'}$$





The resulting sets of parameters for $^3$He and $\alpha$ particles in the incoming and outgoing channels, respectively, are listed in Table 5.1.

Table 5.1

Optical model parameters for $^3$He and $\alpha$ particles [a]

| Particle | Potential | $V_i$ [b] (MeV) | $W_i$ (MeV) | $r_0$ (fm) | $a$ (fm) | $r'_0$ (fm) | $a'$ (fm) |
|---|---|---|---|---|---|---|---|
| $\tau$ | 1 | 110 | 10.5 | *1.08* [c] | *0.80* | *1.78* | *0.60* |
| $\tau$ | 2 | 151.8 | 14.6 | *1.08* | *0.80* | *1.78* | *0.60* |
| $\tau$ | 3 | 197.6 | 21.8 | *1.08* | *0.80* | *1.78* | *0.60* |
| $\alpha$ | 4 | 150 | *17.0* | 1.41 | *0.60* | 1.41 | *0.60* |
| $\alpha$ | 5 | 220 | *17.0* | 1.34 | *0.60* | 1.34 | *0.60* |
| $\alpha$ | 6 | 285 | *17.0* | 1.27 | *0.60* | 1.27 | *0.60* |

[a]) Symbol $\tau$ is used for $^3$He.

[b]) Subscript $i$ is $\tau$ for $^3$He or $\alpha$ for $\alpha$ particles.

[c]) Numbers in italics represent fixed parameters.

### The distorter wave analysis

I have carried out the distorted wave Born approximation (DWBA) analysis of the transfer reaction data assuming that the pickup neutron moved in the spherical Woods-Saxon potential, $V(r) = V_n f(r)$ with the radius $R_n = r_n A^{1/3}$. I have used $r_n = 1.2$ fm and $a_n = 0.65$ fm. The depth $V_n$ was adjusted to give the binding energy equal to the separation energy for the relevant states in $^{25}$Mg.

In the preliminary analysis, I have studied how the shapes and absolute values of $^{26}$Mg($^3$He,$\alpha$)$^{25}$Mg angular distributions depend on the lower cut-off radius and on the choice of optical model parameters. A sample of the results of the calculations is shown in Figure 5.2 for the 1.962 MeV excited state.

This preliminary analysis indicated that deep potentials (with $V_\tau \approx 150$ MeV and $V_\alpha \approx 200$ MeV for $^3$He and $\alpha$ particles, respectively) give nearly identical distributions with and without the lower cut-off radius, in both the shape and absolute values. They also resemble most closely the experimental distributions (see Figure 5.4).

When fitting transfer reaction cross sections, the parameters derived from elastic scattering are often adjusted to optimise the fits. To understand this process and to see how the adjusted parameters relate to the original parameters determined from analysis of elastic scattering data, I have calculated and examined the $\chi^2$ function:

$$\chi^2 = \frac{1}{N} \sum_{i=1}^{N} \left[ \frac{\sigma_{\exp}(\theta_i) - \sigma_{th}(\theta_i)}{\Delta \sigma_{\exp}(\theta_i)} \right]^2$$

where $N$ is the number of the experimental data points, $\sigma_{exp}(\theta_i)$ is the experimental differential cross section measured at the angle $\theta_i$, $\sigma_{th}(\theta_i)$ is the corresponding theoretical cross section, and $\Delta \sigma_{exp}(\theta_i)$ is the error in $\sigma_{exp}(\theta_i)$. The function is multidimensional in the space of the optical model parameters. An example of a two-dimensional $\chi^2$ map is shown in Figure 5.3.





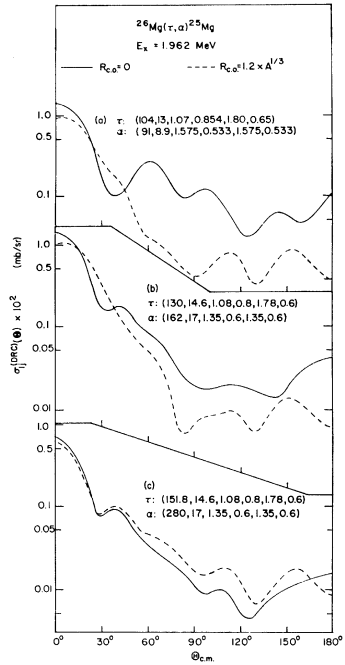

Figure 5.2. The dependence of the $^{26}$Mg($^3$He,$\alpha$)$^{25}$Mg angular distributions on the choice of optical model parameters in the input and output channels and on the lower cut-off radius $R_{c.o.}$. The distributions were calculated using three sets of optical potentials for $^3$He (marked here as $\tau$) and $\alpha$ particles. The parameters ($V$, $W$, $r_0$, $r_0'$, $a$ and $a'$) used in each case are shown in the figure. For the deep potentials, the shapes and the absolute values of the calculated cross sections do not depend strongly on the value of the lower cut-off radius. The calculated shapes resemble also more closely the experimental angular distribution (*cf* Figure 5.4).

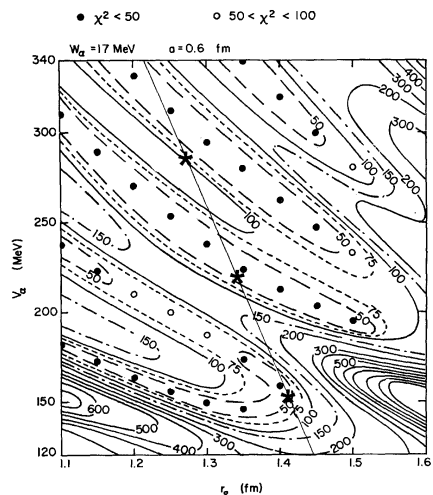

Figure 5.3. An example of a two-dimensional map of the $\chi^2$ function for the $^{26}$Mg($^3$He,$\alpha$)$^{25}$Mg reaction leading to the 0.976 MeV state in $^{25}$Mg. The map was constructed using the set 2 for $^3$He particles (see Table 5.1). Parameters optimising the fits to the reaction cross sections are shown as closed circles. The parameter sets 4-6 (Table 5.1) obtained by fitting elastic ($\alpha,\alpha$) scattering are shown as asterisks. The figure shows that the parameters optimising the fits to the reaction distributions are close to the parameters based on the analysis of elastic scattering. It also shows that the fits to the angular distributions can be optimised using a constant value for $r_0$ and adjusting only the potential depth $V_\alpha$. However, this procedure leads to a strong dependence of the extracted spectroscopic factors on $V_\alpha$.

This example shows that optical model parameters determined by fitting elastic scattering (indicated using the asterisks) are close to parameters that optimize the





fits to reaction differential cross sections (shown as full circles). Thus, the parameters determined using elastic scattering can be only adjusted slightly to optimize the fits to the reaction cross sections.

An alternative way of optimizing the fits to reaction cross sections is to keep geometrical parameters fixed (such as $r_0$ shown in this example of the $\chi^2$ map) and to adjust only the depth of the optical model potential. However, it will be shown later, that this procedure leads to an undesirably strong dependence of the absolute values of the differential cross sections on the choice of the optical model parameters. Thus, it is better to keep the parameters as closely as possible to the values determined by the analysis of elastic scattering.

An example of the DWBA calculations for the $^{26}$Mg($^3$He,$\alpha$)$^{25}$Mg reaction is shown in Figure 5.4. In these calculations, I used parameter set 2 (see Table 5.1) for $^3$He particles and set 5 and 6 for $\alpha$ particles but adjusting the depth $V_\alpha$ to optimise the fits. As expected, only small adjustments were necessary. The figure shows that the angular distribution for 1.611 MeV state cannot be described by using direct neutron pickup reaction.

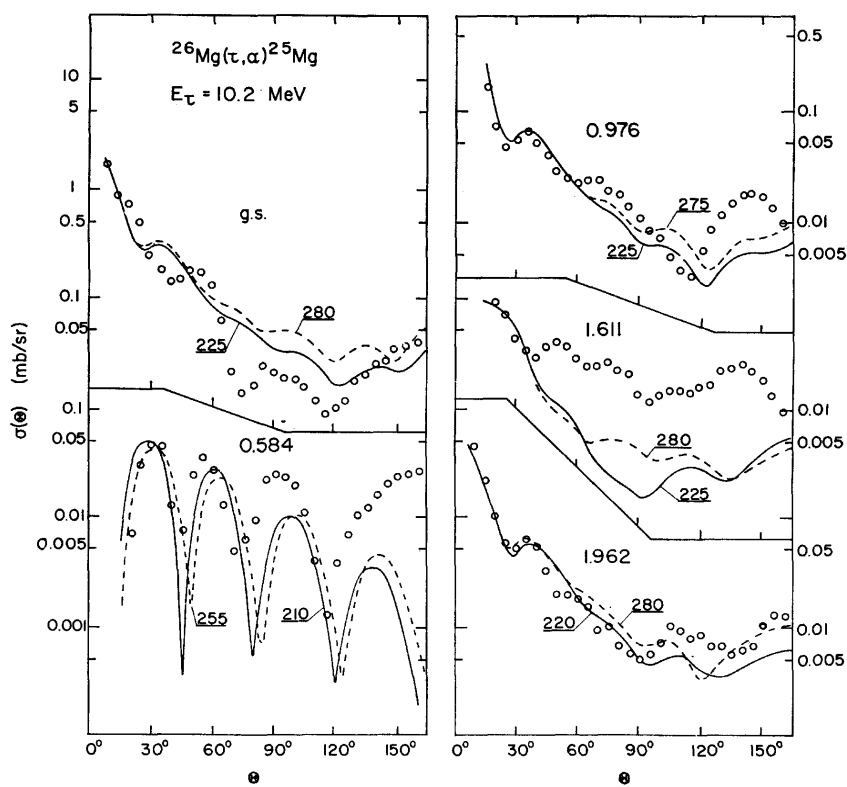

Figure 5.4. Angular distributions for the reaction $^{26}$Mg($^3$He,$\alpha$)$^{25}$Mg are compared with the distorted wave calculations. Parameter set 2 was used for $^3$He particles (see Table 5.1). For $\alpha$ particles, parameter sets 5 and 6 were used but the depth $V_\alpha$ was adjusted to optimise the fits. The adjusted values are shown for each calculated distribution.

As mentioned in the Introduction, the aim of the DWBA analysis was not only to reproduce the shapes of angular distributions and thus to determine the transferred angular momenta but also to extract absolute values of the spectroscopic factors and compare them with model calculations.

Theoretical angular distribution can be expressed as:





$$\sigma_{th}(\theta) = (2J+1)D_0^2 S(l,j)\sigma_{lj}(\theta)$$

where $J$ is the spin of the final state, and $S(l,j)$ is the spectroscopic factor, $l$ and $j$ are orbital momentum and spin of the transferred particle.

The angle-dependent part of the theoretical cross sections, $\sigma_{lj}(\theta)$, gives information about the reaction mechanism. It also gives spectroscopic information in the form of spins and parities of states in the final nucleus.

Spectroscopic factors $S(l,j)$ give additional information about nuclear structure. Experimental values of spectroscopic factors are extracted by comparing the *absolute* values of the experimental and theoretical cross sections at forward angles, that is, at angles where the reaction is dominated by direct transfer process.

It is clear that by comparing the absolute values of the calculated and experimental cross sections at forward angles, only the products $P = D_0^2 S(l,j)$ can be determined. These products can be used to calculate the *ratios* of the spectroscopic factors, $\kappa = S(l,j)/S_{g.s.}(l,j)$, where $S(l,j)$ is the spectroscopic factor for a given excited state and $S_{g.s.}(l,j)$ is the spectroscopic factor for the ground state.

The ratios of the spectroscopic factors give useful and often satisfactory information about nuclear structure but better information is obtained if absolute values of spectroscopic factors can be extracted.

### Relative values of spectroscopic factors

The derived here relative values of the spectroscopic factors are compared with results obtained by other authors in Table 5.2.

As shown in Figure 5.4, angular distribution for the 1.611 MeV state in $^{25}$Mg cannot be described well using direct reaction mechanism. Consequently, the experimental ratios of the spectroscopic factors for this state can be ignored.

Table 5.2

Ratios of the spectroscopic factors for the neutron pickup reactions from $^{26}$Mg

| $E_x$ $^{25}$Mg (MeV) | (p, d) [b] 40 MeV | (d, t) [c] 21.7 MeV | $(\tau, \alpha)$ [c] 33 MeV | $(\tau, \alpha)$ [d] 10.2 MeV | $\kappa_{th}^{(1)}$ [e] | $\kappa_{th}^{(2)}$ [e] |
|---|---|---|---|---|---|---|
| | | $\kappa_{exp}$ [a] | | | | |
| 0.584 | 0.068 | 0.073 | 0.067 | 0.066 | 0.086 | 0.057 |
| 0.976 | 0.044 | 0.049 | 0.042 | 0.111 | 0.148 | 0.117 |
| 1.611 | | <0.098 | 0.100 | 0.042 | 0.002 | 0.002 |
| 1.962 | 0.088 | 0.095 | 0.055 | 0.090 | 0.500 | 0.095 |

[a]) $\kappa_{exp}$ are the experimentally determined ratios of the spectroscopic factors $S(l,j)/S_{g.s.}(l,j)$.

[b]) Reynolds (1966)
[c]) Dehnhard and Yntema (1967)
[d]) My results for the $^{26}$Mg($^3$He,$\alpha$) $^{25}$Mg reaction.

[e]) $\kappa_{th}^{(1)}$ and $\kappa_{th}^{(2)}$ are the theoretical ratios of the spectroscopic factors calculated using a simple rotational model and rotational model with coupling between the rotation and particle motion (Kerman 1956 and Davidson 1965).





Table 5.2 shows a good agreement between the experimental ratios for the 0.584 MeV state. My result for the 1.962 MeV state agree with the results obtained using (p,d) and (d,t) reactions but disagrees with the result of Dehnhard and Yntema (1967) who used ($^3$He,$\alpha$) reaction at 33 MeV. A marked difference between my results and results obtained by other authors is for the 0.976 MeV state. However, all my results, including the 0.976 MeV state, are in excellent agreement with model calculations based on rotational model for $^{25}$Mg with coupling between the rotation and particle motion (Kerman 1956 and Davidson 1965).

### Normalization coefficient $P = D_0^2 S(l, j)$

It is always desirable, but not always possible, to determine the absolute values of spectroscopic factors. However, to find the absolute values one has to know the numerical value of the zero-range coefficient $D_0^2$. Unfortunately, there is a considerable uncertainty about this coefficient for ($^3$He,$\alpha$) reactions.

Normally, when fitting reaction angular distributions, optical model parameters derived from analysis of elastic scattering are used, with a possibly of only minor adjustments. To understand the problem with determining the $D_0^2$ coefficient I have studied the dependence of not only the shapes but also of the absolute values of the calculated differential cross sections on the optical model parameters. I have found that both the shapes of the calculated angular distributions can be reproduced well by varying only one parameter, the depth $V_\alpha$. However, to calculate reliably not only the shapes but also the absolute values of the differential cross sections the parameters should be adjusted as closely as possible around the values determined from analysis of elastic data.

Results of this preferable procedure are shown in Table 5.3, which lists the values of the normalization coefficient $P = D_0^2 S(l, j)$ and the ratios of the spectroscopic factors $\kappa = S(l, j) / S_{g.s.}(l, j)$. Optical model parameters for $^3$He particles are represented by set 2 in Table 5.1. The parameters for $\alpha$ particles are represented by sets 4, 5, and 6, respectively. However, the potential depth in sets 4 and 6 have been adjusted slightly to optimise the fits to the reaction angular distributions.

Table 5.3

The parameters $P = D_0^2 S(l, j)$ and $\kappa = S(l, j) / S_{g.s.}(l, j)$ for the $^{26}$Mg($^3$He,$\alpha$)$^{25}$Mg reaction

| $E$ (MeV) | Set 4 $V_\alpha$ = 160 MeV $r_0$ = 1.41 fm | | Set 5 $V_\alpha$ = 220 MeV $r_0$ = 1.34 fm | | Set 6 $V_\alpha$ = 300 MeV $r_0$ = 1.27 fm | | Average [a] | |
|---|---|---|---|---|---|---|---|---|
| | $P$ | $\kappa$ | $P$ | $\kappa$ | $P$ | $\kappa$ | $\overline{P}$ | $\overline{\kappa}$ |
| 0.000 | 36.57 | 1.000 | 41.56 | 1.000 | 34.91 | 1.000 | 37.68 | 1.000 |
| 0.584 | 1.91 | 0.052 | 2.93 | 0.070 | 2.69 | 0.077 | 2.51 | 0.066 |
| 0.976 | 4.36 | 0.119 | 4.57 | 0.110 | 3.62 | 0.104 | 4.18 | 0.111 |
| 1.611 | 1.63 | 0.045 | 1.68 | 0.040 | 1.41 | 0.040 | 1.57 | 0.042 |
| 1.962 | 3.19 | 0.087 | 3.91 | 0.094 | 3.16 | 0.090 | 3.42 | 0.090 |

[a]) The last two columns give the average values of $P$ and $\kappa$. The listed factors $P$ are in $10^4$ MeV$^2 \cdot$fm$^3$.





___

As mentioned earlier, experimental distributions could be fitted well using fixed parameters for all three sets but adjusting only the depth $V_\alpha$. This is a significant departure from the recommended procedure and while it reproduces the shapes of the distributions it leads to significant dependence of the calculated *absolute* values of the differential cross sections on the discrete values of $V_\alpha$. For instance, the normalization coefficient $P \equiv D_0^2 S(l, j)$ for the ground state is 29.9, 44.9, and 54.9 for $V_\alpha$ = 160, 220, and 270 MeV respectively. However, this type of the calculations does not alter significantly the relative values of the calculated cross sections.

Thus, the calculations show that as far as the ratios of the spectroscopic factors are concerned it does not matter which procedure is used. Referring to the example presented in Figure 5.3, it does not matter whether parameters are adjusted along the diagonal line and close to the values obtained from elastic scattering marked by asterisks or along a vertical line for a fixed value or $r_0$. However, if parameters are kept close to the parameters determined from elastic scattering, the values of $P = D_0^2 S(l, j)$ are significantly less dependent on the choice of optical model parameters, which means that one should be able to determine the absolute values of the spectroscopic factors with a better accuracy.

### Determining the $D_0^2$ coefficient

The advantage of using the $^{26}Mg(^3He,\alpha)^{25}Mg$ reaction is that model calculations for $^{25}Mg$ can be treated with high degree of confidence. The determined average value of the normalization coefficient $P$ shown in Table 5.3 can then be used to extract the experimental value of the zero-range coefficient $D_0^2$, which in turn can be applied to other ($^3He,\alpha$) reactions to extract not only the ratios but also the absolute values of the spectroscopic factors.

$$D_0^2 = \frac{P}{S_{th}(l, j)}$$

where $P$ is the experimentally determined normalization coefficient and $S_{th}(l, j)$ is the model-calculated spectroscopic factor.

Results of my calculations are compiled in Table 5.4. This table includes not only the calculations of $D_0^2$ based on my experimental results but also on the data of Dehnhard and Yntema (1967) at 33 MeV $^3He$ energy.

In the case of the 10.2 MeV data, the $D_0^2$ coefficients were calculated by using the experimentally determined normalization coefficient $P$ = 37.7×10$^4$ MeV$^2$·fm$^3$ and the theoretical spectroscopic factors for the ground state in $^{25}Mg$. This procedure gives then the experimental spectroscopic factors for the excited states, which are compared with the theoretical values.

I have carried out calculations using model-calculated spectroscopic factors $S_{th}^{(1)}(l, j)$ based on a simple rotational model, and $S_{th}^{(2)}(l, j)$ based on a model with coupling between the rotation and particle motion (Kerman 1956 and Davidson 1965).





Table 5.4

Comparison of the experimental spectroscopic factors $S(l, j)$ with model calculations and determination of the zero-range $D_0^2$ coefficients

| $E_t$ $^{25}$Mg (MeV) | $l$ | $S_{exp}(l, j)$ | | $S_{th}^{(1)}(l, j)$ a) | $S_{exp}(l, j)$ | | $S_{th}^{(2)}(l, j)$ a) |
|---|---|---|---|---|---|---|---|
| | | $E_t = 33$ MeV b) $D_0^2 = 23.6$ | $E_t = 10.2$ MeV c) $D_0^2 = 26.9$ | | $E_t = 33$ MeV b) $D_0^2 = 14.4$ | $E_t = 10.2$ MeV c) $D_0^2 = 16.4$ | |
| 0.0 | 2 | 1.4 | 1.4 | 1.4 | 2.3 | 2.3 | 2.3 |
| 0.584 | 0 | 0.093 | 0.093 | 0.12 | 0.15 | 0.15 | 0.13 |
| 0.976 | 2 | 0.059 | 0.16 | 0.20 | 0.098 | 0.26 | 0.27 |
| 1.611 | (4) | 0.140 | 0.058 | 0.0024 | 0.230 | 0.096 | 0.0044 |
| 1.962 | 2 | 0.076 | 0.13 | 0.7 | 0.125 | 0.21 | 0.22 |

a) $S_{th}^{(1)}(l, j)$ and $S_{th}^{(2)}(l, j)$ are the theoretical spectroscopic factors calculated using a simple rotational model and a model with coupling between the rotation and particle motion (Kerman 1956 and Davidson 1965), respectively.

b) The $D_0^2$ coefficients and the corresponding spectroscopic factors $S_{exp}(l, j)$ calculated using the experimental data of Dehnhard and Yntema (1967) and the relevant theoretical spectroscopic factors for the ground state.

c) The $D_0^2$ coefficients and the corresponding spectroscopic factors $S_{exp}(l, j)$ calculated using my experimental data and the relevant theoretical spectroscopic factors for the ground state. The listed $D_0^2$ values are in units of $10^4$ MeV²·fm³. Excellent agreement is between my experimental results and the theoretical values of $S_{th}^{(2)}(l, j)$ and thus the experimentally determined $D_0^2 = 16.4 \times 10^4$ MeV²·fm³.

As can be seen by comparing theoretical and experimental values of the spectroscopic factors my results are in excellent agreement with the rotational model for $^{25}$Mg with coupling between rotation and particle motion. The experimentally derived $D_0^2 = 16.4 \times 10^4$ MeV²·fm³ is in good agreement with the theoretical value of $17 \times 10^4$ MeV²·fm³ calculated using Irvin-Gunn wave functions (Bassel and Drisko, 1967).

## Conclusions

I have measured angular distributions for the $^{26}$Mg($^3$He,$\alpha$)$^{25}$Mg reaction leading to the ground state and first four excited states in $^{25}$Mg. The aim was to study the spectroscopic applicability of ($^3$He,$\alpha$) reactions. I have selected $^{26}$Mg as a target nucleus because theoretical values of the spectroscopic factors for states in $^{25}$Mg are relatively well known and thus they can be used to determine the elusive $D_0^2$ factor for ($^3$He,$\alpha$) reactions.

Experimental data were analysed using distorted wave theory. With the exception of the transition to the 1.611 MeV state in $^{25}$Mg, all measured angular distributions can be described well by direct reaction mechanism over a wide range of angles in the forward direction. Conclusions of this study can be summarised as follows:

1. Spectroscopic factors extracted in the present study are in excellent agreement with theoretical values calculated using rotational model for low-energy states in $^{25}$Mg with coupling between the rotation and particle motions (Kerman 1956 and Davidson 1965).





2. The extracted zero-range coefficient for ($^3$He,$\alpha$) reactions, $D_0^2$ = 16.4×10⁴ MeV²·fm³, is in good agreement with the theoretical value of 17×10⁴ MeV²·fm³ calculated using Irvin-Gunn wave functions (Bassel and Drisko, 1967).

3. The single neutron pickup reaction ($^3$He,$\alpha$) can serve as a useful spectroscopic tool. However, my study suggests the following recommendations.

    a. Deep optical model potentials with $V_\tau$ >100 MeV for $^3$He particles ($\tau$ = $^3$He) and $V_\alpha$ >150 MeV for $\alpha$ particles should be used. Deep potentials give better fits to the angular distributions. They also allow to minimise the dependence of the shape of the distributions and the absolute values of the calculated cross sections on the choice of the lower cut-off radius $R_{c.o.}$. In fact, the lower cut-off radius is unnecessary if deep potentials are used. In addition, deep potentials allow for a better estimation of spectroscopic factors.

    b. The shapes of the calculated angular distributions do not depend strongly on the optical model parameters for $^3$He particles. These parameters can be kept fixed at the values determined from elastic scattering. Parameters in the exit channel can be adjusted to optimise the fits but they should be kept as close as possible to the values determined by fitting elastic scattering angular distributions.

    c. The accuracy of the extracted absolute values of the spectroscopic factors is improved if analysis is carried out for various sets of the optical model potential and the resulting spectroscopic factors are averaged.

    d. To extract the absolute values of the spectroscopic factors it is recommended to use the experimentally determined normalization coefficient $D_0^2$ = 16.4×10⁴ MeV²·fm³. Alternatively, the theoretical value $D_0^2$ = 17×10⁴ MeV²·fm³ can be used.

## References


Bassel, R. H. and Drisko, R. M. 1967, Proc. Symp. On Direct Nuclear Reactions with $^3$He, September, Tokyo (I.P.C.R. Cyclotron Report, Supplement 1) p. 13.

Budzanowski, A., Grotowski, K., Micek, S., Niewodniczanski, H., Sliz, J., Strzalkowski, A., and Wojciechowski, H. 1964, INP Report No. 347; Phys. Lett. **11**:74.

Davidson, J. P. 1965, *Rev. Mod. Phys.* **37**:105.

Dehnhard, D. and Yntema, J. L. 1967, Phys. Rev. **160**:964.

Gibbs, W. R. Madsen V. A., Miller, J. A., Tobocman, W., Cox, E. C. and Mowry, L. 1964, *NASA TN D-2170.*

Glendenning, N. K. 2004, *Direct Nuclear Reactions*, World Scientific Publishing Co. Pte. Ltd., Singapore.

Hamburger, E. W. and Blair, A. G. 1960, Phys. Rev. **119**:777.

Kerman, A. K. 1956, *Mat. Fys. Medd. Dan. Vid. Selsk.* **30**, No. 15

Nilsson, S. G. 1955, *Mat. Fys. Medd. Dan. Vid. Selsk.* **29**, No 16.

Perey, F. G. 1963, *Phys. Rev.* **131**:745.

Reynolds, G. M. 1966, Ph.D. thesis, Univ. of Minnesota, Minneapolis.






Satchler, G. R. 1958, Ann. of Phys. **3**:275.
Satchler, G. R. 1983, *Direct Nuclear Reactions*, Oxford University Press, Oxford.
Tobocman, W. 1961, *Theory of Direct Nuclear Reactions*, Oxford University Press, Oxford.





<div style="text-align:center">

**6**

</div>

# The Discrete Radius Ambiguity of the Optical Model Potential

***Key features:***

1. This study revealed a hitherto unknown discrete radius ambiguity. It explained why some analyses resulted in an unusually large radius for the imaginary part of the optical model potential.

2. I have carried out theoretical analysis of elastic scattering of $^3$He particles from $^{16}$On nuclei using optical model formalism and a computer code, which I have modified and adapted to run on an ANU mainframe computer.


***Abstract:*** Angular distributions for the elastic scattering of $^3$He on $^{16}$O measured at six incident energies between 9.802 and 10.720 MeV were averaged to yield a distribution at 10.25 MeV. The averaged distribution was then analysed using optical model potential. Calculations were carried out using surface or volume absorption, four or six parameters for the central part of the optical potential, and with or without spin-orbit interaction. Detailed investigation of the $\chi^2$ function has led to a discovery of a hitherto unknown discrete radius ambiguity for the real part of the central components of the optical model.


## Introduction

Previous analyses (Yntema, Zeidman, and Bassel 1964; Bray and Nurzynski 1965; Bray, Nurzynski, and Bourke 1968) of elastic scattering of $^3$He particles demonstrated that low-energy data can be fitted satisfactorily using six-parameter, volume absorption potential with standard geometry. However, a study of $^{16}$O($^3$He,$^3$He)$^{16}$O elastic scattering indicated that this geometry is probably inapplicable in the case of the target nuclei with atomic mass A ≤ 16. Some other calculations (Erskine, *et al.* 1965; Fortune et al. 1968; Kellogg and Zurmuhle 1966; Weller, Robertson and Tilley 1968) for carbon and oxygen targets suggest that an unusually large radius $r_0'$ >2 has to be used to fit the experimental data. I have, therefore, decided to look more closely at the $^3$He + $^{16}$O scattering to understand why different sets of geometrical parameters are apparently needed for such light nuclei.

## Experiment

Doubly charged $^3$He particles accelerated to required energies in the Australian National University tandem accelerator were used to bombard an oxygen gas target. Angular distributions for the elastic scattering of $^3$He on $^{16}$O were measured at bombarding energies 9.802, 9.955, 10.210, 10.313, 10.465, 10.720, 11.229, and 11.738 MeV. The quoted energies were calculated at the centre of the oxygen gas cell. The measurements were carried out in 5º steps between 15º and 165º (lab) [8]. Results of measurements are presented in Figure 6.1. The absolute values of the differential cross sections are accurate to about ±2.7%. The various component errors are shown in Table 6.1. The relative errors of the experimental points vary between 1% and 5%.

---

[8] *lab* stand for the *laboratory system*.





Table 6.1
Components of the absolute error

| Quantity | Error (%) |
|----------|-----------|
| beam heating | ±1.5 |
| gas pressure | ±0.53 |
| geometrical factor | ±1.0 |
| current integrator | ±2.0 |

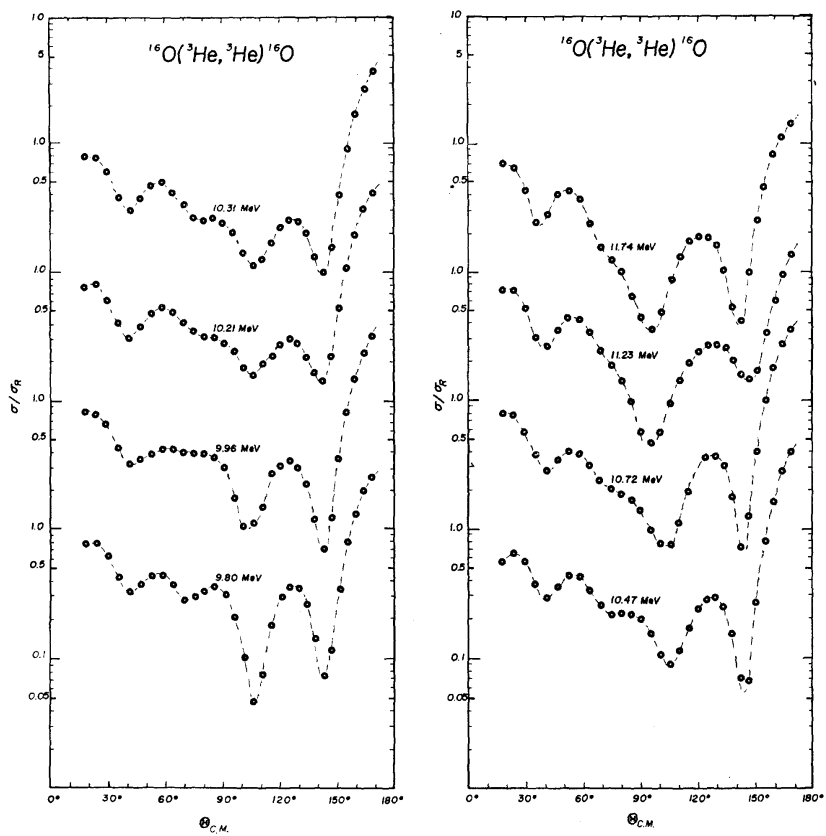

Figure 6.1. The experimental angular distributions for the elastic scattering of $^3$He particles on $^{16}$O. The lines have been drawn to guide the eye.

## Discussion of the experimental results

At the first glance, the overall shape of the elastic angular distributions appears to be similar at all energies. The most striking features are the large backward angle increase in the cross section and a large maximum at 125° (c.m.)[9], which are common to all distributions. The third maximum at about 85° appears somewhat anomalous in that its size fluctuates with energy. At 9.955 MeV, it comes close to the second maximum but is separated at four higher energies. However, at 11.229 and 11.738 MeV the second maximum has almost disappeared.

Figure 6.2 shows the integrated cross sections as a function of energy. The integration over each angular distribution has been performed between 15° and 165° (lab). The experimental points follow roughly the $E^{-2}$ dependence of the Rutherford scattering but they also show some marked deviations from the calculated curve.

---

[9] *c.m.* stand for the *centre of mass* system





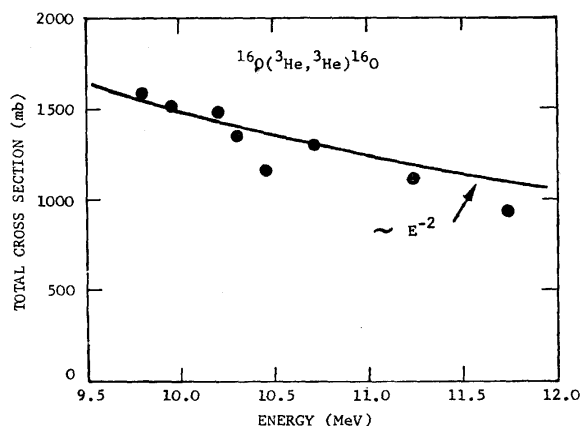

Figure 6.2. The integrated (total) cross sections for the elastic scattering of $^3$He particles from $^{16}$O nuclei as a function of incident energy. The full line represents the $E^{-2}$ dependence.

## Optical model analysis

### *Optical potential*

The optical model potential employed in this analysis had the following form:

$$U(r) = -Vf(r) - iW_k g_k(r) + V_{S.O.} f_{S.O.}(r)\mathbf{s}\cdot\mathbf{l} + V_C(r)$$

where

$$f(r) = \left\{1 + \exp\left[(r - r_0 A^{1/3})/a\right]\right\}^{-1}$$

Index $k$ is $S$ for the volume (Saxon-Woods) absorption or $D$ for the surface (derivative) absorption.

$$g_S(r) = \left\{1 + \exp\left[(r - r_0' A^{1/3})/a'\right]\right\}^{-1}$$

$$g_D(r) = 4a'\left|\frac{d}{dr}g_S(r)\right|$$

$$f_{S.O.}(r) = \left(\frac{\hbar}{Mc}\right)^2 \frac{2}{r}\left|\frac{d}{dr}\left\{1 + \exp\left[(r - r_{0s} A^{1/3})/a_s\right]\right\}^{-1}\right|$$

$$V_C(r) = \frac{Ze^2}{r_C A^{1/3}}\left(3 - \frac{r^2}{r_C^2 A^{1/3}}\right) \text{ for } r \leq r_C A^{1/3}$$

$$V_C(r) = \frac{2Z^2}{r} \text{ for } r > r_C A^{1/3}$$

In my calculations, I used $r_C = 1.4$ fm.

## Analysis

The experimental angular distribution used in the analysis was obtained by averaging measurements at six incident energies between 9.801 and 10.720 MeV. This is a close cluster of experimental distributions and they show similar angular pattern.





The computation was carried out using an automatic search code (Perey 1963), which I have modified and adapted to run on the Australian National University IBM-360/50 computer.

I have carried out extensive calculations using surface or volume absorption, central potential with four or six parameters, and by including or excluding the spin orbit interaction. The parameters were determined by minimising the function

$$\chi^2 = \frac{1}{N} \sum_{i=1}^{N} \left[ \frac{\sigma_{\exp}(\theta_i) - \sigma_{th}(\theta_i)}{\Delta \sigma_{\exp}(\theta_i)} \right]^2$$

where $N$ is the number of the experimental data points, $\sigma_{exp}(\theta_i)$ is the experimental differential cross section measured at the angle $\theta_i$, $\sigma_{th}(\theta_i)$ is the corresponding theoretical cross section, and $\Delta\sigma_{exp}(\theta_i)$ is the error in $\sigma_{exp}(\theta_i)$.

### Results of the optical model analysis

A few examples of fits to the experimental angular distribution are shown in Figure 6.3. The figure has been prepared by copying directly from the computer plots prepared using my plotting routine, which I have added to the optical model code. The outputs show that the fits do not depend strongly on the number of optical model parameters and on whether volume or surface absorption is used, at least for calculations at forward angles.

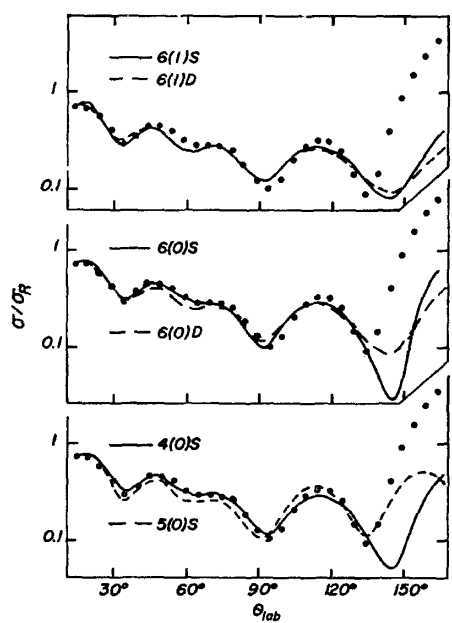

Figure 6.3. Examples of the optical model calculations for the elastic scattering of [3]He particles from [16]O at 10.25 MeV incident energy. The parameters were optimised around $V$ = 150 MeV. In the signature $n_1(n_2)k$, $n_1$ is the number of optical model parameters (both fixed and variable), $n_2$ the number of fixed parameters, and $k$ is $S$ for the volume absorption potential or $D$ for the surface absorption.

### The discrete radius ambiguity

To understand the role of the optical model parameters I have carried out a detailed mapping of the $\chi^2$ function. This function is multidimensional in the space of optical model parameters and as such cannot be presented in a graphic form for all parameters simultaneously. However, the function can be easily studied by plotting





two-dimensional projections. The plots can be further simplified by showing only the points corresponding to the minimum values of the $\chi^2$ function. An example of such plots is presented in Figure 6.4. The figure shows clearly the presence of the hitherto unknown discrete radius ambiguity.

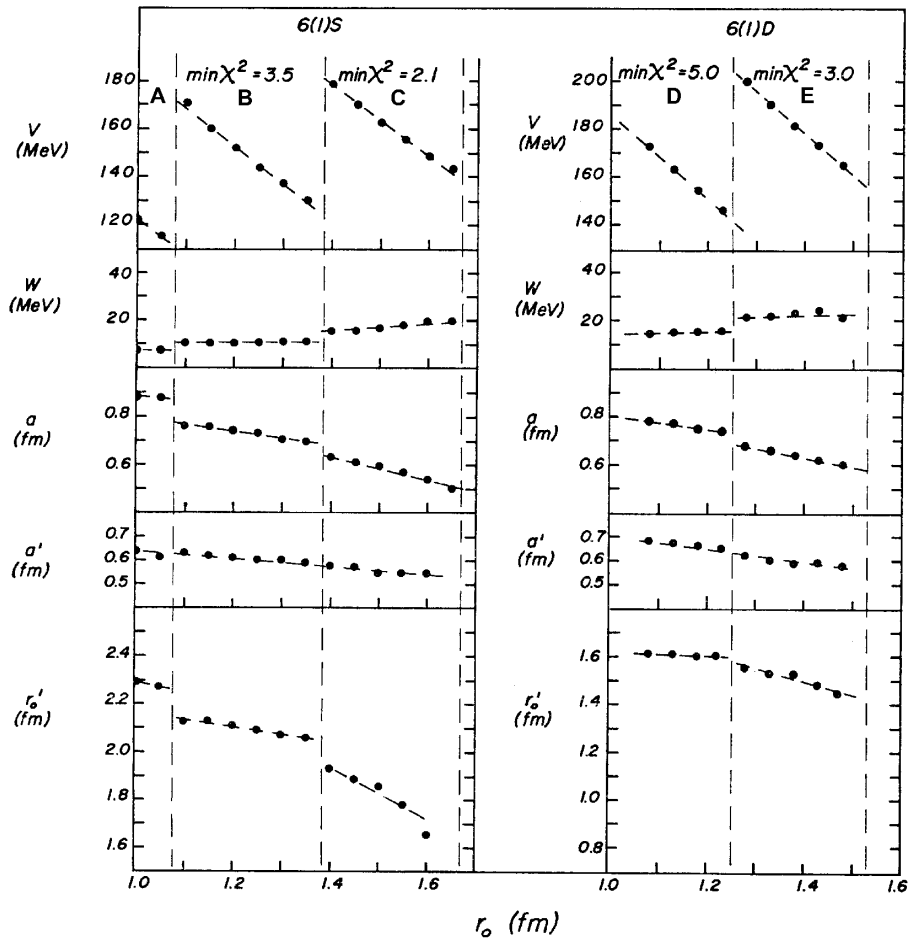

Figure 6.4. An example of discrete radius $r_0$ ambiguities. These results show that it is possible to obtain practically equivalent fits to the elastic scattering data (compare the minimum values of the $\chi^2$ function) using discrete values of the radius $r_0$ parameter for the real part of the central component of the optical model potential.

The figure presents various families of parameter sets that optimise the fits to the elastic scattering of $^3$He from $^{16}$O nuclei. The families are identified using letters *A*, *B*, *C*, *D*, and *E*.

For instance, in the family *B*, any set of parameters selected for a given $r_0$ along the lines indicated by the full circles will fit the data well and will correspond to $\chi^2 = 3.5$. Parameters in the family *C* will produce nearly identical fits (the minimum of $\chi^2$ function is only slightly lower) but their sets of parameters correspond to an entirely different range of $r_0$ parameters. Thus, for instance, if *V* = 150 MeV one can fit the data using any of the two discrete values of $r_0$, either 1.21 fm or 1.60 fm for the volume absorption potential. For the surface absorption, potential one can have either $r_0$ = 1.15 or 1.23 fm.





This study explains also why some searches have led to an unusually large radius $r_0'$ for the imaginary component of the central part of the optical model potential. Figure 6.4 shows that this large value belongs to the family *B*. For the family *C,* the parameter has lower values $r_0'$. Depending on the initial values of the optical model, the automatic search will lead to one of the two equivalent minima in $\chi^2$ function.

## Summary and conclusions

This work describes a detailed study of the optical model for the elastic scattering of $^3$He particles from $^{16}$O nuclei at 10.25 MeV. The analysis of experimental angular distribution was carried out using optical model with six or four parameters for the central part, volume or surface absorption, and with or without spin-orbit interaction. Excellent fits were obtained over a large range of angles of up to about $135^0$ in the centre of mass system.

I have found that angular distributions were insensitive to spin-orbit interactions. Likewise, experimental distributions were reproduced well using either surface or volume absorption. Finally, the shapes of the calculated distributions did not depend strongly on whether six or four parameters were used for the central part of the optical model potential.

Detailed analysis of the $\chi^2$ functions has revealed a hitherto unknown and therefore unexpected discrete radius ambiguity for the real component of the central part of the optical potential. Discrete ambiguities in the depth *V* of the real component of the central part of the optical potential are well known. However, the discrete radius ambiguities have not been previously reported. The demonstrated here discrete radius ambiguity explains why in certain analyses automatic searches lead to unusually large value of the imaginary component of the central part of the optical model potential.

## References


Bray, K. H. and Nurzynski, J. 1965, *Nucl. Phys.* **67**:417.

Bray, K. H., Nurzynski, J. and Bourke, W. P. 1968, *Nucl. Phys.* **A114**:309.

Erskine, J. R., Holland, R. E., Lawson, R. D., MacFarlane, M. H. and Schiffer, J. P. 1965, *Phys. Rev. Lett.* **14**:915.

Fortune, H. T., Gray, T. J., Trost, W. and Fketcher, N. R. 1968, *Phys. Rev.* **173**:1002.

Kellogg, E. M. and Zurmuhle, R. W. 1966, *Phys. Rev.* **152**:890.

Perey, F. G. 1963, *Phys. Rev.* **131**:745.

Weller, H. R., Robertson, N. R. and Tilley, D. R. 1968, *Nucl. Phys.* **A122**:529.

Yntema, J. L., Zeidman, B. and Bassel, R. H. 1964, *Phys. Lett.* **11**:302.






---

**7**

# The $^{27}$Al($^3$He,$\alpha$)$^{26}$Al Reaction at 10 MeV $^3$He Energy

***Key features:***

1. We have carried out the measurements of angular distributions for the $^{27}$Al($^3$He,$\alpha$)$^{26}$Al reaction, and my early version of a computer code for unfolding Gaussian distributions was successfully applied to analyse the particle spectra.

2. Experimental data were analysed using distorted wave Born approximation theory.

3. Spectroscopic factors extracted in this study show that the low-energy states in $^{26}$Al can be described using a simple $jj$ coupling scheme that consists of $(1d_{5/2})^{-2}$ configuration only. Such a configuration gives states with spins ranging from $0^+$ to $5^+$.

4. All distributions were fitted using $l = 2$ angular momentum. Contrary to the results for the (p,d) and (d,t) reactions, no evidence for $l = 0$ was obtained, which is in agreement with the conclusion of Blair and Wagner (1962) that when more than one $l$ value is allowed, the ($^3$He,$\alpha$) reactions show preference for higher values.

***Abstract:*** Angular for $^{27}$Al($^3$He,$\alpha$)$^{26}$Al reaction have been measured at 10 MeV incident $^3$He energy. They were analysed using distorted wave Born approximation theory. Spectroscopic factors extracted in this study show that the low-energy states can be described using a simple $jj$ coupling of two nucleons in $1d_{5/2}$ configuration.

## Introduction

The 1d-2s shell nuclei have been the subject of considerable experimental and theoretical interest, and for many of them the strong coupling Nilsson model (Nilsson 1955) has been successfully applied. The model works particularly well at the beginning of this shell. However, it has been pointed out (Gove 1960; Bar-Touv and Kelson 1965) that the deformation changes abruptly from prolate to oblate in the vicinity of the mass number $A = 27$, and consequently, as discussed in Chapter 3, the strong coupling description seems to be inadequate in this region. Indeed, as shown earlier, the low-energy states of the $^{27}$Al nucleus can be described well assuming the excited-core model. The target and the residual nucleus involved in the reaction discussed in this study are both in this interesting region of the periodic table.

Angular distributions of $\alpha$ particles from $^{27}$Al($^3$He,$\alpha$)$^{26}$Al reaction have been measured previously (Taylor *et al.* 1960) at an incident energy of 5.2 MeV. States below the 2.07/2.08 MeV doublet were studied and the data were found to exhibit $l = 2$ angular momentum transfers only. The excitation of 1.76 MeV state was interpreted as a direct transition in disagreement with the results discussed here and with the results for the (p,d) reaction (Bevington and Anderson 1966).

Measurements for the analogous neutron pickup reaction (d,t) induced by 19 MeV deuterons have been also reported (Vlasov *et al.* 1960). These authors measured angular distributions between about $5^0$ and $30^0$ only and analysed them using a simple plane-wave theory. The group corresponding to the third excited state exhibited an $l = 0$ angular momentum transfer. On the basis of the extracted reduced widths for this state, the authors concluded that the s-wave admixture in the ground state of $^{27}$Al must be small. Their conclusion was later confirmed by the (p,d) data (Bevington and Anderson 1966).





## Experimental details

The experimental setup was similar to that used in the $^{26}$Mg($^3$He,$\alpha$)$^{25}$Mg measurements (see Chapter 5). Target thickness was 170 ± 6 μg/cm$^2$. Angular distributions for the reaction $^{27}$Al($^3$He, $\alpha$)$^{26}$Al were measured at 10 MeV $^3$He energy between 5$^0$ and 165$^0$ (lab) in steps of 5$^0$. Figure 7.1 shows a typical $^{27}$Al($^3$He,$\alpha$)$^{26}$Al spectrum.

As can be seen, peaks corresponding to the first and second excited states were only partly resolved. To extract differential cross sections for these states I have written a program for unfolding Gaussian distributions. The program was optimising the fits to particle spectra by searching for both the widths and positions of the peaks around the positions calculated using nuclear kinematics. Spectral analysis was carried out using an IBM 1620 computer.

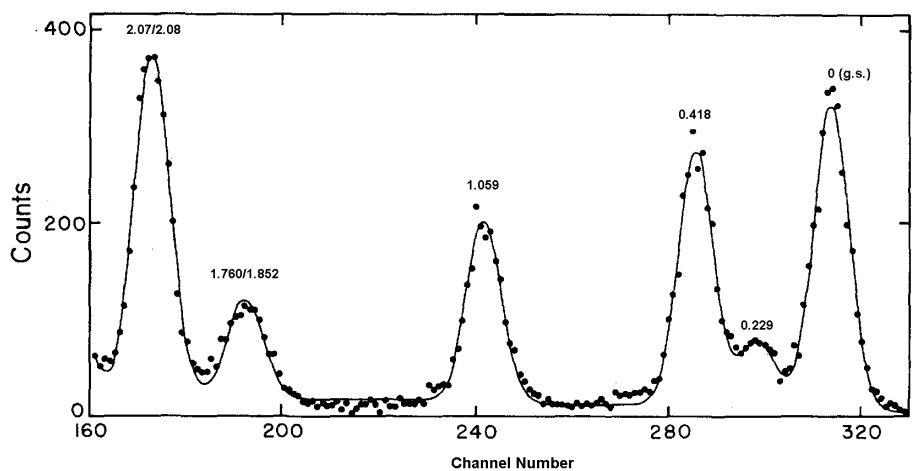

Figure 7.1. Typical spectrum of $\alpha$ particles taken at $\theta = 60^0$ (lab) for the reaction $^{27}$Al($^3$He, $\alpha$)$^{26}$Mg at 10 MeV of $^3$He energy. The continuous line was calculated using my program for unfolding Gaussian distributions and an IBM 1620 computer.

Angular distributions of $\alpha$ particles were measured for the ground state and for 0.229, 0.418, 1.059, 1.760/1.852, and 2.07/2.08 MeV states in $^{26}$Al. Apart from a few small angles, the yield for the 1.852 MeV state was not sufficiently high to extract angular distribution. Angular distributions are shown in Figure 7.3. The relative errors were generally smaller than the size of the displayed data points. The absolute error was mainly due to uncertainties in the measurements of the target thickness and was estimated at about ±4%.

## Theoretical analysis

Angular distributions shown in Figure 7.3 were analysed using the DRC code (Gibbs *at al.* 1964) and the ANU IBM 360/50 computer.

Optical model parameters in the incident, $^3$He + $^{27}$Al, channel were obtained from an earlier work (Bray, Nurzynski, and Satchler 1965) of $^3$He scattering at energies between 5.5 and 10 MeV. A fit to the data at 10 MeV $^3$He energy is shown in Figure 7.2. The corresponding optical model parameters are listed in Table 7.1.

A simple optical model potential was used containing only the central part:

$$U(r) = V_C(r) - Vf(r) - iWg(r)$$





where $V_C(r)$ is the potential due to a uniformly charged sphere of radius $R_C = rcA^{1/3}$, $V$ and $W$ are the real and imaginary well depths, and the form factors $f(r)$ and $g(r)$ are of the Woods-Saxon type:

$$f(r) = \left\{1 + \exp\left[(r - r_0 A^{1/3}) / a\right]\right\}$$

$$g(r) = \left\{1 + \exp\left[(r - r_0' A^{1/3}) / a'\right]\right\}$$

The parameters for the exit, $\alpha$ + $^{26}$Al, channel were suggested by the analysis of elastic scattering by McFadden and Satchler (1966).

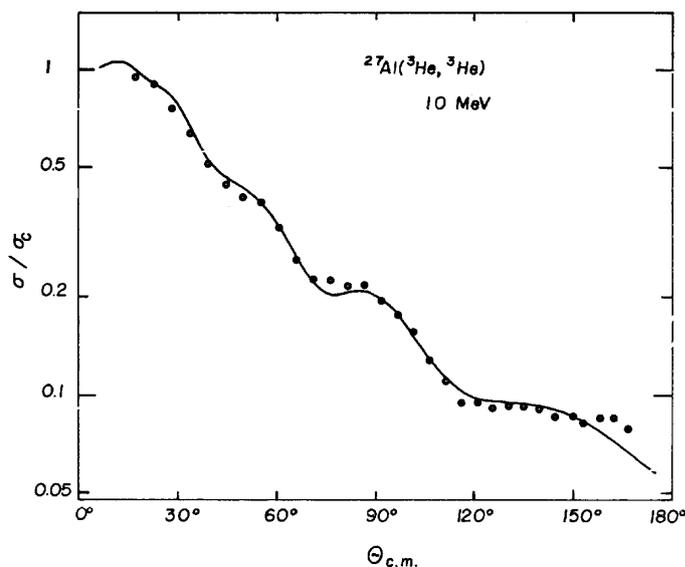

Figure 7.2. The experimental (points) and calculated (line) angular distributions for the elastic scattering of 10 MeV $^3$He particles from $^{27}$Al. The optical model parameters are listed in Table 7.1.

Table 7.1

Optical model parameters for $^3$He and $\alpha$ particles used in the distorted waves analysis of the $^{27}$Al($^3$He,$\alpha$)$^{26}$Al angular distributions

|  | $V$ (MeV) | $W$ (MeV) | $r_0$ (fm) | $a$ (fm) | $r_0'$ (fm) | $a'$ (fm) | $r_C$ (fm) |
|---|---|---|---|---|---|---|---|
| $^3$He potential | 155 | 15 | 1.08 | 0.80 | 1.78 | 0.60 | 1.40 or 1.08 |
| $\alpha$ potential | 190 | 17 | 1.35 | 0.60 | 1.35 | 0.60 | 1.35 |

Theoretical calculations for $^{27}$Al($^3$He,$\alpha$)$^{26}$Al angular distributions are compared with experimental data in Figure 7.3. All the theoretical distributions were fitted assuming a direct pickup of a neutron with the orbital angular momentum $l$ = 2 moving in a Woods-Saxon potential, whose depth was adjusted to give the correct binding energy. The calculations were carried out either with a lower cut-off radius of 1.7 fm or with full integration. As can be seen in Figure 7.3, calculations without a cut-off radius produce good enough fits to the angular distributions to allow for extracting the spectroscopic factors.

The possible $l$ = 0 transition to the 0.418 MeV state was also tried but was found to be in gross disagreement with the observations. The calculated $l$ = 0 and $l$ = 2 were also mixed in various proportions but no improvement to the $l$ = 2 fit was obtained.





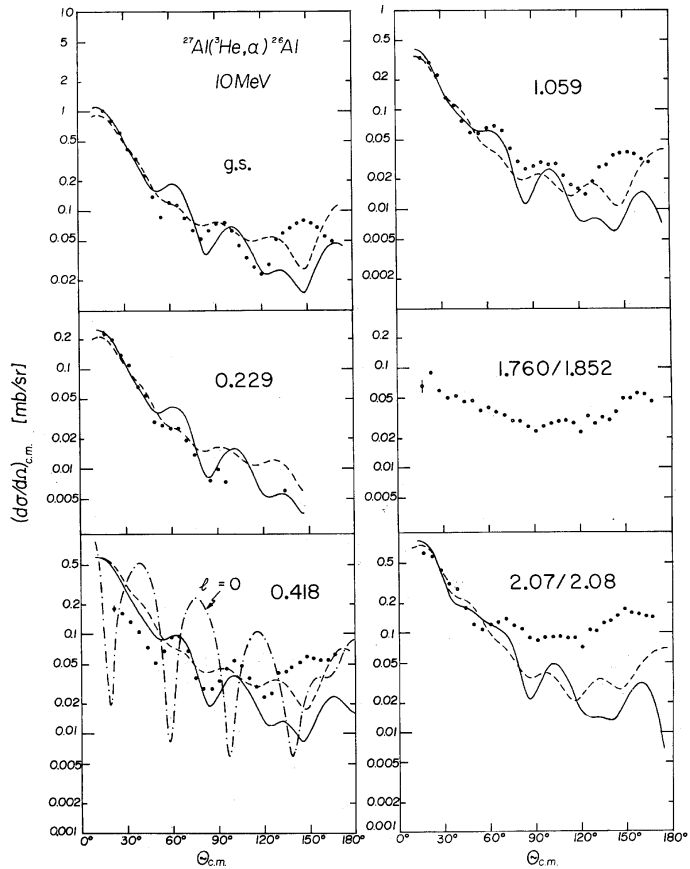

Figure 7.3. Angular distributions of $\alpha$ particles from the $^{27}$Al($^3$He,$\alpha$)$^{26}$Al reaction are compared with theoretical calculations. The curves were calculated using the distorted waves theory of direct nuclear reactions. The full curves correspond to a lower cut-off radius of 1.7 fm. Other curves were calculated without a cut-off radius.

## Discussion

The available information about the spins and parities of the low-energy states in $^{26}$Al are summarised in Table 7.2. The energies quoted are those of Endt and van der Leun (1962). All the spins and parity assignments are from the work of Horvat *et al.* (1963).

Two alternative descriptions of the low-energy states in $^{26}$Al have been proposed. In one of them, the strong coupling Nilsson model was applied (Kelson 1964; Picard and de Pinho 1966; Pyatov 1963; Varshalovich and Peker 1961) whereas the other used the shell model (Bouten, Elliott, and Pullen 1967; Brennan and Bernstein 1960; Ferguson 1964). However, the strong coupling model does not seem to be adequate for $^{27}$Al (Ophel and Lawergren 1964.) In fact, the low-lying states in $^{27}$Al are described remarkably well using core-excited model (see Chapter 3).

Assuming the *jj* coupling for the odd neutron and proton in $^{26}$Al, the most likely relationship is that they are both in the $1d_{5/2}$ configuration coupling to the total spin $J$ = $0^+$, $1^+$, $2^+$, $3^+$, $4^+$, or $5^+$. There is also a possibility that any member of the unpaired neutron and proton is excited to $2s_{1/2}$ or $1d_{3/2}$ orbits and that the resulting wave functions can mix with the $(1d_{5/2})^{-2}$ configuration giving the total spin $J$ = $0^+$, $1^+$, $2^+$, $3^+$, or $4^+$. It should be expected, however, that the nucleus in the ground state, coupled to $J^\pi$ = $5^+$ is in the pure $(1d_{5/2})^{-2}$ configuration.





Table 7.2

Available information on the low-energy states in the $^{26}$Al nucleus

| Energy (MeV) | $J\pi$ | $T$ | ($^3$He, $\alpha$) [a] | | (p, d) [e] | | (d, t) [d] | | ($^3$He, $\alpha$) [f] | | $(S/S_{g.s.})_{th}$ |
|---|---|---|---|---|---|---|---|---|---|---|---|
| | | | $l$ | $\dfrac{\sigma}{\sigma_{g.s.}}$ [b] | $l$ | $\dfrac{S}{S_{g.s.}}$ | $l$ | $\dfrac{S}{S_{g.s.}}$ [e] | $l$ | $\dfrac{S}{S_{g.s.}}$ | |
| 0 | 5⁺ | 0 | 2 | 1 | 2 | 1 | 2 | 1 | 2 | 1 | 1 |
| 0.229 | 0⁺ | 1 | 2 | 0.22 | 2 | 0.15 | | (0.15) | 2 | 0.23 | 0.09 |
| 0.418 | 3⁺ | 0 | | | 0 | 0.16 | 0 | 0.12 | 2 | 0.61 | 0.64 |
| 1.059 | 1⁺ | 0 | 2 | 0.50 | 2 | 0.30 | 2 | 0.29 | 2 | 0.30 | 0.27 |
| 1.760 | 2⁺ | 0 | 2 | 0.57 | | | | | | | |
| 1.852 | (3⁺) | 0 | | | 0 | 0.03 | | | | | |
| 2.07 | 3⁺ | 0 | } 2 | } 2.19 | } 0 | } 0.20 | } (2) | } 0.37 | } 2 | } 0.59 | 0 |
| 2.08 | 2⁺ | 1 | | | | | | | | | 0.46 |

[a]) Taylor *at al.* 1960
[b]) The ratios of the differential cross-sections at the first maximum
[c]) Bevington and Anderson 1966
[d]) Vlasov, *et al.* 1960
[e]) Ratios of spectroscopic factors calculated from the reduced widths assuming that the ratio $S/S_{g.s.}$ for the state 0.229 MeV state is the same as that measured in the (p,d) reaction listed in column 7.
[f]) Our data for 10 MeV $^3$He projectiles. The last column contains theoretical spectroscopic factors calculated using a formula of Macfarlane and French (1960).

In the simplest case, therefore, when the configuration mixing is neglected, the spectroscopic factors for the single-nucleon transfer reaction can be calculated readily using an explicit formula of Macfarlane and French (1960). The results for the transition from $(1d_{5/2})^{-1}$ to $(1d_{5/2})^{-2}$ between $^{27}$Al and $^{26}$Al are shown in the last column of Table 7.2. It has been assumed that the total strength of the $(1d_{5/2})^{-2}$ wave function for $J = 3$ is in the second excited state.

For all states, except the 0.229 MeV state, the agreement with experiment is exceptionally good. The 0.229 MeV state has a relative spectroscopic factor about twice as large as calculated using the $jj$ coupling. The data from other experiments included in the table show similar results. The close agreement between experiment and the simple theory in the case of the 0.418 MeV and the 2.07/2.08 MeV states suggests that the assumption that the 0.418 MeV state contains nearly all of the $J^\pi = 3^+$, $(1d_{5/2})^{-2}$ configuration is substantially correct.

The major difference between our data and the (p,d) and (d,t) experimental results is that the reported $l = 0$ transitions for the 0.418 MeV and the 2.07/2.08 MeV states observed in these reactions were not confirmed by the neutron pickup induced by $^3$He particles. This differences may be a further evidence of the effect, first observed by Blair and Wanger (1962), that when two values of the angular momentum transfer are allowed, the ($^3$He,$\alpha$) reactions tend to proceed via the higher value.

## Conclusions

Apart from the transitions to the 1.760/1.852 MeV levels, the present $^{27}$Al($^3$He,$\alpha$)$^{26}$Al experimental angular distributions show forward peaking characteristic of a direct





pick-up reaction and consequently they have been analysed using the distorted-wave direct transfer theory. Using optical-model parameters, which describe the related elastic scattering results, and $l = 2$ neutron transfer was found to be most consistent with the data in all cases. At large angles it is likely that other reaction mechanisms are contributing to the angular distributions since the cross sections are comparable to those for the 1.760 MeV state, which does not display direct transfer features.

The relative spectroscopic factors calculated from a simple $jj$ coupling scheme in which $^{27}$Al and $^{26}$Al are considered to consist of $(1d_{5/2})^{-1}$ and $(1d_{5/2})^{-2}$ configurations only, are in good agreement with those determined from the distorted-wave analysis.

The predominance of $l = 2$ transitions over $l = 0$ transitions when both are possible and when the latter have been reported for the analogous (p,d) and (d,t) reactions may be further evidence for the preference of $(^3He,\alpha)$ reactions to proceed via the higher $l$ value.

## References


Bar-Touv, J and Kelson, I. 1965, *Phys. Rev.* **139**:B1035.

Bevington, P. R. and Anderson, A. S. 1966, *Bull. Am. Phys. Soc.* **11**:908.

Blair, A. G. and Wanger, H. E. 1962, *Phys. Rev.* **127**:1233.

Bouten, M. C., Elliott, J. P. and Pullen, J. A. 1967, *Nucl. Phys.* **A97**:113.

Bray, K. H., Nurzynski, J. and Satchler, G. R. 1965, *Nucl. Phys.* **67**:417.

Brennan, M. H. and Bernstein, A. M. 1960, *Phys. Rev.* **120**:927.

Ferguson, J. M. 1964, *Nucl. Phys.* **59**:97.

Gibbs, W. R. Madsen V. A., Miller, J. A., Tobocman, W., Cox, E. C. and Mowry, L. 1964, *NASA TN D-2170*.

Gove, H. E. 1960, Proc. Int. Conf. on Nuclear Structure, ed. By D. A. Bromley and E. W. Voght, University of Toronto Press, Toronto, p. 438.

Horvat, P., Kump, P. and Povh, B. 1963, *Nucl. Phys.* **45**:341.

Endt, C. M. and van der Laun, *Nucl. Phys.* **34**:1.

Kelson, I. 1964, *Phys. Rev.* **134**:B267.

Macfarlane, M. H. and French, J. B. 1960, *Rev. Mod. Phys.* **32**:417.

McFadden, L. and Satchler, G. R. 1965, *Nucl. Phys.* **84**:177.

Niewodniczanski, H., Nurzynski, J., and Strzalkowski, A. 1963, *Le Journal de Physique,* **24**:944; also in IFJ Report No. 263,1963.

Nilsson, S. G. 1955, *Mat. Fys. Medd. Dan. Vid. Selsk.* **29**, No 16.

Ophel, T. R. and Lawergren, B. T. 1964, *Nucl. Phys.* **52**:417.

Picard, J. and de Pinho, A. G. 1966, *Nuovo Cim.* **41B**:239.

Pyatov, N. I. 1963, Izv. Akad. Nauk SSSR (ser. Fiz.) **27**:1436.

Taylor, I. J., de Barros, S. Forsyth, P. D., Jaffe, A. A. and Ramavataram, S. 1960, *Proc. Phys. Soc.* **75**:772.

Varshalovich, D. A. and Peker, R. K. 1961, *Izv. Akad. Nauk SSSR* (ser. Fiz.) **25**:287.

Vlasov, N. A., Kalinin, S. P., Ogloblin, A. A. and Chuev, V. I. 1960, *JETP (Sov. Phys.)* **10**:844.








# Configuration Mixing in the Ground State of $^{28}$Si

**Key features:**

1.  Angular distributions for the elastic scattering of $^{3}$He particles from $^{28}$Si and for the pickup reaction $^{28}$Si($^{3}$He,$\alpha$)$^{27}$Si were measured and analysed using optical model and distorted wave theory.

2.  Results of this study show that neutron orbits, $1d_{5/2}$, $2s_{1/2}$, and $1d_{3/2}$ in the ground state of $^{28}$Si are filled to approximately 60%, 30%, and 40%, respectively.

**Abstract:** Angular distributions for the elastic scattering and for the $^{28}$Si($^{3}$He,$\alpha$)$^{27}$Si were measured using 12 MeV $^{3}$He particles. The data were analysed using optical model and distorted wave theory. Spectroscopic factors for low-energy states were extracted. Results of our study are compared with previous investigations of neutron pickup reactions from $^{28}$Si. The extracted spectroscopic factors were used to calculate configuration mixing in the ground state of $^{28}$Se.

## Introduction

In the absence of residual interaction, $^{28}$Si nucleus should have the closed $1d_{5/2}$ subshell for protons and neutrons. However, transitions involved in neutron (Jones, Johnson and Griffiths 1968; Kozub 1968; Swenson, Zurmühle and Fou 1967; Wildenthal and Glaudemans 1967) and proton (Gove *at al.* 1968; Wildenthal and Newman 1968) pickup reactions suggest significant admixture of $2s_{1/2}$ and $1d_{3/2}$ configurations. This should be expected when a residual interaction is included.

The relative contributions of the three configurations can be determined by studying spectroscopic factors. Measurements for neutron pickup have been carried out using (p,d) and ($^{3}$He,$\alpha$) reactions (Jones, Johnson and Griffiths 1968; Kozub 1968; Swenson, Zurmühle and Fou 1967; Wildenthal and Glaudemans 1967) but there are some disagreements between determined spectroscopic factors for low-energy states in $^{27}$Si. The aim of this study was to carry out new measurements for the $^{28}$Si($^{3}$He,$\alpha$)$^{27}$Si reaction, use the new and existing data, and try to determine the configuration mixing in the ground state of $^{28}$Si.

## Experiment

The beam of 12 MeV $^{3}$He$^{++}$ ions was provided by the Australian National University tandem accelerator. Experimental equipment and technique were similar to those described for $^{26}$Mg($^{3}$He,$\alpha$)$^{25}$Mg reaction (Chapter 5).

The silicon targets were produced by vacuum evaporation of natural silicon powder using a VARIAN *e*-gun.[10] Initially, an attempt was made to produce self-supporting targets. These, however, were found to have too short a lifetime under beam bombardment to be of practical use, and consequently a target, which was backed by a thin carbon foil, was used to record the data reported here. This target also contained impurities of both oxygen and tantalum. The tantalum was due to the use of a tantalum dish as a lining inside the *e*-gun crucible in order to facilitate the silicon evaporation. Due to these impurities and the carbon backing the angular range of the

---

[10] Manufactured by Vacuum Products Division, Varian Associates, Palo Alto, California, USA.





data for both the elastic scattering and the ($^3$He,$\alpha$) measurements was restricted mainly to forward angles.

Figure 8.1 shows two examples of particle spectra. Contaminant groups are shown by their chemical symbols. Results of measurements for the elastic scattering of $^3$He particles from $^{28}$Si at 12 MeV are shown in Figure 8.2. Figure 8.3 shows the distributions for the $^{28}$Si($^3$He,$\alpha$)$^{27}$Si reaction.

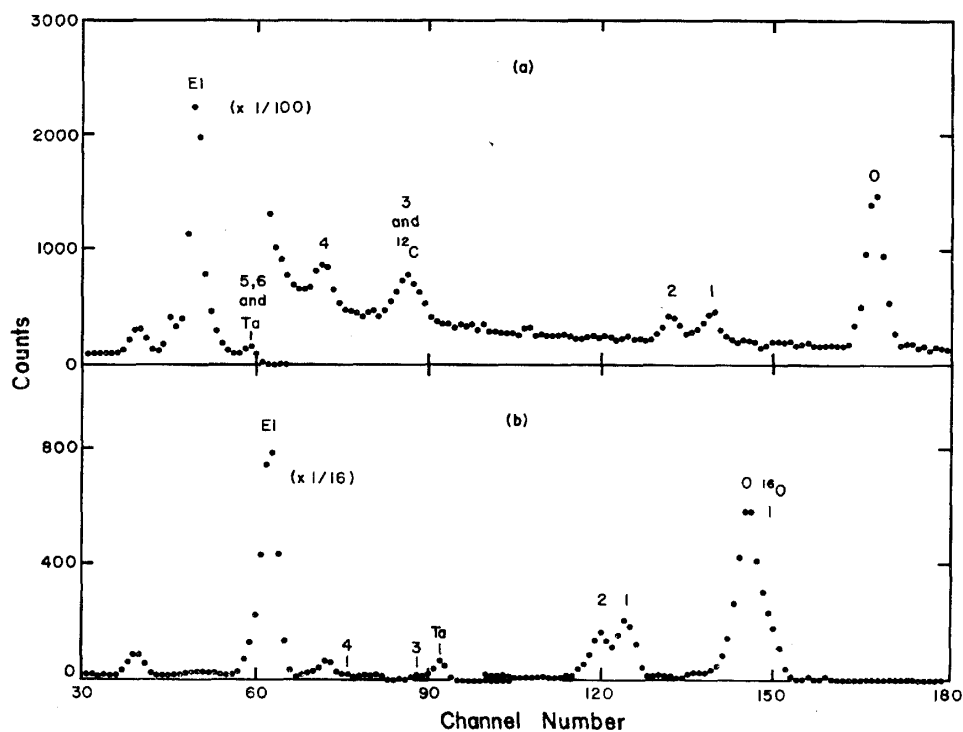

Figure 8.1. Spectra obtained with 12 MeV $^3$He particles incident on a carbon-backed, natural silicon target at (a) 30°(lab) and (b) 60°(lab). The groups labelled 0, 1, 2, etc., correspond to the ground state and the excited states of $^{27}$Si from the reaction $^{28}$Si($^3$He,$\alpha$)$^{27}$Si. Those labelled "$^{16}$O" and "$^{12}$C" were produced by the reactions $^{16}$O($^3$He,$\alpha_0$)$^{15}$O and $^{12}$C($^3$He,$\alpha_0$)$^{12}$C respectively. The group marked "Ta" is the elastic peak from Ta($^3$He,$^3$He)Ta and that marked "El" is from the Si + $^3$He elastic scattering.

## Theoretical analysis

The elastic scattering angular distributions was analysed using a six-parameter Woods-Saxon optical model potential with volume absorption and the computer code (Perey 1963). Samples of the best fits using different sets of parameters are presented in Figure 8.2 and the corresponding optical model parameters are listed in Table 8.1.

Potentials of Table 8.1 were used in the subsequent DWBA analysis of the $^{28}$Si($^3$He,$\alpha$)$^{27}$Si angular distributions. The parameters for the exit channel were the same as in the analysis of the $^{27}$Al($^3$He,$\alpha$)$^{26}$Al reaction and were based on the elastic scattering of McFadden and Satchler (1966). The neutron bound states wave functions were calculated for a real Woods-Saxon potential of radius 1.2A$^{1/3}$ fm and diffuseness of 0.65 fm. The depths were adjusted to give the separation energy for the transferred neutrons. The DWBA calculations were carried out using the DRC computer code (Gibbs *et al.* 1964) as described in Chapter 5.

The distribution corresponding to the 2.17 MeV, $J^\pi = {}^7/_2{}^+$ state was measured but was not analysed because it does not exhibit characteristics of a direct pickup





transition. Direct pickup transition to this state would require a $1f_{7/2}$ admixture in the ground state wave function of $^{28}$Si. Such an admixture is not only unlikely but also evidently absent.

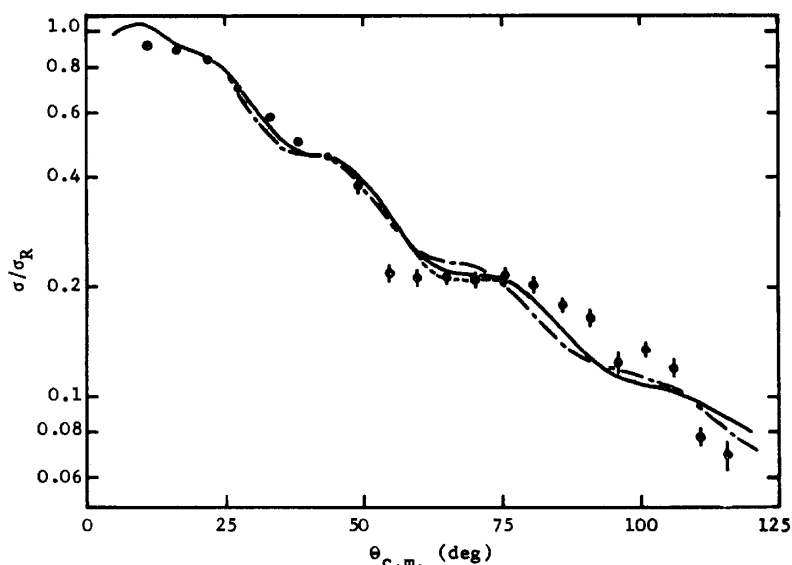

Figure 8.2. Differential cross sections for the elastic scattering of $^3$He from $^{28}$Si at 12 MeV $^3$He energy. The curves show optical model calculations. The full curve is for the set 1 (see Table 8.1). The dashed and the dash-dot curves are for sets 2 and 3, respectively.

Table 8.1

Optical model parameters for the elastic scattering of 12 MeV $^3$He particles from $^{28}$Si

| Parameter set no | $V$ (MeV) | $W$ (MeV) | $r_0$ (fm) | $a$ (fm) | $r_0'$ (fm) | $a'$ (fm) | $\chi^2$ |
|---|---|---|---|---|---|---|---|
| 1 | 108.0 | 16.3 | 1.07 | 0.854 | 1.81 | 0.650 | 10.1 |
| 2 | 145.3 | 23.3 | 1.07 | 0.854 | 1.81 | 0.650 | 9.4 |
| 3 | 155.9 | 18.3 | 1.08 | 0.800 | 1.78 | 0.600 | 10.2 |

The results of the calculation are compared with experimental data in Figure 8.3. The three $l = 2$ angular distributions did not show a preference for any of the three sets of the $^3$He parameters and only the set 1 predictions for $l = 2$ are shown in Figure 8.3. The $l = 0$ transfer for the first excited state is fitted best by set 2 of the optical model potential. All curves in Figure 8.3 were calculated with a zero cut-off radius. DWBA calculations with a lower cut-off radius of 3.5 fm were also tried but did not produce any significant difference in the calculated distributions.

## Discussion

The observed direct pickup transitions to the low-lying states of $^{27}$Si and the relatively large absolute cross sections indicate a significant admixture of $1d_{5/2}$, $2s_{1/2}$, and $1d_{3/2}$ orbits in the ground state wave function of $^{28}$Si.

The contributions of $1d_{5/2}$, $2s_{1/2}$, and $1d_{3/2}$ orbits in the ground state wave function of $^{28}$Si can be estimated by calculating neutron occupation numbers:





$$V_j^2 = \frac{\sum_i S_{lj}^{(i)}}{2j+1},$$

where $S_{lj}^{(i)}$ is the spectroscopic factor of the *i*th state corresponding to the *lj* configuration. The sum is over all states corresponding to the *lj* configuration.

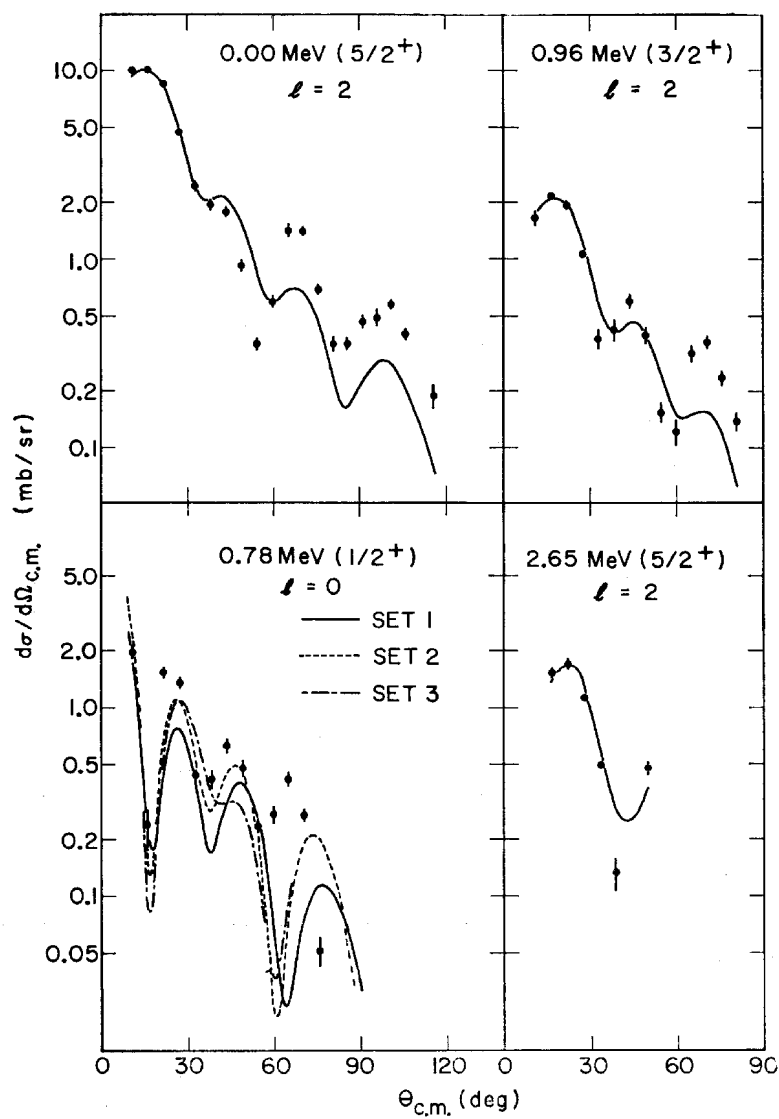

Figure 8.3. The distorted wave calculations are compared with the measured differential cross sections for the reaction $^{28}$Si($^3$He,$\alpha$)$^{27}$Al at 12 MeV incident $^3$He energy. The full curves were calculated using set 1 of the optical model parameters for $^3$He particles (see Table 8.1). Calculations using sets 2 and 3 are also shown for the state 0.78 MeV. All calculations were done using a zero cut-off radius.

Spectroscopic factors found in the present experiment as well as those from other studies of neutron pickup reactions from $^{28}$Si are compiled in Table 8.2. Comparing them with other ($^3$He,$\alpha$) measurements taken at two adjacent energies it may be seen that the 12 MeV results agree better with the 15 MeV data than with those obtained at 10 MeV.





The last column in Table 8.2 contains the average values of the spectroscopic factors. They add up to 5.83, which is close to the expected number of 6.

Table 8.2

Spectroscopic factors for neutron pickup reactions from $^{28}$Si

| $E_x$ (MeV) | $J\pi$ | $l$ | ($^3$He, $\alpha$) [a] 10 MeV | ($^3$He, $\alpha$) [b] 12 MeV | ($^3$He, $\alpha$) [c] 15 MeV | (p, d) [d] 27.6 MeV | (p, d) [e] 33.6 MeV | $S_{av}$ |
|---|---|---|---|---|---|---|---|---|
| 0 | $\frac{5}{2}^+$ | 2 | 1.84 | 2.64 | 2.71 | 2.62 | 2.93 | 2.55 |
| 0.774 | $\frac{1}{2}^+$ | 0 | 0.65 | 0.29 | 0.38 | 0.80 | 0.54 | 0.53 |
| 0.952 | $\frac{3}{2}^+$ | 2 | 0.46 | 0.59 | 0.34 | 0.45 | 0.29 | 0.43 |
| 2.647 | $\frac{5}{2}^+$ | 2 | 1.20 | 0.63 | 0.72 | 0.28 | 0.40 | 0.65 |
| 2.90 [f] | $(\frac{3}{2})^+$ | (2) | | | (1.30) | (1.00) | (0.69) | (1.00) |
| 4.275 | $(\frac{3}{2}, \frac{5}{2})^+$ | 2 | | | | | 0.29 | 0.29 |
| 6.324 | $(\frac{3}{2}, \frac{5}{2})^+$ | 2 | | | | | 0.38 | 0.38 |

[a]) Wildenthal and Glaudemans 1967
[b]) Our results
[c]) Swenson, Zurmühle and Fou 1967
[d]) Jones, Johnson and Griffiths 1968
[e]) Kozub 1968
[f]) Unresolved doublet

Table 8.3

Configuration parameters for the ground state of $^{28}$Si

| Orbit | $n'_{lj}$ | $n_{lj}$ | $V_j^2$ |
|---|---|---|---|
| 1d$_{5/2}$ | 6.00 | 3.20-3.87 | 53-65% |
| 2s$_{1/2}$ | 0.00 | 0.53 | 27% |
| 1d$_{3/2}$ | 0.00 | 1.43-2.10 | 36-53% |

$n'_{lj}$ – the number of neutrons in a given configuration assuming no residual interaction.

$n_{lj}$ – the experimentally determined number of neutrons in a given configuration.

$V_j^2$ – the experimentally determined occupation numbers for a given configuration.

The experimentally determined number of neutrons $n_{lj}$ in a given configuration and the occupation numbers $V_j^2$ are listed is Table 8.3. The range of values for $l$ = 2 configurations are due to uncertainties in spin assignments for the 4.275 MeV and 6.324 MeV states.

Without residual interaction, the orbit 1d$_{5/2}$ would contain 6 neutrons and thus would be 100% full. Our results show that residual interaction causes about 40% of neutrons outside the closed $N$ = 8 shell to occupy orbits 2s$_{1/2}$ and 1d$_{3/2}$.

## Summary and conclusions

Angular distributions for the $^{28}$Si($^3$He,$\alpha$)$^{27}$Si reaction have been measured at 12 MeV incident $^3$He energy. The distributions were analysed using direct DWBA theory. Spectroscopic factors were extracted for transitions to the low-energy states in $^{27}$Si.





Results were compared with other available data for neutron pickup reactions from $^{28}$Si and average values of the spectroscopic factors have been calculated. Using them, configuration components for the ground state wave function of $^{28}$Si have been determined.

Without residual interaction, the six neutrons outside the 1p shell in the ground state of $^{28}$Si would occupy the lowest $1d_{5/2}$ orbit. However, neutron pickup reactions show that the ground state wave function contains a mixture of three configurations, $1d_{5/2}$, $2s_{1/2}$, and $1d_{3/2}$. The determined here neutron occupation numbers show that the $1d_{5/2}$ orbit is only 53-65% full. Neutrons, which are outside the 1p shell occupy also other two orbits in the 1d-2s shell. They fill in 27% of the $2s_{1/2}$ orbit and 36-53% of the $1d_{3/2}$ orbit.

## References


Gibbs, W. R. Madsen V. A., Miller, J. A., Tobocman, W., Cox, E. C. and Mowry, L. 1964, *NASA TN D-2170*.

Gove, H. E., Purser, K. H., Schwartz, J. J., Alford, W. P. and Cline, D. 1968, *Nucl. Phys*. **A116:**369

Jones, G. D., Johnson, R. R. and Griffiths, R. J. 1968, *Nucl. Phys*. **A107**:659.

Kozub, R. L., 1968, *Phys. Rev*. **172**:1078.

McFadden, L. and Satchler, G. R. 1966, *Nucl. Phys*. **84**:177

Perey, F. G. 1963, *Phys. Rev*. **131**:745.

Swenson, L. W., Zurmühle, R. W. and Fou, C. M. 1967, *Nucl. Phys*. **A90**:232.

Wildenthal, B. H. and Glaudemans, P. W. M. 1967, *Nucl. Phys*. **A92**:353.

Wildenthal, B. H. and Newman, E. 1968, *Phys. Rev*. **167**:1027






# 9

# The *j* - dependence for the $^{54}Cr(d,p)^{55}Cr$ Reaction

***Key features:***

1. The aim of this work was to study the *j* - dependence using the reaction $^{54}Cr(d,p)^{55}Cr$.

2. The low-energy states in $^{55}Cr$ belong to $2p_{3/2}$ and $2p_{1/2}$ configurations and we have observed both $j = {}^{3}/_{2}$ and $j = {}^{1}/_{2}$ transitions, which indicated considerable configuration mixing.

3. We have found that *j* - dependence varies with the excitation energy and is more pronounced for low excited states.

4. We have explained the *j* - dependence as being due to the real component of the spin-orbit interaction. The imaginary component has no effect on the *j* -dependence.

5. Using the observed *j* - dependence we have assigned spin values to the ground state and low-energy excited states in $^{55}Cr$.

6. We have extracted spectroscopic factors and calculated occupation probabilities of the orbits $2p_{3/2}$ and $2p_{1/2}$. We have found that the two neutrons outside the closed shell in the ground state of $^{54}Cr$ do not occupy entirely the lower $2p_{3/2}$ orbit, as it would have been expected on the basis of the independent particle shell model, but spend a considerable time also in the $2p_{1/2}$ orbit. On average, there are 1.48 neutrons in the $2p_{3/2}$ orbit and 0.52 in $2p_{1/2}$. These orbits are 37% and 26% full, respectively.

7. We have calculated the centre-of-gravity single particle energies for the configurations $2p_{3/2}$ and $2p_{1/2}$ and found that the spacing between these two energies agrees well with the spacing for the neighbouring nuclei.

***Abstract:*** Differential cross sections for six transitions in the reaction $^{54}Cr(d,p)^{55}Cr$ have been measured at a deuteron energy of 8 MeV. The data have been described qualitatively by the distorted wave Born approximation calculations employing spin-orbit terms in the deuteron and proton optical model potentials to reproduce the observed dependence on the spin transfer, *j*. We have found that the observed *j* - dependence arises from the deuteron and proton real spin-orbit interactions. Spin assignments have been made for the ground state and for the 7 low-lying exited states in $^{55}Cr$. The spectroscopic factors have been determined and used to calculate occupation numbers for the configurations $2p_{3/2}$ and $2p_{1/2}$ as well as the centre-of-gravity single particle energies for these configurations.

## Introduction

I was working late one night in my office. The door was open as usual and my next-door neighbour David Baugh, a visiting research fellow, came in to have a little chat. We talked for a while and then unexpectedly he asked me whether I could help him with the supervision of his PhD student, David Rosalky. He was wondering whether I could suggest a suitable research project for him.

It so happened that I had in mind to look into the *j* - dependence by using $^{54}Cr$ as the target nucleus but I was already too busy with other projects. However, David's request sounded attractive so I decided to work with him and his student.

Usually, direct reactions allow only for the determination of the orbital angular momentum *l* but do not distinguish between states with different values of the total angular momentum, *j*, which belong to the same *l*. However, it has been demonstrated (Lee and Schiffer 1964 and 1967) that in some cases involving (d,p) reactions it is possible to determine not only *l* but also *j* values.





For a spin-zero target nucleus, the spin of the residual nucleus $J$ is equal to the transferred total angular momentum $j$ and the (d,p) angular distributions depend directly upon $J$. The $j$ - dependence, as this property is called, provides a valuable spectroscopic tool for distinguishing between states belonging to the same $l$ but differing in their total angular momentum. Thus, using the $j$ − dependence one should be able to distinguish between states with $J = l + 1/2$ and $J = l - 1/2$.

I have chosen $^{54}$Cr as a target nucleus because it contains two neutrons outside the closed $1f_{7/2}$ shell. Thus, stripping reaction (d,p) should be expected to lead to both $2p_{3/2}$ and $2p_{1/2}$ configurations and thus offer a good opportunity to study the $j$ - dependence. Indeed, earlier measurements (Bock *et al.* 1965a) identified eight states corresponding to the $l$ =1 transfers. The single-neutron stripping reaction should therefore be expected to show clear $j$ - dependence and thus should allow for determining the $j$ values for low-energy states in the residual nucleus of $^{55}$Cr. I also hoped that this reaction could be used to study the conditions for the $j$ − dependence and explain why this feature is not always clear in the (d,p) reactions.

**Experimental procedure and results**

A 94% enriched target of $^{54}$Cr, on a thin carbon backing (15-20 μg/cm$^2$), was bombarded with 8 MeV deuterons from the ANU EN tandem accelerator. The charged reaction products were detected by two movable 1000-micron silicon surface barrier detectors mounted 20° apart in a 51-cm scattering chamber. Experimental set up was similar as that described in Section 5 for the $^{26}$Mg($^3$He,$\alpha$)$^{25}$Mg reaction.

The spectrum from each detector was collected in 2048 channels of an IBM 1800 data acquisition system with a dispersion of about 7 keV/channel. For measurements at forward angles, beam intensities were reduced to minimize pulse pile-up effects and dead time corrections, which were kept below 5%. The best resolution obtained with this arrangement was 23 keV for 11 MeV protons.

A product of the target thickness and solid angle was determined by elastic scattering of 5 MeV alpha particles in the range 20° to 80°. The ratio of the experimental to Rutherford cross sections was constant within experimental errors over this range. The alpha-particle spectra resulting from these measurements revealed the presence of carbon (used as target backing), oxygen and tantalum impurities in the target. The thickness of $^{54}$Cr was found to be 52 ± 3 μg/cm$^2$.

Figure 9.1 shows proton spectra measured at $50^0$ and $120^0$ in the laboratory system. The proton groups arising from the $^{54}$Cr(d,p)$^{55}$Cr reaction are marked as $p_i$. Angular distributions between $15^0$ and 160° (except for angles where $^{16}$O and $^{12}$C contaminant proton peaks interfered) were extracted for the six strongest $l$ =1 transitions and for the elastically scattered deuterons from $^{54}$Cr.





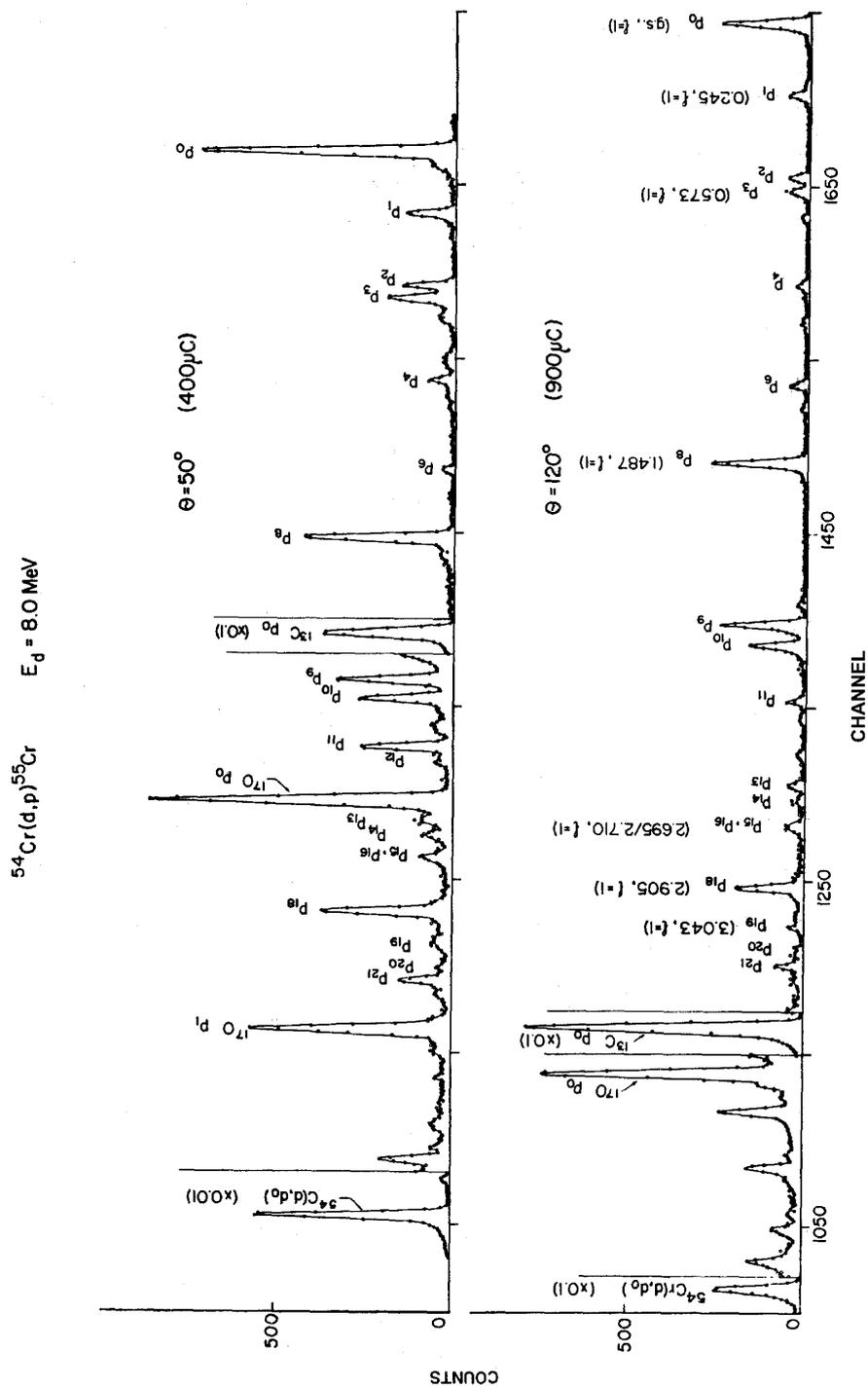

Figure 9.1. Proton spectra from the $^{54}$Cr(d,p)$^{55}$Cr stripping reaction at 8 MeV deuteron bombarding energy measured at 50⁰ and 120⁰ (lab). Proton spectra are numbered according to Table 8 of Macgregor and Brown (1966). The excitation energies of states populated by $l$ = 1 transitions are quoted.

The level at 2.7 MeV excitation energy, which was weakly excited in our work, consisted of an unresolved doublet, 2.695 and 2.710 MeV, (Macgregor and Brown 1966), of which at least one component was formed by an $l$ = 1 transition (Bock *et al.*





1965a). An accurate angular distribution could be obtained for the 2.905 MeV level since this is strongly excited compared to the partially resolved 2.874 MeV state.

The measured angular distributions are shown in Figure 9.3, the error bars represent relative errors which include contributions from counting statistics, background subtraction, the unfolding of partially resolved peaks and errors in beam monitoring. Average values have been taken of repeated measurements weighted according to the inverse square of their errors. The scattering angle could be set with an accuracy of ± 0.2⁰.

**Distorted wave analysis**

Zero-range distorted-wave Born approximation (DWBA) calculations  were carried out using optical model potential as defined in Chapter 6 for the $^{16}O(^{3}He,^{3}He)^{16}O$ scattering. However, in this analysis we have used only surface absorption potential.

We have tried both real and imaginary potentials for the spin-orbit interaction but found that the imaginary component had little effect on the calculated distributions. Thus, we have found that the dependence of the calculated angular distributions on $j$ arises from deuteron and proton real spin-orbit potentials. Consequently, we have used only real component for the spin-orbit interaction.

To understand the dependence of $j$ - dependence on the excitation energy we have calculated a series of angular distributions for $j = ^{1}/_{2}$ and $j = ^{3}/_{2}$ as a function of excitation energies. These distributions are shown in Figure 9.3. It can be seen that theoretical calculations produce different shapes of angular distributions for different $j$ values. However, the most prominent differences between $j = l$ - 1/2 (the left-hand side of the figure) and $j = l$ + 1/2 (the right-hand side) occur at low excitation energies of the residual nucleus ($^{55}Cr$). As the excitation energies increases, it is increasingly more difficult to distinguish between the two $j$ values.

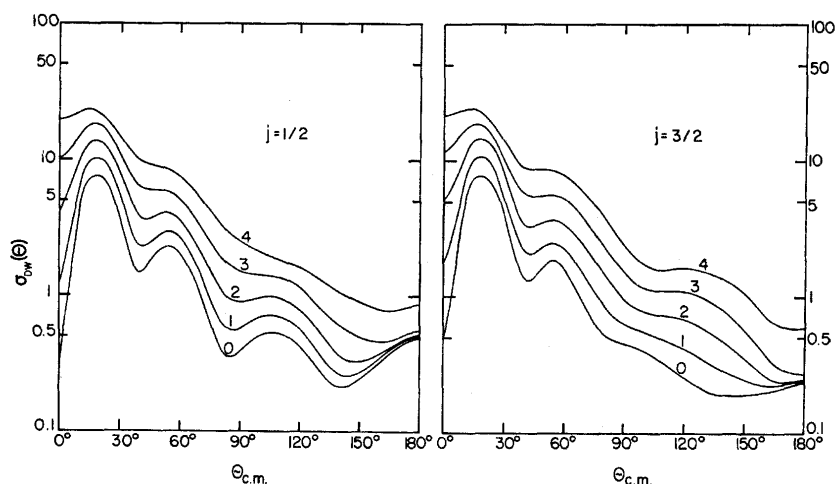

Figure 9.2. The dependence of the calculated angular distributions for $j$ = 1/2 (the left-hand side of the figure) and $j$ = 3/2 (the right-hand side) on the excitation energy in $^{55}Cr$ for the reaction $^{54}Cr(d,p)^{55}Cr$ at 8 MeV deuteron energy. The curves are labelled by the excitation energies with *0* being for the ground state. The figure shows that the $j$ – dependence is clear for the low excitation energies but becomes gradually less pronounced when the excitation energy is increasing.

Results of our DWBA calculations are compared with experimental distributions in Figure 9.3. It can be seen that by comparing theoretical and experimental distributions one can distinguish between different $j$ values and thus one can assign the spin values to the relevant states in the residual nucleus.





Spectroscopic information based on our study of the $^{54}$Cr(d,p)$^{55}$Cr reaction is summarized in Tables 9.1 and 9.2.

Spectroscopic factors for stripping reactions are proportional to the degree a given orbit is empty. Without a residual interaction, orbit 2p$_{1/2}$ in the ground state of $^{54}$Cr would be empty and orbit 2p$_{3/2}$ would contain two neutrons. Our results show that these orbits are 26% and 37% percent full indicating, as expected, significant residual interaction. Orbit 2p$_{1/2}$ contains on average 0.52 neutrons and orbit 2p$_{3/2}$ 1.48 neutrons.

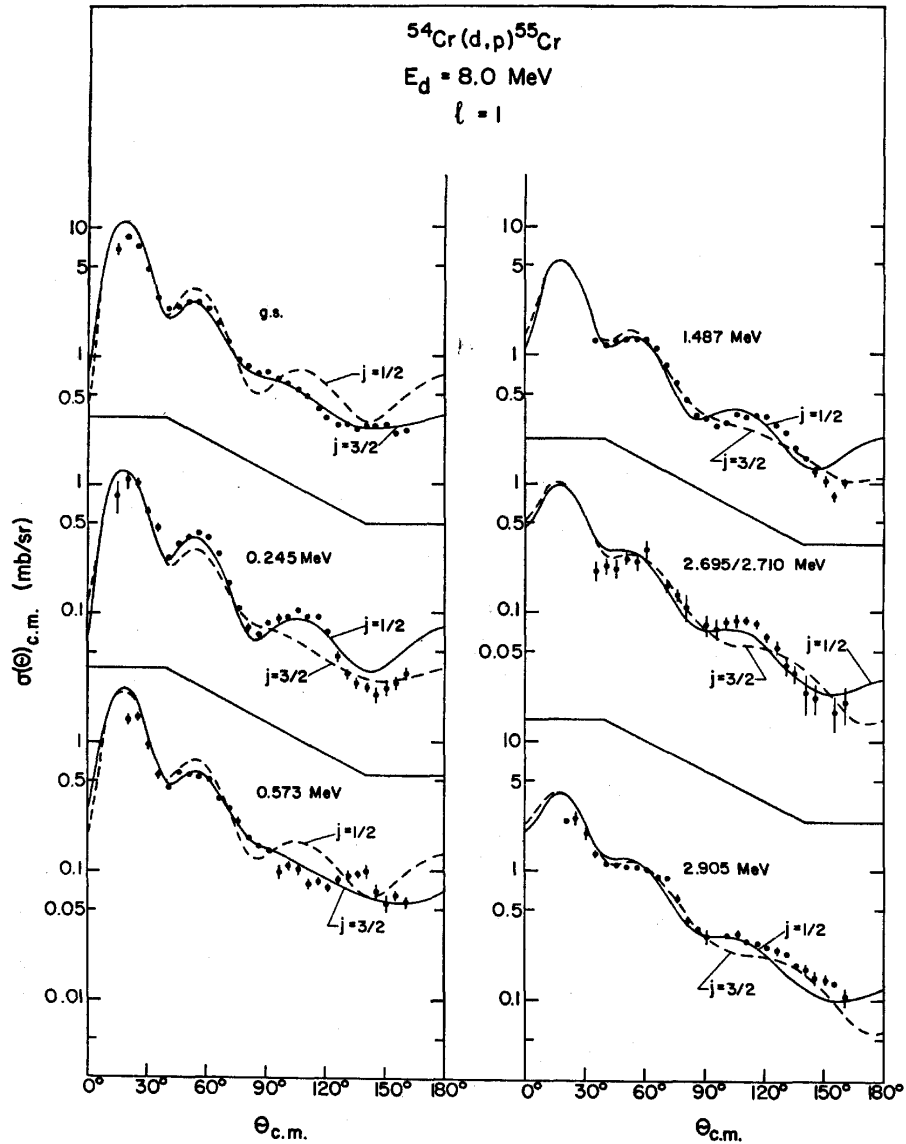

Figure 9.3. Angular distributions of protons from the $^{54}$Cr(d,p)$^{55}$Cr reaction corresponding to the $l = 1$ transitions. The curves are the DWBA predictions for $j = 1/2$ and $j = 3/2$. This figure shows clear the $j$ dependence at low excitation energies. At higher excitation energies, the differences between the two $j$ values are less clear but it is still possible to make a distinction between the two $j$ values by comparing the theoretical calculations with experimental data.





Table 9.1

Spectroscopic information based on the study of the reaction $^{54}$Cr(d,p)$^{55}$Cr at 8 MeV deuteron energy

| Level (MeV) | $j$ | Configuration | $S(l, j)$ [a] $j = 3/2$ | $S(l, j)$ [a] $j = 1/2$ |
|---|---|---|---|---|
| 0.000 | 3/2 | 2p$_{3/2}$ | 0.52 | |
| 0.245 | 1/2 | 2p$_{1/2}$ | | 0.12 |
| 0.573 | 3/2 | 2p$_{3/2}$ | 0.10 | |
| 1.487 | 1/2 | 2p$_{1/2}$ | | 0.34 |
| 2.7 [b] | (1/2) | (2p$_{1/2}$) | | 0.06 |
| 2.905 | 1/2 | 2p$_{1/2}$ | | 0.21 |
| 3.043 | (3/2) | (2p$_{3/2}$) | 0.01 | |
| 3.696 | (3/2) | (2p$_{3/2}$) | | 0.01 |

[a] Spectroscopic factors for the last two states (3.043 and 3.696 MeV) are from Bock *et al.* (1965a).

[b] Unresolved doublet 2.695/2.710 MeV.

Table 9.2

Additional spectroscopic information based on the study of the reaction $^{54}$Cr(d,p)$^{55}$Cr at 8 MeV deuteron energy

| Orbit | $U_j^2$ | $h_j$ | $h_j'$ | $V_j^2$ | $n_j$ | $n_j'$ | $E_j$ |
|---|---|---|---|---|---|---|---|
| 2p$_{3/2}$ | 0.63 | 2.52 | 4.00 | 0.37 | 1.48 | 2.00 | 0.14 |
| 2p$_{1/2}$ | 0.74 | 1.48 | 0.00 | 0.26 | 0.52 | 0.00 | 1.79 |

$U_j^2 = \sum_i S_i(l, j)$ – The experimentally determined vacancy number, i.e. the degree to which a given neutron orbit in the ground state of $^{54}$Cr is empty.

$h_j = (2j+1)U_j^2$ – The experimentally determined average number of neutron holes in a given configuration in the ground state of $^{54}$Cr.

$h_j'$ – The number of expected neutron holes in the absence of the residual interaction.

$V_j^2 = 1 - U_j^2$ – The experimentally determined occupation number, i.e. the degree to which a given neutron orbit in the ground state of $^{54}$Cr is full.

$n_j = (2j+1)V_j^2$ – The experimentally determined average number of neutrons in a given configuration in the ground state of $^{54}$Cr.

$n_j'$ – The expected number of neutrons in the absence of the residual interaction.

$E_j$ (MeV) – Centre-of-gravity energies for states with the total angular momentum $j$.

Spectroscopic factors allow also for calculating the center of gravity energies $E_j$ for states corresponding to configurations 2p$_{1/2}$ and 2p$_{3/2}$.

$$E_j = \frac{\sum_i S_i(l, j)E_i}{\sum_i S_i(l, j)}$$

where $E_i$ are excitation energies for spins $j$, and the sum is over all the states with the same spin $j$.





Results of calculations are shown in Table 9.2. The spacing between the two energies is 1.65 MeV, which is in good agreement with the energy spacing in the neighboring nuclei (see Table 9.3).

Table 9.3

The spacing between the $2p_{1/2}$ and $2p_{3/2}$ centre-of-gravity energies for nuclei in the Cr region

| Nucleus | $^{47}$Ca | $^{49}$Ca | $^{51}$Cr | $^{51}$Cr | $^{53}$Cr | $^{55}$Cr | $^{59}$Ni | $^{61}$Ni |
|---------|-----------|-----------|-----------|-----------|-----------|-----------|-----------|-----------|
| Spacing | 1.86 [a] | 2.03 [b] | 1.60 [c] | 1.64 [d] | 1.68 | 1.65 [e] | 1.90 [f] | 1.20 [f] |

[a] Belote *et al.* 1966; [b] Kashy *et al.* 1964; [c] Robertshaw *et al.* 1968; [d] Delic, G. and Robson, B. A. 1969; [e] Our result; [f] Fulmer *et al.* 1964.

## Summary and conclusions

We have studied the applicability of the $j$ - dependence for the (d,p) reactions in assigning spin values of states in residual nuclei. We have found that this feature is indeed a useful tool in distinguishing between $j = l + 1/2$ and $j = l - 1/2$ values.

By studying the dependence of the shapes of the angular distributions on the excitation energy we have also found that $j$ - dependence is clear at low excitation energies. As the energy of excited states increases the distinction between $j = l + 1/2$ and $j = l - 1/2$ decreases. However, by comparing theoretical and experimental distributions it is still possible to distinguish between different $j$ values even for high-excited states.

We have found that $j$ - dependence is related only to the real component of the spin-orbit interaction. The imaginary component has no influence.

By comparing the DWBA calculations with the experimental angular distributions we have assigned spins $3/2$, $1/2$, $3/2$, $1/2$, and $1/2$ to the ground state, 0.245, 0.573, 1.487, and 2.905 MeV in $^{55}$Cr, respectively. These states belong to the $2p_{3/2}$ and $2p_{1/2}$ configurations.

We have found that neutron orbits $2p_{3/2}$ and $2p_{1/2}$ in the ground state of $^{54}$Cr are 37% and 26% full and thus they contain on average 1.48 and 0.52 neutrons, respectively. We have calculated the centre-of-gravity single particle energies for orbits $2p_{3/2}$ and $2p_{1/2}$ in $^{55}$Cr. The energy spacing between them is in good agreement with the spacing in the neighbouring nuclei.

## References


Bock, R., Duhm, H. H., Martin, S., Rüdel R. and Stock, R.1965a, *Nucl. Phys.* **72**:273

Bock, R., Duhm, H. H., Jahr, R., Santo, R., and Stock, R. 1965b, *Phys. Lett.* **19**:417.

Belote, T. A., Chen, H. Y., Hansen, O. and Rapaport, J. 1966, *Phys. Rev.* **142**:624.

Delic, G. and Robson, B. A. 1969, *Nucl. Phys.* **A134**:470

Kashy, E., Sperduto, A., Enge, H. A. and Buechner, W. W. 1964, *Phys. Rev.* **135**:B865.

Lee, L. L. and Schiffer, J. P. 1964, *Phys. Rev.* **136**:B405.

Lee, L. L. and Schiffer, J. P. 1967, *Phys. Rev.* **154**:1097.

Macgregor, A. and Brown, G. 1966, *Nucl. Phys.* **88**:385

Robertshaw, J. E., Mecca, S., Sperduto, A. and Buechner, W. W. 1968, *Phys. Rev.* **170**:1013






<div align="center">



**Tensor Analyzing Powers for Mg and Si Nuclei**

</div>

***Key features:***

1. Unpolarized deuterons accelerated to 7 MeV were polarized in the elastic scattering from Mg and Si nuclei and the corresponding angular distributions of the tensor analyzing powers $T_{20}(\theta)$, $T_{21}(\theta)$, and $T_{22}(\theta)$ were measured using the $^3$He(d,p)$^4$He polarization analyser.

2. Experimental results were analysed using optical model in combination with the statistical theory of Hauser and Feshbach. The optical model contained not only spin-orbit but also tensor components.

3. Various shapes of the tensor component were examined.

4. Our calculations show that only the distributions of the tensor analyzing powers are sensitive to tensor interaction. Vector analyzing power $iT_{11}$ (not measured but calculated) is sensitive only to the spin-orbit interaction.

5. The tensor interaction potential has been found to be shallow, attractive, and long range.

***Abstract:*** Angular distributions for the elastic scattering cross sections and for the three components of the tensor analyzing powers $T_{20}(\theta)$, $T_{21}(\theta)$, and $T_{22}(\theta)$ were measured for the elastic scattering of 7 MeV deuterons from Mg and Si nuclei. In addition, differential cross sections were measured at 7 MeV for Mg and 7, 8, 9, 10, and 11 MeV for Si targets. All the distributions were analysed using optical model formalism with the inclusion of the statistical theory of Hauser and Feshbach. Angular distributions of the differential cross sections were reproduced well without spin-dependent interactions. However, the analysis of the tensor analyzing powers required not only the spin-orbit but also tensor interaction. Only one component of the tensor interaction, $T_R$, was necessary. Tensor interaction potential was found to be shallow and attractive, and to have a long-range.

## Introduction

It is well known that nuclear potential contains not only central but also spin dependent components. Measurements of differential cross sections serve as a useful tool to study the central part of nuclear interaction. However, the best way to study the spin-dependent components is by carrying out the polarization measurements. In particular, tensor components of the interaction potential can be studied by studying tensor analyzing powers.

Interaction of spin $^1/_2$ particles with spinless target nuclei involves only spin-orbit forces. However, interaction of spin 1 particles, such as deuterons, involves also tensor dependent forces. It is to study these tensor-dependent forces that we have carried out measurements of tensor analyzing powers for the elastic scattering of deuterons from Mg and Si target nuclei.

## Experimental procedure and results

Angular distributions of the tensor analyzing powers $T_{20}(\theta)$, $T_{21}(\theta)$, and $T_{22}(\theta)$ were measured for the elastic scattering of 7 MeV unpolarized deuterons from Mg and Si targets. In addition, to support the theoretical analysis of experimentally determined tensor moments (tensor analyzing powers), measurements of differential cross





sections of the elastically scattered deuterons were carried out at 7 for Mg and 7, 8, 9, 10, and 11 MeV for Si. All the measurements were carried out using deuterons accelerated in the EN electrostatic tandem accelerator in the Department of Nuclear Physics of the Australian National University.

### The targets

In order to reduce the data collection time, thick targets were used for the polarization measurements. However, their thickness had to be kept within a reasonable limit to avoid an excessive energy spread of elastically scattered deuterons. The thickness of the Mg target was $(44.2 \pm 4.5)$ $\mu g/cm^2$ or $(420 \pm 46)$ keV for 7.0 MeV deuterons. The thickness of Si target was $(58.5 \pm 6.2)$ $\mu g/cm^2$ or $(440 \pm 49)$ keV.

The Mg target was prepared by rolling a natural magnesium strip, which was initially about 0.2 mm thick. This simple method allowed for the production of thin Mg foils with the required thickness of about 0.025 mm.

However, due to the very brittle nature of silicone, the preparation of the thick Si targets was more complex (Djaloeis and Nurzynski 1972). Briefly, the idea was to glue a silicone disk to a glass slide and then to grind the silicone to a suitable thickness.

The surface of one side of a circular silicon disk with the diameter of about 22 mm and thickness of about 1.5 mm was smoothed by grinding it with fine-grained $Al_2O_3$ powder (grade 500) on a smooth piece of glass. Kerosene was used to wet the powder. A glass slide located on top of a hot plate was then heated and after the temperature had reached approximately 80°C a 'Lakeside 70 thermal glue'[11] was evenly spread on the glass slide. Next, after cleaning in alcohol, the smoothed side of the disk was firmly attached to the glass slide by means of the thermal glue while the glass slide was still on the hot plate. To make the surfaces of the slide and the disk parallel to each other, the disk was pressed firmly against the glass slide while the glue was still in its molten state.

The glass slide-glue-disk combination was then quickly removed from the hot plate, placed on an even surface and similar pressure was again applied to the silicone disk until the combination cooled down and the glue hardened. This procedure was necessary to avoid the formation of a wedge-shaped finish between the silicone disk and the glass slide.

Having completed this stage, the other side of the disk was ground down to a thickness of approximately 0.025 mm by using a succession of grinding powders of decreasing grain size. Once the required thickness was approximately obtained the combination was again heated to melt the glue. The target was pushed, with extreme care, toward the edge of the glass slide until it became free. The target was then immersed in alcohol for a few hours until all the adhering glue became dissolved, thus leaving the target free of contaminants. Finally, the target was stored in a dry and clean atmosphere to allow complete evaporation of the alcohol.

This procedure was, however, long and tedious and the rate of success was low. From the twelve disks used only two acceptable targets were finally produced.

---

[11] Manufactured by Hugh Courtright and Co., Chicago, Ill., USA.





Initially, the thickness of the resulting targets was measured using a Mitutoyo micrometre, accurate to 2.5 μm. Later, Rutherford scattering of low-energy α particles was used to find the thickness of the targets more accurately.

Angular distributions for elastic scattering differential cross sections were measured over a wider range of angles than for the distributions of the tensor analyzing powers. The measurements of the differential cross sections could be carried out with thinner targets, which were prepared by evaporation using VARIAN *e*-gun. However, again the preparation of thin silicone targets had to be done with special care.

We have found that the stability of the target depended critically on the distance between the crucible and the supporting glass slide. With a short distance, the deposited wafer of Si tended to crack. When the glass slide was placed far from the crucible, the evaporation time was too long. We have found that the optimal distance was 8 cm.

The temperature was also critical. Any attempt to speed up the evaporation resulted in damage to the material already deposited. The best method was a slow heating up to the melting point of the silicone ($1420^0$C) followed directly by a slow continuous evaporation. The temperature was measured using a Leeds and Northrup optical pyrometer.

### *Experimental setup*

Schematic diagram of the experimental setup is shown in Figure 10.1. Figures 10.2 and 10.3 show the horizontal and vertical view of the apparatus, respectively.

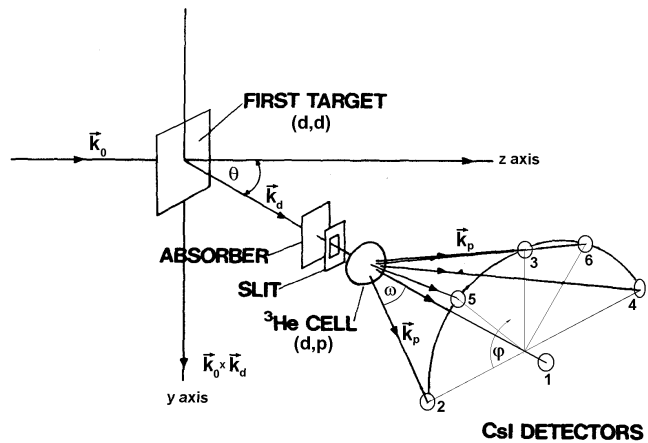

Figure 10.1. Schematic diagram of the experimental setup used in the measurements of the tensor analyzing powers in the elastic scattering of deuterons from Mg and Si targets. Deuterons were polarized by scattering from the first target (Mg or Si) and the corresponding tensor analyzing powers were measured by using the $^3$He(d,p)$^4$He polarimeter. The scattered deuterons were decelerated by Mylar foil to the energy of about 430 keV before using them to induce the resonance reaction $^3$He(d,p)$^4$He.

Unpolarized deuterons were scattered from Mg or Si targets. The scattered (and now polarized) deuterons were decelerated to around 430 keV using Mylar foil. The decelerated deuterons were directed to the $^3$He cell to induce the $^3$He(d,p)$^4$He reaction leading though a $^3/_2{}^+$ resonance in $^5$Li (see Figure 10.4) which served as the polarization analyser. In general, an operating pressure in the $^3$He cell was 5 atm, which was sufficiently high to stop up to 800 keV deuterons.





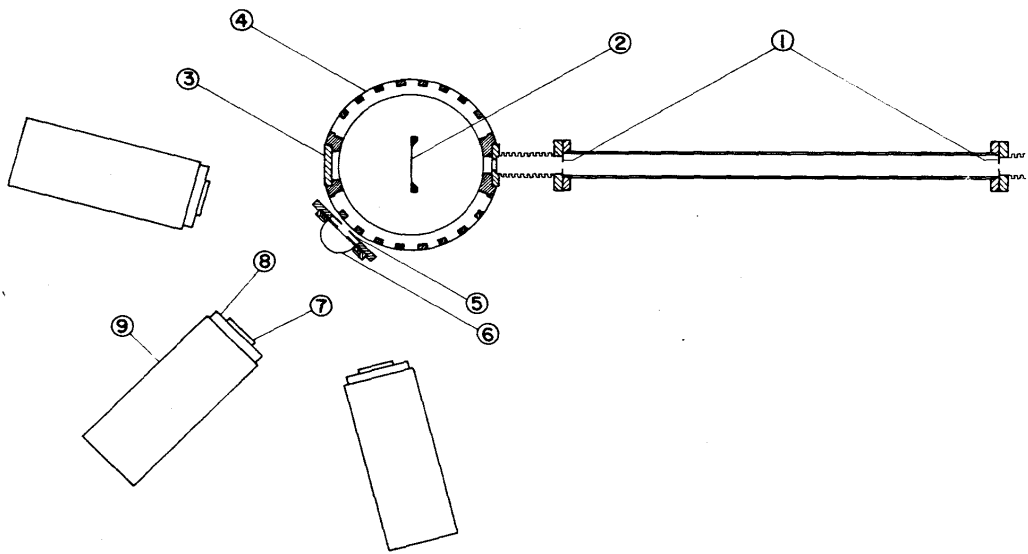

Figure 10.2. The horizontal view of the apparatus. 1. Beam collimator and anti-scattering baffle. 2. Solid target (Mg or Si). 3. Beam stop. 4. Havar foil. 5. A defining slit. 6. The $^3$He cell. 7. The CsI crystal (only three CsI detectors out of six are shown here). 8. Lucite light pipe. 8. Photomultiplier tube.

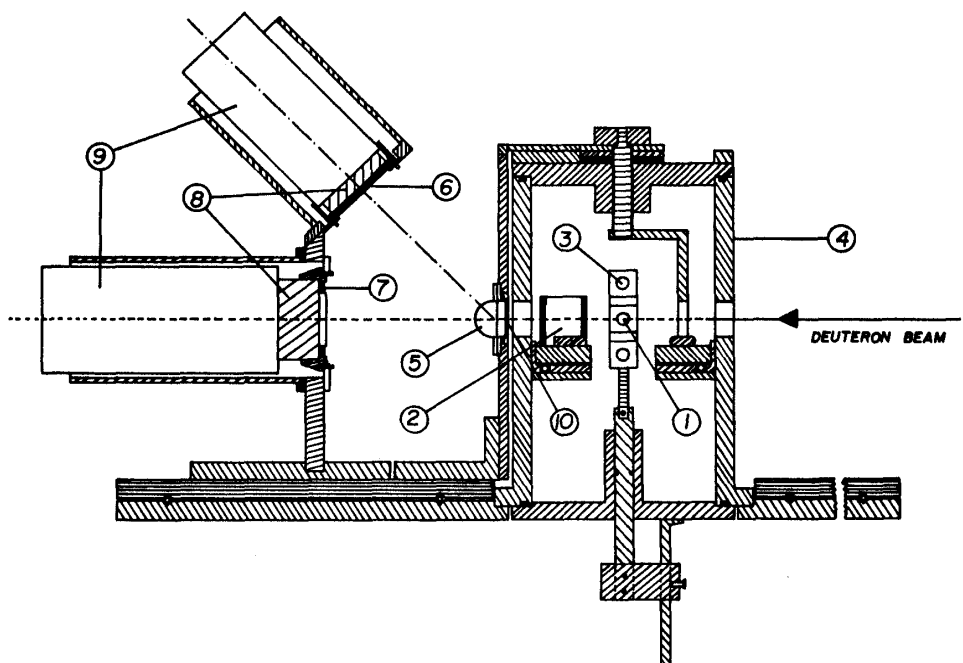

Figure 10.3. A vertical cross section of the apparatus. 1. Solid target (Mg or Si). 2. Collimator. 3. Gold target. 4. Scattering chamber. 5. $^3$He cell. 6. Square CsI crystal. 7. Annular CsI crystals. 8. Lucite light pipes. 9. Photomultiplier tubes. 10. Slit for the $^3$He cell.





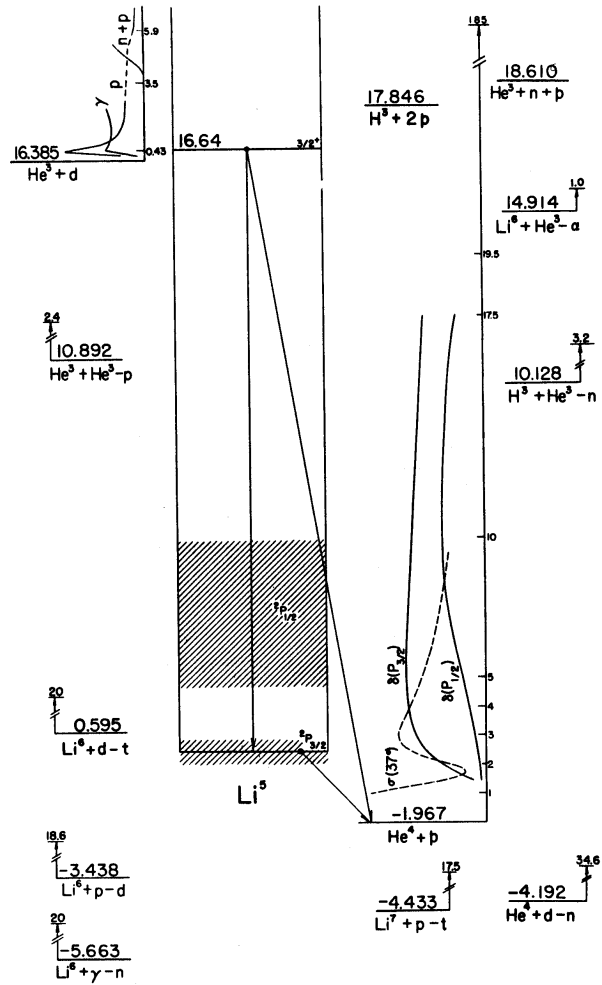

Figure 10.4. The energy level diagram for $^6$Li (Lauritsen and Ajzenberg-Selove 1962). The $^3/_2{}^+$ resonance at 16.64 MeV is formed by 430 keV deuterons interacting with $^3$He. The corresponding $^3$He(d,p)$^4$He reactions serves as a polarization analyser.

Protons from the $^3$He(d,p)$^4$He resonance reaction were detected using 6 scintillation counters placed at azimuthal angles of $0^0$, $45^0$, $90^0$, $135^0$, and $180^0$. Square CsI crystals were used for all detectors except for the detector positioned in the direct line of the incident deuterons (i.e. deuterons scattered from Mg or Si targets). This detector was equipped with an annular CsI crystal to prevent the detection of protons coming directly from the target.

Measurements of the tensor analyzing powers could be carried out with only 4 detectors (see below). The extra two detectors (at $\varphi = 45^0$ and $135^0$) were added to increase data collection efficiency.

The CsI crystals were attached to Lucite light pipes and mounted on 5 cm Dumont 6392 photomultiplier tubes. The pulses from photomultipliers were fed through Franklin double-delay-line preamplifiers, amplifiers, and directed via analogue-to-digital converters to an online IBM-computer.

The data collection time was between about 2 and 30 hours depending on the reaction angle. The collection time for the differential cross sections was significantly shorter.





### The $^3$He(d,p)$^4$He polarimeter

The differential cross section for the $^3$He(d,p)$^4$He reaction at the resonant energy of 430 keV can be expressed by the following formula (McIntyre 1965; Welton 1963):

$$\sigma(\omega, \varphi) =$$

$$\sigma_0(\omega)\left\{1 - f\left[\frac{1}{2\sqrt{2}}(3\cos^2\omega - 1)T_{20}(\theta) - \sqrt{3}(\sin\omega\cos\omega\cos\varphi)T_{21}(\theta) + \frac{\sqrt{3}}{2}(\sin^2\omega\cos 2\varphi)T_{22}(\theta)\right]\right\}$$

where $\omega$ is the reaction angle for the $^3$He(d,p)$^4$He reaction, $\varphi$ is the azimuthal angle between the first reaction plane (for the elastic scattering) and the second reaction plane (for the $^3$He(d,p)$^4$He reaction), and $T_{2q}(\theta)$ are the tensor analyzing powers for the first reaction (the elastic scattering). They are defined in the right-handed coordinate system with the $z$-axis along the incident beam $\vec{k}_0$ and the $y$-axis along the $\vec{k}_0 \times \vec{k}_d$ vector as shown if Figure 10.1. The energy dependence of the factor $f$ has been studied by Brown, Christ, and Rudin (1966) in the energy range of 300-1700 keV.

It is clear that by a proper combination of measurements at various azimuthal angles $\varphi$ the components of tensor moments $T_{2q}(\theta)$ can be determined.

The reaction angle for the $^3$He(d,p)$^4$He analyser for detectors 2-6 was chosen to correspond to $\cos^2\omega = 1/3$ ($\omega \approx 54.74^0$ in the centre of mass system or $\approx 52.55^0$ in the laboratory system). For the detector 1, $\omega = 0$. The equation for the differential cross section for the $^3$He(d,p)$^4$He reaction is given by simpler expressions:

*For detector 1*

$$\sigma_1 = \sigma_0\left[1 - f\frac{1}{\sqrt{2}}T_{20}(\theta)\right]$$

*For detectors 2-6*

$$\sigma_i = \sigma_0\left[1 + f\left(\sqrt{\frac{2}{3}}(\cos\varphi)T_{21}(\theta) - \frac{1}{\sqrt{3}}(\cos 2\varphi)T_{22}(\theta)\right)\right] \qquad i = 2,3,4,5,6$$

More explicitly, for detectors 2-6, we have the following relations:

Detector 2 ($\varphi = 0^0$): $\qquad \sigma_2 = \sigma_0\left[1 + f\left(\sqrt{\frac{2}{3}}T_{21}(\theta) - \frac{1}{\sqrt{3}}T_{22}(\theta)\right)\right]$

Detector 3 ($\varphi = 90^0$) $\qquad \sigma_3 = \sigma_0\left[1 + f\frac{1}{\sqrt{3}}T_{22}(\theta)\right]$

Detector 4 ($\varphi = 180^0$): $\qquad \sigma_4 = \sigma_0\left[1 - f\left(\sqrt{\frac{2}{3}}T_{21}(\theta) + \frac{1}{\sqrt{3}}T_{22}(\theta)\right)\right]$

Detector 5 ($\varphi = 45^0$): $\qquad \sigma_5 = \sigma_0\left[1 + f\frac{1}{\sqrt{3}}T_{21}(\theta)\right]$





Detector 6 ($\varphi = 135^0$):     $\sigma_6 = \sigma_0 \left[ 1 - f \frac{1}{\sqrt{3}} T_{21}(\theta) \right]$

It follows that the analyzing powers are given by the following relations:

$$T_{20}(\theta) = \frac{1}{f} \sqrt{2} \left( 1 - \frac{\sigma_1}{\sigma_0} \right)$$

$$T_{21}(\theta) = \frac{1}{f} \frac{\sqrt{3}}{2\sqrt{2}} \left( \frac{\sigma_3 - \sigma_4}{\sigma_0} \right)$$

$$T_{21}(\theta) = \frac{1}{f} \frac{\sqrt{3}}{2} \left( \frac{\sigma_5 - \sigma_6}{\sigma_0} \right)$$

$$T_{22}(\theta) = \frac{1}{f} \sqrt{3} \left( \frac{\sigma_3}{\sigma_0} - 1 \right)$$

$$T_{22}(\theta) = \frac{1}{f} \sqrt{3} \left( 1 - \frac{\sigma_2 + \sigma_4}{2\sigma_0} \right)$$

$$\sigma_0 = \frac{\sigma_2 + 2\sigma_3 + \sigma_4 + \sigma_5 + \sigma_6}{6}$$

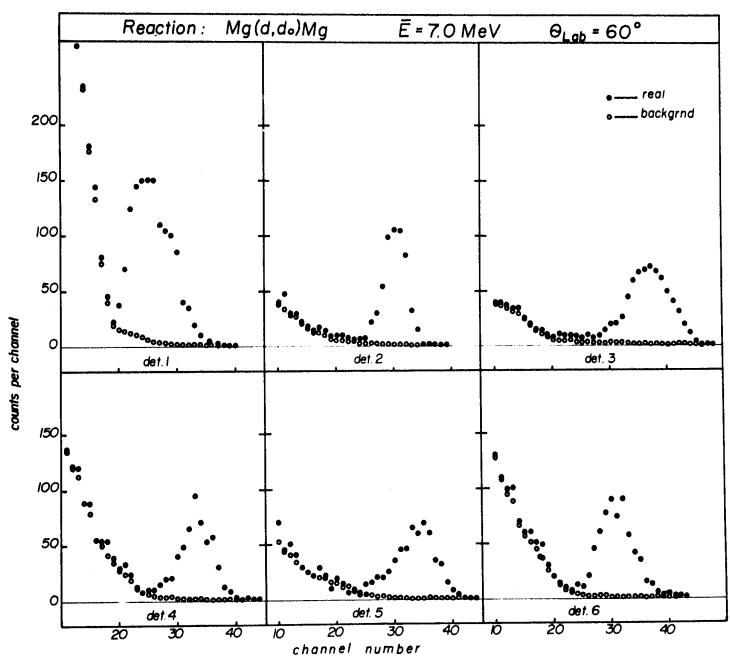

Figure 10.5. Examples of proton spectra from the reaction $^3$He(d,p)$^4$He induced by polarized deuterons following the elastic scattering from Mg target. Similar spectra are also for the Si target.

In practice, the analyzing powers are determined by measuring the ratios of the number of counts with a given target (Mg or Si in our case) to the number of counts with gold target for each reaction angle $\theta$. The above expressions are the same except that now $\sigma_i$ are replaced with the relevant ratios $R_i$ for each detector, and





$\sigma_0$ by $R_0$, which is determined by taking an average value for all 5 detectors as shown in the last equation. This procedure simplifies the procedure of measuring the analyzing powers. It also eliminates geometrical asymmetries.

A sample of proton spectra for the $^3$He(d,p)$^4$He reaction detected by the six scintillation counters is shown in Figure 10.5. This figure shows that even though the $^3$He(d,p)$^4$He reaction angle $\omega$ is the same for all six detectors the number of counts depends on the azimuthal angle $\varphi$, as it should if the incident deuterons are polarized. The observed azimuthal asymmetries were used to calculate tensor moments $T_{2q}$ for deuterons scattered from the first target (Mg or Si).

Results of measurements are shown in Figures 10.7 and 10.10 for the tensor analyzing powers, and in Figures 10.6, 10.8, and 10.9 for the differential cross sections, together with results of theoretical calculations.

**Theoretical analysis**

*The theory*

It is well known that direct processes induced by low-energy deuterons bombarding light targe nuclei can be influenced by compound nucleus formation. It has been therefore decided to include the statistical corrections in these calculations using statistical theory as developed by Wofenstein (1949) and Hauser and Feshbach (1952). The theoretical differential cross section $\sigma_{th}(\theta)$ for the differential cross sections can be then written as:

$$\sigma_{th}(\theta) = \sigma_{OM}(\theta) + R\sigma_{HF}(\theta)$$

where $\sigma_{OM}(\theta)$ is the theoretical differential cross section calculated using optical model, $\sigma_{HF}(\theta)$ is the statistical, Hauser-Feshbach cross section, and $R$ is the reduction factor (Hodgson and Wilmore 1967).

Likewise,

$$\left[T_{2q}(\theta)\right]_{th} = \left[1 - \frac{R\sigma_{HF}(\theta)}{\sigma_{\exp}(\theta)}\right]\left[T_{2q}(\theta)\right]_{OM}$$

Theoretical analysis of the experimental differential cross sections $\sigma(\theta)$ and tensor moments $T_{2q}(\theta)$ was carried out using optical model potential containing not only central and spin-orbit ($\vec{S}\cdot\vec{L}$) components but also tensor terms $T_R$ and $T_L$:

$$T_R = \frac{(\mathbf{S}\cdot\mathbf{r})^2}{r^2} - \frac{2}{3}$$

$$T_L = (\mathbf{L}\cdot\mathbf{S})^2 + \frac{1}{2}(\mathbf{L}\cdot\mathbf{S}) - \frac{2}{3}\mathbf{L}^2$$

The potential used in the calculations had the following form:

$$U(r) = V_C(r) - Vf_V - iW_D g + V_S\left(\frac{\hbar}{m_\pi c}\right)^2 \frac{1}{r}\frac{df_S}{dr}(\mathbf{S}\cdot\mathbf{L}) + V_R h_R^k T_R + V_L h_L^k T_L$$





where $V_C(r)$ is the Coulomb potential, $V$, $W_D$, $V_S$, $V_R$, $V_L$ are the depths of the relevant components, and $f$, $g$ and $h$ are the radial form factors. The form factors $f$ and $g$ are the usual Woods-Saxon and derivative of Woods-Saxon type, respectively.

If $f_i$ is defined as

$$f_i \equiv f(r, r_i, a_i) = \left\{ 1 + \exp\left[ \left( r - r_i A^{1/3} \right) / a_i \right] \right\}^{-1}$$

then

$$f_V \equiv f(r, r_V, a_V)$$

$$f_S \equiv f(r, r_S, a_S)$$

$$g = 4 a_W \left| \frac{d}{dr} f(r, r_W, a_W) \right|$$

The form factors $h$ can have various forms. We have tried the following three options:

(i) Derivative of Woods-Saxon ($D$):

$$h_R^D = -4 a_R \frac{d}{dr} f(r, r_R, a_R)$$

$$h_L^D = -4 a_L \frac{d}{dr} f(r, r_L, a_L)$$

(ii) Thomas ($T$):

$$h_R^T = -\left( \frac{\hbar}{m_\pi c} \right)^2 \frac{1}{r} \frac{d}{dr} f(r, r_R, a_R)$$

$$h_L^T = -\left( \frac{\hbar}{m_\pi c} \right)^2 \frac{1}{r} \frac{d}{dr} f(r, r_L, a_L)$$

(iii) Gaussian ($G$)

$$h_i^G = \exp(-x_i^2)$$

with $x_i = (r - r_i A^{1/3}) / a_i$ and $i = R$ or $L$.

The quality of fits was examined both visually and by calculating the function:

$$\chi^2 = \frac{1}{N} \left[ N_\sigma \chi_\sigma^2 + \sum_{q=0}^{2} N_q \chi_q^2 \right]$$

where

$$N = N_\sigma + \sum_{q=0}^{2} N_q$$

$$\chi_\sigma^2 = \frac{1}{N_\sigma} \sum_{i=1}^{N_\sigma} \left[ \frac{\sigma_{\exp}(\theta_i) - \sigma_{th}(\theta_i)}{\Delta \sigma_{\exp}(\theta_i)} \right]^2$$





$$\chi_q^2 = \frac{1}{N_q} \sum_{i=1}^{N_q} \left[ \frac{\left[T_{2q}(\theta_i)\right]_{exp} - \left[T_{2q}(\theta_i)\right]_{th}}{\Delta \left[T_{2q}(\theta_i)\right]_{exp}} \right]^2$$

$N_\sigma$ is the number of experimental points for the differential cross sections, $\sigma_{exp}(\theta_i)$ measured at angles $\theta_i$; $N_q$ is the number of experimental points $[T_{2q}(\theta_i)]_{exp}$; $\Delta\sigma_{exp}(\theta_i)$ and $\Delta[T_{2q}(\theta_i)]_{exp}$ are the experimental uncertainties of $\sigma_{exp}(\theta_i)$ and $[T_{2q}(\theta_i)]_{exp}$, respectively.

### *Analysis of the Mg(d,d)Mg data*

Calculations for the differential cross sections $\sigma(\theta)$ for the scattering from Mg target are shown in Figure 10.6. Similar fits were obtained for the elastic scattering from Si. It can be seen that theoretical calculations reproduce well the experimental angular distribution.

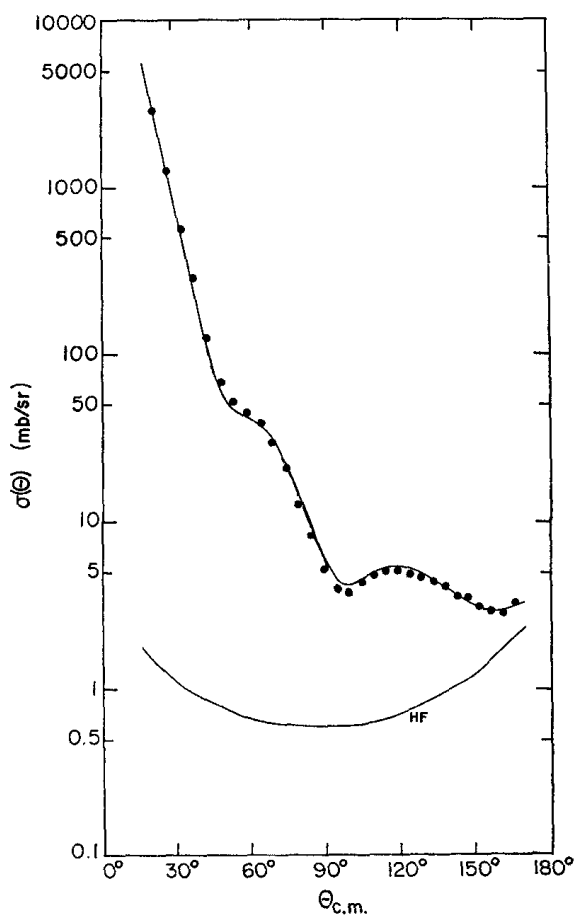

Figure 10.6. The differential cross sections $\sigma(\theta)$ for the elastic scattering of 7 MeV deuterons from Mg. The curve marked HF is the calculated Hauser-Feshbach contribution to the elastic scattering. The continuous curve going through the data points is the optical model fit, which included the Hauser-Feshbach contribution. Optical model potential parameters used in this analysis are listed in Table 10.1 as the set $\sigma$-1.

We have spent a considerable amount of computation time trying to understand influence of various optical model parameters on the calculated distributions of the tensor analyzing powers for both target nuclei. Examples of the calculated curves for





the tensor analyzing powers for Mg are shown in Figures 10.7. The corresponding potentials are listed in Table 10.1

The calculated shapes for the $T_{20}$ tensor analyzing power are similar for all parameter sets. The $\chi^2$ value is the lowest for the Gaussian shape (THF-G set; $\chi^2 = 6.82$) and the highest for the spin-orbit interaction without tensor components (THF-S1 set, $\chi^2 = 9.75$).

The best fit for the $T_{21}$ tensor analyzing power is obtained by using the derivative shape of the $T_R$ component of the optical model potential (set THF-D1;$\chi^2 = 6.68$). Adding the $T_L$ component does not improve the fit (set THF-DD1; $\chi^2 = 6.85$). Changing the shape to Thomas or Gaussian results in higher $\chi^2$ values (10.27 and 17.37 respectively).

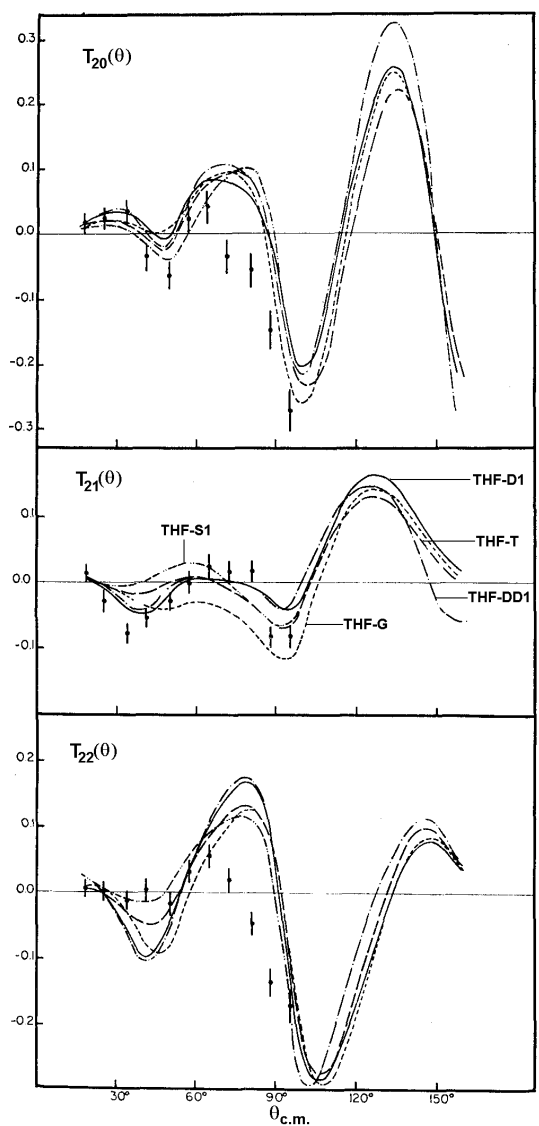

Figure 10.7. Tensor analyzing powers for the Mg(d,d)Mg scattering at 7 MeV deuteron energy. Experimental points are compared with theoretical calculations. The parameter sets are listed in Table 10.1. The curves belonging to the respective potentials used in the calculations are identified for the $T_{21}(\theta)$ angular distribution.





Table 10.1

Parameters of the optical model potential used in the analysis of differential cross sections and tensor analyzing powers for the elastic scattering of 7 MeV deuterons from Mg

| Set | Central | | $S \cdot L$ | $T_R$ | | $T_L$ | |
|---|---|---|---|---|---|---|---|
| | Re | Im | | par | shape | par | shape |
| $\sigma$-1 | $V = 72.92$ $r_V = 1.050$ $a_V = 0.968$ | $W_D = 17.84$ $r_W = 1.549$ $a_W = 0.565$ | | | | | |
| THF-S1 | $V = 72.92$ $r_V = 1.050$ $a_V = 0.968$ | $W_D = 11.0$ $r_W = 1.575$ $a_W = 0.565$ | $V_S = 25.0$ $r_S = 0.75$ $a_S = 0.40$ | | | | |
| THF-D1 | $V = 72.92$ $r_V = 1.050$ $a_V = 0.968$ | $W_D = 11.0$ $r_W = 1.575$ $a_W = 0.565$ | $V_S = 25.0$ $r_S = 0.75$ $a_S = 0.40$ | $V_R = -0.35$ $r_R = 2.5$ $a_R = 2.0$ | D | | |
| THF-T | $V = 72.92$ $r_V = 1.050$ $a_V = 0.968$ | $W_D = 11.0$ $r_W = 1.575$ $a_W = 0.565$ | $V_S = 25.0$ $r_S = 0.75$ $a_S = 0.40$ | $V_R = -10.0$ $r_R = 2.5$ $a_R = 2.0$ | T | | |
| THF-G | $V = 72.92$ $r_V = 1.050$ $a_V = 0.968$ | $W_D = 11.0$ $r_W = 1.575$ $a_W = 0.565$ | $V_S = 25.0$ $r_S = 0.75$ $a_S = 0.40$ | $V_R = -0.35$ $r_R = 2.5$ $a_R = 2.0$ | G | | |
| THF-DD1 | $V = 72.92$ $r_V = 1.050$ $a_V = 0.968$ | $W_D = 11.0$ $r_W = 1.575$ $a_W = 0.565$ | $V_S = 25.0$ $r_S = 0.75$ $a_S = 0.40$ | $V_R = -0.35$ $r_R = 2.5$ $a_R = 2.0$ | D | $V_L = +0.70$ $r_L = 0.80$ $a_L = 0.40$ | D |

The depths $V$, $W_D$, $V_R$, and $V_L$ are in MeV. $V_S$ is in Mev·fm$^2$ and the geometrical parameters $r$ and $a$ are in fm.

Re – the real part of the central potential; Im – the imaginary part; par – parameters; shape – shape of the tensor parts of the optical model; D – derivative shape; T – Thomas shape; G – Gaussian shape. Italicised numbers represent fixed parameters during the search.

The $\sigma$-1 potential was used to fit only the experimental differential cross sections $\sigma(\theta)$. Potentials labelled as *THF* were used to analyse not only the differential cross sections but also the tensor analyzing powers. They included calculations of the compound nucleus contributions.

The best fit to the $T_{22}$ tensor analyzing power is by using the spin-orbit interaction without the tensor components (set THF-S1; $\chi^2 = 19.23$). However, this is mainly due to one experimental point at $\theta_{c.m.} = 40.43^0$. When $T_R$ component is included, the best fit is for the Thomas shape (set THF-T; $\chi^2 = 31.75$).

The overall results indicate that tensor component $T_R$ in the optical model potential is necessary, in addition to the spin-orbit component, to describe the observed tensor analyzing powers. Adding tensor component $T_L$ makes practically no difference. There is also no compelling need for shapes other than the more conventional derivative shape.

### Analysis of the Si(d,d)Si data

Calculations of the differential cross sections for Si are compared with the experimental results in Figure 10.8 and the corresponding parameters of the optical model potential are listed in Table 10.2. Similar fits were obtained for $V \approx 100$ MeV.





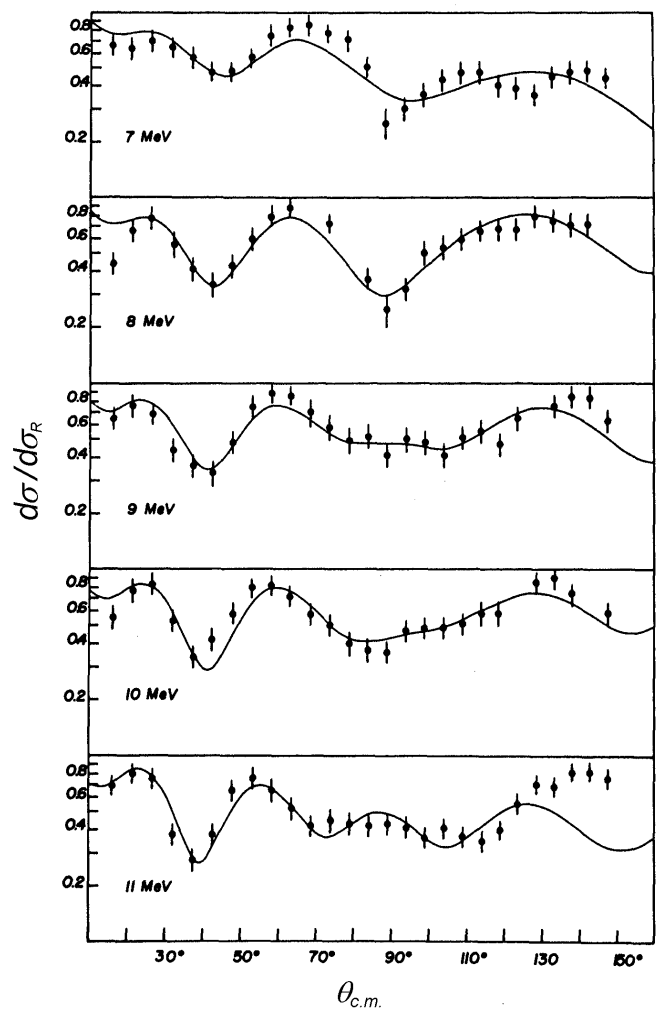

Figure 10.8. Examples of fits to the differential cross sections for the elastic scattering of 7-11 MeV deuterons from Si. The parameters of the optical model potential are listed in Table 10.2

Table 10.2

Optical model parameters generating theoretical distributions displayed in Figure 10.8

| $E_d$ (MeV) | $V$ (MeV) | $r_V$ (fm) | $a_V$ (fm) | $W_D$ (MeV) | $r_W$ (fm) | $a_W$ (fm) |
|---|---|---|---|---|---|---|
| 7 | 65.5 | 1.05 | 0.877 | 10.6 | 1.539 | 0.758 |
| 8 | 66.3 | 1.05 | 1.013 | 15.7 | 1.701 | 0.485 |
| 9 | 62.0 | 1.05 | 0.970 | 15.2 | 1.638 | 0.555 |
| 10 | 65.6 | 1.05 | 0.932 | 18.6 | 1.565 | 0.484 |
| 11 | 61.8 | 1.05 | 0.944 | 20.0 | 1.485 | 0.551 |

We have also carried out calculations of the differential cross sections by including the spin-orbit and tensor components. Results are shown in Figure 10.9 for 7 MeV deuterons for two sets of parameters OM-1 corresponding to $V = 70.3$ and OM-2 for $V = 116.0$ MeV. The parameters are listed in Table 10.3. Surprisingly, the fit for $V = 116.0$ MeV is worse than for the shallower potential.





Finally, we have also included Houser-Feshbach corrections. The calculations, labelled as HF-D are also shown in Figure 10.9 and the corresponding optical model parameters are listed in Table 10.3

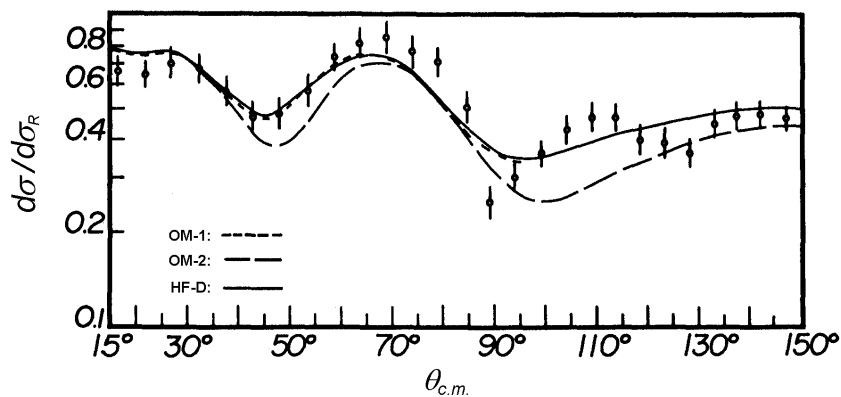

Figure 10.9. Optical model calculations with spin-dependent components for the differential cross sections of 7 MeV deuterons scattered elastically from Si nuclei. One set of calculations (HF-D) includes also Hauser-Feshbach corrections. The corresponding optical model parameters are listed in Table 10.3.

Table 10.3

Optical model parameters for the calculations with the central, spin-orbit, and tensor components

| Set | Central potential | | Spin-orbit | $T_R$ tensor |
|---|---|---|---|---|
| OM-1 | $V = 70.3$ MeV; $r_V = 1.050$ fm; $a_V = 0.791$ fm; | $W = 6.0$ MeV $r_W = 1.675$ fm $a_W = 0.795$ fm | $V_S = 20.0$ MeV·fm² $r_S = 0.75$ fm $a_S = 0.40$ fm | $V_R = -0.35$ MeV $r_R = 2.50$ fm $a_R = 1.00$ fm |
| OM-2 | $V = 116.0$ MeV; $r_V = 1.050$ fm; $a_V = 0.810$ fm; | $W = 9.5$ MeV $r_W = 1.500$ fm $a_W = 0.746$ fm | $V_S = 20.0$ MeV·fm² $r_S = 0.75$ fm $a_S = 0.40$ fm | $V_R = -0.35$ MeV $r_R = 2.50$ fm $a_R = 0.50$ fm |
| HF-D | $V = 68.0$ MeV; $r_V = 1.050$ fm; $a_V = 0.840$ fm; | $W = 8.5$ MeV $r_W = 1.425$ fm $a_W = 0.919$ fm | $V_S = 25.0$ MeV·fm² $r_S = 0.75$ fm $a_S = 0.40$ fm | $V_R = -0.35$ MeV $r_R = 2.50$ fm $a_R = 0.75$ fm |

Figure 10.10 compares optical model calculations for the angular distributions of the tensor analyzing powers with experimental results. Two sets of curves are displayed. They both belong to the potential set HF-D of Table 10.3 but with or without tensor interaction. Shown are also calculations for the vector analyzing power $iT_{11}(\theta)$. Calculations using the two other sets (OM-1 and OM-2) of Table 10.3 produced similar results.

Figure 10.10 shows that angular distributions for the tensor analyzing powers are sensitive to tensor interaction, and thus this interaction is essential in the analysis of the experimental angular distributions of the tensor analyzing powers. In contrast, the vector analyzing power is not sensitive to tensor interaction. The calculated distributions for this component are nearly identical whether tensor interaction is included or not. Thus, while measurements of the vector tensor analyzing powers can yield information about the spin-orbit interaction, the study of tensor analyzing powers opens an opportunity to learns about tensor forces.





The $T_R$ component used in these calculations had a derivative shape. The parameters optimising the fits are similar to the parameters used in the analysis of the Mg(d,d)Mg scattering, except for $a_R$ which for Si has a significantly smaller value than for Mg.

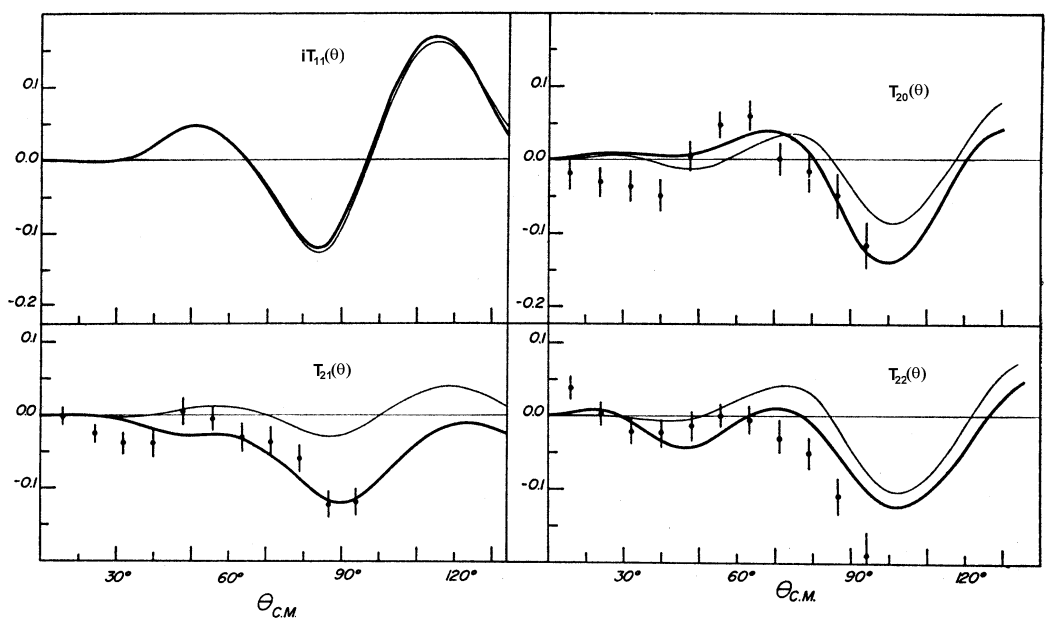

Figure 10.10. Measured angular distributions of the tensor analyzing powers for Si(d,d)Si elastic scattering are compared with theoretical distributions. Optical model parameters used in the calculations are listed in Table 10.3 as set HF-D. Similar results, published but not shown here, have been also obtained for sets OM-1 and OM-2. The fine line was calculated using central and spin orbit components only. The thicker solid line was calculated by including also the $T_R$ component of the tensor interaction. Calculations for the vector analyzing power $iT_{11}$ are also shown. As can be seen, they are insensitive to the tensor interaction.

## Summary and conclusions

In this work, we have studied nuclear interaction in the deuteron-nucleus system. We have measured angular distributions of the differential cross sections and of the three components of the tensor analyzing powers, $T_{20}(\theta)$, $T_{21}(\theta)$, and $T_{22}(\theta)$ in the elastic scattering of 7 MeV deuterons from Mg and Si. A beam of unpolarized deuterons was polarized by nuclear forces experienced in the elastic scattering and its tensor analyzing powers were measured using the resonance reaction $^{3}$He(d,p)$^{4}$He, which served as the polarization analyser. Being a 'double-scattering' experiment (*cf* Chapter 1), the yield of the detected protons was low, and the data collection time was between 2-30 hours per angle, depending on the scattering angle from the first target (Mg or Si).

Experimental results were analysed using the optical model procedure in combination with the Hauser-Feshbach statistical theory. We have found that the inclusion of the spin-orbit interaction did not result in any significant improvement of the fits of the differential cross sections. We have also found that a combination of the central and spin-orbit interactions alone was inadequate in reproducing the observed tensor analyzing powers angular distributions. Addition of the tensor interaction $T_R$ was necessary to improve the fits to the tensor analyzing powers. The tensor potential was found to be shallow, attractive and to have a long range. An inclusion of the second component $T_L$ of the tensor interaction turned out to be unnecessary.





Measurements of vector analyzing power yield useful information about spin-orbit interaction. Tensor components of nuclear interaction can be studied by measuring tensor analyzing powers.

## References


Brown, L., Christ, H. A. and Rudin, H. 1966, *Nucl. Phys.* **79**:459.

Djaloeis, A. and Nurzynski, J. 1972, Report ANU-P/532.

Hauser, W. and Feshbach, H. 1952, *Phys. Rev.* **87**:366.

Hodgson, P. E. and Wilmore, D. 1967, *Proc. Phys. Soc.* **90**:361.

Lauristen, T. and Ajzenberg-Selove, F. 1962, *Energy Levels of Light Nuclei: Nuclear Data Sheet*, Printing and Publishing Office, National Academy of Sciences, Washington 25, DC.

McIntyre, L. C. 1965, PhD Thesis, University of Wisconsin

Welton, T. A. 1963, *Fast Neutron Physics Part II*, ed. by J.B. Marion and J. L. Fowler, Interscience Publisher, New York.

Worfenstein, L. 1949, *Phys. Rev.* **75**:1664.






_______________________________________________________________________

## 11
# Optical Model Potential for Tritons

***Key features:***

1. I have carried out a detailed examination of the optical model potential for tritons interacting with a wide-range of target nuclei ($40 \leq A \leq 207$).

2. The analysis shows a preference for the surface absorption potential.

3. The unusually shallow depth of the real central component, $V \approx 25$ MeV, for the volume absorption potential obtained in earlier calculations, has been explained as being associated with strong attenuation of the radial wave functions inside the nucleus.

4. With proper care, it is possible to discriminate between nearly equivalent discrete sets of parameters. Out of the 16 sets found in this analysis only one family was identified as giving the best fits to the experimental angular distributions using either four- or six-parameter potentials. This family corresponds to the volume integral $J_R \approx 440$ MeV·fm$^3$ or to the real potential depths $V \approx 150$ MeV.

5. Volume integrals have been found to depend on the mass number of the target nuclei.

6. Formulae for the mass dependence of the optical model parameters have been derived. Geometrical parameters depend weakly on the mass number of the target nucleus and can be fixed.

7. The dependence of the real and imaginary potential depths on the symmetry parameter $\varepsilon = (N - Z)/A$. Clear linear dependence has been found for the four- and six-parameter potentials. However, the gradient of the relevant functions depends on the parameterisation of the optical model potential and is most likely dominated by the dependence on the mass of the target nuclei.

***Abstract:*** Elastic scattering of 20 MeV tritons from target nuclei with $A \geq 40$ has been analysed using four- or six-parameter optical model potentials. A total of 16 parameter families have been identified and studied. Formulae for the mass dependence have been derived. Limitations of the conventional optical model are examined. The dependence of the potential depths on the symmetry parameter $\varepsilon = (N - Z)/A$ has been investigated.

## Introduction

The interaction of tritons with nuclei has not been explored sufficiently well. Acceleration of tritons is avoided because of the problems with nuclear radiation and consequently there have not been enough experimental data to support a systematic theoretical investigation. Tritium half-life is relatively long (12.3 years) and its mobility is high. When deposited in various places along the beam line (ion source, slits, Faraday cups, etc.) tritium can spread quickly everywhere in the system. The use of tritons as projectiles to induce nuclear reactions is therefore unwelcome and unpopular.

The most extensive measurements of triton elastic scattering were carried out at 20 MeV incident energy at Los Alamos (Hafele, Flynn, and Blair,1967; Flynn, *et al.* 1969). Results of measurements were analysed using Woods-Saxon volume absorption with or





without the addition of a surface-peaked isospin component (Hafele, Flynn, and Blair,1967; Flynn, *et al.* 1969; Urone, *et al.* 1971a).

I have decided to investigate the interaction of tritons with atomic nuclei for three reasons. First, such information was needed in connection of my study of both (d,t) and (p,t) reactions. It is well known that the description of reactions involving mass-three particles depends significantly on the parameterisation of the optical model potential generating the relevant distorted waves (Baer, et al. 1970; Barnard and Jones 1968a, 1968b) and yet, the interaction in the triton channel was often approximated by using potentials derived from $^3$He scattering.

Another reason was the puzzling results in the previous analysis (Hafele, Flynn, and Blair,1967; Flynn, *et al.* 1969), which yielded unusually shallow potential depths of around 25 MeV for the real central component.

Finally, there was also the question whether there was a preference for any type of a form factor for the imaginary part of the optical model potential for tritons. Glover and Jones (1966) carried out an analysis for 12 MeV tritons scattered from light nuclei and concluded that the quality of fits did not depend on the type of the form factor of the imaginary component. However, considering the more thoroughly investigated $^3$He scattering, and in particular data extending to large angles, it appeared that there was good evidence in favour of the surface-peaked absorption (Cage, *et al.* 1972; Chang, *et al.* 1973; Fulmer and Hafele 1972; Marchese 1972; Siegel 1971; Urone, *et al.* 1971a, 1972). Unfortunately, surface absorption has not been studied sufficiently well for tritons.

## The analysis

### Interaction potential

The main part of this work is a study of the surface absorption interaction. However, I have also carried out calculations using volume absorption with an aim of trying to understand why this type of the potential leads to an unusually low value for the potential depth of the real component.

Thus, most of my analysis was done using the following form of the interaction potential:

$$U(r) = -Vf(r, r_0, a) - i4a'W_D \left| \frac{d}{dr} f(r, r_0', a') \right| + V_C(r)$$

where

$$f(r, r_0, a) = \left\{ 1 + \exp\left[ (r - r_0 A^{1/3})/a \right] \right\}^{-1}$$

For the imaginary component, $r_0$ is replaced by $r_0'$ and $a$ by $a'$. $V_C(r)$ is the Coulomb potential due to a uniformly charged sphere of radius $1.4A^{1/3}$.

For the limited calculations with volume absorption, the form factor

$$4a' \left| \frac{d}{dr} f(r, r_0', a') \right|$$

was replaced by

$$f(r, r_0', a') = \left\{ 1 + \exp\left[ (r - r_0' A^{1/3})/a' \right] \right\}^{-1}$$





and $W_D$ by $W$.

The calculations were carried out using computer code JIB-3 (Perey 1963), which I have modified and adapted to run on the ANU UNIVAC-1108 computer.

The fits to the data were optimised by minimising the function $\chi^2$ defined as:

$$\chi^2 = \frac{1}{N} \sum_{i=1}^{N} \left[ \frac{\sigma_{\exp}(\theta_i) - \sigma_{th}(\theta_i)}{\Delta \sigma_{\exp}(\theta_i)} \right]^2$$

where $N$ is the number of the experimental data points for a given angular distribution, $\sigma_{exp}(\theta_i)$ is the experimental differential cross section measured at the angle $\theta_i$, $\sigma_{th}(\theta_i)$ is the corresponding theoretical cross section, and $\Delta \sigma_{exp}(\theta_i)$ is the error in $\sigma_{exp}(\theta_i)$.

### Investigating the unusually shallow potential

The unusually shallow potential ($V \approx 25 MeV$) was reported for the four-parameter volume-absorption potential, i.e. for the potential with identical geometrical parameters, $r_0 = r_0'$ and $a = a'$ for the real and imaginary components (Hafele, Flynn, and Blair,1967; Flynn, *et al.* 1969.) I have decided to investigate this parameterisation to try to understand the reasons for such a shallow potential.

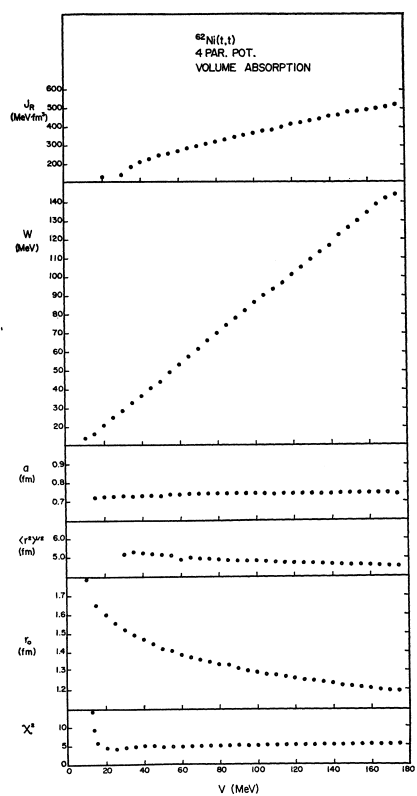

Figure 11.1. Example of the grid search for the four-parameter optical model potential with volume absorption.[12] The figure displays the strong continuous ambiguity of the optical model potential. The plot of the $\chi^2$ function shows that for $V \geq 20$ MeV, any value of $V$ gives equivalent fit to the elastic scattering angular distributions as long as remaining parameters follow the indicated lines.

---

[12] $J_R$ and $\left\langle r^2 \right\rangle_R^{1/2}$ are the volume integral and root-mean-square radius, respectively.





First, I have carried out an automatic search and I have reproduced the earlier results. Next, I have carried out a grid search to see whether there are perhaps less deep minima in the $\chi^2$ function, which might have been skipped in the automatic search. Results of the grid search are shown in Figure 11.1.

The calculations revealed a strong continuous ambiguity extending over a wide range of values for the real potential depths. The $\chi^2$ function displays no clear minima. Indeed, the function has a constant value for $V \geq 20$ MeV but rises sharply for $V \leq 15$ MeV. Any potential depths greater than around 20 MeV will result in equivalent fits to the experimental angular distributions. This continuous ambiguity means that there is no way of determining the preferred set of discrete parameters for the nuclear potential. In this sense, the four-parameter potential gives no useful information about the triton-nucleus interaction.

These results should be compared with the results based on the identical procedure using the surface absorption potential (see Figure 11.2).

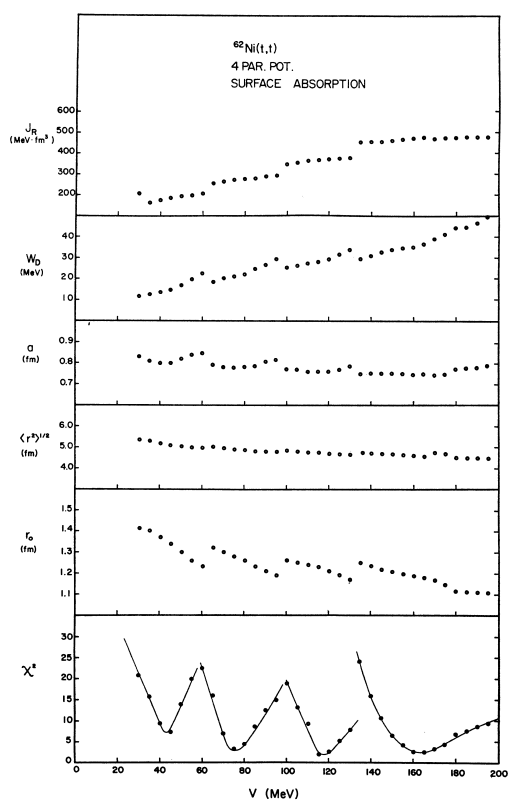

Figure 11.2. Example of the grid search for the four-parameter optical model potential with the *surface* absorption. In contrast with the results presented in Figure 11.1, the volume absorption potential leads to distinct discrete values of parameter sets corresponding to well-defined minima in the $\chi^2$ function.

This figure shows the well-known characteristics of the optical model. Now the $\chi^2$ function has a series of clear and distinct minima, which give a well-defined series of discrete sets of parameters, which optimise the fits to experimental distributions.

It has been shown (Drisko, Satchler and Bassel 1963) that the discrete series of parameters correspond to a different number of oscillations of partial wave functions inside the nucleus. I have therefore decided to investigate this aspect and see





whether there are differences between the volume and surface absorption potentials. Figure 11.3 shows some examples of the calculated wave functions.

The figure shows clear differences in the behaviour of the radial wave functions inside the nucleus for potentials with either volume or surface absorption. Radial wave functions for the volume absorption potential are strongly attenuated inside the nucleus. In contrast, the wave functions for the potential with surface absorption display clear and only weakly attenuated oscillations.

I have also carried out calculations using six-parameter volume absorption potential. It can be also seen in Figure 11.3 that for this parameterisation the attenuation of the wave functions inside the nucleus is much weaker than for the potential with four parameters. The six-parameter potential leads to the usual discrete sets of parameters. It is also worth pointing out that the number of oscillations inside the nucleus depends not just on the discrete set of parameters as pointed out by Drisko, Satchler and Bassel (1963) but also on the $L$ value of the radial wave function.

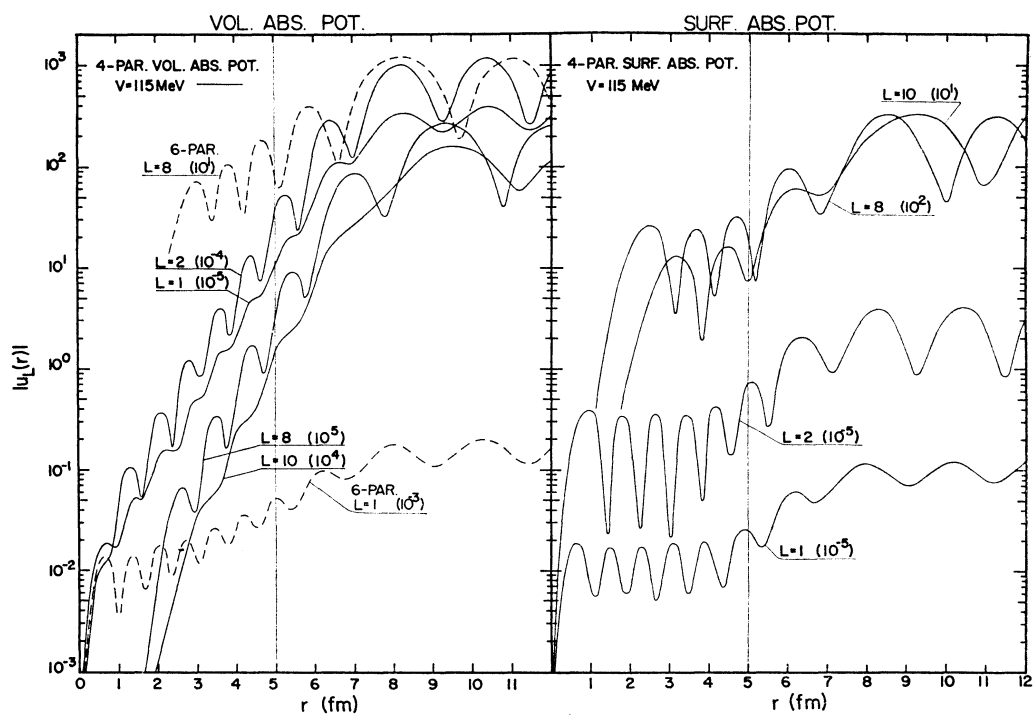

Figure 11.3. Examples of the radial wave functions $|u_L(r)|$ calculated for sets of parameters corresponding to $V$ = 115 MeV for the four-parameter potentials with volume or surface absorptions (the left- and right-hand side of the figure, respectively). The parameter sets are as shown in Figures 11.1 and 11.2. Two examples of radial wave functions for the six-parameter volume absorption potential are also shown on the left-hand side of the figure.

### Fits to the experimental angular distributions

The remaining calculations have been carried out using only surface absorption potentials. In these calculations, I have identified 9 sets of discrete sets of parameters for the four-parameter potential and 7 sets for the six-parameter potential. They all give acceptable fits to the experimental angular distributions.

Examples of fits obtained using set 4 (i.e. with $V \approx 150$ MeV) for the four-parameter potential with surface absorption are shown in Figure 11.4. Excellent fits are





obtained when all four parameters are allowed to vary. However, calculations with fixed geometry give also satisfactory results.

Six parameter potentials give nearly identical results. The quality of fits cannot be distinguished visually but only by the minima in the $\chi^2$ function, which are slightly deeper than for the four-parameter potentials.

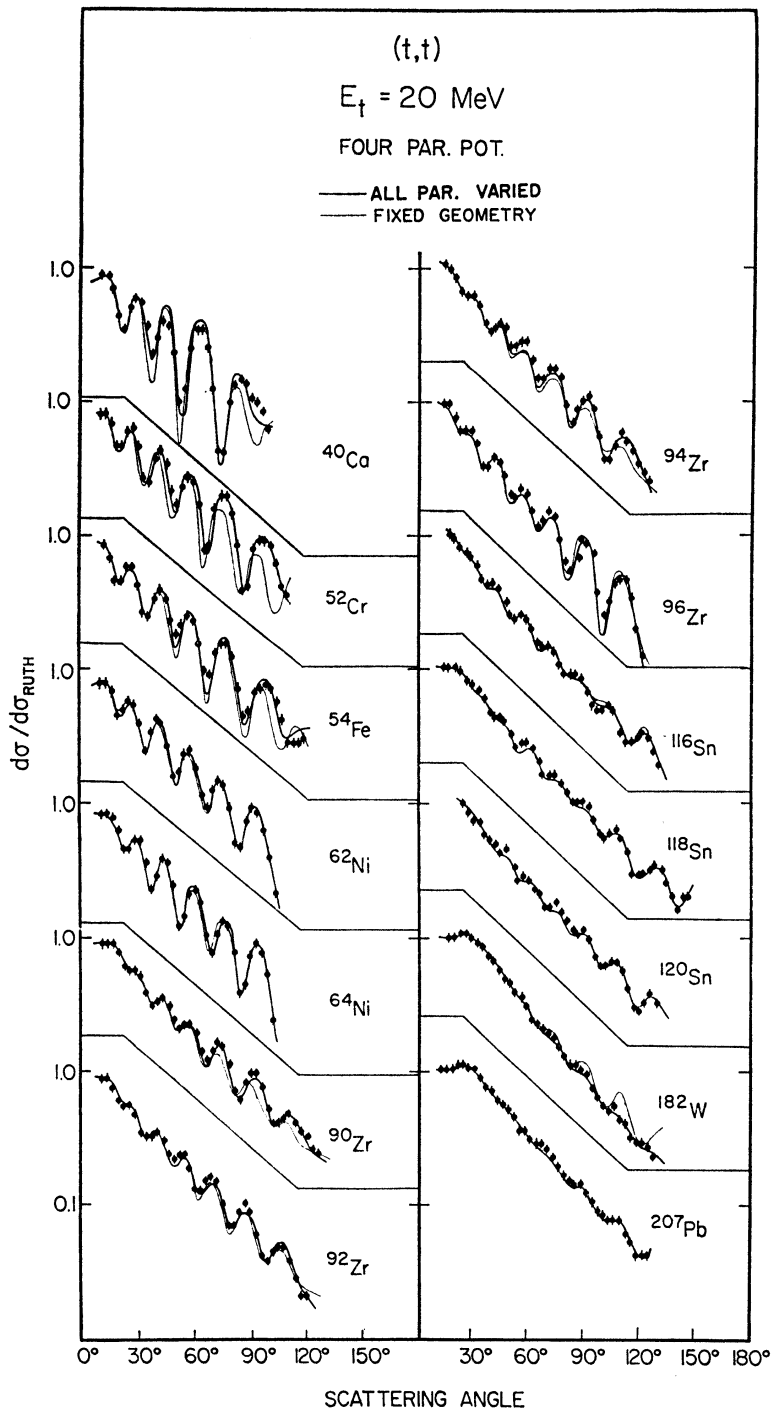

Figure 11.4. Examples of fits to the 20 MeV triton data obtained using the four-parameter optical model potential with surface absorption. Results of the calculations represented by thicker lines were obtained by searching for all parameters. Results represented by thinner lines are for calculations with fixed geometrical parameters $r_0$ = 1.20 fm, and $a$ = 0.755 fm. Both sets of calculations are for $V \approx 150$ MeV.





### *Selecting the best set of parameters*

The next step in the analysis was to try to discriminate between various sets of parameters and see whether the number of sets can be reduced. Visual examination of the fits was ineffective and examination of individual minima of the $\chi^2$ for all the angular distributions and parameter sets was too tedious. (The total number of individual minima was 240.) Accordingly, I have averaged the minimum values of the $\chi^2$ function for all target nuclei for each set of parameters. Results are presented in Figure 11.5. Each family is identified not only by the family number but also by the approximate value of the volume integral $J_R$ calculated using the following expression:

$$J_R = \frac{4\pi V R^3}{3 A A_i} \left[ 1 + \left( \frac{\pi a}{R} \right)^2 \right]$$

where $A_i$ is the atomic mass number of the projectile.

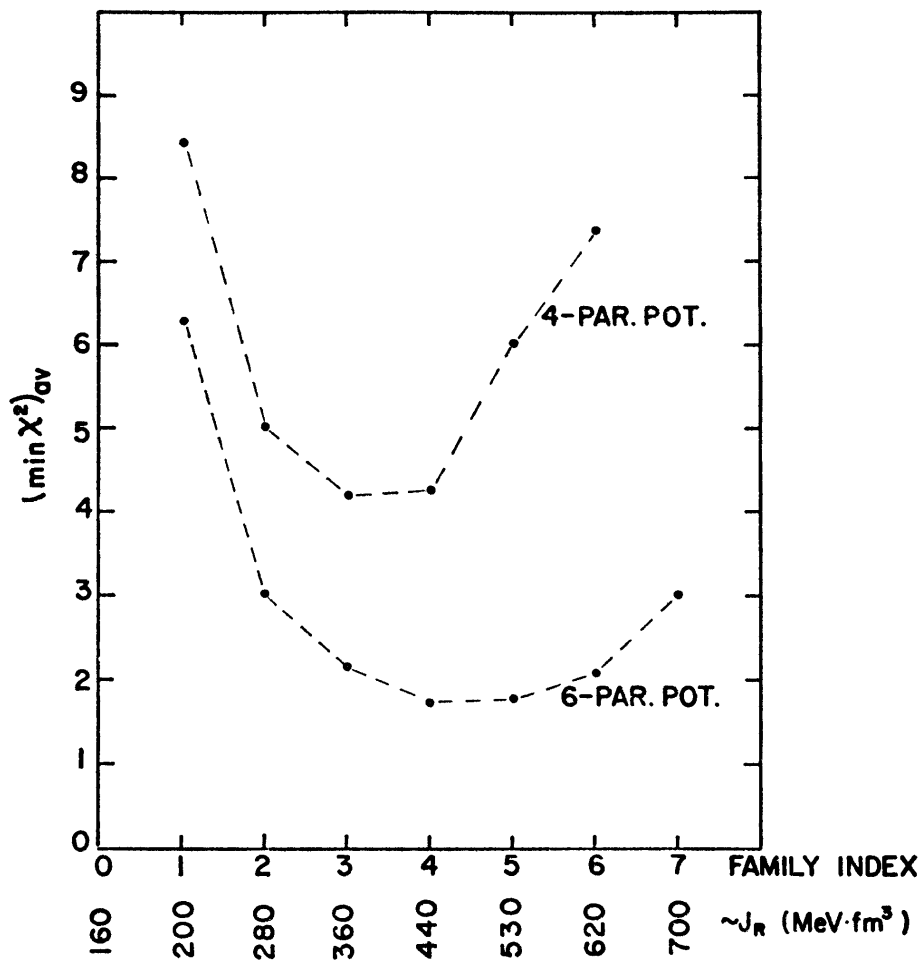

Figure 11.5. Plots of the minima of the $\chi^2$ function averaged over the atomic mass of the target nuclei calculated for each set (family) of the optical model parameters. Only sets with no fixed parameters were used in this calculation. This figure shows that by a careful study of the minima in the $\chi^2$ function it is possible to eliminate many of the possible discrete sets of parameters and select only one or two sets (families). In my analysis, the preferable families of parameters are 3 and 4 for the 4-parameter potential, and 4 and 5 for the 6-parameter potential. The best common set is the family 4, which corresponds to $J_R \approx 440$ MeV·fm$^3$ or $V \approx 150$ MeV.





It can be seen that the number of acceptable families can be limited to two for the four-parameter potential. The best families are 3 and 4 corresponding $J_R \approx 360$ and 440 MeV·fm³ or to $V \approx 100$ and 150 MeV. For the six-parameter potential, the best sets are 4 and 5, but families 3 and 6 give also acceptable fits. The best common family is the family 4 corresponding to $J_R \approx 440$ MeV•fm³ or $V \approx 150$ MeV.

The problem of resolving discrete optical model ambiguities for ³He particles has been discussed in a number of publications. In the analysis of 33 and 53 MeV ³He scattering on ⁵⁷Fe (Marchese 1973) the number of acceptable families was reduced to two, corresponding to $J_R \approx 330$ and $\approx 440$ MeV·fm³. Most results suggested the preference for the $J_R \approx 450$ MeV · fm³ family but some calculations (Chang, *et al.* 1973; Fulmer and Hafele 1972; Fulmer and Hafele 1973a, 1972b) carried out for data collected at sufficiently high energy and in a wide range of angles revealed a unique family with $J_R \approx 330$ MeV · fm³. Only one set of parameters was also found in the analysis of 217 MeV ³He particles scattered from targets ranging from ⁶Li to ²⁰⁸Pb (Willis *et al.* 1973), namely a set corresponding to $J_R \approx 255$ MeV · fm³.

The analysis of 139 MeV α-particle scattering on ¹²C yielded a unique set corresponding to $J_R \approx 353$ MeV · fm³ (Smith *et al.* 1973). Similar results were also reported for α-particle scattering on ²⁴Mg (Duhm 1968; Sinnh *et al.* 1969; Yang *et al.* 1973) and on ⁹⁰Zr (Paans, Put and Malfliet 1973) isotopes.

In general, results obtained so far seem to show some preference for the $J_R \approx 360$ MeV · fm³ family of parameters. However, parameter sets I have identified in this work were also applied in my study of (p,t) reactions on a range of Se isotopes at 33 MeV proton energy. Significantly better fits to the reaction distributions were obtained using sets corresponding to $J_R \approx 450$ MeV · fm³ (i.e. with $V \approx 150$ MeV) rather than to $J_R \approx 350$ MeV · fm³ (i.e. with $V \approx 100$ MeV). Thus, the analysis of our (p,t) results confirmed my conclusion based on the analysis of the (t,t) elastic scattering as discussed above.

### The mass dependence

In fitting nuclear reaction angular distributions, it is often desirable to know the dependence of the optical model parameters on the mass number $A$ of the target nucleus. Parameters derived here displayed an approximately linear dependence on $A$. Tables 11.1 and 11.2 show the formulae obtained by fitting linear functions to the parameters determined from the optical model analysis. The examples are for the families 3 and 4.

Table 11.1

Mass dependence of the optical model parameters for the four-parameter potential with surface absorption

| | Family 3 | | Family 4 | |
|---|---|---|---|---|
| | all parameters varied | fixed geometry | all parameters varied | fixed geometry |
| $V$ | $107.5 + 0.119A$ | $115.7 + 0.089A$ | $156.1 + 0.032A$ | $151.8 + 0.034A$ |
| $W_D$ | $28.03 - 0.0197A$ | $32.09 - 0.055A$ | $36.22 - 0.034A$ | $36.20 - 0.054A$ |
| $r_0$ | $1.25 - 0.00026A$ | $1.20$ | $1.21 - 0.00013A$ | $1.20$ |
| $a$ | $0.735 + 0.00028A$ | $0.755$ | $0.711 + 0.00036A$ | $0.755$ |
| $J_R$ | $371.3 - 0.116A$ | $364.5 - 0.0501A$ | $496.1 - 0.438A$ | $500.9 - 0.416A$ |





Table 11.2

Mass dependence of the optical model parameters of the six-parameter potential with surface absorption

| | Family 3 | | Family 4 | |
|---|---|---|---|---|
| | all parameters varied | fixed geometry | all parameters varied | fixed geometry |
| $V$ | $107.5+0.111A$ | $110+0.0896A$ | $159.2-0.026A$ | $146.6+0.027A$ |
| $r_0$ | $1.25-0.00021A$ | $1.23$ | $1.21+0.00004A$ | $1.23$ |
| $a$ | $0.716+0.00019A$ | $0.720$ | $0.696+0.00014A$ | $0.720$ |
| $W_D$ | $28.5-0.0384A$ | $27.3-0.031A$ | $34.05-0.046A$ | $30.76-0.031A$ |
| $r_0'$ | $1.22-0.00039A$ | $1.15$ | $1.16-0.00033A$ | $1.15$ |
| $a'$ | $0.742+0.00092A$ | $0.850$ | $0.796+0.00028A$ | $0.850$ |
| $J_R$ | $379.4-0.087A$ | $361.4-0.0015A$ | $494.5-0.369A$ | $478.6-0.244A$ |

It can be seen that geometrical parameters depend only weakly on $A$. Furthermore, in contrast with the suggestion of Marchese *et al.* (1972), $J_R$ is not constant. For sets with no fixed parameters, $J_R$ decreases at the rate of $\approx 0.4$ MeV $\cdot$ fm$^3$ per a.m.u. for the family 4 and at $\approx 0.1$ MeV $\cdot$ fm$^3$ per a.m.u. for the family 3. The 0.5 MeV $\cdot$ fm$^3$ rate was reported for $J_R \approx 330$ MeV $\cdot$ fm$^3$ for $^3$He particles (Fulmer and Hafele 1973a, 1973b).

Figure 11.6 shows the dependence of $J_R$ on $A$ for various families with no fixed parameters. The gradient $dJ_R/dA$ varies between families. The difference between the adjacent values of $J_R$ is about 80 MeV $\cdot$ fm$^3$ for $A \approx 120$ and it decreases with the increasing mass of the target nucleus.

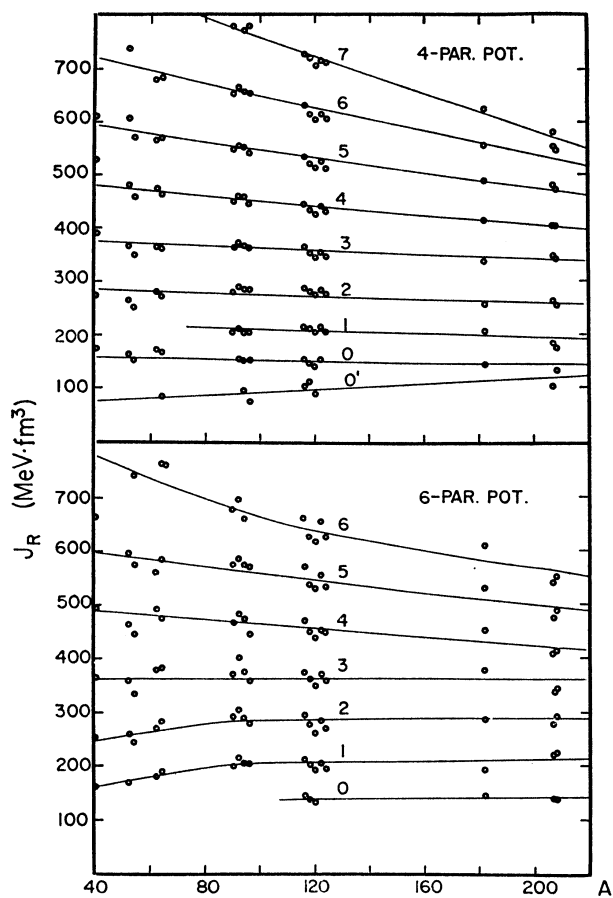

Figure 11.6. Examples of the dependence of the volume integral, $J_R$, on the atomic mass number $A$ of the target nuclei. The family index is used to label separate groups of points. The lines are drawn as a visual aid. The displayed points were calculated for sets with no fixed parameters.





*The symmetry term for tritons*

It has been long recognised that the depth of the real part of the optical model potential depends on a symmetry parameter $\varepsilon = (N - Z)/A$ (Green and Sood 1958; Lane 1958). Lane (1962a, 1962b) has shown that this dependence might be related to an isospin term in the optical model potential. If the potential is calculated as an average sum of the two-body forces, then the real part of the optical model potential can be expressed as:

$$V = V_0 + \frac{\mathbf{t} \cdot \mathbf{T}}{A} V_1$$

where $\mathbf{t}$ and $\mathbf{T}$ are the isotopic spins of the incident particle and the target nucleus, respectively.

By averaging this potential over all allowed values of the total isotopic spin the mean optical model potential can be written in terms of the symmetry parameter $\varepsilon = (N - Z)/A$. For instance, for protons:

$$V = V_0 - \frac{1}{4} \frac{N - Z}{A} V_1$$

The early evidence for such dependence was provided by comparing neutron and proton scattering (Melkanoff 1956; Melkanoff, Nodvik and Saxon 1957) and by shell model calculations (Green 1956; Ross, Mark and Lawson 1956; Ross, Lawson and Mark 1956). It has been also suggested that the symmetry term is complex (Satchler 1967).

Attempts have been made to determine the $\mathbf{t} \cdot \mathbf{T}$ interaction for mass three projectiles using elastic scattering (Becchetti and Greenlees 1971; Drisko *et al.* 1967; Flynn *at al.* 1969; Hafele *et al.* 1967; Urone *et al.* 1972). The common feature of all the previous analyses was the presence of a volume absorption term in the optical model potential. The results are inconclusive and do not form a consistent pattern.

Most analyses seem to indicate zero or weak dependence of the real component on the symmetry parameter, although a strong symmetry term was reported by Becchetti and Greenlees (1971) for [3]He particles.

Comparison of [3]He and triton angular distributions resulted in an imaginary symmetry term similar to that employed in analyses of the ([3]He, t) reaction (Urone *et al.* 1971). The result was, however, obtained by averaging over a wide range of values differing by as much as a factor of about 30. It is also worth noting that, owing to the possible complex multistep mechanism (Schaeffer and Bertsch 1972; Schaeffer and Glendenning 1973; Toyama 1972), the usual DWBA analysis of the reaction ([3]He, t) can hardly serve as a point of reference in assessing the magnitude of the symmetry term. The apparent agreement is therefore most probably accidental.

Considering previous attempts based usually on volume absorption potentials, it was interesting to see whether surface absorption would produce better results. For consistency with the previously published studies I have adopted the recommended method of deriving the symmetry potential from the elastic scattering (Satchler 1969). In this method potentials corresponding to a fixed average geometry are used and compared with a linear dependence on the symmetry parameter $\varepsilon = (N - Z)/A$. Plots of





$V$ and $W_D$ for the two families of parameters are shown in Figure 11.7. The straight lines represent the least-squares fits of the functions:

$$V = V_0 + \varepsilon V_1$$

$$W_D = W_0 - \varepsilon W_1$$

The factor 1/4 is now absorbed in $V_1$ and $W_1$. The relevant numerical values of $V_0$, $V_1$, $W_0$, and $W_1$ are listed in Figure 11.7.

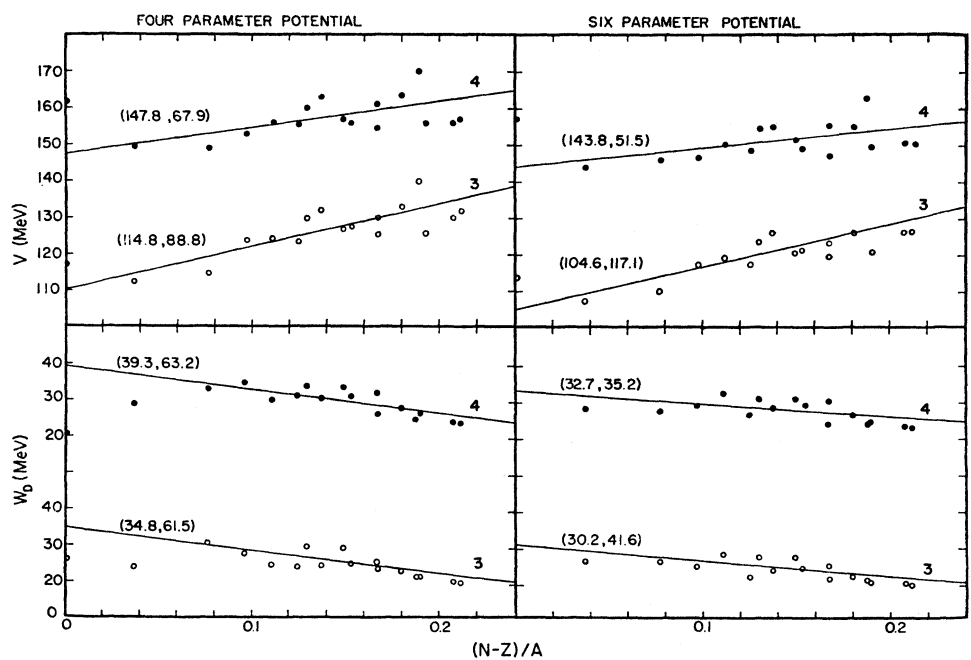

Figure 11.7. The dependence of the real and imaginary potential depths, $V$ and $W_D$, on the symmetry parameter $\varepsilon = (N - Z)/A$ for the elastic scattering of 20 MeV tritons. The potential depths correspond to a fixed averaged geometry $r_0 = 1.20$ fm, $a = 0.755$ fm for the four-parameter potential and $r_0 = 1.23$ fm, $a = 0.72$ fm, $r_0' = 1.15$ fm, $a' = 0.85$ fm for the six-parameter potential. The straight lines are the least-squares fits of the functions $V = V_0 + \varepsilon V_1$ and $W_D = W_0 - \varepsilon W_1$ to the relevant points. The numerical values of $(V_0, V_1)$ and $(W_0, W_1)$ in MeV are displayed on the left-hand side of each fitted line, and the family index on the right-hand side. The optical model with surface absorption was used in these calculations.

The figure shows clear linear dependence of both $V$ and $W_D$ on the symmetry term $\varepsilon$. However, the derived values $V_1$ and $W_1$ do not appear to represent the strengths of the isospin potential because they depend on the parameterisation of the optical model potential.

This simple method of determining isospin interaction, though used in the past, have been criticised by many authors (see for instance Hodgson 1971 and referenced therein). It has been pointed out that not only the depths of nuclear potentials but also geometrical parameters may depend on the isospin interaction, and that it is either difficult or impossible to separate geometrical effects from the purely isospin effects. Isospin interaction may have different components, which are neglected in this simple linear representation. The assumption that nuclear radius is proportional to $A^{1/3}$ is also





questionable. Isospin interaction depends critically on the relative distribution of protons and neutrons, which is neglected in the simple description given by the optical model. There could be also nuclear structure effects that are not accounted for in these simple linear representations of the dependence of potential depths on the symmetry parameter $\varepsilon$.

The average symmetry potentials found in my analysis are $V_1$ = 88.8 MeV and $W_1$= 50.4 MeV.

## Summary and conclusions

A detailed study of the optical model description of the elastic scattering of tritons at 20 MeV incident energy has been carried out. In this work, I have concentrated on studying the surface absorption potential. However, I have also carried out some calculations with volume absorption.

I have shown that the unusually shallow potential reported for the four-parameter optical model potential with volume absorption is associated with a strong attenuation of radial wave functions inside the nucleus. Surface absorption potential gives better description of triton-nucleus interaction.

The analysis yielded nine sets of parameters for the four-parameter potential and seven for the six-parameter potential, or a total of 16 sets. By comparing the averaged minima of the $\chi^2$ functions it was possible to reduce the number of acceptable parameter sets to only two for the four-parameter potential and 2-4 for the six-parameter potential. Only one group of parameters results in the best fits for both types of the optical potential. This group corresponds to the volume integral $J_R \approx$ 440 MeV·fm$^3$ or to $V \approx$ 150 MeV.

In general, volume integral decreases with the mass number. However, for one group of parameters $J_R$ is almost constant. This group corresponds to $J_R \approx$ 360 MeV·fm$^3$ or to $V \approx$ 100 MeV and it gives almost identical fits to the experimental angular distributions as the group corresponding to $V \approx$ 150 MeV.

Formulae for the mass dependence of the optical model parameters have been derived. They show that the dependence of the geometrical parameters on the mass of the target nucleus is weak and consequently, these parameters can be fixed.

A study of the dependence of $V$ and $W_D$ on the symmetry term $\varepsilon = (N - Z) / A$ has been studied. In contrast to most previous analyses for mass-3 projectiles, linear dependence on $\varepsilon$ has been demonstrated. However, the gradient of the linear functions depends on the parameterisation of the optical model. The observed linear dependence on the symmetry term represents most likely mainly the mass dependence of the optical model parameters.

## References

Baer, H. W., Kraushaar, J. J., Moss, C. E., King, N. S. P. and Green, R. E. L. 1970, *Phys. Rev. Lett.* **25**:1035.

Barnard, R. W. and Jones, G. D. 1968a, *Nucl. Phys.* **A106**:497.

Barnard, R. W. and Jones, G. D. 1968b, *Nucl. Phys.* **A108**:641.

Cage, M. E, Clough, D. L., Cole, A. J., England, J. B. A., Pyle, G. J., Rolph, P. M., Watson, L. H. and Worledge, D. H. 1972, *Nucl. Phys.* **A183**:449






Chang, H. H. *et al,* 1973, University of Colorado, Nuclear Physics Laboratory Technical Report COO-535-693:14

Drisko, R. M., Satchler, G. R. and Bassel, R. H. 1963*, Phys.* Lett. 5:347.

Drisko *et al,* R. M. 1967, *Proc. Int. Conf. on Nuclear Structure*, Tokyo, p. 347.

Duhm, H. H. 1968, *Nucl. Phys.* **A118**:563.

Flynn, E. R., Armstrong, D. D., Beery, J. G. and Blair, A. G.1969, *Phys. Rev.* **182**:1113.

Fulmer, C. B. and Hafele, J. C. 1972, *Phys. Rev.* **C5**:1969.

Fulmer, C. B. and Hafele, J. C. 1973a, Phys. Rev. **C7**:631;

Fulmer, C. B. and Hafele, J. C. 1973a, Phys. Rev. **C8**:172

Glover, R. N. and Jones, A. D. 1966, *Nucl. Phys.* **81**:268.

Green, A. E. S. 1956, *Phys. Rev.* **102**:1325

Green A. E. S. and Sood, P. C. 1958, *Phys. Rev.* **111**:1147

Hafele, J. C., Flynn, E. R. and Blair, A. G. 1967, *Phys. Rev.* **155**:1238.

Lane, A. M. 1958, *Proc. Int. Conf. on Nuclear Physics*, Paris, p. 32.

Lane, A. M. 1962a, *Phys. Rev. Lett.* **8**:171.

Lane, A. M. 1962b, *Nucl. Phys.* **35**:676.

Marchese, C. J. 1972, *Nucl. Phys.* **A191**:627.

Marchese**,** C. J., Clarke, N. M. and Griffiths, R. J. 1973, *Nucl. Phys.* **A202:**421.

Melkanoff, M. A., Moszkowski, S. A., Nodvik, J. and Saxon, D. S. 1956, *Phys. Rev.* **101**:507

Melkanoff, M. A., Nodvik, J. and Saxon, D. S. 1957, *Phys. Rev.* **106**:793

Hodgson, P. E. 1971, *Nuclear Reactions and Nuclear Structure*, Clarendon Press, Oxford.

Paans, A. M. J., Put, L. W. and Malfliet, R. A. R. L. 1973, *Proc. Int. Conf. Nuclear Physics*, Munich, ed. J. de Boer and H. J. Mang (North-Holland, Amsterdam,) p. 340

Perey, F. G. 1963, *Phys. Rev.* **131**:745.

Ross, A. A., Mark, H. and Lawson, R. D. 1956, *Phys. Rev.* **102**:1613.

Ross, A. A. Lawson, R. D. and Mark, H. 1956, *Phys. Rev.* **104**:401.

Satchler, G. R. 1967, *Nucl. Phys.* **A91**:75.

Satchler, G. R. 1969, Isospin in Nuclear Physics, ed. D. H. Wilkinson (North-Holland, Amsterdam,) p. 389.

Schaeffer, R. and Bertsch, G. F. 1972, *Phys. Lett.* **38B**:159.

Schaeffer, R. and Glendenning, N. K. 1973, *Nucl. Phys.* **A207**:321.

Siegel, G. R. 1971, Ph.D. thesis, Washington University, St. Louis.

Singh, P. P., Malmin, R. E., High, M. and Devins, D. W. 1969, *Phys. Rev. Lett.* **23**:1124.

Smith, S. M., Tebell, G., Cowley, A. A., Goldberg, D. A., Pugh, H. G., Reichart, W. and Wall, N. S. 1973, *Nucl. Phys.* **A207**:273.

Toyama, M. 1972, *Phys. Lett.* **38B**:147.

Urone, P. P., Put, L. W., Chang, H. H. and Ridley, B. W. 1971a, *Nucl. Phys.* **A163**:225.

Urone, P. P., Put, L. W., Ridley, B. W. and Jones, G. D. 1971b, *Nucl. Phys.* **A167**:1383.

Urone, P. P., Put, L. W., Chang, H. H. and Ridley, B. W. 1972, *Nucl. Phys.* **A186**:344.

Willis, N., Brissaud, I., Le Bornec, Y., Tatischeff, B. and Duhamel, G. 1973, *Nucl. Phys.* **A204**:454.

Yang *et al,* G. G. 1973, *Proc. Int. Conf. Nuclear Physics*, Munich, ed. J. de Boer and H. J. Mang (North-Holland, Amsterdam,) p. 338








# The $^{54}$Cr(d,t)$^{53}$Cr and $^{67,68}$Zn(d,t)$^{66,67}$Zn Reactions at 12 MeV

***Key features:***

1. We have measured a total of 35 angular distributions for the neutron pickup reaction induced by 12 MeV deuterons: 5 for the $^{54}$Cr(d,t)$^{53}$Cr reaction, 13 for $^{67}$Zn(d,t)$^{66}$Zn and 17 for $^{68}$Zn(d,t)$^{67}$Zn.

2. Particle identification was done by using a $\Delta E$-$E$ particle identification technique.

3. We have carried out theoretical analysis of the experimental differential cross sections using direct reactions theory and a computer code, which I have modified and adapted to run on an ANU computer.

4. We have demonstrated *j*-dependence for *l* = 1 and 3 transitions. However, the observed *j*-dependence could not be reproduced theoretically even if spin-orbit potentials were used in the calculations for deuterons and tritons.

5. We have assigned spins and parities to states in residual nuclei and extracted spectroscopic factors.

6. We have extracted information about configuration mixing and compared it with theoretical calculations using pairing theory and a computer code I have written for this purpose.

***Abstract:*** Angular distributions for the (d,t) reaction on $^{54}$Cr and $^{67,68}$Zn target nuclei have been measured using 12 MeV deuterons. Experimental results were analysed theoretically using direct nuclear reactions theory. Spin assignments have been made to states in residual nuclei and spectroscopic factors have been extracted. The *j*-dependence has been observed but it could not be reproduced theoretically. Configuration mixing for the f-p shell has been studied using experimental spectroscopic factors. Results are compared with predictions of the pairing theory.

## Introduction

The nuclei $^{54}$Cr, $^{67}$Zn, and $^{68}$Zn have 2, 6, and 10 neutrons outside the closed $N = 28$ shell. I have chosen these nuclei to extend the study of nuclear spectroscopy in the f-p shell. Without residual interaction, the 2p$_{3/2}$ orbit should be half full in the ground state of $^{54}$Cr; there should be one 1f$_{5/2}$ neutron hole in $^{67}$Zn; and orbits 2p$_{3/2}$ and 1f$_{5/2}$ should be full in $^{68}$Zn. Residual interaction causes configuration mixing and one should expect to observe states belonging to configurations 2p$_{3/2}$, 1f$_{5/2}$, 2p$_{1/2}$ and possibly even to 1g$_{9/2}$ for all these target nuclei.

The (d,t) neutron pickup reaction has not been explored well enough in this region. Fulmer and Daehnick (1964, 1965) examined the *l* =1 transitions in $^{55}$Fe, $^{59}$Ni, and $^{63}$Ni nuclei using the (d,t) reactions and reported a deep minimum at backward angles for *j* = 1/2 similar to that observed in (d,p) reactions. They also observed some indication of differences at forward angles between two *j* values for *l* = 3 transitions in $^{59}$Ni but their measurements for this transition did not extend beyond about 60$^0$. If confirmed, this feature could serve as a tool in assigning not only orbital angular momenta to excited states but also the spin values. There was also no previous attempt to reproduce theoretically the observed (d,t) *j*-dependence.





## Experimental procedure

Measurements of angular distributions were carried out using the ANU EN tandem accelerator. The beam intensity on the target varied between 15 nA and 350 nA. The experimental arrangement was similar to that used in the $^{26}$Mg($^{3}$He,$\alpha$)$^{25}$Mg measurements (see Chapter 5). However, in order to separate triton groups from deuterons and protons and thus to obtain clear triton spectra, $\Delta E$-$E$ detector telescope assemblies, made of silicon surface barrier detectors, were used. Care has been taken to select suitable thickness of the $\Delta E$ detectors to optimise the mass resolution. The $\Delta E$ detectors were 40-100 $\mu$m thick and $E$ detectors 2 $\mu$m thick. To ensure that the $\Delta E$ detectors were fully depleted, a bias voltage of at least 10% higher than required for the total depletion was applied. We have used three such telescope assemblies in our measurements. The electronics of the particle identification system is shown in Figure 12.1.

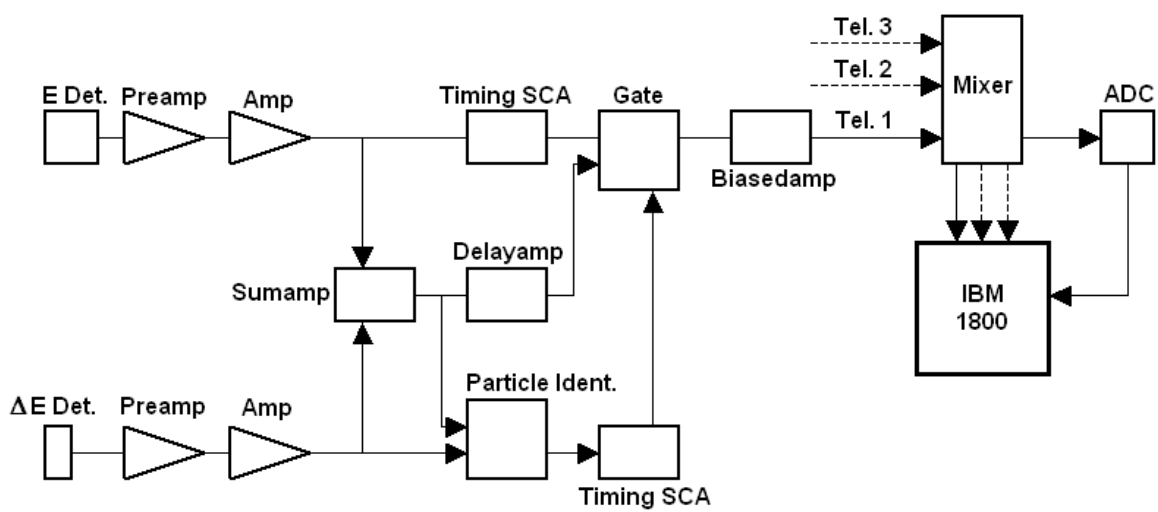

Figure 12.1. The electronics diagram of the particle identification system used in the measurements of angular distributions for the (d,t) reaction at 12 MeV incident deuteron energy. Preamp – Preamplifier; Amp – Amplifier; Sumamp – Summing Amplifier; Timing SCA – Timing Single Channel Analyser; Dalayamp – Delay Amplifier; Particle Ident. – Particle Identifier; Gate – Linear Gate and Slow Coincidence; Biasedamp – Biased Amplifier; Mixer – Mixer and Routing Unit; ADC – Analogue-to-Digital Converter; IBM 1800 – IBM 1800 computer.

The output pulses from the particle identifier, representing the mass spectrum, were fed into a timing single channel analyser (TSCA). The threshold levels of the TSCA were set so that only the triton pulses were allowed to pass through a linear gate and slow coincidence unit. Pulses from the linear gates for all three particle telescopes were fed into a mixer and routing system. The mixed unipolar pulses were then fed into an analogue-to-digital converter, which was interfaced with the IBM 1800 computer. The total energy pulses were separated according to their associated routing signals and stored in different memory locations in the computer. A typical particle spectrum is shown in Figure 12.2.

The targets were about 100 $\mu$g/cm$^2$ thick. They were supported by an approximately 30 $\mu$g/cm$^2$ carbon backing. Measurements of the angular distributions were carried out for a wide range of angles, from 12.5$^0$ to 150$^0$ in steps of 2.5$^0$.





A total of 35 angular distributions have been measured: 5 for the $^{54}$Cr(d,t)$^{53}$Cr reaction, 13 for $^{67}$Zn(d,t)$^{66}$Zn and 17 for $^{68}$Zn(d,t)$^{67}$Zn. They are presented in Figures 12.4 to 12.8.

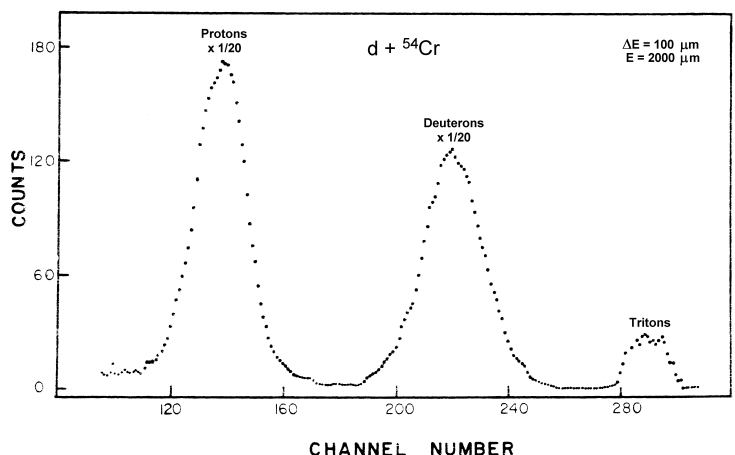

Figure 12.2. A typical mass spectrum for the interaction of the 12 MeV deuterons with $^{54}$Cr leading to the (d,d'), (d,p), and (d,t) reactions.

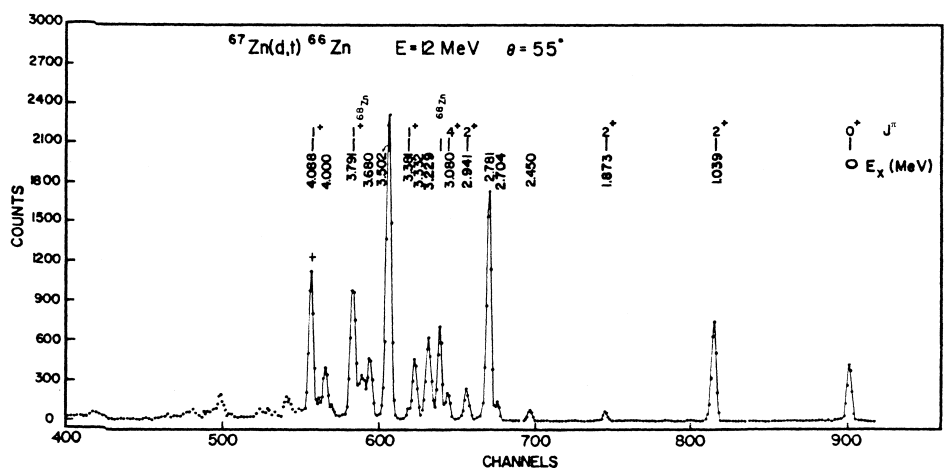

Figure 12.3. An example of a triton spectrum.

## Theoretical analysis

Theoretical analysis of the data was carried out using the computer code DWUCK (Kunz 1966), which I have modified and adapted to run on the Australian National University UNIVAC 1108 computer. This program allows for the calculation of the finite range and non-locality corrections (see the Appendix E).

The optical model potential was made of the central part with the surface absorption (see Chapter 5). Deuteron parameters (i.e. the parameters for the incident channel) were derived from the mass dependent formulae of Perey and Perey (1963). The triton potential (for the outgoing channel) was based on my analysis of the 20 MeV triton scattering (see Chapter 11). The potential depths $V$ and $W_D$ were adjusted using the gradients of $dV/dE$ = -0.15 and $dW_D/dE$ = -0.50 determined from $^3$He scattering (Chang and Ridley 1971) to match the energy of tritons from the (d,t) reactions at 12 MeV.





Both deuteron and triton potentials did not contain the spin-orbit interactions. We have carried out calculations with spin-orbit interaction in the incident and outgoing channels but its effect on the calculated angular distributions was negligible. In some cases, spin-orbit interaction resulted in making the fits worse.

Theoretical calculations are compared with experimental results in Figures 12.4 to 12.8. The corresponding spectroscopic information is summarised in Tables 12.1 to 12.3. They contain spin and parity assignments and experimentally extracted spectroscopic factors.

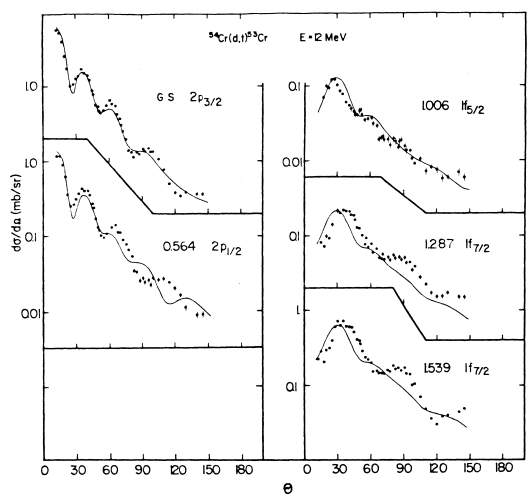

Figure 12.4. Angular distributions for the reaction $^{54}$Cr(d,t)$^{53}$Cr at 12 MeV deuteron energy compared with theoretical calculations.

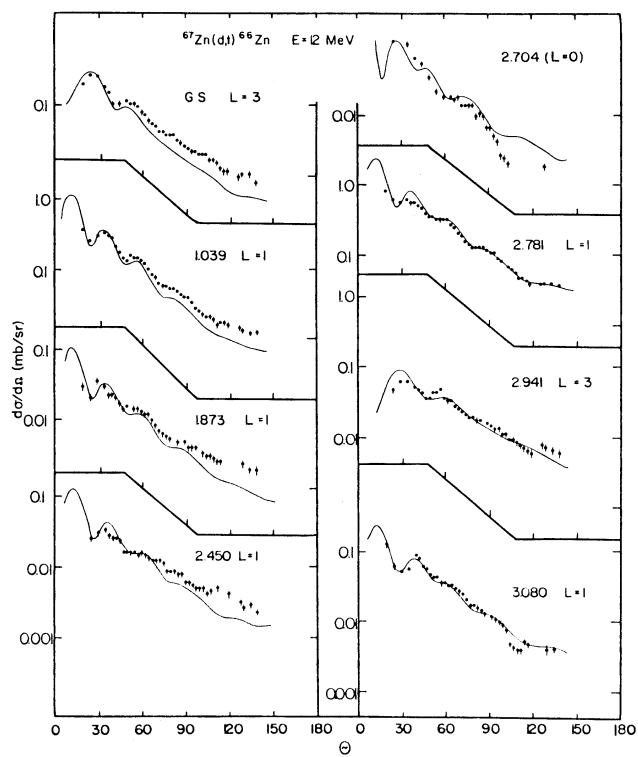

Figure 12.5. Angular distributions for the reaction $^{67}$Zn(d,t)$^{66}$Zn at 12 MeV deuteron energy compared with theoretical calculations.





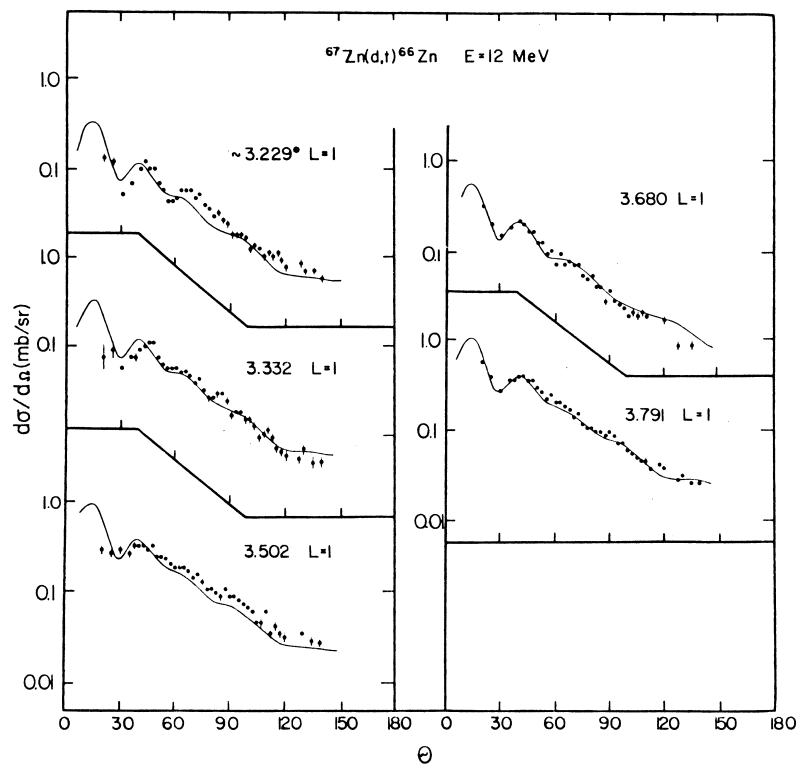

Figure 12.6. Angular distributions for higher excited states for the reaction $^{67}$Zn(d,t)$^{66}$Zn at 12 MeV deuteron energy compared with theoretical calculations.

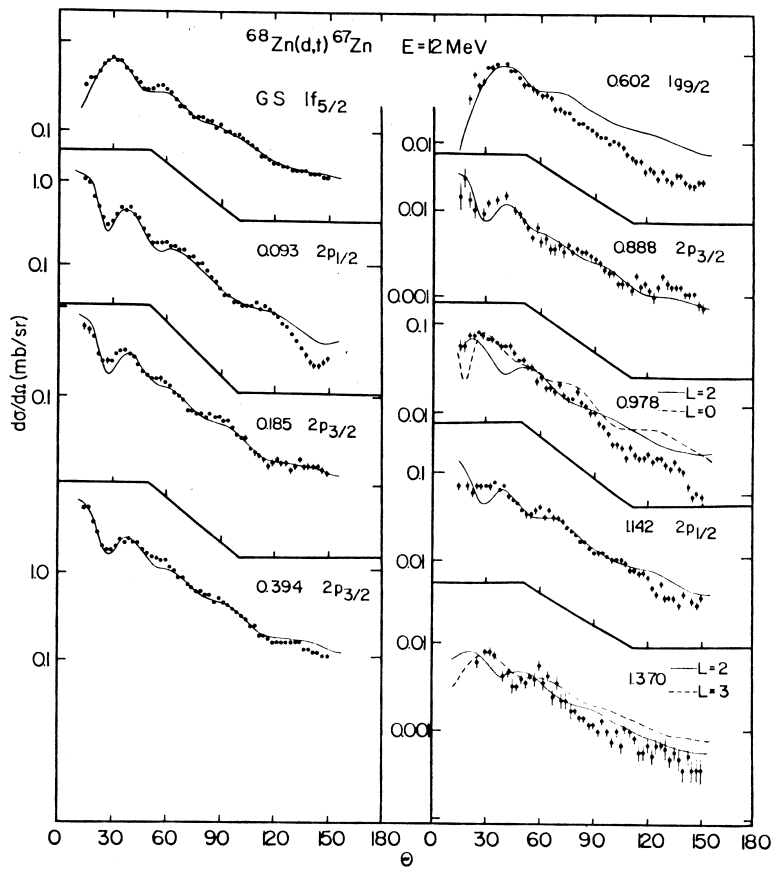

Figure 12.7. Angular distributions for the reaction $^{68}$Zn(d,t)$^{67}$Zn at 12 MeV deuteron energy compared with theoretical calculations.





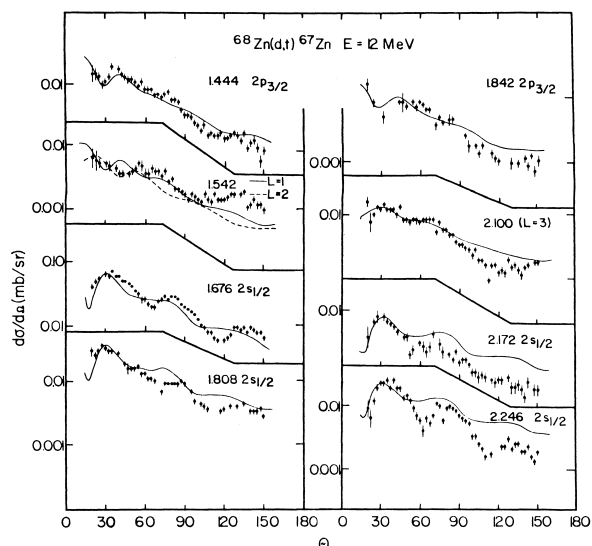

Figure 12.8. Angular distributions for higher excited states for the reaction $^{68}$Zn(d,t)$^{67}$Zn at 12 MeV deuteron energy compared with theoretical calculations.

Table 12.1

Summary of the results for the single neutron pickup reactions to states in $^{53}$Cr

| $E_x$ (MeV) | $J^\pi$ | | | (d,t) [a] | (d,t) [b] | (p,d) [c] | ($^3$He,α) [d] |
|---|---|---|---|---|---|---|---|
| | | $l$ | $j$ | $S$ | $S$ | $S$ | $S$ |
| 1 | 3/2 | | | | 0.61 | 0.83 | 1.10 |
| | 1/2 | | | 0.24 | 0.22 | 0.31 | 0.26 |
| 1.006 | 5/2⁻ | 3 | 5/2 | 0.54 | 0.31 | 0.51 | 0.49 |
| 1.287 | 7/2⁻ | 3 | 7/2 | 0.68 | 0.61 | 0.70 | 0.45 |
| 1.539 | 7/2⁻ | 3 | 7/2 | 2.30 | 1.80 | 3.20 | 3.00 |

$E_x$ – Excitation energy of levels in $^{53}$Cr. $J^\pi$– Adopted spins and parities.
$l$ – the angular momentum of transferred neutron.
$j$ – the spin of transferred neutron ($j = l \pm 1/2$.) $S$ – spectroscopic factors.
a) Our results. [b]) Fitz *et al.* 1967. [c]) Whitten 1967. [d]) David *et al.* 1969.

Table 12.2

Summary of the results for the single neutron pickup reactions to states in $^{66}$Zn

| $E_x$ (MeV) | $J^\pi$ | (d,t) | | (p,d) | |
|---|---|---|---|---|---|
| | | $l$ | $S$ | $l$ | $S$ |
| 0.000 | 0⁺ | 3 | 0.27 | 3 | 0.44 |
| 1.039 | 2⁺ | 1 | 0.10 | 1 | 0.10 |
| 1.873 | 2+ | 1 | 0.01 | 1 | 0.01 |
| 2.450 | 4⁺ | (1) | 0.02 | (1) | 0.02 |
| 2.704 | (2,3)⁻ | (0) | 0.06 | | |
| 2.781 | (1,2)⁺ | 1 | 0.55 | 1 | 0.52 |
| 2.941 | 4⁺ | 3 | 0.22 | 3 | 0.47 |
| 3.080 | | 1 | 0.06 | 1 | 0.10 |
| 3.229 | 1⁺ | 1 | 0.12 | 1 | 0.15 |
| 3.332 | (1,2)⁺ | 1 | 0.10 | 1 | 0.14 |
| 3.502 | | 1 | 0.39 | 1 | 0.30 |
| 3.680 | | 1 | 0.28 | 1 | 0.30 |
| 3.791 | 1⁺ | 1 | 0.56 | 1 | 0.80 |

$E_x$ – Excitation energy of levels in $^{53}$Cr. $J^\pi$– Adopted or possible spins and parities.
$l$ – the angular momentum of transferred neutron. $S$ – spectroscopic factors.
(d,t) – Our results. (p,d) – McIntyre 1966.





Table 12.3

Summary of the results for the single neutron pickup reactions to states in $^{67}$Zn

| $E_x$ (MeV) | $J^\pi$ | (d,t) | | | (p,d) | | |
|---|---|---|---|---|---|---|---|
| | | $l$ | $j$ | $S$ | $l$ | $j$ | $S$ |
| 0.000 | 5/2⁻ | 3 | 5/2 | 3.87 | 3 | 5/2 | 3.80 |
| 0.093 | 1/2⁻ | 1 | 1/2 | 0.58 | 1 | 1/2 | 0.40 |
| 0.185 | 3/2⁻ | 1 | 3/2 | 0.20 | 1 | 3/2 | 0.19 |
| 0.394 | 3/2⁻ | 1 | 3/2 | 1.90 | 1 | 3/2 | 1.76 |
| 0.602 | 9/2⁺ | 4 | 9/2 | 0.88 | (4) | | 0.90 |
| 0.888 | 3/2⁻ | 1 | 3/2 | 0.03 | (1) | | 0.04 |
| 0.978 | (5/2⁺) | (0,2) | | 0.39 ᵃ⁾ | (1) | | 0.06 |
| 1.142 | 1/2⁻ | 1 | 1/2 | 0.19 | 1 | 1/2 | 0.18 |
| 1.370 | | (2,3) | | 0.02 ᵃ⁾ | | | |
| 1.444 | 3/2⁻ | 1 | 3/2 | 0.05 | | | |
| 1.542 | (3/2⁻) | (1,2) | (3/2) | 0.03 ᵇ⁾ | | | |
| 1.676 | 1/2⁺ | 0 | 1/2 | 0.32 | | | |
| 1.808 | (1/2⁺) | (0) | (1/2) | 0.19 | | | |
| 1.842 | (3/2⁻) | 1 | (3/2) | 0.05 | | | |
| 2.100 | | (3) | | 0.42 | | | |
| 2.172 | (1/2⁺) | (0) | (1/2) | 0.10 | | | |
| 2.246 | (1/2⁺) | (0) | (1/2) | 0.28 | | | |

$E_x$ – Excitation energy of levels in $^{67}$Zn.    $J^\pi$– Probable spins and parities.
$l$ – the angular momentum of transferred neutron.
$j$ – the spin of transferred neutron ($j = l \pm 1/2$.)    $S$ – spectroscopic factors.
(d,t) – Our results.   (p,d) – McIntyre 1966.   ᵃ⁾ For $l = 2$.   ᵇ⁾ For $l = 1$.

## *The j-dependence*

I have selected the nucleus $^{54}$Cr for our study because it is a good candidate for studying the *j*-dependence in the (d,t) reactions. Spins and parities of the low-energy states in $^{53}$Cr are well known and they belong to both $l = 1$ and 3 orbital angular momenta.

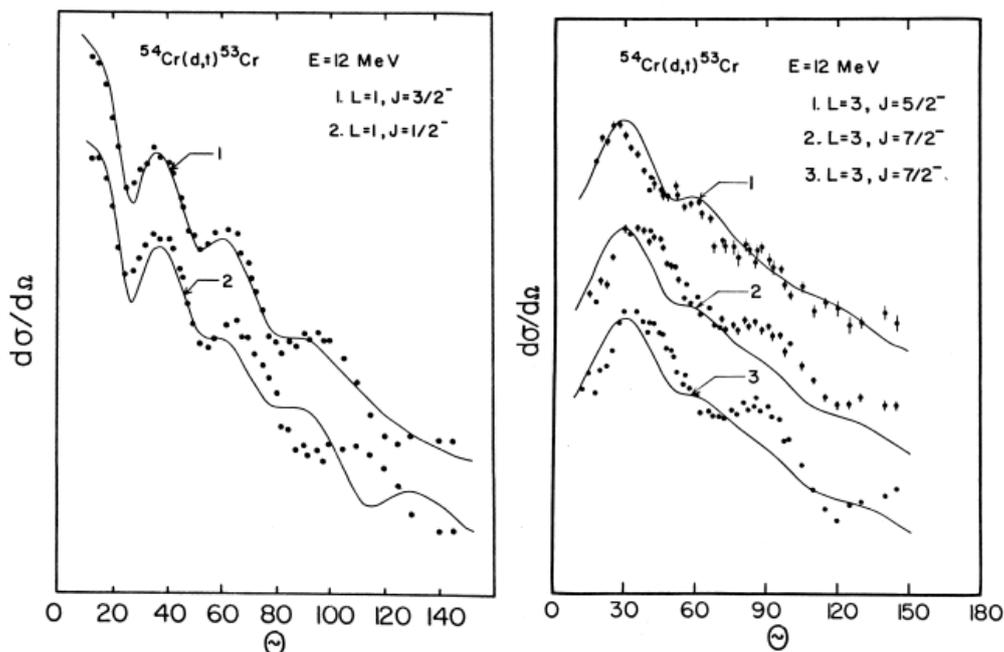

Figure 12.9. Experimentally observed *j* - dependence for the $^{54}$Cr(d,t)$^{53}$Cr reaction for $l = 1$ and 3 transitions. (Compare the differential cross sections at backward angles.)





The ground state and the first excited state have spins $^3/_2$ and $^1/_2$, respectively, belonging to $l = 1$, whereas the third and the fourth excited states have spins $^5/_2$ and $^7/_2$ belonging to $l = 3$. The fourth excited state also has spin $^7/_2$. Thus, the reaction $^{54}$Cr(d,t)$^{53}$Cr offers a good opportunity to study the $j$-dependence for both $l = 1$ and 3 transitions.

Figure 12.9 compares angular distributions for $l = 1$ and 3. It can be seen that for $l = 1$, the last maximum and minimum for the $^3/_2$ angular distribution is shifted to larger angles for the $^1/_2$ distribution. For $l = 3$, the distributions corresponding to $j = ^7/_2$ have more pronounced structure at large reaction angles than the distribution corresponding to $j = ^5/_2$. However, these experimentally observed signatures were not reproduced theoretically even when spin-orbit potentials were used for deuterons and tritons.

### Configuration mixing

Using spectroscopic factors for the pickup reactions one can calculate configuration mixing in the ground state wave function of the target nucleus caused by a residual interaction. Experimentally, occupation numbers, $V_j^2$, and the centre-of-gravity energies, $E_j$, are given by:

$$V_j^2 = \frac{\sum_i S_i(l, j)}{2j + 1}$$

$$E_j = \frac{\sum_i S_i(l, j) E_x^{(i)}(j)}{\sum_i S_i(l, j)}$$

where $S_i(l, j)$ is the spectroscopic factor for $(l, j)$ pickup to state $i$, and $E_x^{(i)}(j)$ is the excitation energy of the $i$th state of the residual nucleus with the total angular momentum $j$.

The sum is over all states belonging to $(l, j)$ configuration. The sum of all spectroscopic factors should be equal to the number of neutrons outside the closed shell.

$$n = \sum_{ij} S_i(l, j)$$

Measurements for the $^{68}$Zn(d,t)$^{67}$Zn reaction show clearly that the ground state of $^{68}$Zn is a mixture of all four configurations, 2p$_{3/2}$, 1f$_{5/2}$, 2p$_{1/2}$, and 1g$_{9/2}$, located outside the $N = 28$ closed neutron shells.

Experimentally determined occupation numbers, $V_j^2$, and centre-of-gravity single particle energies, $E_j$, are listed in Table 12.4. They are compared with theoretical values calculated using the pairing theory of Kisslinger and Sorensen (1960, 1963) and a computer code I have written for this purpose.





According to this theory

$$V_j^2 = \frac{1}{2}\left\{1 - \frac{\varepsilon_j - \lambda}{\left[(\varepsilon_j - \lambda)^2 + \Delta^2\right]^{1/2}}\right\}$$

$$E_j = \left[(\varepsilon_j - \lambda)^2 + \Delta^2\right]^{1/2} - \Delta$$

where $\varepsilon_j$, $\lambda$, and $\Delta$ are the single particle energies, the chemical potential and the gap parameter, respectively, all defined by Kisslinger and Sorensen (1960, 1963). The single particle energies are calculated using their relations. The parameters $\lambda$, and $\Delta$ were determined by solving the gap equations of Kisslinger and Sorensen (1960, 1963).

Table 12.4

Occupation numbers and centre-of-gravity energies for ground state of $^{68}$Zn

| $lj$ | $\varepsilon_j$ | $n_{jl}$ | $n'_{jl}$ | $V_j^2$ | | $E_j$ | |
|------|------|------|------|------|------|------|------|
| | | | | Exp. | Theory | Exp. [a]) | Theory |
| 2p$_{3/2}$ | -0.043 | 2.76 | 4.00 | 0.69 | 0.85 | 0.44 | 0.58 |
| 1f$_{5/2}$ | 0.371 | 5.23 | 6.00 | 0.87 | 0.79 | 0.21 | 0.32 |
| 2p$_{1/2}$ | 2.742 | 0.94 | 0.00 | 0.47 | 0.16 | 0.35 | 0.54 |
| 1g$_{9/2}$ | 2.758 | 1.07 | 0.00 | 0.11 | 0.10 | 0.60 | 0.55 |

$j$ – single particle configuration.

$\varepsilon_j$ – calculated single particle energies.

$n_{jl}$ – The experimentally determined average number of neutrons in the configuration $jl$.

$n'_{jl}$ – The number of neutrons expected in the absence of the residual interaction.

Exp. – experimental values as derived from the $^{68}$Zn(d,t)$^{67}$Zn reaction at 12 MeV.

Theory – theoretical values calculated using the pairing theory of Kisslinger and Sorensen (1960, 1963) and my computer code.

Results of our study show that the population of the configuration 1f$_{5/2}$ is close to the expected value of 6. However, the population of the 2p$_{1/2}$ configuration is significantly higher than that predicted by the pairing theory and the population of 2p$_{3/2}$ slightly lower.

About 52% of all neutrons outside the closed shell $N$ = 28 occupy the 1f$_{5/2}$ orbit, which is close to the value of 60% that would be expected if there were no residual interaction. However, the orbit p$_{1/2}$, which would have been empty without residual interaction, is now nearly 50% full. Even the remote orbit 1g$_{9/2}$ has 11% of its allowed population filled in. Both orbits, 2p$_{1/2}$ and 1g$_{9/2}$, are populated mainly at the expense of the 2p$_{3/2}$ configuration, which donates 31% of its neutrons to other orbits. The 1f$_{5/2}$ donates only 13% of its neutrons.





## Summary and conclusions

We have measured a total of 35 angular distributions of the differential cross sections for neutron pickup (d,t) reaction using 12 MeV deuterons and target nuclei belonging the f-p shell ($^{54}$Cr, $^{67,68}$Zn). We have carried out theoretical analysis, extracted spectroscopic factors, and assigned spins and parities to the corresponding states in the final nuclei.

We have observed $j$ - dependence for both $l$ = 1 and 3 orbital angular momenta. However, the experimentally observed $j$ - dependence could not be reproduced using the direct reaction theory even when spin-orbit interaction was included in the calculations in the incident and outgoing channels.

Using the experimentally determined spectroscopic factors we have calculated the occupations numbers for orbits in the ground state of $^{68}$Zn. In the absence of a residual interaction, neutrons would occupy only configurations $2p_{3/2}$ and $1f_{5/2}$. However, we have found that orbits $2p_{3/2}$ and $1f_{5/2}$ are only 69% and 87% full, respectively. Neutrons located outside the closed $N$ = 28 shell spend also considered time in orbits $2p_{1/2}$ and $1g_{9/2}$. These orbits are 47% and 11% full, respectively.

## References


Chang, H. H. and Ridley, B. W. 1971, University of Colorado Report, COO-535-635:55.

David, P., Duhm, H. H., Bock, R. and Stock, R. 1969, *Nucl. Phys.* **A128**:47

Fitz, W., Hegar, R., Jahr, R. and Santo., R. 1967, *Z. Phys.* **202**:109.

Fulmer, R. H. and Daehnick, W. W. 1964, *Phys. Rev. Lett.* **12**:455.

Fulmer, R. H. and Daehnick, W. W. 1965, *Phys. Rev.* **B139**:579.

Kisslinger, L. S. and Sorensen, R. A. 1960, *Kgl. Danske Videskab. Selskab. Mat. Fys. Medd.* 32, No. 9.

Kisslinger, L. S. and Sorensen, R. A. 1963, *Rev. Mod. Phys.* **35**:853.

Kunz, P. D. 1966, University of Colorado Report COO-536-606.

McIntyre, L. C. 1966, *Phys. Rev.* **152**:1013.

Perey, C. M. and Perey, F. G. 1963, *Phys. Rev.* **132**:755.

Whitten, Jr, C. A. 1967, *Phys. Rev.* **156**:1228.






<div style="text-align:center">

**13**

# A Study of the $^{76,78}$Se(p,t)$^{74,76}$Se Reactions at $E_p$ = 33 MeV

</div>

**Key features:**

1. We have observed a total of 53 states in $^{74}$Se and $^{76}$Se (28 states in $^{74}$Se and 25 in $^{76}$Se) and we have assigned excitations energies to all of them.

2. We have found that the (p,t) reaction is strongly selective – it leads mainly to the ground states and the first excited states in the residual nuclei. Other states are excited with significantly smaller intensity.

3. We have measured a total of 38 angular distributions of the differential cross sections for the reactions $^{76,78}$Se(p,t)$^{74,76}$Se. Measurements were carried out using a $\Delta E$-$E$ particle identification technique.

4. We have analysed the angular distributions using direct reactions theory.

5. We have assigned a total of 35 spin-parity values to states in $^{74}$Se and $^{76}$Se.

6. We have found that the calculated absolute values of the differential cross sections depend strongly on the assumed transfer configuration. We conclude that better description of the ground state wave functions of the target nuclei is needed to yield more reliable values of the theoretical spectroscopic factors.

**Abstract:** The $^{76,78}$Se(p,t)$^{74,76}$Se reactions have been studied at a proton energy of 33 MeV using detector telescopes and particle identification techniques. Twenty-eight states up to the excitation energy of 4.64 MeV in $^{74}$Se and twenty-five states up to 4.43 MeV in $^{76}$Se were observed. Angular distributions were measured for many of these states in the range of 15°-90° and the results were compared with the distorted waves direct transfer calculations. Many $J^\pi$ assignments were made on the basis of the theoretical analysis of the data and a comparison of the angular distributions with empirical shapes for transitions to states with well-known $J^\pi$. Enhancement coefficients were calculated for the simple two-neutron pickup configurations. Results indicate that the (p,t) reaction is sensitive to assumed configurations in the ground state wave functions of the target nuclei and that better description of the relevant wave functions is needed to yield more reliable values for the theoretical spectroscopic factors.

## Introduction

The study of the (p,t) reaction was carried out in conjunction with the study of the (p,d) reaction on Se isotopes (see Chapter 14). Only slight modification of the electronic system was necessary to measure angular distribution simultaneously for both reactions.

The $^{76}$Se and $^{78}$Se isotopes have 14 and 16 neutrons outside the closed shell $N$ = 28, respectively. In the absence of the residual interaction, these neutrons would fill in the orbits 2p$_{3/2}$, 1f$_{5/2}$, and 2p$_{1/2}$, and there would be 2 neutrons in the 1g$_{9/2}$ orbit in $^{76}$Se and 4 in $^{78}$Se.

Nuclei with neutrons in the 1g$_{9/2}$ region have long resisted a complete and reliable description in terms of any one model. The large number of valence neutrons and available orbits complicate the calculations in this region. There was also not a great deal of experimental information available on many of these nuclei at the time of our study.





One of the aims of our study was to provide information on the low-spin states in $^{74}$Se and $^{76}$Se. Investigation of the single-step direct transfer mechanism to reproduce two-neutron transfer data in this area was also of interest. It was also interesting to see whether the existing simple assumptions about the configurations in the ground state wave functions could result in reliable predictions of the absolute values of the differential cross sections for these reactions.

These reactions served also as a good test of the of the optical model parameterisation for tritons as discussed in Chapter 11. We have used my sets of parameters successfully in the analysis of our results for the (d,t) reaction induced by 12 MeV deuterons (Chapter 12). However, we had to scale down the parameters to match the low energy of tritons in this reaction. The (p,t) measurements were carried out at a significantly higher energy and thus the parameters I have derived earlier could be tested with a better accuracy.

### The experimental method

Measurements were performed using the 33 MeV proton beam from the ANU cyclograaff facility consisting of the CNI-30 cyclotron injecting 26 MeV H⁻ beam into the EN electrostatic tandem accelerator. Experimental arrangement was the same as described for (d,t) measurements (see Chapter 12) but the electronic system was modified to allow for simultaneous measurements of both the (p,d) and (p,t) reactions (see Figure 13.1).

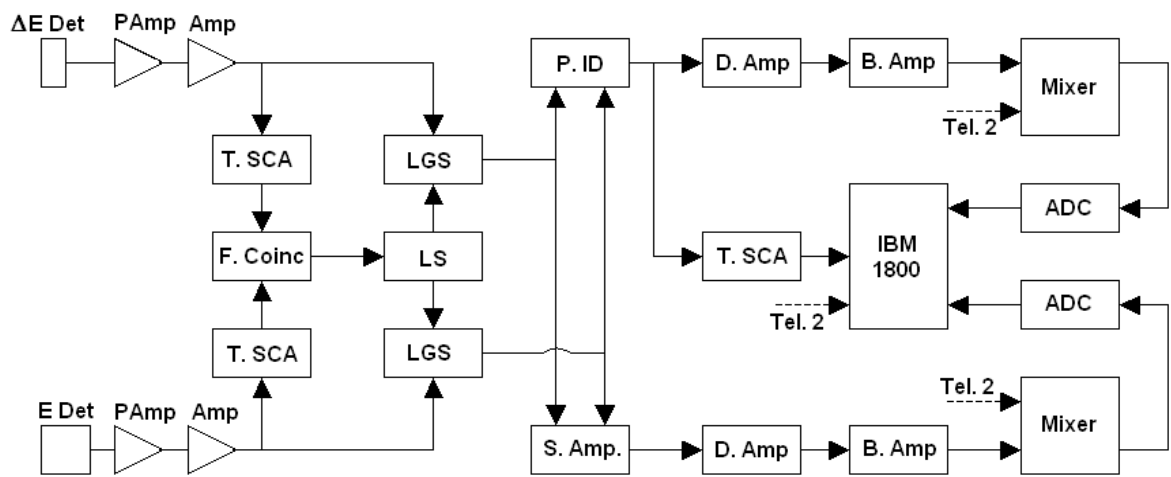

Figure 13.1. A diagram of electronic system used in measurements of the (p,d) and (p,t) reactions on Se isotopes. P Amp – Pre-amplifier; Amp – Amplifier; T. SCA – Timing Single Channel Analyser; F. Coinc – Fast Coincidence unit; LGS – Linear Gate and Stretcher; LS – Logic Shaper and Delay; P. ID – Particle Identifier; S. Amp. – Summing Amplifier; D. Amp – Delay Amplifier; B. Amp – Biased Amplifier; Mixer – Mixer and Routing unit; ADC – Analogue to Digital Converter; IBM 1800 – IBM 1800 computer.

Typical mass spectrum for the (d,t) reaction is shown in Figure 13.2. The deuteron spectra resulting from the (p,d) reaction were recorded simultaneously with the triton spectra from the (p,t) reaction by setting windows on the deuteron and triton peaks in the mass spectrum. Coincidences were required between each window and the mixed total energy spectra. The setting of the mass windows, the establishment of the appropriate coincidence conditions and the subsequent routing of the deuteron and triton energy spectra for the $\Delta E$-$E$ telescopes into different computer areas were all achieved by digital means. This was done by the data acquisition program *Routed Window* in the IBM 1800 computer.





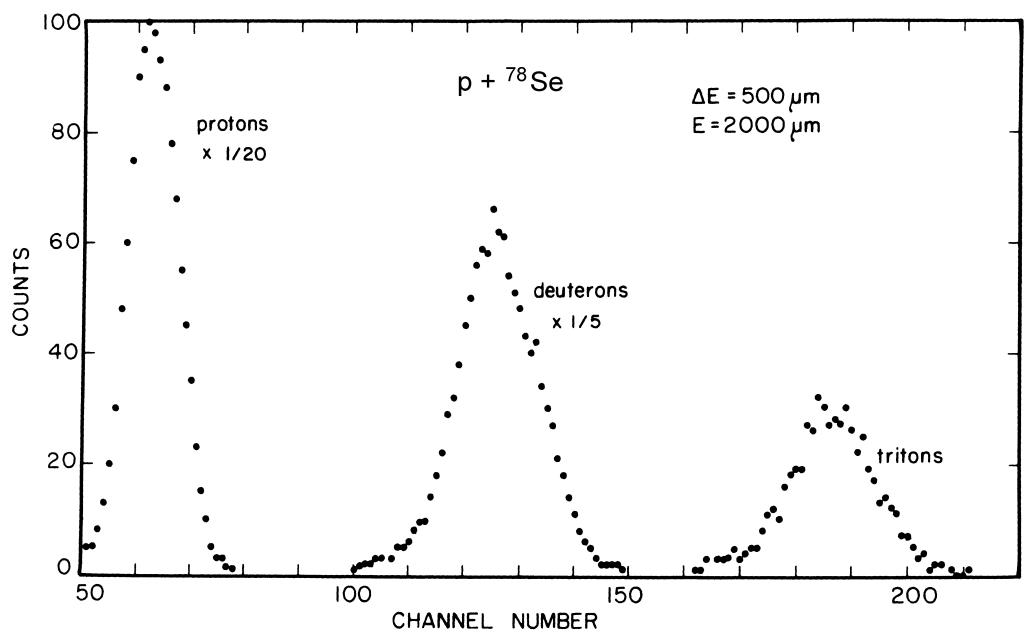

Figure 13.2. A typical mass spectrum for the interaction of protons with $^{78}$Se leading to (p,p'), (p,d), and (p,t) reactions induced by 33 MeV protons.

The data were stored in the data acquisition area of the IBM 1800 computer and dumped into the computer buffer after each run. When the buffer was full or the measurement was finished, the spectra were transferred to a demountable disc pack for storage.

Each surface-barrier detector telescope was made of a 500 μm $\Delta E$ counter and a 2000 μm $E$ counter. The energy resolution obtained varied from 60 to 85 keV.

Typical particle spectra for both isotopes are shown in Figures 13.3 and 13.4. Levels were observed for up to around 4.5 MeV excitation energy. The energy calibrations were obtained using several well-known low-lying states in $^{74}$Se and $^{76}$Se. The quoted excitation energies are accurate to around 10 keV for strongly excited states and up to 25 keV for some weakly excited states at high excitation. We have observed 28 states in $^{74}$Se and 25 in $^{76}$Se.

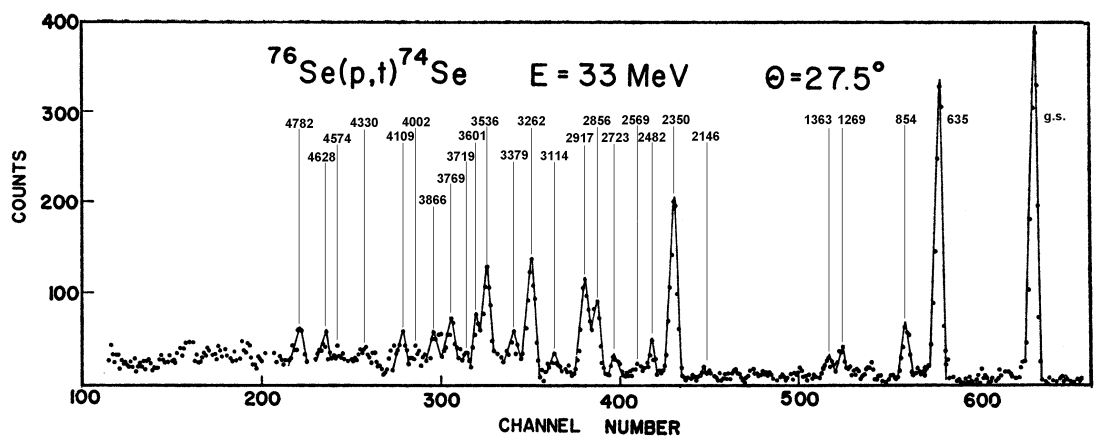

Figure 13.3. An example of a spectrum of tritons from the $^{76}$Se(p,t)$^{74}$Se reaction.





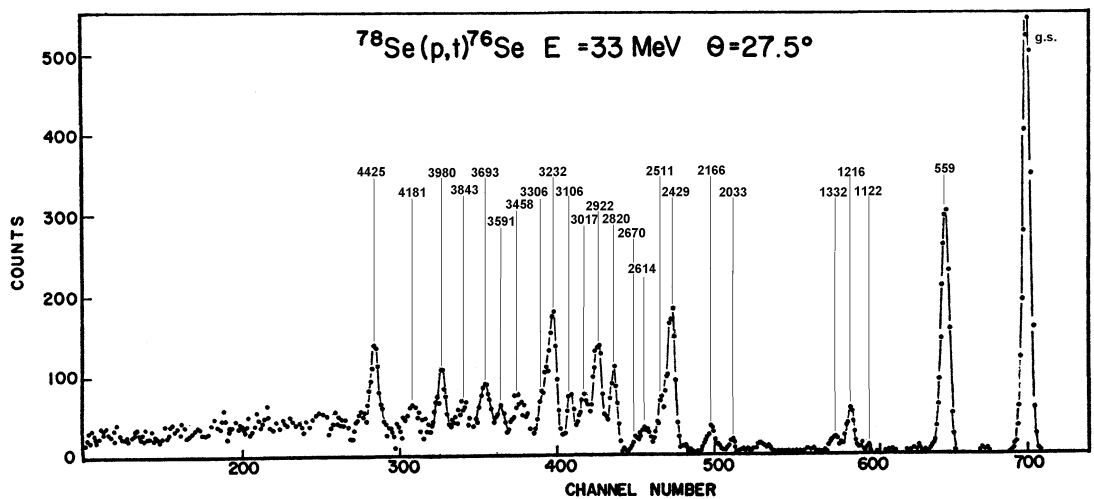

Figure 13.3. An example of a spectrum of tritons from the $^{78}$Se(p,t)$^{76}$Se reaction. These two spectra show that the (p,t) reaction is dominated by just a few transitions, mainly to the ground state and to the first excited state.

The targets were 400 μg/cm$^2$ thick and were made by vacuum evaporation from enriched material onto 30 μg/cm$^2$ carbon backings. Because Se sublimes rapidly under beam bombardment, a second thin carbon layer was evaporated onto the Se. These sandwich targets were able to withstand beams of up to 100 nA, being the highest currents used during the measurements. The target stability was constantly monitored by a Si(Li) detector at 90°.

Angular distributions for most states were measured from 15° to 90°. The solid angle-target thickness product, which was required to convert the relative cross sections into absolute values, was determined from the Rutherford scattering at proton energy of 4.0 MeV. The errors in the absolute cross-sections are estimated to be typically around 15 %.

Even though we were able to assign excitation energies to a total of 53 states in $^{74}$Se and $^{76}$Se we could measure angular distributions for only 38 of them. These distributions are presented in Figures 13.4 and 13.5. For the remaining states, triton peaks could be observed only at certain angles and thus angular distributions could not be measured.

## Theoretical analysis

Theoretical analysis was carried out using direct reactions distorted waves Born approximation theory and the computer code (Kunz 1966), which I modified and adapted to run on the ANU UNIVAC 1108 computer. For protons, we have used parameter sets of Becchetti and Greenlees (1969). Their potentials contained both volume and surface absorptions. For tritons, we have tested parameters from four sources: Baer et al. (1973); Becchetti and Greenlees (1970); Flynn *et al.* 1969; and Nurzynski (1975). We have found that my sets of parameters produced, in general, the best fits to the experimental angular distributions. Results of the distorted wave analysis are presented in Figures 13.4 and 13.5.





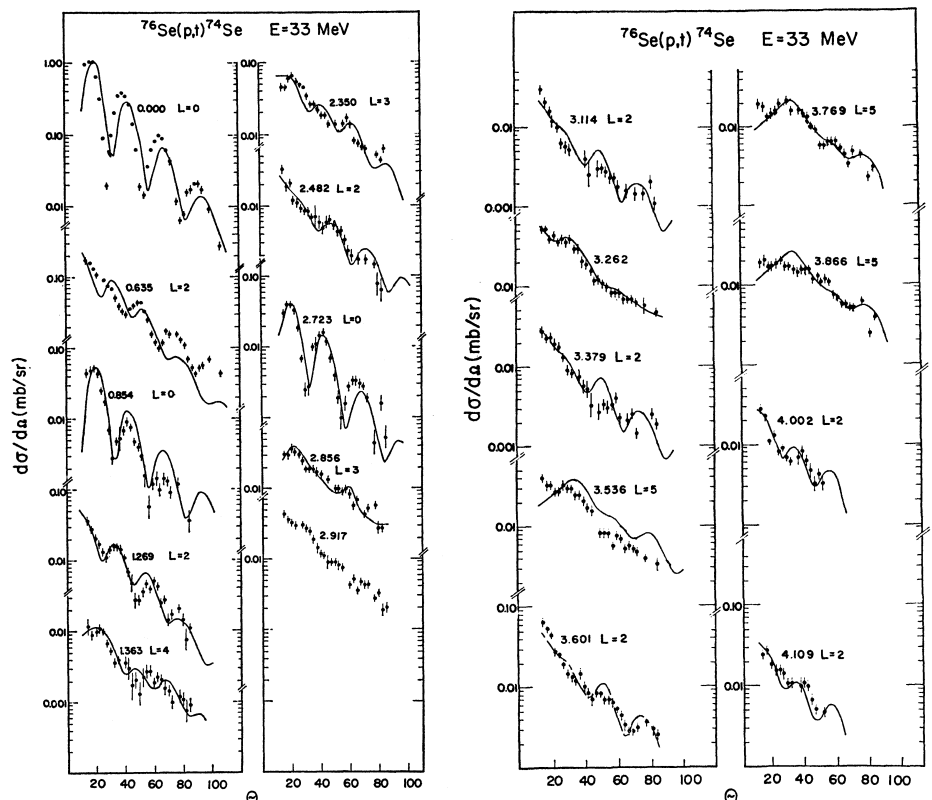

Figure 13.4. Measured angular distributions of the differential cross sections for the reaction $^{76}$Se(p,t)$^{74}$Se induced by 33 MeV protons are compared with the theoretical calculations.

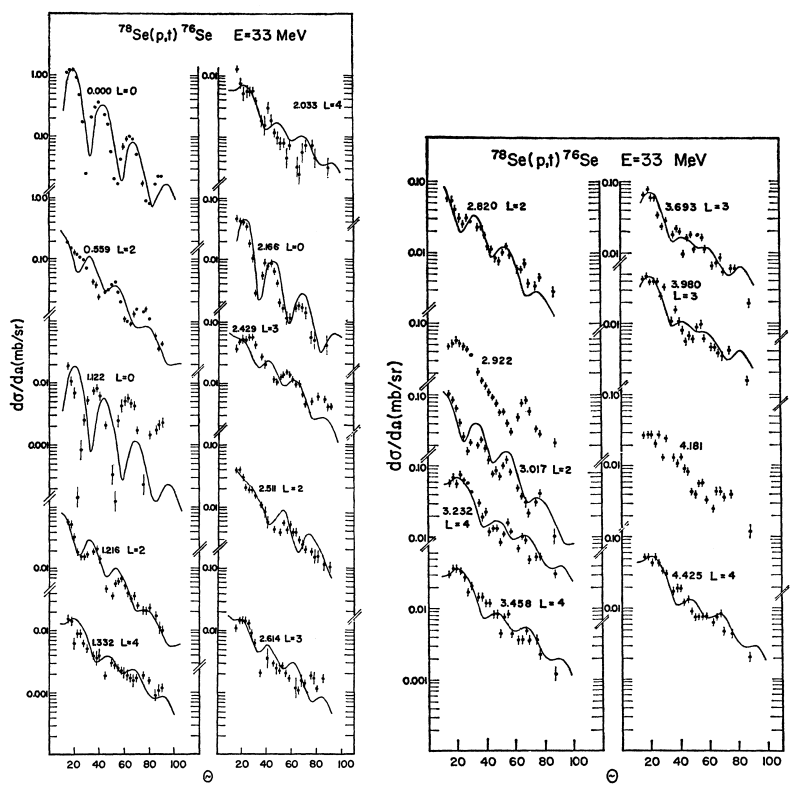

Figure 13.5. Measured angular distributions of the differential cross sections for the reaction $^{78}$Se(p,t)$^{76}$Se induced by 33 MeV protons are compared with the theoretical calculations.





### Selection rules for the (p,t) reactions

Let us consider a reaction A(a,b)B involving a transfer of two nucleons. Each transferred nucleon is described by a set of three quantum numbers, $j_i$, $l_i$, and $s_i$ (the total angular momentum, orbital angular momentum, and spin), where $i$ = 1 or 2. The total angular momentum for the transferred pair can be written as:

$$\boldsymbol{J} = \boldsymbol{j}_1 + \boldsymbol{j}_2 = \boldsymbol{L} + \boldsymbol{S}$$

where

$$\boldsymbol{L} = \boldsymbol{l}_1 + \boldsymbol{l}_2$$

$$\boldsymbol{S} = \boldsymbol{s}_1 + \boldsymbol{s}_2$$

For a pair of nucleons, $S$ = 0 or 1. In particular, for two neutrons, $S$ = 0.

The conservation of the angular momentum gives the following relations between the angular momentum $J_A$ of the target nucleus and $J_B$ of the residual nucleus:

$$\boldsymbol{J}_A = \boldsymbol{J}_A + \boldsymbol{L} + \boldsymbol{S} = \boldsymbol{J}_A + \boldsymbol{J}$$

which means that

$$\left| J_A - J_B \right| \leq J \leq J_A + J_B$$

For even-even target nuclei $J_A$ = 0 and therefore

$$J_B = J$$

or

$$J_B = L \pm 1 \quad \text{for the S = 1 pair of nucleons}$$

$$J_B = L \qquad \text{for the S = 0 pair of nucleons}$$

The conservation of parity is fulfilled if

$$\Delta \pi = (-)^L$$

Table 13.1
Excitation energies and spin-parity assignments to states in $^{74}$Se based on the study of the $^{76}$Se(p,t)$^{74}$Se at 33 MeV proton energy

| $E_x$ | $J^\pi$ | $E_x$ | $J^\pi$ | $E_x$ | $J^\pi$ | $E_x$ | $J^\pi$ | $E_x$ | $J^\pi$ |
|-------|---------|-------|---------|-------|---------|-------|---------|-------|---------|
| 0.000 | $0^+$ | 2.101 | | 2.856 | $(3^-)$ | 3.601 | $(2^+)$ | 4.330 | |
| 0.635 | $2^+$ | 2.146 | | 2.917 | | 3.719 | | 4.574 | |
| 0.854 | $0^+$ | 2.350 | $3^-$ | 3.114 | | 3.769 | $(5^-)$ | 4.628 | |
| 1.269 | $2^+$ | 2.482 | $(2^+)$ | 3.262 | | 3.866 | $(5^-)$ | 4.782 | |
| 1.363 | $4^+$ | 2.569 | | 3.379 | $(2^+)$ | 4.002 | $(2^+)$ | | |
| 1.839 | | 2.723 | $0^+$ | 3.536 | $(5^-)$ | 4.109 | $(2^+)$ | | |

$E_x$ – Excitation energy (in MeV) based on our measurements.
$J^\pi$ – Spin-parity assignments based on our measurements. In cases of insufficient data at all angles, only excitations energies have been assigned.





Table 13.2
Excitation energies and spin-parity assignments to states in $^{76}$Se based on the study of the $^{78}$Se(p,t)$^{76}$Se at 33 MeV proton energy

| $E_x$ | $J^\pi$ | $E_x$ | $J^\pi$ | $E_x$ | $J^\pi$ | $E_x$ | $J^\pi$ | $E_x$ | $J^\pi$ |
|---|---|---|---|---|---|---|---|---|---|
| 0.000 | $0^+$ | 2.166 | $0^+$ | 2.670 | | 3.106 | | 3.693 | $(3^-)$ |
| 0.559 | $2^+$ | 2.347 | | 2.820 | $2^+$ | 3.232 | $(4^+)$ | 3.843 | |
| 1.122 | $0^+$ | 2.429 | $3^-$ | 2.922 | | 3.306 | | 3.980 | $(3^-)$ |
| 1.216 | $2^+$ | 2.511 | $(2^+)$ | 3.017 | $(2^+)$ | 3.458 | $(4^+)$ | 4.181 | |
| 2.033 | $(4^+)$ | 2.614 | $(3^-)$ | 3.106 | | 3.591 | | 4.425 | $(4^+)$ |

By fitting the angular distributions for the (p,t) reactions one can find the $L$ value for the transferred pair of neutrons. If an even-even target nucleus is used (as in our measurements) one can then assign the $J^\pi$ values (the spin and parity) to states in the residual nucleus. Thus, for $L$ = 0, 1, 2, 3, 4, 5, etc. $J^\pi$ = $0^+$, $1^-$, $2^+$, $3^-$, $4^+$, $5^-$ etc., respectively.

Table 13.1 and 13.2 summarises our assignments of excitation energies (for the study of particle spectra) and $J^\pi$ values (from fitting the angular distributions) to states in $^{74}$Se and $^{76}$Se.

### The absolute values of the differential cross sections

In general, theoretical calculations reproduce the measured shapes of angular distributions sufficiently well. However, as we have anticipated, we have experienced a problem with reproducing the absolute values of the differential cross sections.

The relationship between the experimental and theoretical differential cross sections is as follows:

$$\left(\frac{d\sigma}{d\Omega}\right)_{exp} = \frac{D_0^2\left(\frac{1}{2}\pi\Delta^2\right)^{\frac{3}{2}}\varepsilon S^2}{2L+1}\left(\frac{d\sigma}{d\Omega}\right)_{th}$$

where $\left(\frac{d\sigma}{d\Omega}\right)_{exp}$ is the experimental differential cross section, $D_0^2$ is the zero-range coefficient (see the Appendix E), $\Delta$ = 1.7 fm is the root mean square radius of the triton, $\varepsilon$ the enhancement factor, $S$ is the theoretical spectroscopic factor, and $\left(\frac{d\sigma}{d\Omega}\right)_{th}$ is the theoretical cross section as calculated by the DWBA code.

We expected a problem with reproducing the absolute values of differential cross sections because the (p,t) reactions involve a number of unknown quantities. For a start, to calculate the $D_0^2$ coefficient one has to know not only the wave function for tritons but also the interaction potential between proton and the pair of neutrons.

To compare the absolute values of the experimental and theoretical differential cross sections, one has to calculate spectroscopic factors for the states involved in the





transition of two neutrons. Each of the two neutrons can be picked up from different orbits in the ground state of the target nuclei. Thus, a series of spectroscopic factors would have to be calculated using various combinations of configurations for the two neutrons. Such calculations would be complex but they were even impossible because the wave functions for the ground states of the target nuclei were unknown. In our calculations, we used information, which was available to us at the time of our study.

For $D_0^2$ we used the value of 23.5 MeV$^2 \cdot$fm$^3$ determined by Broglia, *et al.* (1972). To calculate the spectroscopic factors $S$, we have assumed that the two neutrons are picked up from the same orbit, i.e. that the transitions occur only between neutron configurations $j^n \rightarrow j^{n-2}$ ($n$ even). Under this assumption, the spectroscopic factors are given by the following formulae (Bassani, Hinz and Kavaloski 1964):

$$S = \frac{n}{2} \frac{2j + 3 - n}{2j + 1} \qquad \text{for L = 0}$$

$$S = \frac{n(n-2)(2L+1)}{(2j-1)(2j+1)} \quad \text{for L = 2, 4, 6, ...}$$

Using these spectroscopic factors, we have tested theoretical predictions of the absolute values of the differential cross sections. The degree of the reliability of the theoretical predictions of the absolute values of the differential cross section is given by the enhancement factor $\varepsilon$, which should be equal to one if the theory gives correct predictions. Table 13-3 lists examples of the enhancement factors calculated for states corresponding to unambiguous spin-parity assignments.

Table 13.3
Enhancement coefficients

| Nucleus | States | | | | | |
|---------|--------|--------|--------|--------|--------|--------|
|  | $0_1^{+\,a)}$ | $2_1^{+\,b)}$ | $0_2^{+\,b)}$ | $2_2^{+\,b)}$ | $4_1^+$ | $0_3^{+\,b)}$ |
| $^{74}$Se | 1.47 | 4.13 | 1.25 | 0.86 | 0.34$^{b)}$ | 1.12 |
| $^{76}$Se | 1.89 | 4.47 |  | 1.39 | 2.00$^{c)}$ | 1.03 |

$^{a)}$ Assumed transfer: $(2p_{3/2})^4 \rightarrow (2p_{3/2})^2$ ; $^{b)}$ Assumed transfer: $(1f_{5/2})^6 \rightarrow (1f_{5/2})^4$

$^{c)}$ Assumed transfer: $(1g_{9/2})^4 \rightarrow (1g_{9/2})^2$

It can be seen that in general $\varepsilon \neq 1$. Furthermore, the values of enhancement factors depend on the excited state.

The only quantity in the theoretical cross section that depends on the excited state is the spectroscopic factor. Consequently, the different values of $\varepsilon$ indicate that the assumed transfer configurations must be incorrect.

We have tried various other options for transfer configurations and found that $\varepsilon$ depends strongly on the assumed configurations. For instance, for the transfer to the ground state ($0_1^+$) in $^{74}$Se, assuming $(2p_{1/2})^2 \rightarrow (2p_{1/2})^0$ transfer results in $\varepsilon = 3.65$, as





compared with $\varepsilon = 1.47$ for the $(2p_{3/2})^4 \rightarrow (2p_{3/2})^2$ transfer. If assumed transfer is $(1f_{5/2})^6 \rightarrow (1f_{5/2})^4$ then $\varepsilon = 23.56$.

Clearly, the absolute values of the calculated differential cross sections depend strongly on the assumed configurations involved in the transfer of the two neutrons and better description of the ground state wave functions is needed to yield more reliable values for the theoretically calculated spectroscopic factors.

## Summary

The $^{76,78}$Se(p,t)$^{74,76}$Se reactions were studied at the proton energy of 33 MeV using particle telescopes to detect the outgoing tritons. Many new states were observed and their excitation energies were calculated. Many new $J^\pi$ assignments were made for states in the residual nuclei. Angular distributions were measured for most of the observed states and were compared with distorted wave calculations. Spin-parity assignments were made from a comparison of the data with both the distorted wave calculations and the empirical shapes of angular distributions for well-known states.

Enhancement coefficients were calculated for several well-known states using various options for configurations in the ground state wave functions of the target nuclei. The calculated absolute values of the differential cross sections were found to depend strongly on the assumed configurations indicating a need for a better description of the relevant wave functions.

## References

Bassani, G., Hinz, N. M, and Kavaloski, C. D. 1964, Phys. Rev. **136B**:1006.

Baer, H. W., Kraushaar, J. J., Moss, C. E., King, N. S. P., Green, R. E. L., Kunz, P. D. and Rost, E. 1973, *Ann. of Phys.* **76**:437

Becchetti, F. D. and Greenlees, G. W. 1969, *Phys. Rev.* **182**:1190

Becchetti, F. D. and Greenlees, G. W. 1970, Proc. Third Int. Symp. Polarization Phenomena in Nuclear Reactions, Madison, ed. H. H. Barschall and W. Haeberli (Univ. of Wisconsin Press) p. 682

Broglia, R. A., Riedel, C. and Udagawa, T. 1972, *Nucl. Phys.* **A184**:23

Flynn, E. R., Armstrong, D. D., Beery, J. B. and Blair, A. G. 1969, *Phys. Rev.* **182**:1113

Kunnz, P. D. 1966, University of Colorado Report COO-536-606.

Nurzynski, J. 1975, *Nucl. Phys.* **A246**:333





_______________________________________________________________________

**14**

# Single-neutron Transfer Reactions on $^{76,\,78,\,80,\,82}$Se Isotopes Induced by 33 MeV Protons

**Key features:**

1. The Se isotopes are in an interesting region of the periodic table where orbits outside the closed $N = 28$ shell are almost full. Model interpretation of these nuclei is difficult and there was a need to support theoretical work by experimental investigation.

2. We have observed a total of 120 states in $^{75,77,79,81}$Se isotopes and we have assigned excitation energies to these states.

3. We have measured a total of 88 angular distributions for the single neutron pickup reactions $^{76,\,78,\,80,\,82}$Se(p,d)$^{75,77,79,81}$Se induced by 33 MeV protons.

4. We have analysed the distributions using the distorted wave theory. We have determined orbital angular momenta for the relevant states in residual nuclei and extracted corresponding spectroscopic factors.

5. Using the experimental spectroscopic factors, we have calculated the occupation numbers for orbits in the f-p shell and compared them with the theoretical predictions using the pairing theory. We have found a good agreement between experimental and theoretical values.

6. In addition to the expected $l_n = 1, 3$, and 4 transitions we have also observed a number of anomalous $l_n = 2$ transitions. We have carried out calculations using the Coriolis coupling model. We have found that in general, the model can account for the presence of the positive parity states belonging to anomalous transitions.

**Abstract:** Single-neutron transfer reactions on even-even Se nuclei have been studied using 33 MeV incident protons from the ANU cyclograaff facility. A total of 120 levels have been observed below the 4 MeV excitation energy in the $^{75,77,79,81}$Se isotopes. Angular distributions for 88 states were extracted and analysed with the distorted wave theory. Coriolis coupling calculations have been carried out for low-level spin states in all four isotopes.

## Introduction

The Se isotopes presented an interesting and challenging case for research. The $^{76,78,80,82}$Se nuclei contain a large number of neutrons outside the closed $N = 28$ shell, ranging from 14 for $^{76}$Se to 20 for $^{82}$Se. Without the residual interaction, orbits $2p_{3/2}$, $1f_{5/2}$, and $2p_{1/2}$ would be fully occupied and orbit $1g_{9/2}$ would be filling in. With the residual interactions, orbits $2p_{3/2}$ and $1f_{5/2}$ are expected to be about 90% full for $^{76}$Se, orbit $2p_{1/2}$ about 85% full, and orbit 40% full. Thus, orbits $2p_{3/2}$, $1f_{5/2}$, and $2p_{1/2}$ would contain, on average less neutrons than expected in the absence of the residual interaction, but orbit $1g_{9/2}$ would contain more.[13] With the increasing mass number, the distribution of neutrons between various orbits would come progressively closer to the distribution corresponding to the independent particle shell model distribution

_______________________________________________________________________

[13] Without the residual interaction, orbits $2p_{3/2}$, $1f_{5/2}$, and $2p_{1/2}$ would be 100% full and orbit $1g_{9/2}$ only 20% full for $^{76}$Se.





until for $^{82}$Se, orbits $2p_{3/2}$, $1f_{5/2}$, and $2p_{1/2}$ would be nearly 100% full and orbit $1g_{9/2}$ nearly 80% full.

The large number of neutrons outside the closed shell $N = 28$ makes the model description of Se isotopes difficult. Theoretical descriptions need to be tested by experimental results and at the time of our study little experimental information was available.

The only spectroscopic information available from single neutron transfer reactions to the odd Se nuclei was based on the $^{76,78,80}$Se(d,p)$^{77,79,81}$Se (Lin 1965) and $^{76}$Se(d,t)$^{75}$Se (Sanderson 1973) reactions. Lin also identified some levels in $^{77,79,81}$Se using the (d,t) reaction but did not measure the corresponding angular distributions.

Single neutron stripping reactions populate vacant neutron configurations. Consequently, the (d,p) reactions are not expected to have significantly strong transitions to $2p_{3/2}$, $1f_{5/2}$, and $2p_{1/2}$, orbits, particularly for heavier Se isotopes. One can, however, expect transitions to the $1g_{9/2}$ orbit ($l = 4$) or to orbits outside the closed $N = 50$ shell ($l = 0$ and 2 transitions). Indeed, Lin observed many $l = 2$ transitions but only a small number of $l = 1$, 3, and 4.

On the other hand, the single neutron pickup reactions are expected to have strong transitions from the well-populated configurations in the target nuclei. Consequently, the (p,d) reactions should show many $l = 1$ and 3 transitions. As the $1g_{9/2}$ orbits are filling in with the increasing mass number, the number of $l = 4$ transitions is expected to increase for heavier Se isotopes.

Nuclei in the Se mass region cannot be fully described by any known model. It was originally thought that the level structure of the even Se nuclei should be well explained by the vibrational model (Scharff and Weneser 1955). However, Barrette *et al.* (1974) have shown that with the exception of $^{74}$Se, the ratios of the electric quadrupole transitions probabilities, $B(E2)$, in the even Se isotopes do not comply with the simple vibrational description.

Lieder and Draper (1970), McCauley and Draper (1971), Wyckhoff and Draper (1973) and Nolte *et al.* (1977) investigated the even-even Ge, Se and Kr nuclei using heavy ion reactions and found quasi-rotational bands with spins of up to 10$^+$. The large moment of inertia determined for these bands indicate that the ground states of these nuclei may be deformed. Coupling a neutron to such a deformed core should stabilize the deformation and one could therefore expect the odd nuclei in this mass region to be also deformed.

The anharmonic vibrator model is also able to describe such quasi-rotational bands as have been shown in this region in a study of heavy ion reactions. Holzwarth and Lie (1972) and Lie and Holzwarth (1975) used such a model to describe $^{76}$Se and $^{78}$Se and obtained good agreement between the calculated and experimental levels below 2.5 MeV. They also calculated quadrupole moments and $B(E2)$ values which agree reasonably well with the experimental results.

Another problem that needs to be solved and explained is the presence of the low-lying $^5/_2{}^+$ and $^7/_2{}^+$ levels in all odd $N$, even $Z$ nuclei throughout the $39 < N < 49$ region. They are unlikely to be single particle states since the $2d_{5/2}$ and the $1g_{7/2}$ orbits should be filling in at $N > 50$. Similar $^5/_2{}^+$ and $^7/_2{}^+$ low-lying states are also found in odd-Z, even-N nuclei in the $1g_{9/2}$ mass region. The presence of such states in all of





these nuclei is somewhat surprising and has led to several theoretical attempts to explain their origins.

The first attempt to explain these states was made by Flowers (1952) who used the seniority-coupling model. His calculations predict that the $7/2^+$ level is never the ground state. This is in conflict with the experimental results, which show, for example, that $^{79}$Se has a ground state spin of $7/2^+$.

Kisslinger and Sorensen (1960) coupled the $1g_{9/2}$ quasi-particle to the neighbouring $2^+$ one-phonon state (QPC model). However, their calculations could not explain the ground state spins of $5/2^+$ and $7/2^+$ in $^{75}$Se and $^{79}$Se, respectively. Their calculations were later improved by Sherwood and Goswami (1966) who included the quasihole-phonon coupling (EQPC model) and by Goswami and Nalcioglu (1968) who included the quadrupole-quadrupole interaction. Again, these extended calculations failed to explain fully the presence of $5/2^+$ and $7/2^+$ states, which often lie below the $9/2^+$ state.

Because of the inability of the spherical shell model with appropriate residual interactions to describe adequately the *odd proton* nuclei in this region, Scholz and Malik (1968) extended their successful use of Coriolis coupling model in the $f_{7/2}$ shell to some of these nuclei. The model correctly predicted spins and parities for the low-lying states with the right energy spacing for all the nuclei they studied.

Since the situation is analogous to the low-lying positive parity states that occur in the $1g_{9/2}$ *odd neutron* nuclei, Coriolis coupling model has been applied with some success to $^{75}$Se by Sanderson (1963) and to a number of odd neutron nuclei in the $1g_{9/2}$ mass region ($73 < A < 87$) by Heller and Friedman (1974, 1975). These authors obtained spins, parities and level spacing for the low-lying positive parity states that are in good agreement with the experimental data. Their calculations predict a prolate deformation in this mass region.

One of the aims of our study was to extend the study of the low-lying $5/2^+$ and $7/2^+$ states for Se nuclei. Available experimental evidence suggests that Se nuclei might be deformed and that such states could be interpreted as arising from Coriolis coupling. If such is the case then these states should have significant single particle components and thus should be excited in the (p,d) reaction.

**Experimental method**

Experimental arrangement was the same as for the (p,t) reactions (see Chapter 13). Target enrichments were $^{76}$Se 86%, $^{78}$Se 96%, $^{80}$Se 94% and $^{82}$Se 97%. Examples of deuteron spectra are shown in Figure 14.1. Careful checks were made for each isotope to ensure that identified peaks did not arise from Se impurities. A total of 120 levels have been identified and the corresponding excitation energies have been assigned to all of them. The quoted energies are accurate to around 10 keV for strongly excited states and up to 25 keV for weaker states at high excitations.

We have measured a total of 88 angular distributions, 25 for $^{76}$Se, 16 for $^{78}$Se, 18 for $^{80}$Se, and 29 for $^{82}$Se. They are shown in Figures 14.2 – 14.11. They served as a basis for assigning orbital angular momenta to the respective states in residual nuclei and to derive spectroscopic factors.

Absolute cross sections are accurate to around 15% and were determined from the Rutherford scattering of 4 MeV protons. As mentioned in Chapter 13, (p,d) and (p,t) reactions were measured simultaneously. As the cross sections for the (p,d) reaction





were in general higher than for the two-neutron (p,t) transfer, good statistical accuracy for the (p,t) reaction resulted in excellent accuracy for the (p,d) data.

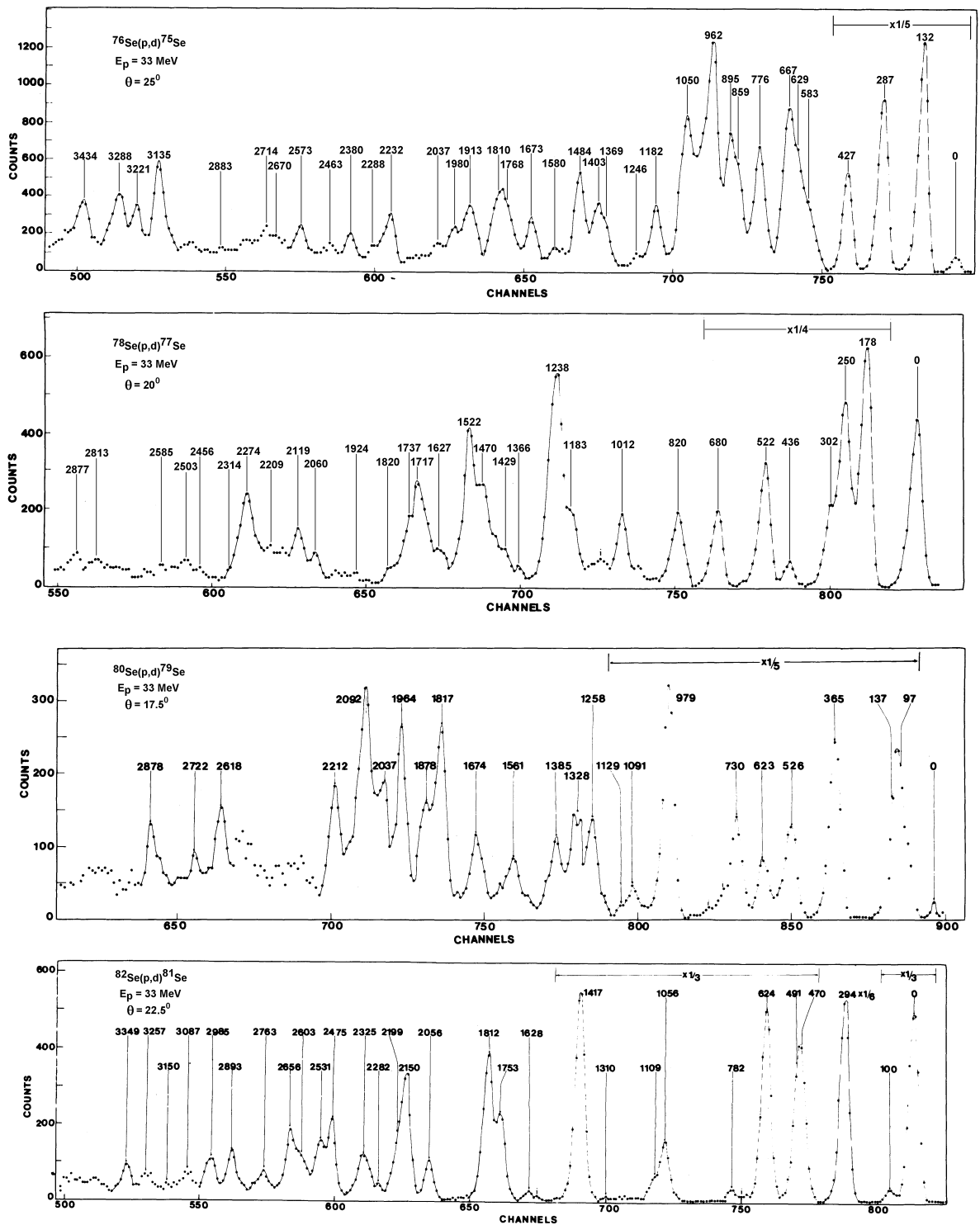

Figure 14.1. Examples of deuteron spectra for the reactions $^{76,78,80,82}$Se(p,d)$^{75,77,79,81}$Se induced by 33 MeV protons.





**The distorted wave analysis**

The distorted wave calculations were carried out using the code DWUCK (Kunz 1966) which included standard finite-range and non-locality corrections. The proton optical-model parameters were derived from the sets given by Becchetti and Greenlees (1969). While many sets of deuteron parameters were tried, those that gave the best fits to the data were obtained from the elastic scattering on Zn at 25.9 MeV (Perey and Perey 1966). The transferred neutron was assumed to be bound in a Woods-Saxon well and the customary separation energy prescription was used. Calculated angular distributions are compared with the experimental results in Figures 14.2 – 14.7.

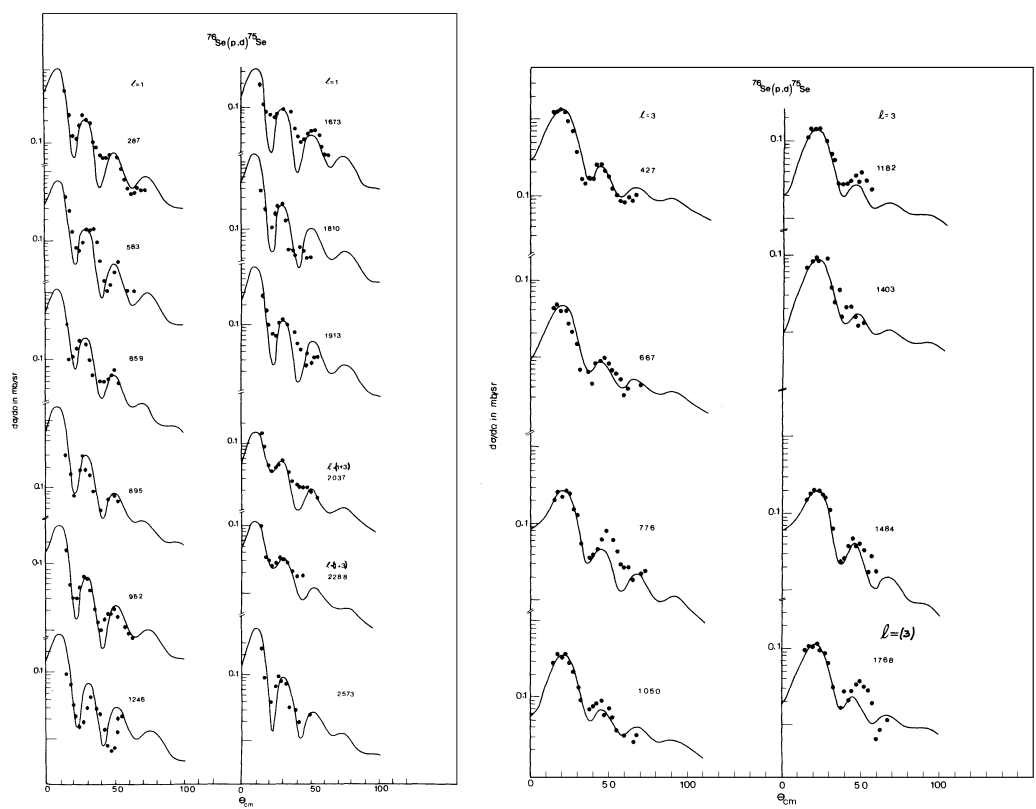

Figure 14.2. Deuteron angular distributions for the reaction $^{76}$Se(p,d)$^{75}$Se induced by 33 MeV protons compared with the distorted wave calculations for $l_n = 1$ and $l_n = 1 + 3$ (the left-hand side of the figure) and for $l_n = 3$ (the right-hand side).

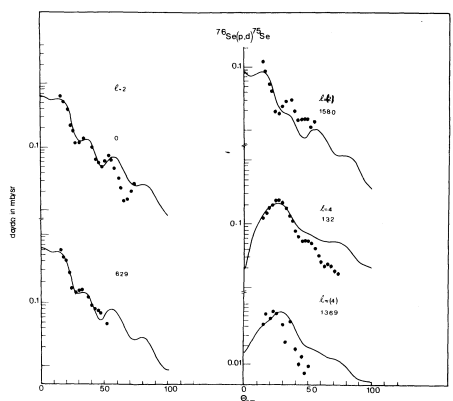

Figure 14.3. Deuteron angular distributions for the reaction $^{76}$Se(p,d)$^{75}$Se induced by 33 MeV protons compared with the distorted wave calculations for the even $l_n$ values.





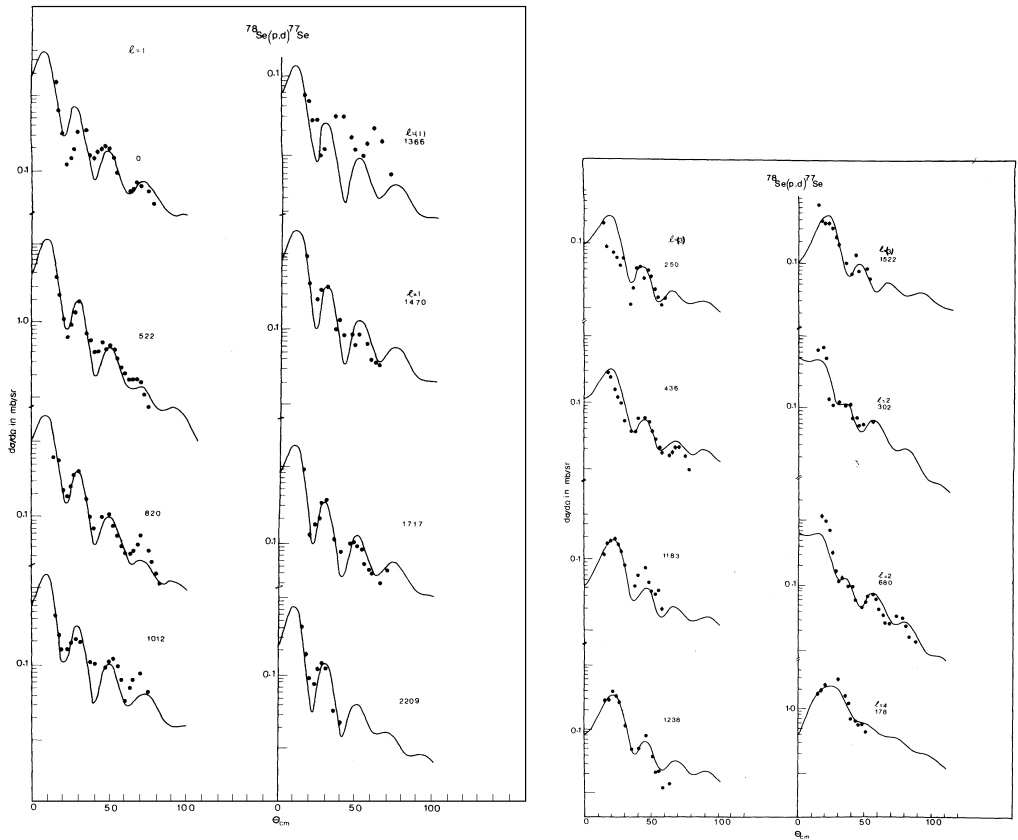

Figure 14.4. Deuteron angular distributions for the reaction $^{78}$Se(p,d)$^{77}$Se induced by 33 MeV protons compared with the distorted wave calculations for $l_n$ = 1 (the left-hand side of the figure) and for $l_n$ = 2, 3, and 4 (the right-hand side).

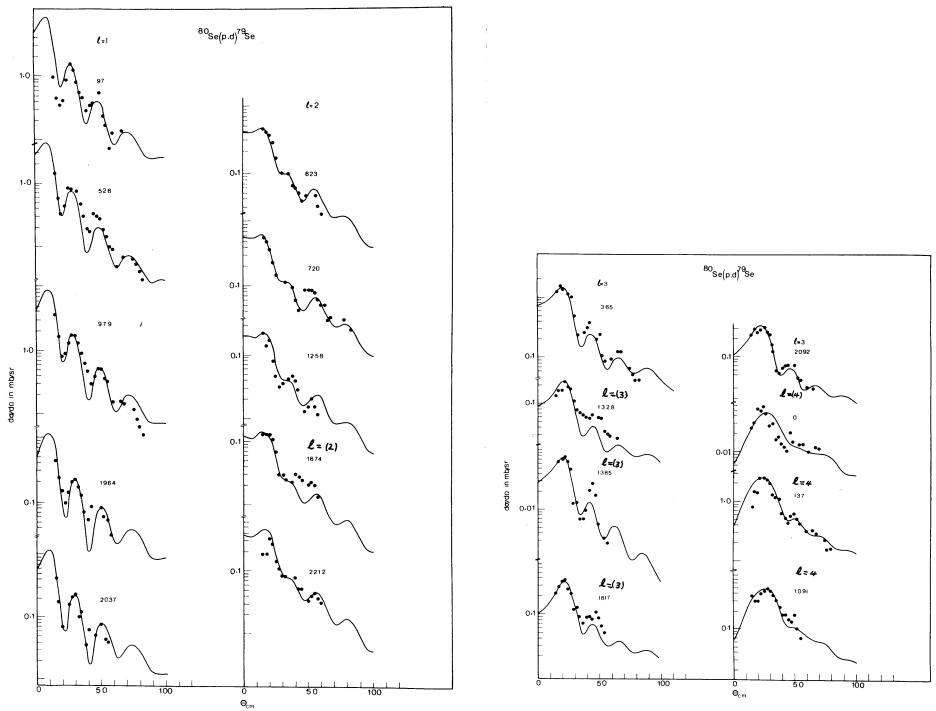

Figure 14.5. Deuteron angular distributions for the reaction $^{80}$Se(p,d)$^{79}$Se induced by 33 MeV protons compared with the distorted wave calculations for $l_n$ = 1 and 2 (the left-hand side of the figure) and for $l_n$ = 3 and 4 (the right-hand side).





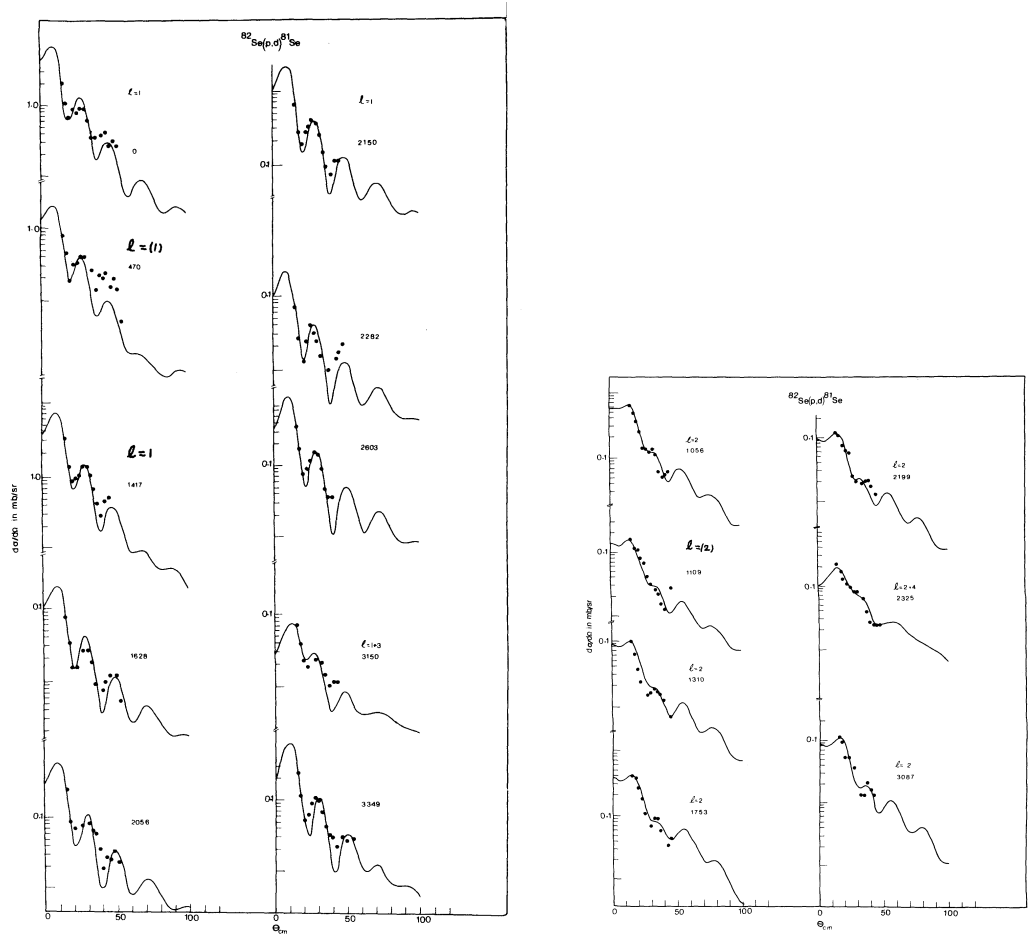

Figure 14.6. Deuteron angular distributions for the reaction $^{82}$Se(p,d)$^{81}$Se induced by 33 MeV protons compared with the distorted wave calculations for $l_n = 1$ and $l_n = 1 + 3$ (the left-hand side of the figure) and for the even $l_n$ values (the right-hand side).

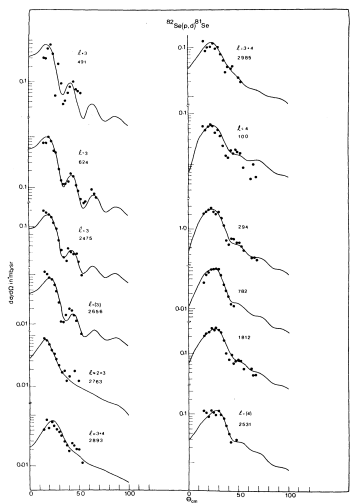

Figure 14.7. Deuteron angular distributions for the reaction $^{82}$Se(p,d)$^{81}$Se induced by 33 MeV protons compared with the distorted wave calculations for $l_n = 3$, 4, 2+3, and 3+4.

In general, the main transfer peak in the $l_n = 2$, 3, 4 angular distributions is well reproduced by the calculations while for $l_n = 1$ the calculations often tend to over-emphasize the depth of the first minimum in the data. With the exception of $^{81}$Se, the calculations had difficulty in reproducing the data for all values of $l_n$ at larger angles.





Table 14.1

Spectroscopic information for $^{75}$Se from the reaction $^{76}$Se(p,d)$^{75}$Se induced by 33 MeV protons

| $E_x$ | $l_n$ | $S$ | $E_x$ | $l_n$ | $S$ | $E_x$ | $l_n$ | $S$ | $E_x$ | $l_n$ | $S$ |
|---|---|---|---|---|---|---|---|---|---|---|---|
| 0 | 2 | 0.12 | 776 | 3 | 0.53 | 1369 | (4) | 0.15 | 1913 | (1) | 0.15 |
| 132 | 4 | 3.97 | 859 | 1 | 0.15 | 1403 | (3) | 0.22 | 2037 | (1+3) | 0.04,0.07 |
| 287 | 1 | 2.34 | 895 | 1 | 0.15 | 1484 | 3 | 0.32 | 2288 | (1+3) | 0.04,0.06 |
| 427 | 3 | 2.15 | 962 | 1 | 0.58 | 1580 | (2) | 0.03 | 2573 | 1 | 0.09 |
| 583 | 1 | 0.13 | 1050 | 3 | 0.64 | 1673 | (1) | 0.10 | | | |
| 629 | 2 | 0.15 | 1182 | 3 | 0.26 | 1768 | (3) | 0.25 | | | |
| 667 | 3 | 0.88 | 1246 | 1 | 0.04 | 1810 | 1 | 0.23 | | | |

$E_x$ – Excitation energy in keV; $l_n$ – Orbital angular momentum; $S$ – Spectroscopic factor

Table 14.2

Spectroscopic information for $^{77}$Se from the reaction $^{78}$Se(p,d)$^{76}$Se induced by 33 MeV protons

| $E_x$ | $l_n$ | $S$ | $E_x$ | $l_n$ | $S$ | $E_x$ | $l_n$ | $S$ | $E_x$ | $l_n$ | $S$ |
|---|---|---|---|---|---|---|---|---|---|---|---|
| 0 | 1 | 0.61 | 436 | 3 | 0.64 | 1012 | 1 | 0.26 | 1470 | 1 | 0.48 |
| 178 | 4 | 4.39 | 522 | 1 | 1.7 | 1183 | (3) | 0.43 | 1522 | (3) | 0.81 |
| 250 | (3) | 4.12 | *680* | *2* | *0.17* | 1238 | 3 | 0.93 | 1717 | 1 | 0.45 |
| *302* | *(2)* | *0.16* | 820 | 1 | 0.45 | 1366 | (1) | 0.17 | 2209 | 1 | 0.19 |

Table 14.3

Spectroscopic information for $^{79}$Se from the reaction $^{80}$Se(p,d)$^{79}$Se induced by 33 MeV protons

| $E_x$ | $l_n$ | $S$ | $E_x$ | $l_n$ | $S$ | $E_x$ | $l_n$ | $S$ | $E_x$ | $l_n$ | $S$ |
|---|---|---|---|---|---|---|---|---|---|---|---|
| 0 | (4) | 0.36 | *720* | *2* | *0.45* | 1385 | (3) | 0.72 | 2092 | 3 | 2.02 |
| 97 | (1) | 3.82 | 979 | 1 | 5.17 | 1674 | (2) | 0.09 | 2212 | (2) | 0.36 |
| *137* | *4* | *9.26* | 1096 | 4 | 1.98 | 1817 | (3) | 1.93 | | | |
| 365 | 3 | 7.19 | 1258 | 2 | 0.14 | 1964 | 1 | 0.72 | | | |
| *623* | *2* | *0.325* | 1328 | (3) | 1.17 | 2037 | 1 | 0.76 | | | |





Table 14.4

Spectroscopic information for $^{81}$Se from the reaction $^{82}$Se(p,d)$^{81}$Se induced by 33 MeV protons

| $E_x$ | $l_n$ | $S$ | $E_x$ | $l_n$ | $S$ | $E_x$ | $l_n$ | $S$ | $E_x$ | $l_n$ | $S$ |
|---|---|---|---|---|---|---|---|---|---|---|---|
| 0 | 1 | 1.32 | *1056* | *2* | *0.21* | 2056 | 1 | 0.16 | 2656 | 3 | 0.46 |
| 100 | (4) | 0.19 | *1109* | *2* | *0.05* | 2150 | 1 | 0.74 | 2763 | (2+3) | 0.03,0.07 |
| 294 | 4 | 5.61 | *1310* | *2* | *0.05* | 2199 | 2 | 0.05 | | | |
| 470 | (1) | 0.60 | 1417 | 1 | 2.25 | 2282 | 1 | 0.09 | 2893 | (3+4) | 0.23, 0.14 |
| 491 | (3) | 1.74 | 1628 | 1 | 0.05 | 2325 | (2+4) | 0.07,0.12 | 2985 | (3+4) | 0.21,0.16 |
| 624 | 3 | 2.37 | 1753 | 2 | 0.16 | 2531 | 4 | 0.35 | *3087* | *(2)* | *0.16* |
| 782 | 4 | 0.14 | 1812 | 4 | 1.07 | 2603 | 1 | 0.30 | 3150 | (1+3) | 0.05,0.07 |

Table 14.5

Occupation numbers and centre-of-gravity energies for the ground states in Se isotopes

| Isotope | $lj$ | $n_{jl}$ | $n'_{jl}$ | $V_j^2$ | | $E_j$ | |
|---|---|---|---|---|---|---|---|
| | | | | Exp. | Theory | Exp. | Theory |
| $^{76}$Se | 2p$_{1/2}$+2p$_{3/2}$ | 6 | 4.2 | 0.70 | 0.75 | 0.70 | 0.96 |
| | 1f$_{5/2}$ | 6 | 5.4 | 0.90 | 0.91 | 0.76 | 1.01 |
| | 1g$_{9/2}$ | 2 | 4.2 | 0.42 | 0.41 | 0.18 | 0.03 |
| $^{78}$Se | 2p$_{1/2}$+2p$_{3/2}$ | 6 | 4.3 | 0.72 | 0.80 | 0.85 | 1.20 |
| | 1f$_{5/2}$ | 6 | 6.0 | 1.00 | 0.94 | 0.61 | 1.35 |
| | 1g$_{9/2}$ | 4 | 4.4 | 0.44 | 0.55 | 0.17 | 0.01 |
| $^{80}$Se | 2p$_{1/2}$+2p$_{3/2}$ | 6 | 5.8 | 0.96 | 0.87 | 0.76 | 1.48 |
| | 1f$_{5/2}$ | 6 | 6.0 | 1.00 | 0.96 | 1.00 | 1.73 |
| | 1g$_{9/2}$ | 6 | 5.3 | 0.53 | 0.70 | 0.29 | 0.11 |
| $^{82}$Se | 2p$_{1/2}$+2p$_{3/2}$ | 6 | 5.8 | 0.97 | 0.92 | 1.28 | 1.70 |
| | 1f$_{5/2}$ | 6 | 5.8 | 0.96 | 0.99 | 1.24 | 1.99 |
| | 1g$_{9/2}$ | 8 | 7.6 | 0.76 | 0.76 | 0.76 | 0.26 |

$j$ – single particle configuration.

$n_{jl}$ – The experimentally determined average number of neutrons in the configuration $jl$.

$n'_{jl}$ – The number of neutrons expected in the absence of the residual interaction.

Exp. – experimental values as derived from the (p,d) reactions of Se isotopes induced by 33 MeV protons.

Theory – theoretical values calculated using the pairing theory of Kisslinger and Sorensen (1960, 1963) and my computer code.

Spectroscopic factors were extracted from the data by comparing the theoretical calculations with the experimental distributions. A summary of spectroscopic information derived in this study (excitation energies, orbital angular momenta, and spectroscopic factors) is given in Tables 14.1-14.4.





As expected, we have observed many $l_n = 1$ and 3 transitions as well as an increasing number of $l_n = 4$ transitions with the increasing atomic number $A$ of Se isotopes. All transitions belong to the pickup of neutrons from the $2p_{1/2}$, $2p_{3/2}$, $1f_{5/2}$, and $1g_{9/2}$ orbits.

However, we have also observed a number of anomalous $l_n = 2$ transitions. The states belonging to these anomalous transitions will be discussed later.

Using the experimentally determined spectroscopic factors we have calculated neutron occupation numbers for configurations in the f-p shell and compared them with the theoretical calculations using the pairing theory and the computer code I have written. (The procedure is described in Chapter 12.) As the spins of states in the relevant residual nuclei are generally unknown, we had to consider an average occupation numbers for configurations $2p_{1/2}$ and $2p_{3/2}$.

$$(V_{2p}^2)_{av}^{(th)} = \frac{V_{1/2}^2 + 2V_{3/2}^2}{3}$$

$$(V_{2p}^2)_{av}^{(\exp)} = \frac{\sum_i S_i(l_n = 1)}{6}$$

where $(V_{2p}^2)_{av}^{(th)}$ and $(V_{2p}^2)_{av}^{(\exp)}$ are the theoretical and experimental average occupation numbers, respectively, for the configurations $2p_{1/2}$ and $2p_{3/2}$, $S_i(l_n = 1)$ is the spectroscopic factor for $l_n = 1$ transition to state $i$, and the sum is over all $l_n = 1$ states in a given Se isotope. Results of our calculations are presented in Table 14.5.

To try to account for $l_n = 2$ transitions we have carried out the Coriolis calculations. Coriolis coupling model was used by Sanderson and Summers-Gill (1976) to reproduce energy levels for anomalous low-energy positive parity states as well as various E2 and M1 transition rates in $^{75}$Se. In addition, Heller and Friedman (1974, 1975) have reproduced moments and transition rates for several nuclei in this mass region using the same model. It was therefore worthwhile to extend such calculations to other Se isotopes.

Coriolis coupling model assumes that the odd neutron moves in the deformed potential of the rotating, axially symmetric doubly even core. The coupling of this neutron to the deformed core then gives rise, in the case of Se, to positive parity energy spectra associated with strong mixing between rotational bands built on Nilsson states arising from the $1g_{9/2}$ orbital.

We have carried out our calculations using computer code of Caseten and Newton (1968). We have used the Nilsson and pairing-model parameters derived from experimental information by Sanderson. The deformation parameter $\beta$ was varied from -0.3 to 0.3 but no attempt was made to do an exhaustive search on other parameters.

Table 14.6 compares the calculated level energies and (p,d) spectroscopic factors for low-lying positive parity states with those obtained from experiment. A positive deformation of $\beta = 0.275$ was found to give the closest agreement with the data for all isotopes and is the same as determined Sanderson for $^{75}$Se.

In general, the model generates sets of low-lying $^5/_2{}^+$, $^7/_2{}^+$ and $^9/_2{}^+$ states whose excitation energies are in close vicinity of the corresponding experimental values.





However, the calculated spectroscopic factors, while of the correct order of magnitude, are not always close to the experimental values.

Table 14.6

Results of the Coriolis coupling calculations compared with the experimental data for $^{75, 77, 79, 81}$Se

| Nucleus | $J^\pi$ | $E_{exp}$ | $E_{th}$ | $S_{exp}$ | $S_{th}$ | Nucleus | $J^\pi$ | $E_{exp}$ | $E_{th}$ | $S_{exp}$ | $S_{th}$ |
|---------|---------|-----------|----------|-----------|----------|---------|---------|-----------|----------|-----------|----------|
| $^{75}$Se | $5/2^+$ | 0 | 58 | 0.12 | 0.44 | $^{77}$Se | $7/2^+$ | 162 | 162 | | 0.04 |
| | $7/2^+$ | 112 | 364 | | 0.03 | | $9/2^+$ | 178 | 174 | 5.39 | 4.00 |
| | $9/2^+$ | 132 | 0 | 3.97 | 2.87 | | $5/2^+$ | 302 | 325 | 0.19 | 0.30 |
| $^{79}$Se | $7/2^+$ | 0 | 0 | 0.14 | 0.05 | $^{81}$Se | $7/2^+$ | 100 | 100 | 0.19 | 0.04 |
| | $9/2^+$ | 137 | 101 | 3.56 | 3.95 | | $9/2^+$ | 294 | 236 | 5.61 | 3.79 |
| | $1/2^+$ | | 527 | | 0.12 | | $1/2^+$ | | 622 | | 0.25 |
| | $5/2^+$ | 623 | 569 | 0.12 | 0.27 | | $3/2^+$ | | 782 | | 0.36 |
| | | | | | | | $5/2^+$ | 1056 | 1238 | 0.21 | 0.22 |

As discussed in the Introduction, selenium isotopes present an interesting but difficult case for nuclear structure interpretation. They are in the mass region where a variety of nuclear models have been tried and have failed to provide a satisfactory description of the features observed experimentally. Nuclei in this region are soft and consequently have a rich band structure with a co-existence of states belonging to various shapes. Multiple band crossing and band mixing is quite common. Both neutron and proton rotation-aligned bands are present. There is also evidence for a shape transition at around $N = 40$.

Theoretical treatment of such nuclei is difficult. As a possible alternative to the mentioned theoretical interpretation of nuclei in this region, dynamic deformation theory has been suggested (Kumar *et al.* 1977, Kumar 1978). It includes shape co-existence, shape transition and pair fluctuations. The theory has been used to interpret the structure of the of $^{70,72,74}$Ge nuclei (Kumar 1978).

**Summary and conclusions**

The odd neutron nuclei $^{75,77,79,81}$Se have been studied using the (p,d) reaction at the proton energy of 33 MeV. Angular distributions for the emitted deuterons have been measured from 15° to around 60° using particle identification telescopes, and the experimental results analysed using the distorted wave procedure. From the distorted wave calculations assignments of $l_n = 1, 2, 3$ and 4 have been made. Spectroscopic factors were also extracted from the angular distributions. Many new states were identified.

Using the experimentally determined spectroscopic factors, occupation numbers have been calculated and compared with the predictions of the pairing theory. Experimentally determined occupation numbers are in close agreement with the calculated values.





The presence of the low-energy $^5/_2{}^+$ and $^7/_2{}^+$ states in the Se region has long attracted theoretical attention. It has been shown here that the existence of these states can be explained reasonably well by assuming Coriolis coupling model.

## References


Barrette, J., Marrette, M., Lamoureux, G., Monara, S. and Markiza, S. 1974, *Nucl. Phys.* **A235**:154.

Becchetti, F. D. and Greenlees, G. W. 1969, *Phys. Rev.* **182**:1190.

Casten, R. and Newton, C. S. 1968, Niels Bohr Institute Computer Centre Report GAP 15-16.

Flowers, B. H. 1952, *Proc. Roy. Soc.* **215**:398.

Goswami, A. and Nalcioglu, O. 1968, *Phys. Lett.* **30B**:397.

Heller, S. L. and Friedman, J. N. 1974, Phys. Rev. **C10**:1509.

Heller, S. L. and Friedman, J. N. 1975, Phys. Rev. **C12**:1006

Holzwarth, G. and Lie, S. G. 1972, *Z. Phys.* **249**:332.

Kisslinger, L. S. and Sorensen, R. A. 1960, *Kgl. Danske Videskab. Selskab. Mat. Fys. Medd.* **32**, No. 9.

Kisslinger, L. S. and Sorensen, R. A. 1963, *Rev. Mod. Phys.* **35**:853.

Kumar, K., Remaud, B., Auger, P., Vaagen, J. S., Rester, A. C., Foucher, R. and Hamilton, J. H. 1977, *Phys. Rev.* **C16**:1235.

Kumar, K. 1978, *J. of Phys.* **G4**:849.

Kunz, P. D. 1966, University of Colorado Report COO-536-606.

Lieder, R. and Draper, J. 1970, *Phys. Rev.* **C2**:531.

and 1975, *Phys. Rev.* **12C**:1035.

Lin, E. K. 1965, *Phys. Rev.* **139**:B340.

McCauley, D. G. and Draper, J. E. 1971, *Phys. Rev.* **C4**:475

Nolte, E., Kutschera, W., Shida, Y. and Moringa, H. 1977, *Phys. Lett.* **33B**:294.

Perey, C. M. and Perey, F. G. 1969, *Phys. Rev.* **152**:923.

Sanderson, N. 1973, *Nucl. Phys.* **A216**:173.

Sanderson, N. E. and Summers-Gill, R. G. 1976, *Nucl. Phys.* **A261**:93

Scharff, G. and Weneser, J. 1955, Phys. Rev. 98:212.

Scholz, W. and Malik, F. B. 1968, *Phys. Rev.* **176**:1355.

Sherwood, A. and Goswami, A. 1966, *Nucl. Phys.* **89**:465

Wyckhoff, W. G. and Daraper, J. E. 1973, *Phys. Rev.* **C8**:796.






<div style="text-align: center">

**[15](#)**

**Analysis of Polarization Experiments**

</div>

***Key features:***

1. I have written a computer program, which carries out global analysis of particle spectra and converts them into angular distributions of differential cross sections and analyzing powers.

2. Programming language: FORTRAN-77.

3. Input parameters have to be defined for only one particle spectrum. The program employs nuclear kinematics to trace the positions of peaks for all remaining spectra taken at different angles and performs Gaussian analysis or channel-by-channel integration, depending on the selected option.

4. Computing time: On the VAX VMS 3.6 machine, the program takes less than 1.5 minutes to fit Gaussian distributions to about 200 particle groups and convert them to differential cross sections and polarization distributions.

5. Output is in both tabulated and graphic forms. The program contains interactive options to increase the flexibility of its use and to allow for a prompt and easy assessment of results.

6. The program can analyse up to 20 particle groups corresponding to different excitation energies and produce angular distributions of the differential cross sections and analyzing powers for all of them. Up to 200 particle spectra can be analysed, each containing up to 2048 channels, obtained using up to 8 detectors. Thus, the program can analyse up to 32,000 spectral peaks in one session. These imposed limitations are more than adequate to analyse any experimental results but they can be easily increased.

***Abstract:*** Program LORNA executes *global* analysis of particle spectra generated in nuclear reactions induced by polarized projectiles. Peak positions have to be defined for only one spectrum. The program employs nuclear kinematics to trace the positions at different reactions angles. Spectra corresponding to different signs of the source polarization, taken at various reaction angles by an array of up to 8 detectors and containing up to 20 particle groups are analysed and converted to angular distributions of differential cross-sections and the analyzing powers.

## Introduction

When I visited the Max-Plank-Institut für Kerphysik to carry out research on deuteron polarization I was surprised that there was no suitable program available for calculating angular distributions of differential cross sections and polarizations from particle spectra obtained in polarization experiments. It was therefore necessary for me to write a suitable computer code to support my work.

Measurements of the angular distributions of particles from nuclear reactions are carried out using an array of detectors. Results of measurements are usually contained in a great number of particle spectra taken at various angles. Each spectrum may be composed of a number of particle groups corresponding to various excitation energies of residual nuclei. For reactions induced by polarized projectiles a set of particle spectra has to be taken for each reaction angle with pairs of detectors positioned on each side of the incident beam of polarized particles and corresponding to different





signs of beam polarization. For instance, for measurements of deuteron polarization using cyclotrons, a set of eight spectra is required for each reaction angle to determine all analyzing powers. Information contained in such an assembly of raw data has to be untangled and presented in the form of angular distributions suitable for further theoretical analysis.

The aim in writing the computer code LORNA was to combine various stages of such data reduction calculations into one and easy to execute *global* analysis of particle spectra, in which the massive amount of data recorded by particle detectors is analysed *en block* and converted to final angular distributions of differential cross sections and analyzing powers.

My aim was to minimise the number of input parameters and thus make the job simple and easy for the user. An important part of achieving this aim was to incorporate nuclear kinematics calculations in the program. I also wanted the user to be able to interact with the program and to have opportunity of assessing the results of the calculations quickly and easily. This requirement made it necessary to incorporate an interactive component in the program and to supply suitable visual displays. The program also had to produce results in tabulated form, which could be readily available to further theoretical analysis.

By using the basic information supplied by the user, the program LORNA calculates kinematics calibration coefficients for each set of spectra. The user is asked to select one spectrum only and to indicate groups of particles for which the angular distributions should be calculated by the program. Employing nuclear reaction kinematics calculations, the program traces automatically the indicated peaks in all spectra, performs the desired integration, subtracts background, carries out all other necessary arithmetical manipulations (error calculation, averaging, lab to centre-of-mass (c.m.) conversion, etc.) and produces angular distributions of the differential cross-sections and analyzing powers in a final form (both numerical and graphic).

Calculations can be carried out either for gas of solid targets. Peak intensities can be extracted either through the channel-by-channel integration or by fitting symmetric Gaussian distributions. Calculated differential cross-sections are expressed in the absolute units (mb/sr) in the laboratory and centre-of-mass systems of reference. For the elastic scattering, Rutherford scattering calculation is included in the program to allow for the customary $\sigma(\theta)/\sigma_R(\theta)$ presentation of the data.

Procedure for extracting analyzing powers from the particle spectra depends on the projectile spin $S$ and on a particular experimental configuration employed in the measurements. Program LORNA has been written for $S = 1$ particles. Provision has been made for other spins and suitable modification of the program can be made without changing the general calculation procedure adopted in the program. Being written for polarized ions, the program can be also used for reactions induced by unpolarized particles.

In order to simplify the user's interaction with the computer, graphic display has been incorporated in the program both in the input and output stages. However, the program can be used without graphic display and, therefore, it is adaptable to a wider range of computer installations.

There are no strict limitations on the number of spectra sets, which can be analysed by the program, on the size of the spectra, and on the maximum number





of particle detectors used in measurements. The selected limits of 200, 2048 and 8, respectively, can be changed easily by changing the appropriate parameter definitions at the beginning of the main part of the program and in some subroutines.

Programs with broad application and thus containing great number of options and pathways are often difficult to use. My aim in writing program LORNA was to make it as simple as possible and yet to allow for sufficient flexibility in order to make it applicable to different experimental environments.

**Program arithmetic**

***Differential cross sections***

Expressions for the differential cross sections have the following forms:

*For solid targets*

$$\sigma(\theta) = k_s \frac{JMqN\tau}{C\Omega tf} \quad \text{(in mb/sr)}$$

where

$J$ – Jacobian of transformation from the *lab* to *c.m.* system

$M$ – atomic mass of the target material

$q$ – projectile charge state

$N$ – number of counts

$\tau$ – dead time correction factor

$C$ – total beam charge in $\mu$C

$\Omega$ – detector solid angle in sr

$t$ – target thickness in mg/cm2

$f$ – observed fraction of the total charge distribution of the projectile

$k_s = 2.66018 \times 10^{-7}$

*For gas targets*

$$\sigma(\theta) = k_g \frac{JTqN\tau \sin\theta_L}{CGnpf}$$

where

$T$ – gas temperature in $^0K$

$\theta_L$ – laboratory scattering angle

$n$ – gas atomicity[14]

$p$ – gas pressure in Torr

$G$ – gas factor[15] in cm

$k_g = 1.65986 \times 10^{-6}$

---

[14] The number of atoms in a molecule.
[15] $G_{00}$ in Silverstein (1959).





### The analyzing powers

Polarization of an assembly of $S = 1$ particles can be described using either Cartesian or spherical spin moments, which in turn are defined using basic angular momentum operators. In the reference frame where the $z$-axis is aligned with the quantization axis $\vec{s}$ (see Figure 15.1) deuteron polarization is described using two non-zero quantities $p_z$ and $p_{zz}$. If $N_+$, $N_0$, and $N_-$ represent the fractional populations along the symmetry axis, for the deuteron spin projections *up*, *zero*, or *down*, respectively, then the asymmetry in the fractional populations relative to the field direction (i.e. along the symmetry axis) is given by

$$p_z = \frac{N_+ - N_-}{N_+ + N_- + N_0}$$

This quantity is known as the vector polarization.

Asymmetry in the plane perpendicular to the field direction is given by

$$p_{zz} = \frac{N_+ + N_- - 2N_0}{N_+ + N_- + N_0}$$

which is known as tensor polarization. The quantities $p_z$ and $p_{zz}$ are the expectation values of the spin operators $S_z$ and $S_{zz}$ given by the following matrices

$$S_x = \begin{bmatrix} 1 & 0 & 0 \\ 0 & 0 & 0 \\ 0 & 0 & -1 \end{bmatrix} \quad S_{zz} = \begin{bmatrix} 1 & 0 & 0 \\ 0 & -2 & 0 \\ 0 & 0 & 1 \end{bmatrix}$$

(see the Appendix F).

The relevant spherical spin operators are defined as

$$\tau_{10} = \sqrt{\frac{3}{2}} S_x \text{ and } \tau_{20} = \frac{1}{\sqrt{2}} S_{zz}$$

If we write their expectation values as $t_{10}$, for the vector polarization, and $t_{20}$, for the tensor polarization, still in the reference frame with the z-axis along the quantization axis, then we have the following relations between the Cartesian and spherical definitions of polarizations:

$$t_{10} = \sqrt{\frac{3}{2}} p_z \text{ and } t_{20} = \frac{1}{\sqrt{2}} p_{zz}$$

According to the Madison Convention (Barshal and Haeberlie 1971), measured polarization is defined in the right-handed frame of references with the $z$-axis in the direction of the incident momentum $\vec{k}_{in}$ and the $y$-axis in the direction of $\vec{k}_{in} \times \vec{k}_{out}$, where $\vec{k}_{out}$ is the momentum for the outgoing particles. The rotation from the system defined by the quantization axis to the new coordinate system complicates the description of the polarization and introduces expressions that depend on $\beta$ and $\varphi$ defined in Figure 15.1.





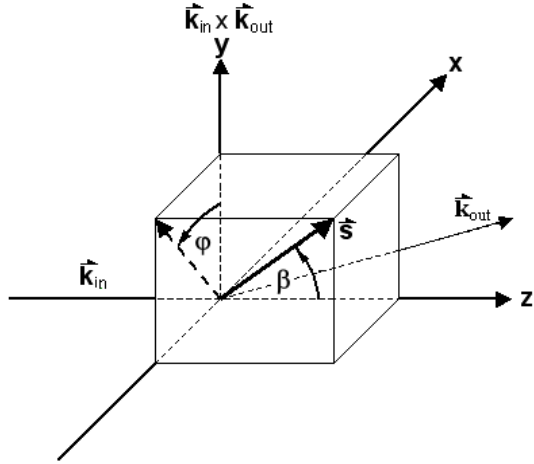

Figure 15.1. The relation between the quantization axis $\vec{s}$ and the right-handed system of coordinates (xyz) as defined by the Madison Convention.

Using the spherical components of polarization, the differential cross section for reactions induced by polarized the $S = 1$ particles is then given by

$$\sigma(\theta, \varphi) = \sigma_0(\theta)[1 + A + B + C + D]$$

with

$$A = \sqrt{2}(\sin \beta)(\cos \varphi)t_{10}iT_{11}(\theta)$$

$$B = \frac{1}{2}(3\cos^2 \beta - 1)t_{20}T_{20}(\theta)$$

$$C = \sqrt{\frac{3}{2}}(\sin 2\beta)(\sin \varphi)t_{20}T_{21}(\theta)$$

$$D = -\sqrt{\frac{3}{2}}(\sin^2 \beta)(\cos 2\varphi)t_{20}T_{22}(\theta)$$

Where $iT_{11}(\theta)$, $T_{20}(\theta)$, $T_{21}(\theta)$, $T_{22}(\theta)$ are the parameters describing the polarization of the outgoing particles. They are also known as the analyzing powers. The quantities $t_{10}$ and $t_{20}$ describe the beam polarization (in the symmetry axis system of reference) produced by the polarized ion source.

To measure the analyzing powers, the required angles $\beta$ and $\varphi$ are selected using a spin rotation device such as a Wien filter. The analyzing powers $iT_{11}(\theta)$, $T_{20}(\theta)$, $T_{21}(\theta)$, $T_{22}(\theta)$ can be measured by using a pair of symmetrically positioned detectors on the left and the right hand sides of the beam, corresponding to angles $\varphi$ and $\varphi + \pi$ and by changing the beam polarization in the source between positive and negative. Four counting rates are recorded for each reaction angle $\theta$: $N_L^+, N_L^-, N_R^+, N_R^-$. Using these counting rates one can calculate the ratios:

$$L = \frac{N_L^+ - N_L^-}{N_L^+ + N_L^-} \text{ and } R = \frac{N_R^+ - N_R^-}{N_R^+ + N_R^-}$$

These terms can be expressed as:





$$L =$$

$$+ \sqrt{2}(\sin \beta)(\cos \varphi)t_{10}iT_{11}(\theta) + \frac{1}{2}(3\cos^2 \beta - 1)t_{20}T_{20}(\theta)$$

$$+ \sqrt{\frac{3}{2}}(\sin 2\varepsilon)(\sin \varphi)t_{20}T_{21}(\theta) - \sqrt{\frac{3}{2}}(\sin^2 \beta)(\cos 2\varphi)t_{20}T_{22}(\theta)$$

$$R =$$

$$- \sqrt{2}(\sin \beta)(\cos \varphi)t_{10}iT_{11}(\theta) + \frac{1}{2}(3\cos^2 \beta - 1)t_{20}T_{20}(\theta)$$

$$- \sqrt{\frac{3}{2}}(\sin 2\beta)(\sin \beta)t_{20}T_{21}(\theta) - \sqrt{\frac{3}{2}}(\sin^2 \beta)(\cos 2\varphi)t_{20}T_{22}(\theta)$$

By selecting suitable values for $\beta$ and $\varphi$, these left-right measurements can be used to determine all four analyzing powers $iT_{11}(\theta)$, $T_{20}(\theta)$, $T_{21}(\theta)$, and $T_{22}(\theta)$:

$$iT_{11}(\theta) = \frac{1}{t_{10}\sqrt{2}}\frac{L-R}{2} \text{ for } (\beta, \varphi) = (90^0, 0^0)$$

$$T_{20}(\theta) = \frac{1}{t_{20}}\frac{L+R}{2} \text{ for } \beta = 0^0$$

$$T_{21}(\theta) = \frac{\sqrt{2}}{t_{20}\sqrt{3}}\frac{L-R}{2} \text{ for } (\beta, \varphi) = (45^0, 90^0)$$

$$T_{22}(\theta) = \frac{\sqrt{2}}{t_{20}\sqrt{3}}\frac{L+R}{2} + \frac{1}{\sqrt{6}}T_{20}(\theta) \text{ for } (\beta, \varphi) = (90^0, 90^0)$$

Measurements with purely vector-polarized beam have to be carried out to determine $iT_{11}(\theta)$. To determine $T_{22}(\theta)$ one must first determine $T_{20}(\theta)$.

The errors associated with the inaccuracies in $\beta$ and $\varphi$ are given by:

$$\Delta iT_{11}(\theta) = \frac{1}{2}[(\Delta \beta)^2 + (\Delta \varphi)^2]iT_{11}(\theta)$$

$$\Delta T_{20}(\theta) = \frac{3}{2}(\Delta \beta)^2 T_{20}(\theta) + \sqrt{\frac{3}{2}}(\cos 2\varphi)(\Delta \beta)^2 T_{22}(\theta)$$

$$\Delta T_{21}(\theta) = \left[2(\Delta \beta)^2 + \frac{1}{2}(\Delta \varphi)^2\right]T_{21}(\theta) + \frac{\sqrt{2}}{3}(\Delta \varphi + \Delta \beta \Delta \varphi)iT_{11}(\theta)$$

$$\Delta T_{22} = \left[\frac{1}{2}(\Delta \beta)^2 + 2(\Delta \varphi)^2\right]T_{22}(\theta) - \sqrt{\frac{3}{2}}(\Delta \varphi)^2 T_{20}(\theta)$$

It can be seen that measurements of $iT_{11}(\theta)$, $T_{20}(\theta)$, and $T_{22}(\theta)$ are insensitive to small deviations in the angles $\beta$ and $\varphi$. The errors depend quadratically on $\Delta \beta$ and $\Delta \varphi$, and consequently deviations of up to about $4^0$ result in errors less than 1% for these analyzing powers. This is important, because precise determination of $\beta$ and $\varphi$





is difficult. Only $\Delta T_{21}(\theta)$ depends linearly on $\Delta \varphi$ and to determine this component with an accuracy of 1% the accuracy of $\Delta \varphi \leq 2^0$ is required.

For experiments carried out with cyclotrons, the angle $\beta$ is fixed by the cyclotron's magnetic field, which means that the angle $\beta$ is determined by the orientation of the plane containing particle detectors (see below). For the fixed vertical quantization axis, it is convenient to express the differential cross sections in terms of the Cartesian spin moments:

$$\sigma(\theta,\varphi) = \sigma_0(\theta)\left[1 + \frac{3}{2}(\cos\varphi)p_z A_y(\theta) + \frac{1}{2}(\sin^2\varphi)p_{zz}A_{xx}(\theta) + \frac{1}{2}(\cos^2\varphi)p_{zz}A_{yy}(\theta)\right]$$

where $p_z$ and $p_{zz}$ are the source polarizations, as defined earlier; $A_y$, $A_{xx}$, and $A_{yy}$ are the polarizations (i.e. the analyzing powers) of outgoing particles. The relations between spherical and Cartesian spin moments is as follows:

$$A_y(\theta) = \frac{2}{\sqrt{3}}iT_{11}(\theta)$$

$$A_{xx}(\theta) = \sqrt{3}T_{22}(\theta) - \frac{1}{\sqrt{2}}T_{20}(\theta)$$

$$A_{yy}(\theta) = -\sqrt{3}T_{22}(\theta) - \frac{1}{\sqrt{2}}T_{20}(\theta)$$

To determine all three analyzing powers, $A_y(\theta)$, $A_{xx}(\theta)$, and $A_{yy}(\theta)$, one has to carry out measurements not only in the horizontal plane but also in the vertical plane. A convenient way to do it is to have a rotating target chamber and to record *eight* counting rates for each reaction angle $\theta$ by using pairs of detectors in the horizontal and vertical planes.

First the target chamber is in the horizontal position and measurements are carried out as before using symmetrically positioned detectors on the left and right hand side of the beam. The left detector corresponds to $\varphi = 0^0$ and the right to $\varphi = \pi$. Four sets of counting rates are collected: $N_L^+, N_L^-, N_R^+$, and $N_R^-$.

Having done that, the target chamber is rotated to a vertical position without changing the reaction angle $\theta$. The detectors are now in the up and down positions and they correspond to $\varphi = \pm\pi/2$. Four more counting rates are recorded, $N_U^+, N_U^-, N_D^+$, and $N_D^-$, for the same reaction angle $\theta$. Using this set of eight counting rates, one can determine the analyzing powers $A_y(\theta)$, $A_{xx}(\theta)$, and $A_{yy}(\theta)$.

We now have the following relations:

$$L = \frac{N_L^+ - N_L^-}{N_L^+ + N_L^-} = +\frac{3}{2}p_z A_y(\theta) + \frac{1}{2}p_{zz}A_{yy}(\theta)$$

$$R = \frac{N_R^+ - N_R^-}{N_R^+ + N_R^-} = -\frac{3}{2}p_z A_y(\theta) + \frac{1}{2}p_{zz}A_{yy}(\theta)$$

$$U = \frac{N_U^+ - N_U^-}{N_U^+ + N_U^-} = \frac{1}{2}p_{zz}A_{xx}(\theta)$$





$$D = \frac{N_D^+ - N_D^-}{N_D^+ + N_D^-} = \frac{1}{2} p_{zz} A_{xx}(\theta)$$

These relations give:

$$A_y = \frac{1}{3 p_z}(L - R) \text{ for } (\beta, \varphi) = (90^0, 0^0)$$

$$A_{xx} = \frac{1}{p_{zz}}(U + D) \text{ for } (\beta, \varphi) = (90^0, 90^0)$$

$$A_{yy} = \frac{1}{p_{zz}}(L + R) \text{ for } (\beta, \varphi) = (90^0, 0^0)$$

The errors associated with the vertical position of the reaction chamber are:

$$\Delta A_y = \frac{1}{2} A_y (\Delta \varphi)^2$$

$$\Delta A_{xx} = (A_{xx} - A_{yy})(\Delta \varphi)^2$$

$$\Delta A_{yy} = (A_{yy} - A_{xx})(\Delta \varphi)^2$$

## Polarimeters

Measurements of the analyzing powers must be accompanied by the measurements of the beam polarization to determine the quantities $t_{10}$ and $t_{20}$ or $p_z$ and $p_{zz}$. This is done by using a reaction for which analyzing powers are known. The device measuring beam polarization is known as a polarimeter. The choice of a polarimeter is dictated by the type and energy of the incident particles and by the so-called figure of merit, which is the product of the differential cross section and polarization. This product should preferably be large for the energy region of interest. A well-known and widely used tensor polarimeter for the medium-energy deuterons is based on the $^3\text{He}(\vec{d}, p)^4\text{He}$ reaction with the detection of protons at $\theta = 0^0$. Examples polarimeters for the vector analyzing powers are $^4\text{He}(\vec{d}, d)^4\text{He}$ and $^{12}\text{C}(\vec{d}, d)^{12}\text{C}$ scattering.

Program LORNA incorporates the polarimeter spectra in the calculations of the analyzing powers.

## Nuclear reactions kinematics

The program uses the non-relativistic nuclear kinematics formulae (see for instance Marion 1960) to trace the positions of peaks in particle spectra.

The parameters defining nuclear reactions kinematics are shown in Figure 15.2.





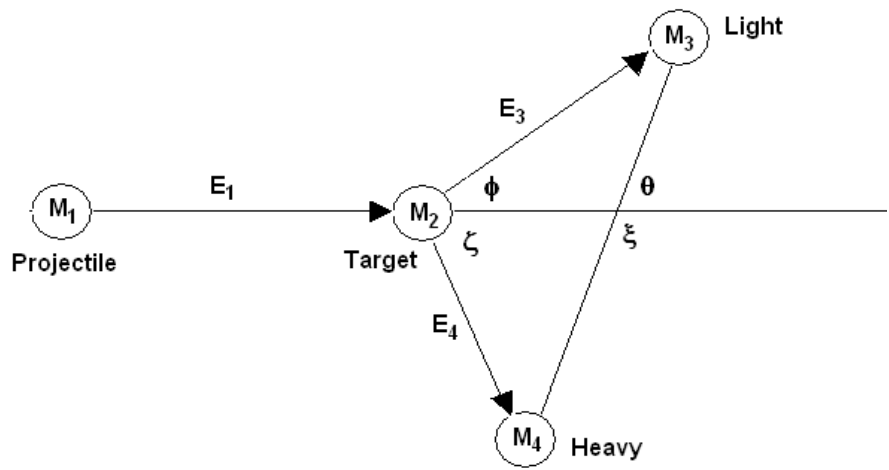

Figure 15.2. A diagram defining the quantities used in the nuclear reactions kinematics calculations. Angles $\phi$ and $\zeta$ are in the laboratory system while angles $\theta$ and $\xi$ in the centre-of-mass system.

The energy of the light product is given by:

$$E_3 = E_T\left[A + D + 2(AC)^{1/2}\cos\theta\right] = E_T B\left[\cos\phi \pm \left(D/B - \sin^2\phi\right)^{1/2}\right]^2$$

where

$$E_T = E_1 + Q = E_3 + E_4$$

$$Q = \left(M_1 + M_2 - M_3 - M_4\right)c^2$$

$$A = \frac{M_1 M_4}{\left(M_1 + M_2\right)\left(M_3 + M_4\right)}\frac{E_1}{E_T}$$

$$B = \frac{M_1 M_3}{\left(M_1 + M_2\right)\left(M_3 + M_4\right)}\frac{E_1}{E_T}$$

$$C = \frac{M_2 M_3}{\left(M_1 + M_2\right)\left(M_3 + M_4\right)}\left(1 + \frac{M_1 Q}{M_2 E_T}\right)$$

$$D = \frac{M_2 M_4}{\left(M_1 + M_2\right)\left(M_3 + M_4\right)}\left(1 + \frac{M_1 Q}{M_2 E_T}\right)$$

The plus sign in the $E_3$ equation is used if $B<D$. If $B>D$, then the maximum emission angle of the light product is given by

$$\phi_{\max} = \sin^{-1}(D/B)^{1/2}$$

The intensity ratio for light products is given by:

$$\frac{\sigma(\theta)}{\sigma(\phi)} = \frac{I(\theta)}{I(\phi)} = \frac{\sin\phi\; d\phi}{\sin\theta\; d\theta} = \frac{\sin^2\phi}{\sin^2\theta}\cos(\theta - \phi) = \frac{(AC)^{1/2}(D/B - \sin^2\phi)^{1/2}}{E_3/E_T}$$





## Input and output parameters

Every effort has been made to make the use of the program simple and easy. Many default parameters are supplied to assist the user who, however, can overrule them.

The program contains an interactive part to allow for a greater flexibility of the use of the program and for an easy assessment of results. The program has an option for either channel-by-channel integration or Gaussian analysis. In both cases, the necessary parameters have to be defined only for one particle spectrum.

For the channel-by-channel integration, one needs to define channel limits for each peak. For the Gaussian integration one needs to give approximate peak positions and the full width at half maximum.

The program uses the user-supplied initial nuclear kinematic parameters to trace the positions of peaks in all the spectra and perform the necessary calculations. Background calculations are included in the program.

Examples of the output are presented in Figures 15.3-15.6. They were copied directly from the computer-generated plots.

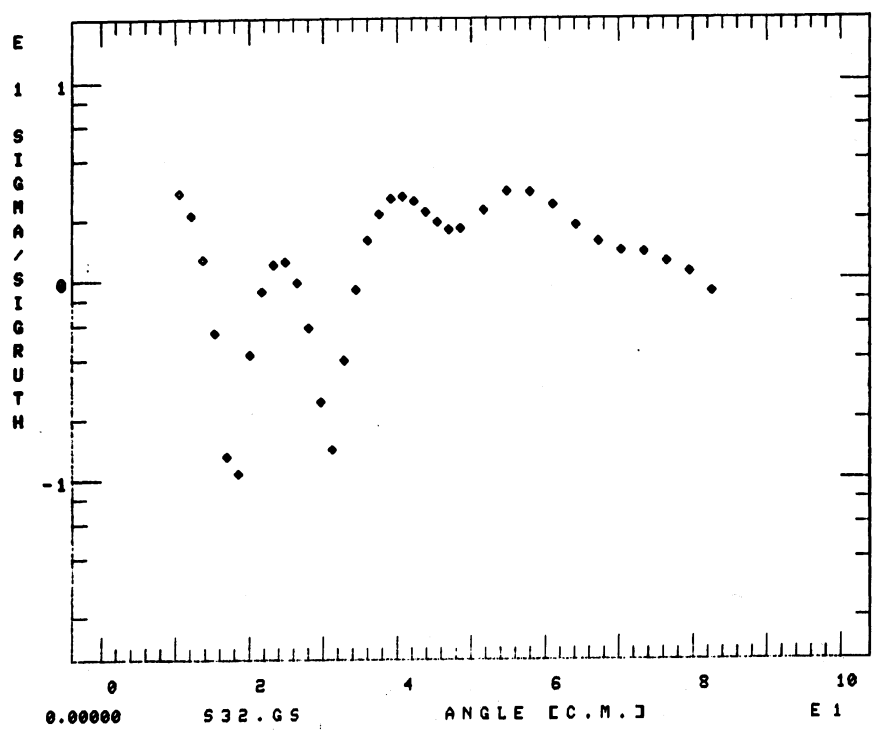

Figure 15.3 An example of the differential cross sections $\sigma(\theta)/\sigma_R(\theta)$ for the reaction $^{32}\text{S}(\vec{d},\vec{d})^{32}\text{S}$ leading to the ground state in $^{32}\text{S}$ calculated by the program LORNA from 198 particle spectra in about 1.5 minutes using the VAX VMS 3.6 computer. The angle is in the centre of mass system. The figure has been copied directly from the computer output. The first label for the horizontal scale shows the excitation energy. Here, the excitation energy is 0. The label E1 on a linear scale means that the displayed numbers should be multiplied by $10^1$. (For instance, in this example, the number 4 on the horizontal scale means $4 \times 10^1$.) The label E1 on a logarithmic scale means that the largest displayed number is $10^1$. The numbers -1, 0, and 1 on the vertical scale mean $10^{-1}$, $10^0$, and $10^1$, respectively.





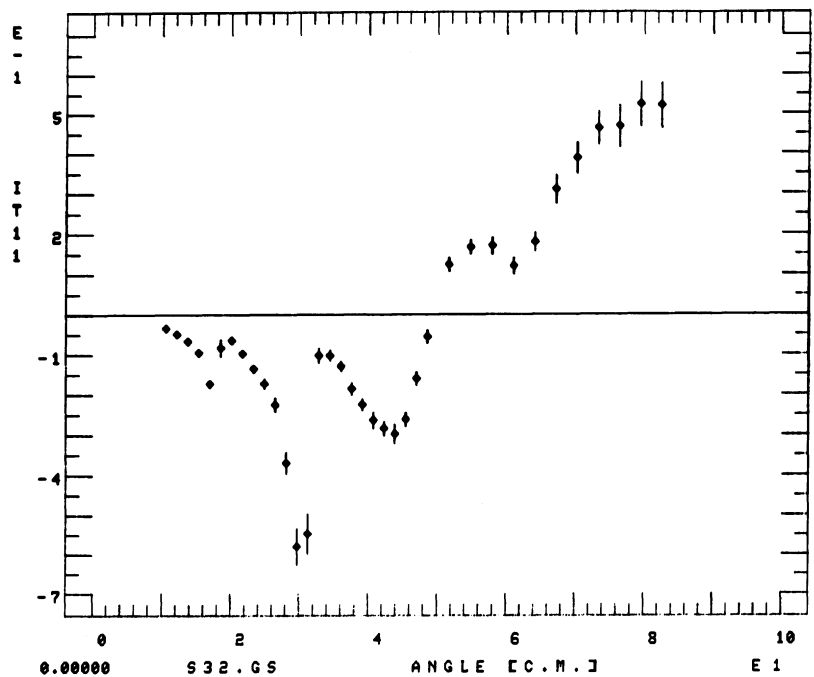

Figure 15.4 An example of the angular distribution of the vector analyzing power $iT_{11}(\theta)$ for the reaction $^{32}S(\vec{d},\vec{d})^{32}S$ leading to the ground state in $^{32}S$ calculated by the program LORNA. The numbers on the vertical scale should be multiplied by $10^{-1}$ as indicated by the label E-1. See also the caption to Figure 15.3.

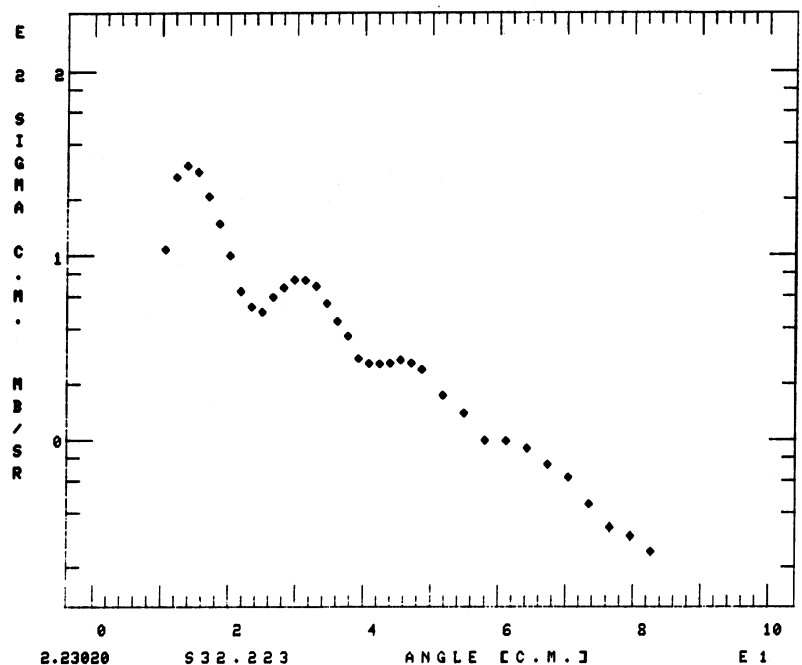

Figure 15.5 The differential cross sections $\sigma(\theta)$ for the reaction $^{32}S(\vec{d},\vec{d}')^{32}S^*$ leading to the 2.23020 MeV excited state in $^{32}S$ calculated by the program LORNA. The numbers 0, 1, and 2 on the vertical scale mean $10^0$, $10^1$, and $10^2$ as indicated by the label E2. See also the caption to Figure 15.3.





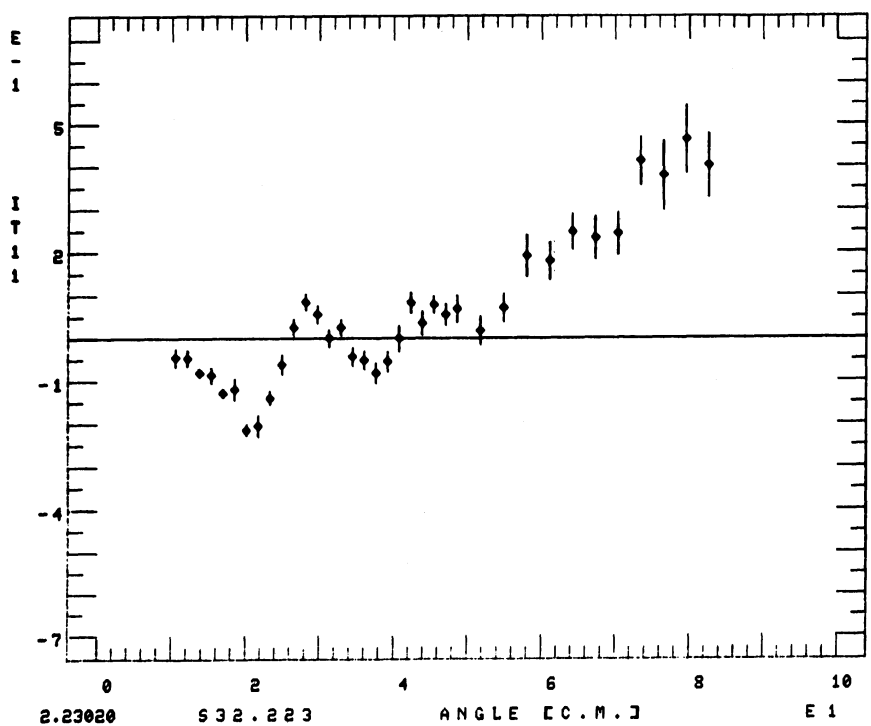

Figure 15.6 The angular distribution of the vector analyzing power $iT_{11}(\theta)$ for the reaction $^{32}\text{S}(\vec{d},\vec{d}')^{32}\text{S}^*$ leading to the 2.23020 MeV excited state in $^{32}\text{S}$ calculated by the program LORNA. The numbers on the vertical scale should be multiplied by $10^{-1}$ as indicated by the label E-1. See also the caption to Figure 15.3.

Output is in both tabulated and graphic forms. Results can be displayed on the screen or plotted. The program calculates differential cross sections, analyzing powers, and the errors. In the case of the differential cross section for the elastic scattering, the program calculated also the Rutherford scattering cross sections. The tabulated outputs contain both the list of the differential cross sections, $\sigma(\theta)$, and the ratios of the differential cross sections to Rutherford cross sections, $\sigma(\theta)/\sigma_R(\theta)$. In the graphic form the user has a choice to display any of the two.

As mentioned earlier, the program has been written specifically for the $S = 1$ projectiles. However, the discussed formalism can be appropriately changed for $S = 1/2$ particles. For larger spins, the program can be extended using for instance formulae presented by Darden (1971) for $S = 3/2$ or the general formalism discussed by Haeberli (1974).

## Acknowledgement

I want to thank my wife Lorna for tolerating the long and late hours of my absence during the development and testing of this computer code.

## References

Barshal, H. H. and Haeberlie, W. (eds) 1971, *Madison Convention, Proceedings of the 3rd International Symposium on Polarization Phenomena in Nuclear Physics, Madison, 1970*, University of Wisconsin Press, Madison, WS, p. xxv

Darden, S. E. 1971, *Polarization Phenomena in Nuclear Reactions*, eds H. H. Barshal and W. Haeberli, University of Wisconsin Press, Madison, Wisconsin, p. 39.





Haeberli, W. 1974, Nuclear Spectroscopy and Reactions, ed. J. Cerny, Academic Press, New York, p. 151 and references therein.

Marion, J. B. 1960, *Nuclear Data Tables*, Part 3, National Academy of Sciences, National Research Council, Washington, DC.

Silverstein, E. A. 1959, *Nucl. Instr. Meth.* **4**:53.

Schwandt, P. and Haeberlie, W. 1968, Nucl. Phys. **A110**:585.







# Reorientation Effects in Deuteron Polarization

***Key features:***

1. This study represents the first systematic survey of the deuteron polarization with the aim of investigating reorientation effects in the scattering of polarized deuterons.

2. Experimental part of this study was carried out using polarized deuterons supplied by the C-LASKA Karlsuhe Lambshift polarized ion source. Polarized deuterons were accelerated to 52 MeV using the Karlsruhe isochronous cyclotron.

3. Conversion of particle spectra to the angular distributions of the differential cross sections $d\sigma(\theta)/d\Omega$ and vector analyzing powers $iT_{11}(\theta)$ was carried out in Heidlberg using the MPI VAX-780 computer and the computer code LORNA described in Chapter 15.

4. Theoretical analysis was carried out using the coupled channels formalism and the computer code ECIS79, which I have modified and adapted to run at the Australian National University. All the theoretical calculations were done using the ANU UNIVAC 1100/82 computer.

5. The deuteron-nucleus interaction containing an imaginary term of the spin-orbit potential has been found to be the most suitable in describing the experimental data.

6. In the single case of a strong oblate deformation (for $^{28}$Si) experimental distribution of the vector analyzing power $iT_{11}(\theta)$ for the first exited state $2_1^+$ is distinctly different than the distributions for all the remaining isotopes. This result suggests that measurements of the $iT_{11}(\theta)$ analyzing power for the $2_1^+$ states could be used to identify nuclei with strong oblate deformations.

7. With the exception of $^{28}$Si, all other nuclei studied here are either prolate or nearly spherical. Experimental results show that the $iT_{11}(\theta)$ distributions to the $2_1^+$ states do not depend strongly on the degree of the quadrupole deformation for all these nuclei.

8. I have carried out theoretical analysis using rotational and vibrational models. With some small exceptions, they all produced similar shapes for the $iT_{11}(\theta)$ distributions for the $2_1^+$ states indicating that reorientation effects are weak at this energy and for these range of nuclei.

***Abstract:*** The differential cross sections, $d\sigma(\theta)/d\Omega$, and vector analyzing powers, $iT_{11}(\theta)$, were measured for the elastic and inelastic scattering of 52 MeV polarized deuterons from $^{20}$Ne, $^{22}$Ne, $^{26}$Mg, $^{28}$Si, $^{32}$S, $^{34}$S, $^{36}$Ar and $^{40}$Ar nuclei. Coupled channels analysis was carried out using an axially symmetric rotational model with either prolate or oblate quadrupole deformations for each isotope. Calculations using harmonic vibrator model were also carried out. In general, reorientation effects were found to be weak. A global optical model potential containing an imaginary spin-orbit component was found to be the most suitable in describing the experimental data at this energy. Our results indicate that vector analyzing power, $iT_{11}(\theta)$, can be used to identify nuclei with oblate deformation but it cannot distinguish between prolate and spherical nuclei.

## Introduction

Reorientation process is associated with self-coupling of an excited state and it involves transitions between various magnetic substates, $IM \rightarrow IM'$ (see the





Appendix K). It has been pointed out that reorientation of electric quadrupole moment for the 2+ states may play an important role in scattering of protons and deuterons (Kurepin, Lombard and Raynal 1973; Raynal1976). For deformed nuclei, the 2+ - 2+ matrix element, which is proportional to the quadrupole moment, $Q_{2^+}$, may have a strong influence on the analyzing powers. The effect is energy dependent and is expected to be stronger at lower projectile energies. However, due to contributions from indirect processes, such as energy fluctuations and resonance scattering, interpretation of the low-energy data may be difficult. However, it was reported that reorientation effects may be noticeable at such high energies as 65 MeV for protons and 56 MeV for deuterons (Hatanaka *et al.* 1981).

Prior to our investigation, no systematic study of reorientation effects in scattering of polarized particles has been undertaken. However, from the available results, confined mainly to energies below 25 MeV, certain patterns could be noticed.

In nearly all cases, the reorientation process was found to affect relatively strongly the analyzing powers for scattering to the first, $2_1^+$, excited states in even-even nuclei and to a certain extend also the inelastic scattering differential cross sections. Scattering to the second, $2_2^+$, states is influenced less strongly (Clement *et al.* 1980).

For oscillating analyzing powers, the reorientation process is claimed to shift oscillations to smaller angles and damp the oscillation amplitude for prolate quadrupole deformations. For oblate shapes, the effect is the opposite: oscillations are shifted towards larger angles and their amplitude is increased.

Such behaviour was demonstrated for 24.5 MeV protons scattered from $^{152}$Sm (Kurepin, Lombard and Raynal 1973), 9.4 MeV deuterons scattered from $^{28}$Si (Clement *et al.* 1978), 20.5 MeV deuterons scattered from $^{154}$Sm (Clement *et al.* 1981), 20 MeV deuterons scattered from $^{24}$Mg, $^{28}$Si and $^{54}$Cr (Clement *et al.* 1980) and 18 MeV deuterons scattered from $^{32}$S (Clement *et al.* 1981).

It should be noted that for $^{32}$S, reorientation effects were also studied using energy averaged results corresponding to deuteron incident energy of $E_d$ = 9.7 MeV (Clement *et al.* 1978). At this energy, experimental data could not be fitted using either prolate or oblate shapes. Good agreement between theory and experiment was, however, obtained assuming the harmonic vibration model. On the other hand, at $E_d$ = 18 MeV, distinction between oblate and prolate shapes for $^{32}$S is clear with a prolate quadrupole deformation giving a definitely better description of the experimental results. Thus, one set of data describes $^{32}$S as a harmonic vibrator while the other shows it to be deformed with a prolate shape. It is clear, therefore, that even in cases for which agreement between theory and experiment is good, discrimination between nuclear shapes and models should be taken with caution.

At very low energies, $\leq 10$ MeV, strong dependence on the sign of the quadrupole deformation parameter, $\beta_2$, was found not only for the inelastic but also for the elastic scattering (Clement *et al.* 1978). At these energies, experimental results had to be averaged over a range of incident energies in order to minimise effects of resonance scattering.

Less clear reorientation effects were reported for 20.3 MeV protons scattered from $^{28}$Si (Blair et al. 1970), 12.3 MeV deuterons scattered from $^{54,56,58}$Fe (Brown *et al.* 1973), 10 MeV deuterons scattered from $^{24}$Mg (Clement *et al.* 1978) and 15 MeV deuterons scattered from $^{56}$Fe (Raynal 1976). For 20.3 MeV protons and for 10 MeV





and 15 MeV deuterons, coupled channels calculations predict strong dependence of the analyzing powers on the sign of $\beta_2$ However, agreement between theory and experiment is far from satisfactory. At 12.3 MeV deuteron energy, theoretical predictions for prolate and oblate shapes are shown only for $^{58}$Fe. For this nucleus fits to the elastic scattering favour an oblate deformation while the distributions for the $2^+_1$ inelastic scattering are described better assuming a prolate shape.

At higher energies, a small but clear angular shift was reported for 65 MeV protons scattered from $^{24}$Mg (Hatanaka *et al.* 1981). The displayed fit to the analyzing powers for the $2^+_1$ state is exceptionally good for a prolate shape. However, fits to other angular distributions are not shown. A more extensive coupled channels analysis was carried out later for 65 protons scattered from $^{24}$Mg, $^{28}$Si and $^{32}$S (Kato *et al.* 1985). Results of this analysis indicated that good fits could be obtained only for the elastic scattering. Fits to the inelastic scattering, and in particular to the $2^+_1$ analyzing powers, are far from satisfactory. The discrepancy between theory and experiment is particularly strong for the $2^+_1$ state in $^{24}$Mg. These calculations put in doubt the earlier results of Hatanaka *et al.* (1981).

An attempt to distinguish between prolate and oblate shapes was also made for 65 MeV protons scattered from $^{28}$Si (Nakamura *at al.* 1978). However, the predicted small angular shift for the two shapes and the poor quality of fits make the results inconclusive.

For the 56 MeV deuterons, reorientation effects were reported for $^{24}$Mg and $^{28}$Si (Hatanaka *et al.* 1981). At this energy, the angular dependence of the vector analyzing powers $iT_{11}(\theta)$ for the $2^+_1$ states is distinctly different from that observed at lower energies. The region sensitive to the sign of $\beta_2$ was found to be confined to angles between about 20° and 60°. By comparing coupled channels calculations with the experimental data, it was possible to distinguish between prolate and oblate shapes with more clear discrimination being for $^{28}$Si. However, here again, theoretical predictions for the two signs of $\beta_2$ are shown only for the $iT_{11}(\theta)$ distributions for the $2^+_1$ states. The aim of our study was to explore more systematically this higher energy region for the *sd*-shell nuclei.

**Experimental procedure**

Measurements of angular distributions of differential cross sections $d\sigma(\theta)/d\Omega$ and vector analyzing powers $iT_{11}(\theta)$ for the elastic and inelastic scattering of 52 MeV polarized deuterons were carried out using a Lamb-shift polarized ion source (Bechtold *at al.* 1976; see also the Appendix G) and the Karlsruhe isochronous cyclotron. The beam intensity on the target was about 5 nA and the overall energy resolution of detected deuterons about 270 keV, which was mainly due to the beam resolution.

The following targets were used in the measurements: $^{20}$Ne (99.9% enriched), $^{22}$Ne (99.7%), $^{26}$Mg(99.7%), Si (with 92.2% of $^{28}$Si), S (as SH$_2$ with 95% of $^{32}$SH$_2$), $^{34}$S(, containing 89.8% of $^{34}$SH$_2$), $^{36}$Ar (99.5%) and Ar (containing 99.59% $^{40}$Ar).

Measurements of $d\sigma(\theta)/d\Omega$ and $iT_{11}(\theta)$ were carried out in the range of angles of about 10°-80° (lab), in steps of 1.5° for up to about 50° and in steps of 3° for larger angles. The method of measurement is described in Chapter 15. The detector





system consisted of *six* $\Delta E$-$E$ solid-state counter telescopes placed at symmetric angles with respect to the beam direction. The detectors were mounted on two remotely controlled, moveable tables inside a large scattering chamber. The angular distance between the adjacent detector telescopes on each table was 3° and the angular resolution of the detector slit systems was ±0.5°. The thickness of the detectors was 1500 μm or 2000 μm for the $\Delta E$ detectors, and 7000 μm (2000 μm + 5000 μm) for the $E$-detectors. The usual pulse multiplication method was used for the particle identification (see Chapter 2). During the measurements, the beam polarization was flipped every few minutes according to a preset value of the beam current integration counts (see Chapter 15).

The absolute value of the beam polarization was monitored using the $^{12}$C$(\vec{d},d\,)^{12}$C elastic scattering at $\theta = 47°$(lab). This angle corresponds to an optimum value of the figure of merit $\left[iT_{11}(\theta)\right]^2\left[d\sigma(\theta)/d\Omega\right]$ with the analyzing power $iT_{11}(47^0) = 0.318\pm0.035$ being known from the double scattering measurements (Seibt and Weddigen 1980). Using this known value of $iT_{11}(47^0)$ for the $^{12}$C$(\vec{d},d\,)^{12}$C the beam polarization $p_z$ was calculated by program LORNA (see Chapter 15) during the data reduction and used to calculate the $iT_{11}(\theta)$ for the studied reactions.

The beam polarimeter was mounted downstream, outside the main scattering chamber and it was followed by a Faraday cup, which was used in the measurements of the integrated charge. The target of the beam polarimeter consisted of a large polyethylene foil. Deuterons from the $^{12}$C$(\vec{d},d\,)^{12}$C reaction were detected by two NaI(Tl) detectors placed symmetrically in respect of the deuteron beam. The thickness of the NaI(Tl) crystals was chosen in such a way as to allow for a clear separation of deuterons from the high-energy protons from the $^{12}$C$(\vec{d},p\,)^{13}$C reaction. In addition, a thin Al foil was mounted in front of each crystal to suppress the $Z \geq 2$ particles. The average beam polarization, $p_z$, during the measurements was 0.46 ±0.05 and its stability was within about 2% over a long period of data collection.

Deuteron spectra from the main detectors and from the beam monitor were stored on magnetic tapes and were analysed off-line using an MPI VAX-780 computer system. All data reduction calculations are carried out using my computer code LORNA (Nurzynski 1985; see also Chapter 15).

As described in Chapter 15, the program performs a global analysis of particle spectra and converts them to angular distributions of differential cross sections and analyzing powers for up to 20 particle groups for each target nucleus. Typically, about 200 spectra were taken for each target. Program LORNA converted them to angular distributions in about 1.5 min for each excitation energy. Calculations of the errors of the experimental data include statistical uncertainties, background subtraction and beam polarization errors.

All experimental distributions are shown together with theoretical calculations in the next section. However, as mentioned earlier, the $iT_{11}(\theta)$ analyzing powers for the $2^+_1$ states are expected to show dependence on the sign of $\beta_2$. It is, therefore, interesting to compare them separately in one figure. Figure 16.1 shows that, except for one isotope, $^{28}$Si, all $iT_{11}(\theta)$ distributions for the $2^+_1$ states display similar features. These results will be discussed further later.





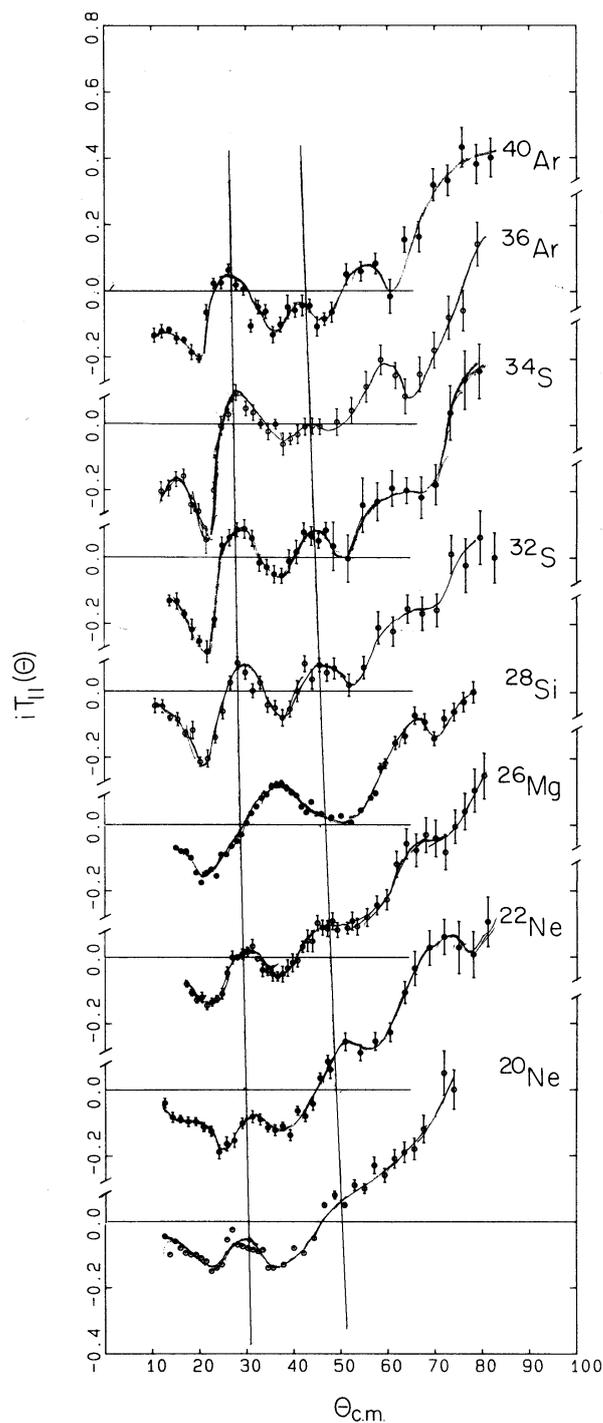

Figure 16.1. Vector analyzing power $iT_{11}(\theta)$ for the first excited states $2_1^+$ measured using 52 MeV polarized deuterons. The lines are to guide the eye for the relevant maxima.

## Coupled channels formalism

The principles of the coupled channels formalism are described in the Appendix E. However, it is useful to mention here some points related to the calculations of our data.

The Schrödinger equation for the interaction between incident and target nucleus can be written as:





$$[T - V(r,\xi) + H(\xi)]\psi(r,\xi) = E\psi(r,\xi)$$

where

$$T = -\frac{\hbar^2}{2\mu}\nabla_r^2$$

is the kinetic energy operator of the incident particle, $V(r,\xi)$ is the interaction potential between the incident particle and target nucleus, $H(\xi)$ is the intrinsic Hamiltonian for the target nucleus, and $\xi$ are the nuclear coordinates.

The intrinsic nuclear states $\phi_i(\xi)$ are the solutions of the equation:

$$H(\xi)\chi_i(\xi) = \varepsilon_i\chi_i(\xi)$$

where $i$ labels the intrinsic states.

The intrinsic wave functions form a complete orthogonal set and consequently the total wave function $\psi(r,\xi)$ can be expressed as a sum of these functions:

$$\psi(r,\xi) = \sum_i \varphi_i(r)\chi_i(\xi)$$

where the sum is over all states of the target nucleus, discrete and continuum.

Substituting it in the original Schrödinger equation we get a set of coupled equations for all channels:

$$(T - E - \varepsilon_i)\varphi_i(r) = \sum_k V_{ik}\varphi_k(r)$$

where

$$V_{ik} = \int \varphi_i^*(\xi)V(r,\xi)\varphi_k(\xi)d\xi$$

The interaction potential $V(r,\xi)$ depends on the way the target nucleus is described. For example, for a simple axially symmetric quadrupole deformation, the potential $V(r,\xi)$ will have the form:

$$V(r,\xi) = V(r - R(\theta,\phi)) \approx V(r - R_0) - \beta_2 R_0 Y_2^0(\theta,\phi)\frac{dV}{dr}$$

where

$$R(\theta,\phi) = R_0(1 + \beta_2 Y_2^0(\theta,\phi))$$

$\beta_2$ is the quadrupole deformation parameter, and $Y_2^0(\theta,\phi)$ is the spherical harmonic function.

For a simple vibrational model

$$R(\theta,\phi) = R_0\left(1 + \sum_m \alpha_m^* Y_2^{m*}(\theta,\phi)\right)$$

where $\alpha_m^*$ are the dynamical distortion parameters that create or annihilate vibrational phonon of angular momentum 2 and $z$ component $m$. The mean-square deformation parameter is given by:





$$\beta_2^2 = \left\langle 0 \left| \sum_m |\alpha_m|^2 \right| 0 \right\rangle$$

The potential $V(r, \xi)$ is then given by:

$$V(r, \xi) = V(r - R) \approx V(r - R_0) - \sum_m \alpha_m^* Y_2^{m*}(\theta, \phi) R_0 \frac{dV}{dr}$$

In practice, even with coupled channels formalism, only a small number of coupled equations are used. In the case discussed here, i.e. for the 0+- 2+ excitations, there are six coupled equations, one for 0+ and five for 2+. The sign of the quadrupole moment is related to the sign of the 2+- 2+ nuclear matrix element, which is proportional to the quadrupole moment $Q_{2^+}$. For spherical nuclei, the matrix element is zero.

The reorientation effect for the inelastic scattering can be studied by performing three sets of calculations for each target nucleus: two for negative and positive deformations and one for vibrational model.

**Theoretical analysis**

I have carried out the coupled channels analysis our data using the well-known computer code ECIS79 (Raynal, 1972, 1081), which I have modified and adapted to run at the Australian National University. All the calculations were performed using the ANU UNIVAC 1100/82 computer.

For each target isotope four distributions, $d\sigma(\theta)/d\Omega$ and $iT_{11}(\theta)$, for the ground and for the first $2_1^+$ excited states, were fitted simultaneously. In order to see whether theoretical fits are sensitive to the sign of the quadrupole deformation, independent searches of potential parameters were carried out using either positive or negative $\beta_2$ parameters for each target isotope. In all these calculations, an axially symmetric rotational model containing quadrupole and hexadecapole deformations was assumed for all nuclei. The central and the spin-orbit components of the deuteron-nucleus interaction potential were assumed to be described by the same deformation parameters.

The optical model parameter search was first carried out using as the starting values the potential parameters derived earlier by Mairle *et al.* (1980) for the 52 MeV deuteron elastic scattering. Unfortunately, searches based on any of their four sets of parameters failed to fit the measured by us angular distributions.

The potential $F'$ of Daehnick, Childs and Vrcelj (1980), containing an imaginary spin-orbit component, was tried next and was found to give a significantly better description of the experimental results. These authors carried out an extensive global optical model analysis of deuteron scattering at energies 11.8, 17, 34, 52, and 79.5 MeV in the mass range of $A$ = 27 - 232. Their potential $F'$ contains five nuclear components and the usual Coulomb potential for a uniformly charged sphere with the radius $R_c = r_c A^{1/3}$. The nuclear components are defined by a total of 13





parameters. The potential I used in my calculations was similar to their potential but the total number of parameters for nuclear components was 15.[16]

$$U(r) = V_C(r) + U_0(r) + U_{s.o.}(r)$$

where $V_C(r)$ is the Coulomb potential, $U_0(r)$ and $U_{s.o.}(r)$ are the central and spin-orbit components of the optical model potential.

$$U_0(r) = -Vf(r, r_0, a_0) - i4a_D W_D \frac{d}{dr} f(r, r_D, a_D) - W_S f(r, r_S, a_S)$$

$$U_{s.o.}(r) = \lambdabar_\pi^2 \frac{1}{r} \left[ V_{so} \frac{df(r, r_{rso}, a_{rso})}{dr} + i W_{so} \frac{df(r, r_{iso}, a_{iso})}{dr} \right] \mathbf{S} \cdot \mathbf{L}$$

where

$$\lambdabar_\pi^2 = \left( \frac{\hbar}{m_\pi c} \right)^2 = 2\,fm^2$$

$$f(r, r_i, a_i) = \frac{1}{1 + e^{x_i}} \text{ with } x_i = \frac{r - r_i A^{1/3}}{a_i} \text{ and } i = 0, D, S, rso, \text{ or } iso$$

Table 16.1
The original set of parameters for the potential $F'$

|  |  |  |  |
|---|---|---|---|
|  |  | $W_D$ | $(12 + 0.031E)e^\beta$ |
| $V$ | $88.0 - 0.283E + 0.88ZA^{1/3}$ | $W_S$ | $(12 + 0.031E)(1 - e^\beta)$ |
| $r_V$ | $1.17$ | $r_D = r_S$ | $1.376 - 0.01\sqrt{E}$ |
| $a_V$ | $0.717 + 0.0012E$ | $a_D = a_S$ | $0.52 + 0.07A^{1/3} - 0.04\sum_i e^{-\mu_i}$ |
| $V_{so}$ | $5.0$ | $W_{so}$ | $0.37A^{1/3} - 0.03E$ |
| $r_{rso}$ | $1.04$ | $r_{iso}$ | $0.80$ |
| $a_{rso}$ | $0.60$ | $a_{iso}$ | $0.25$ |

$\beta = -(E/100)^2$, $\mu_i = [(M_i - N)/2]^2$, $M_i$ is the magic number 8, 20, 28, 50, 82, 126, N – neutron number, E – deuteron energy in the laboratory system in MeV. $r_c = 1.3\,\text{fm}$.

In the initial series of calculations, I have analysed the experimental results for each nucleus using either positive or negative quadrupole deformation parameter $\beta_2$ and searching on all 15 parameters in groups of up to 10 parameters at a time. Later, in

---

[16] Their potential has the same *geometric* parameters for the surface and volume absorption; hence the total number of parameters is 13. In my calculations, I have relaxed this restriction and assumed that the geometric parameters can have different values for these two components and thus the total number of parameters was increased to 15.





the final stages of parameter optimisation, I have searched on all 15 parameters starting with the values determined earlier.

My calculations have shown that all 15 parameters fluctuated around certain smoothly varying, mass-dependent values. Close examination of all these parameters suggested that many of them could be fixed at their original values and that the 15-parameter search could be reduced to a search on only *four* parameters.

In particular, the analysis involving searches on all 15 parameters indicated that (a) the depth of the central volume absorption potential, $W_S$, should have a fixed value of 2.07 MeV for all target isotopes; (b) in contrast with the original potential $F'$, the central surface and volume absorption potentials should have different radius parameters, and (c) all other parameters, except for $V$, $a_0$, $W_D$ and $r_D$, could be assumed to have the values given by Daehnick, Childs and Vrcelj (1980). In particular, I have found no compelling evidence for altering the parameters of the spin-orbit components from their original values.

Taking into account the results of the 15 parameters search, the analysis was repeated by searching on only four parameters, $V$, $a_0$, $W_D$ and $r_D$. In general, the resulting fits were found to be similar to those obtained by searching on all 15 parameters. However, the searching procedure was not only considerably easier and faster but also it eliminated some spurious parameter fluctuations.

The four individually adjusted parameters for each target isotope and for each sign of the $\beta_2$ parameter, are shown in Figure 16.2. They were found to be close to their original values, which are also shown in the same figure. Deformation parameters, $\beta_2$ and $\beta_4$, used in the coupled channels calculations are listed in Table 16.2.

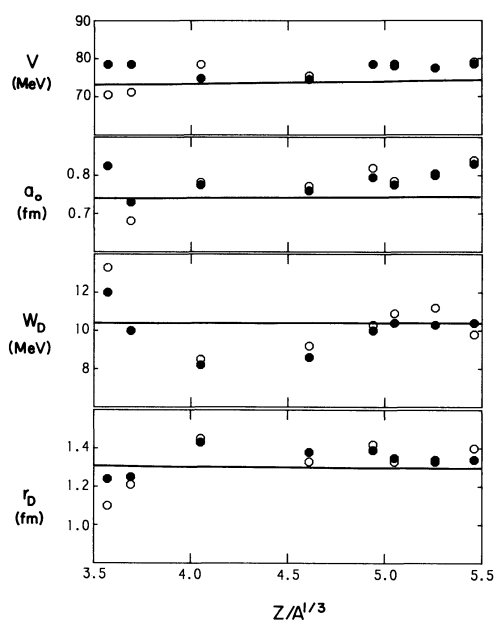

Figure 16.2. The four, out of 15, optical model parameters that had to be adjusted to optimise the fits to the experimental angular distributions of the differential cross sections and vector analyzing powers for the ground states and first excited states of the *sd*-shell nuclei. The full circles correspond to the full lines in Figures 16.3-16.6 for the rotational model with $\beta_2 < 0$ for $^{28}$Si and $\beta_2 > 0$ for all the remaining nuclei. The open circles are the parameters for the reversed signs of the deformation parameters $\beta_2$. The full lines show the original parameter values of the potential $F'$ as determined by Daehnick, Childs and Vrcelj (1980) and as listed in Table 16.1.





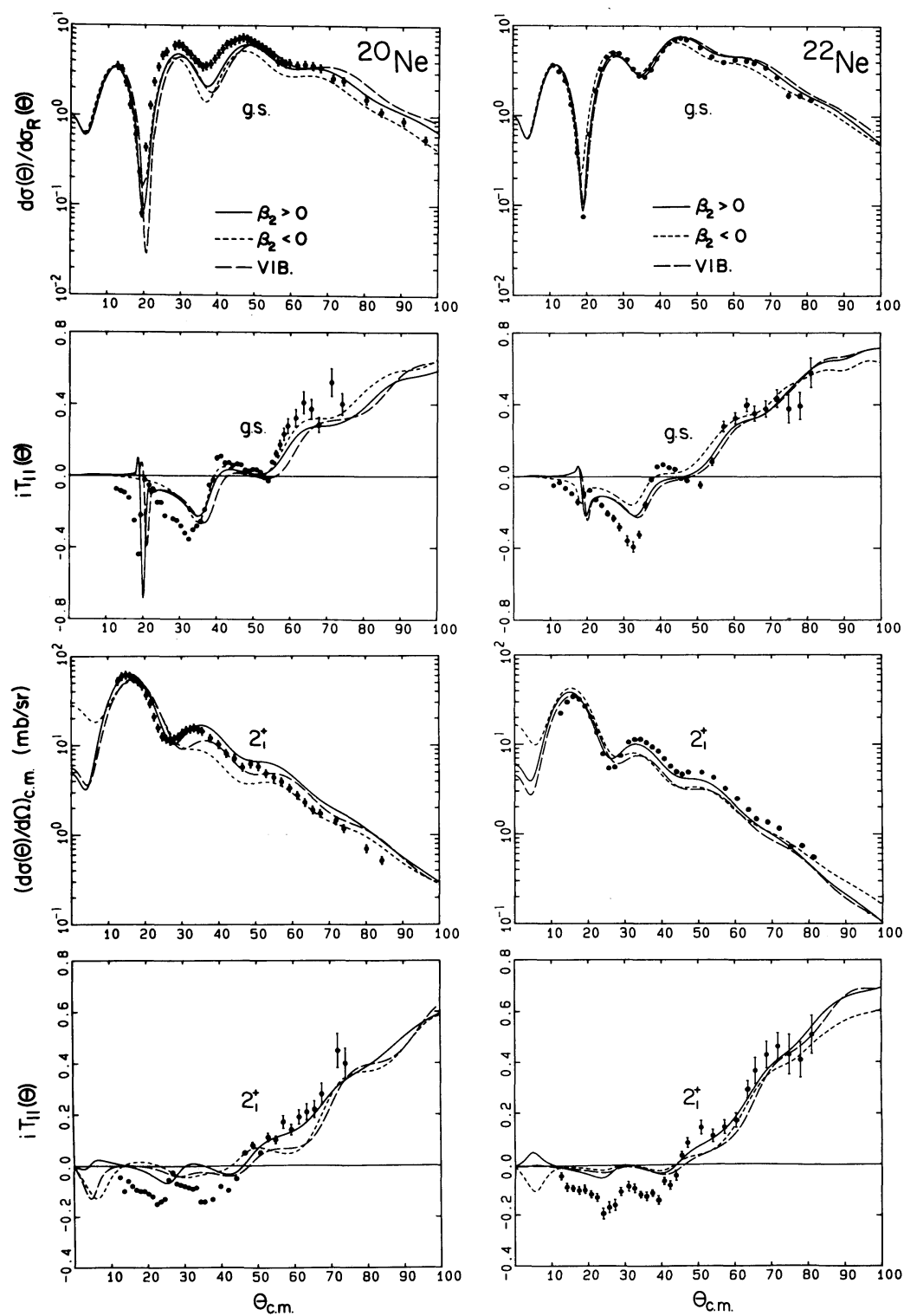

Figure 16.3. The experimental results (dots) for the elastic and inelastic scattering of 52 MeV polarized deuterons from $^{20}$Ne and $^{22}$Ne are compared with the theoretical calculations. Errors smaller than the experimental points are not shown. The coupled channels calculations for an axially symmetric rotational model with prolate or oblate deformations are shown as full and dotted lines, respectively. The dashed lines correspond to calculations using a harmonic vibration model.





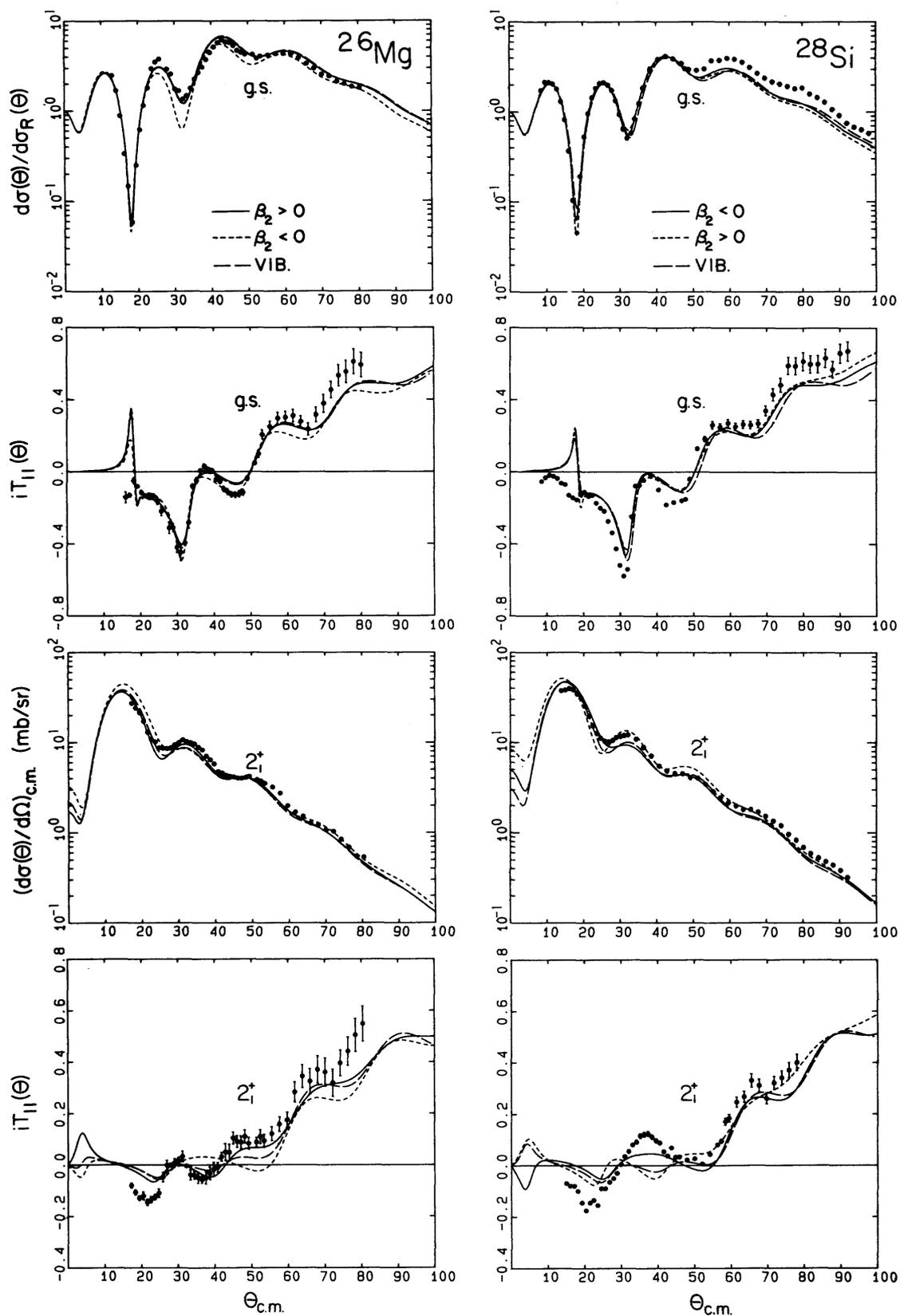

Figure 16.4. Results for $^{26}$Mg and $^{28}$Si. See the caption to Figure 16.3. However, here the continuous line for $^{28}$Si corresponds to an oblate deformation.





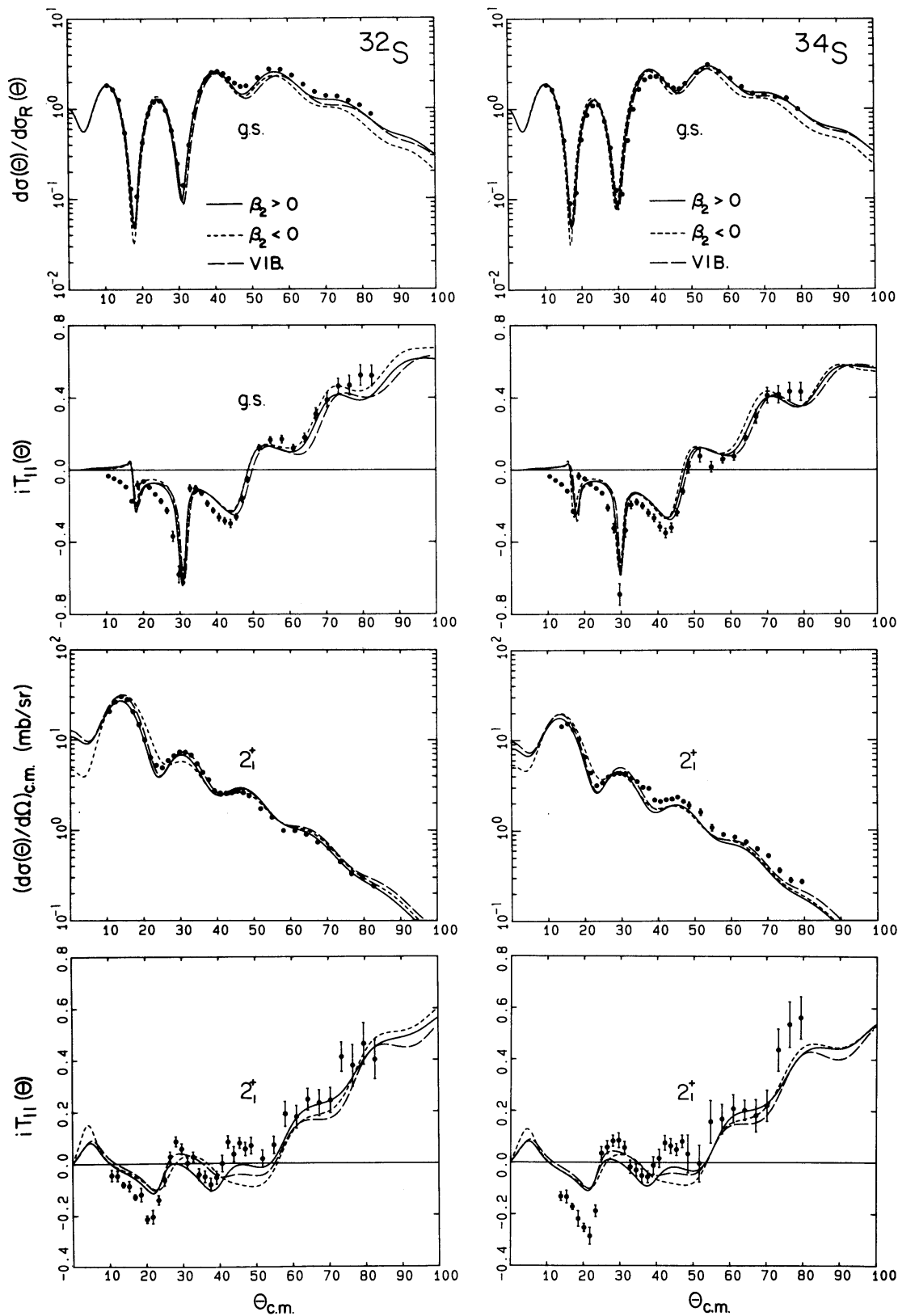

Figure 16.5. Results for $^{32}$S and $^{34}$S. See the caption to Figure 16.3.





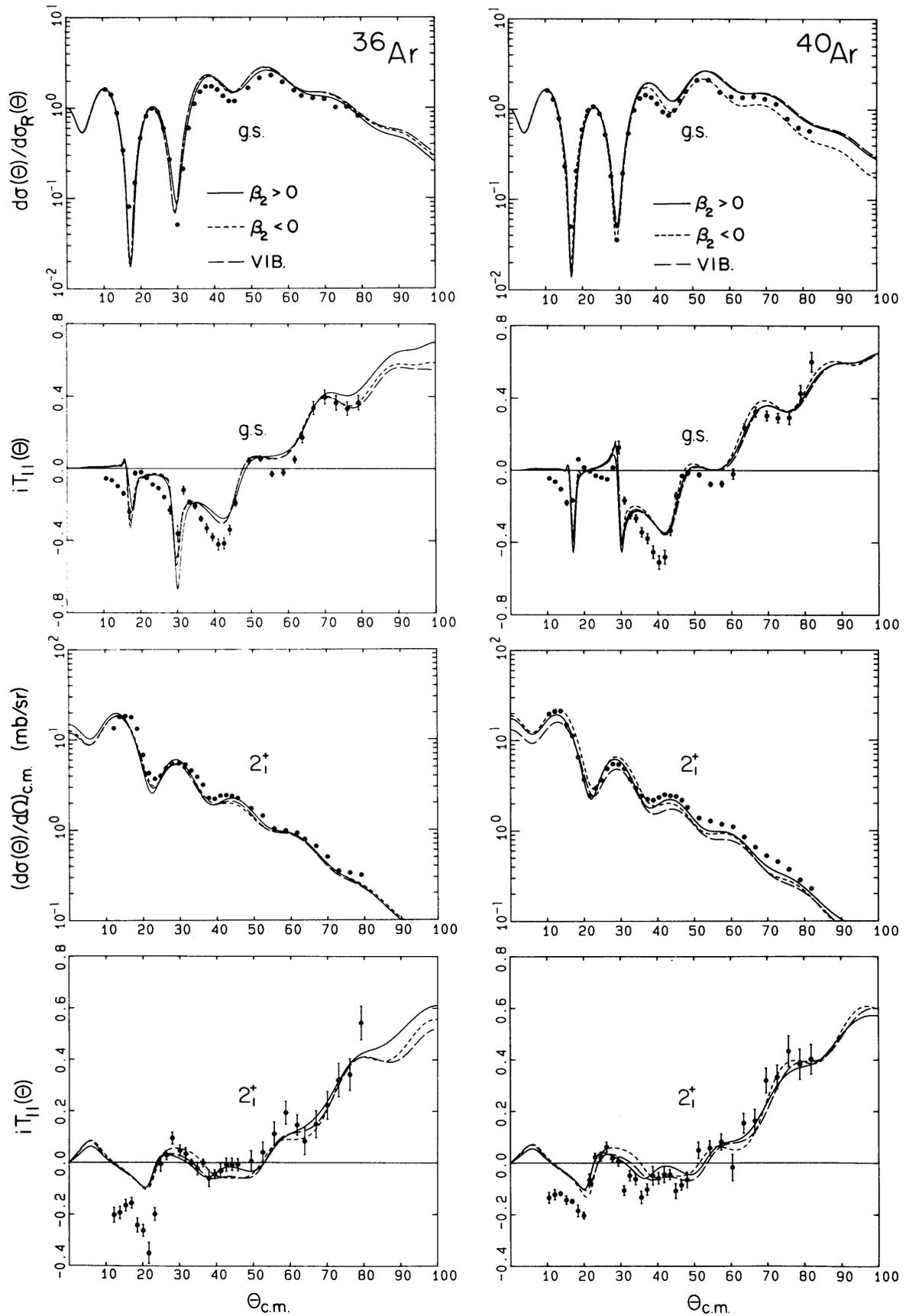

Figure 16.6. Results for $^{36}$Ar and $^{40}$Ar. See the caption to Figure 16.3.





Results of the four-parameter search for each isotope were taken as representing the best theoretical predictions at this deuteron energy. They are displayed in Figures 16.3-16.6 for $\beta_2 > 0$ and $\beta_2 < 0$. With the exception of $^{28}$Si, the full and dotted lines are for the calculations using prolate and oblate deformations, respectively. For $^{28}$Si the representation is reversed because this nucleus has an oblate deformation.

Finally, in order to see to what extent theoretical predictions are model-dependent, the four-parameter search was also carried out assuming a harmonic vibrational model for each isotope. Results of the calculations are shown as dashed lines in Figures 16.3-16.6, and the corresponding coupling parameters $\beta_{02}$ are listed in Table 16.2. In these calculations, the final parameters $V$, $a_0$, $W_D$ and $r_D$ were found to be the same as the parameters corresponding to the full lines in Figures 16.3-16.6.

Table 16.2

Parameters $\beta_2$, $\beta_4$ and $\beta_{02}$ used in the coupled channels analysis of the 52 MeV polarized deuteron scattering from the $sd$-shell nuclei

| | $^{20}$Ne | $^{22}$Ne | $^{26}$Mg | $^{28}$Si | $^{32}$S | $^{34}$S | $^{36}$Ar | $^{40}$Ar |
|---|---|---|---|---|---|---|---|---|
| $\beta_2$ [a] | +0.50 | +0.37 | +0.30 | -0.34 | +0.27 | +0.20 | +0.18 | +0.17 |
| $\beta_4$ [a] | +0.05 | +0.05 | -0.03 | +0.08 | -0.20 | -0.20 | +0.10 | +0.10 |
| $\beta_2$ [b] | -0.53 | -0.44 | -0.35 | +0.34 | -0.33 | -0.24 | -0.18 | -0.20 |
| $\beta_4$ [b] | +0.05 | +0.05 | -0.03 | +0.08 | -0.20 | -0.20 | +0.10 | +0.10 |
| $\beta_{02}$ [c] | 0.50 | 0.37 | 0.30 | 0.34 | 0.27 | 0.20 | 0.17 | 0.17 |
| $Q_{2+}$ [d] | -23(3) | -19(4) | -21(2) | +17(3) | -15(2) | +04(3) | +11(6) | +01(4) |

[a] Deformation parameters associated with the full lines in Figures 16.3-16.6. The corresponding optical model parameters are shown as full circles in Figure 16.2.
[b] Deformation parameters associated with the dotted lines in Figures 16.3-16.6. The corresponding optical model parameters are shown as open circles in Figure 16.2.
[c] Coupling parameters used for harmonic vibrational model calculations. The resulting theoretical angular distributions are shown as dashed lines in Figures 16.3-16.6.
[d] Quadruple moments $Q_{2+}$ in $e$ fm$^2$. The listed values include the uncertainty of the listed values. For instance, -23(3) means -23±3 $e$ fm$^2$ (Stone 2001).

## Summary and conclusions

Measurements of the differential cross sections and analyzing powers were carried out using a 52 MeV beam of deuteron polarized by the C-LASKA Karlsuhe Lambshift source and accelerated by the Karlsruhe isochronous cyclotron. Particle spectra were stored on magnetic tapes and analysed using my computer code LORNA (see Chapter 15) and the VAX-780 computer at the Max-Plank Institute für Kernphysik, Heidelberg. This has resulted in a total of 32 angular distributions of the differential cross sections and vector analyzing powers for the elastic and inelastic scattering on $^{20}$Ne, $^{22}$Ne, $^{26}$Mg, $^{28}$Si, $^{32}$S, $^{34}$S, $^{36}$Ar and $^{40}$Ar nuclei. Theoretical analysis of our experimental results was carried using the Australian National University UNIVAC 1100/82 computer and the coupled channels code ECIS79 (Raynal, 1972, 1981), which I have modified and adapted to run at ANU.

A compilation prepared by Stone (2001) shows that only $^{28}$Si nucleus has a relatively large positive quadrupole moment $Q_{2+}$ =16±3 $e$fm$^2$. Another nucleus that might have





a relatively large positive quadrupole moment is $^{36}$Ar, for which $Q_{2^+} = 11 \pm 6\ e$fm$^2$. However, the experimental error is also large and thus the moment could be close to zero. The quadrupole moments for $^{34}$S and $^{40}$Ar are practically zero. Thus, the heaviest isotopes in this study are either spherical or almost spherical. The remaining nuclei have relatively large quadrupole moments.

Our experimental survey of the vector analyzing powers for the $2_1^+$ states in the *sd*-shell nuclei presented in Figure 16.1 shows that the shapes of the $iT_{11}(\theta)$ distributions for all nuclei except one ($^{28}$Si) have similar oscillation pattern. This group contains both prolate ($^{20}$Ne, $^{22}$Ne, $^{26}$Mg and $^{32}$S) and nearly spherical ($^{34}$S, $^{36}$Ar, and $^{40}$Ar) nuclei. This survey therefore indicates that the underlying process of deuteron polarization at 52 MeV helps to identify only nuclei with strong oblate shapes; it does not allow for a distinction between prolate, spherical or nearly spherical nuclei.

In my coupled channels analysis of our results I have initially used the optical model parameters found earlier by Mairle *et al.* (1980) for 52 MeV deuterons. These parameters served as the starting values but any attempt to optimise the fits to our data by searching around these values was unsuccessful. I have found that the global potential $F'$ of Daehnick, Childs and Vrcelj (1980), which contains an imaginary spin-orbit component produced significantly better results. The slightly modified version of this potential contains 5 components with the total of 15 parameters.

Extensive search on all 15 parameters indicated that 10 of them could be kept at the values defined by the original potential $F'$. The depth of the volume absorption potential $W_S$ had to be lowered to 2.07 MeV and was found to have the same value for all the investigated isotopes. The search was then reduced to only 4 parameters, $V$, $a_0$, $W_D$ and $r_D$.

Considering the calculations for the $iT_{11}(\theta)$ to the $2_1^+$ states, as presented in Figures 16.3-16-6, it is clear that the best distinction between prolate and oblate quadrupole shapes is for $^{26}$Mg and $^{28}$Si nuclei. For $^{26}$Mg, the prolate and oblate deformation produce clear out-of-phase oscillation for angles around $30^0$-$60^0$. For $^{28}$Si, clear out-of-phase results are for angles around $25^0$-$45^0$.

Some dependence on the sign of the quadrupole deformation can be also seen for $^{32}$S and $^{34}$S nuclei. Calculations for these nuclei show a phase shift at angles around $30^0$ and an out-of-phase feature at around $45^0$. In both cases, the calculated curves for the negative deformation in this region of angles is shifted to higher angles as compared with the curves calculated for the positive deformation.

Similar, but less pronounced, features are also observed for $^{36}$Ar and $^{40}$Ar nuclei. However, for $^{20}$Ne and $^{22}$Ne there are no clear differences between calculations for the opposite signs of quadrupole deformation.

Fits to $^{20,22}$Ne are relatively poor and the agreement between the experimental and theoretical results is only marginally better for prolate shapes. Experimental results for this pair of isotopes were found to be the most difficult to fit when using the coupled channels procedure. The easiest to fit were the experimental results for $^{32,34}$S and for $^{36,40}$Ar isotopes. Figures 16.5 and 16.6 show that the assumption of prolate shapes gives only a little better fit for these nuclei.





It is also interesting to compare rotational and vibrational model calculations. As mentioned earlier, the basic difference between the two models is in the $2^+$- $2^+$ matrix elements, which vanish for the harmonic vibrator. Results presented in Figures 16-3-16-6 show that, in general, there are no significant differences between the rotational and the vibrational model calculations. Even for $^{26}$Mg and $^{28}$Si nuclei, calculations using prolate shapes are similar to those obtained using harmonic vibrator. This is in contrast with the results at lower incident energies where for strongly deformed nuclei, differences between rotational and vibrational model calculations are generally better pronounced, even in cases when the fits to the experimental data are poor.

In summary, this study shows that, in general, reorientation effects at 52 MeV deuteron energy for the *sd*-shell nuclei appear to be weak. The experimental $iT_{11}(\theta)$ distributions for nuclei with positive prolate deformations are found to have similar features. Only in a single case of a strong oblate deformation, i.e. for $^{28}$Si, a distinctly different experimental $iT_{11}(\theta)$ distribution for the $2_1^+$ state was observed. Furthermore, for only two isotopes, $^{26}$Mg and $^{28}$Si, distinctly different $iT_{11}(\theta)$ predictions for the two signs of $\beta_2$ were obtained. For $^{20}$Ne and $^{22}$Ne, an assumption of large oblate deformations did not result in altering significantly the calculated $iT_{11}(\theta)$ distributions for the $2_1^+$ states. However, the overall fits to all four distributions for each of these two isotopes appear to favour the correct sign of $\beta_2$.

Finally, my analysis has shown that the best description of our experimental data required a potential with both real and imaginary spin orbit components. The potential is described by 15 parameters but 10 of them have the same values as found earlier by Daehnick, Childs and Vrcelj (1980). Of the remaining five parameters, the depth of the central volume absorption potential had to be lowered to 2.07 MeV and was constant for all target nuclei. Only four parameters $V$, $a_0$, $W_D$ and $r_D$ had to be adjusted individually, but generally only slightly, to optimise the theoretical description of the data.

## References


Bechtold, V., Friedrich, L., Finken, D., Strassner, G. and Ziegler, P. 1976, *Proc. Fourth Int. Symp. on Polarization Phenomena in Nuclear Reactions*, Zürich, 1975, ed. W. Grüebler and V. König (Birkhauser, Basel) p. 849

Blair, A.G., Glashausser, C., de Swiniarski, R., Goudergues, J., Lombard, R., Mayer, B., Thirion, J. and Vaganov, P. 1970, *Phys. Rev.* **C1**:444.

Brown, R.C., Debenham, A.A., Griffith, J.A.R., Karban, O., Kocher, D.C. and Roman, S. 1973, Nucl. Phys. **A208**:589.

Clement, H., Frick, R., Graw, G., Merz, F., Schiemenz, P., Seichert, N. and Hsun, S. T. 1980, *Phys. Rev. Lett.* **45**:599.

Clement, H., Frick, R., Graw, G., Oelrich, I., Scheerer, H.J., Schiemenz, P., Seichert, N. and Hsun, S. T. 1981, *Proc. Fifth Int. Symp. on Polarization Phenomena in Nuclear Physics*, Santa Fe, NM, 1980, ed. G. G. Ohlsen, R.E. Brown, N. Jarmie, W.W. McNaught and G.M. Hale, AIP Conf. Proc. No. 69 (AIP, New York,) p. 376

Clement, H., Graw, G., Kretschmer, W. and Stack, W. 1978, *J. Phys. Soc. Jpn. Suppl.* **44**:570

Daehnick, W. W., Childs, J. D. and Vrcelj, Z. 1980, *Phys. Rev.* **C21**:2253.

Hatanaka, K., Nakamura, M., Imai, K., Noro, T., Shimizu, H., Sakamoto, H., Shirai, J., Matsusue, T. and Nisimura, K. 1981, *Phys. Rev. Lett.* **46**:15






Kato, S., Okada, K., Kondo, M., Hosono, K., Saito, T., Matsuoka, N., Hatanaka, K., Noro, T., Nagamachi, S., Shimizu, H., Ogino, K., Kadota, Y., Matsuki, S. and Wakai, M. 1985, *Phys. Rev.* **C31**:1616.

Kurepin, A. B., Lombard, R.M. and Raynal, J. 1973, *Phys. Lett.* **45B**:184.

Mairle, G., Knopfle, K. T., Riedesel, H. and Wagner, G. J. 1980, *Nucl. Phys.* **A339**:61

Nakamura, M., Sakaguchi, H., Imai, K., Hatanaka, K., Goto, A., Noro, T., Ohtani, F., Kobayashi, S., Hosono, K., Kondo, M., Kato, S., Ogino K. and Kadota, Y. 1978, *J. Phys. Soc. Jpn. Suppl.* **44**:557.

Nurzynski, J., 1985, *Comp. Phys. Commun.* **36**:295.

Raynal, J. 1972, *Computing as a language of physics* (IAEA, Vienna,) p. 281.

Raynal, J. 1976, Proc. Forth Int. Symp. on Polarization Phenomena in Nuclear Reactions, Zurich, 1975, ed. W. Grüebler and V. König (Birkhauser, Basel,) p. 271

Raynal, J. 1981, Phys. Rev. **C23**:2571.

Seibt, E. and Weddigen, C., 1980, *Nucl. Instr. Meth.* **100**:61.

Stone, N. J. 2001, *Tables of Nuclear Magnetic Dipole and Electric Quadrupole Moments*, Oxford Physics, Clarendon Laboratory, Oxford, UK.





---



# Two-step Reaction Mechanism in Deuteron Polarization

***Key features:***

1. We have measured angular distributions of the differential cross sections $\sigma_0(\theta)$ and analyzing powers $iT_{11}(\theta)$, $T_{20}(\theta)$, $T_{21}(\theta)$ and $T_{22}(\theta)$ for the elastic scattering of 12 MeV polarized deuterons from $^{76,78,80,82}$Se isotopes.

2. Our measurements revealed an unusual behaviour of the analyzing powers: the amplitudes *increased* with the increasing mass number *A* of the target nuclei. This is contrary to the normal behaviour, which is characterized by the *decreasing* amplitudes.

3. I have carried out coupled channels analysis of our experimental results and I have found that this unusual behaviour can be explained as being due to the contributions from an indirect, two-step, elastic scattering via the first $2_1^+$ excited states in the target nuclei. Thus, the observed elastic scattering is made of two components: the normal direct scattering (d,d) and the two-step scattering (d,d')2$^+$(d',d).

4. I have also studied other two-step contributions: (d,t)(t,d) and (d,p)(p,d) via the 2p$_{1/2}$ or 1g$_{9/2}$ configurations. I have found that their share in the reaction mechanism is negligible.

5. This study represents the first clear demonstration of the unusual mass-dependence of deuteron polarization and a demonstration of the importance of second-order interaction in the elastic scattering.

***Abstract:*** Angular distributions of the differential cross sections $\sigma_0(\theta)$ and of $iT_{11}(\theta)$, $T_{20}(\theta)$, $T_{21}(\theta)$ and $T_{22}(\theta)$ analyzing powers have been measured for the $(\vec{d},d)$ scattering from $^{76,78,80,82}$Se isotopes at 12 MeV. An unusual mass-dependence of the analyzing powers was observed. Coupled channels analysis explained the experimental results as being due to contributions of the two-step elastic scattering (d,d')2$^+$(d',d) via the first $2_1^+$ excited states in the target nuclei. Two-step processes (d,t)(t,d) and (d,p)(p,d) via the 2p$_{1/2}$ or 1g$_{9/2}$ configurations have been also investigated but have been found to have negligible influence on the measured distributions.

## Introduction

While visiting the Laboratorium für Kernphysik in Zürich, Switzerland, I proposed a study of deuteron polarization using Se isotopes. Normally, one should expect a predictable mass-dependent behaviour of the analyzing powers. However, I thought that it would be interesting to see whether Se isotopes would reveal some new, unexpected features.

As discussed in Chapter 14, selenium isotopes present an interesting case where neutron configurations 2p$_{3/2}$, 1f$_{5/2}$, and 2p$_{1/2}$ are nearly completely occupied and where virtually only one configuration, 1g$_{9/2}$, is filling in as the mass of the selenium isotope increases. Our results for the (p,d) reactions indicated that the neutron occupation numbers for the 1g$_{9/2}$ increased continuously from 40% for $^{76}$Se to 80% for $^{82}$Se. These results were in close agreement with the calculations based on the pairing theory.





The idea behind the proposed measurements of deuteron polarization was to see whether this systematic closing of the f-p shell might be reflected in the angular distributions of the analyzing powers. Selenium isotopes belong to a very limited group of nuclei that can be used for such a study.

## Experimental method

Angular distributions of the differential cross sections $\sigma_0(\theta)$, vector analyzing power $iT_{11}(\theta)$ and all three tensor analyzing powers $T_{20}(\theta)$, $T_{21}(\theta)$ and $T_{22}(\theta)$ for the elastic scattering of 12 MeV polarized deuterons from the $^{76,78,80,82}$Se isotopes were measured using the ETHZ atomic beam polarized ion source and EN Van de Graaff tandem accelerator. The method of polarization measurements was discussed in Chapter 15.

A diagram of the reaction chamber is shown in Figure 17.1. The beam entered the chamber from the left through a collimator, passed through the target and was collected in a Faraday cup, which was equipped with an electrostatic suppressor electrode. The scattered particles were collimated by rectangular slit system consisting of identical slits with antiscattering baffles places between the defining apertures.

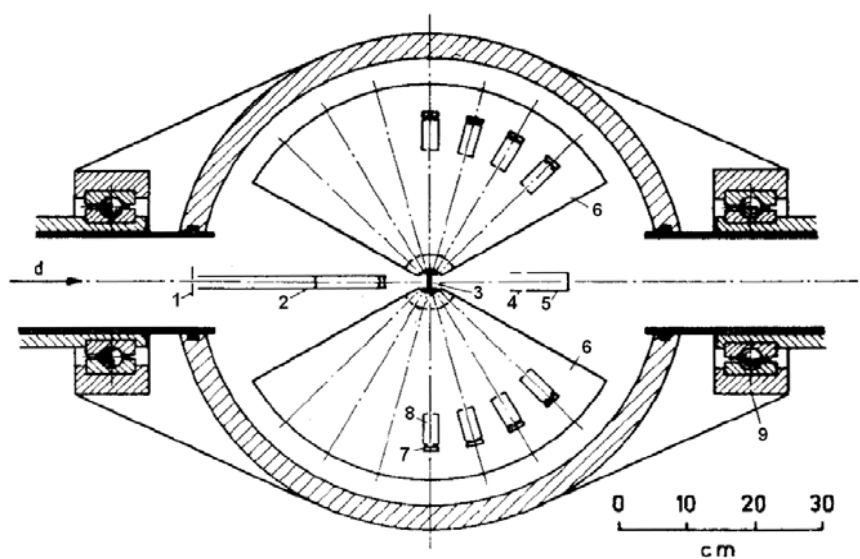

Figure 17.1. Cross section of the reaction chamber used in the measurements of deuteron polarization on Se isotopes. 1 − insulated slits; 2 − collimator tube; 3 − Se target; 4 − suppressor electrode; 5 − Faraday cup; 6 − turntables; 7 − solid state detectors; 8 − detector slit assembly; 9 − bearings for rotating the chamber about the beam axis.

The detectors were mounted on two plates that could be rotated independently around the target. On each turntable, there was room for up to seven detectors placed at positions of $15^0$ apart. With this system, angular distributions could be measured for reaction angles $\theta$ between $20^0$ and $160^0$. Each turntable could be adjusted from outside to the desired position within $0.1^0$. In general, only four detectors were used on each turntable because the electronic system had been designed to accommodate a maximum of only eight detectors. The whole chamber could be also rotated along the beam axis.





The electronic system was relatively simple and is shown in the diagram in Figure 17.2.

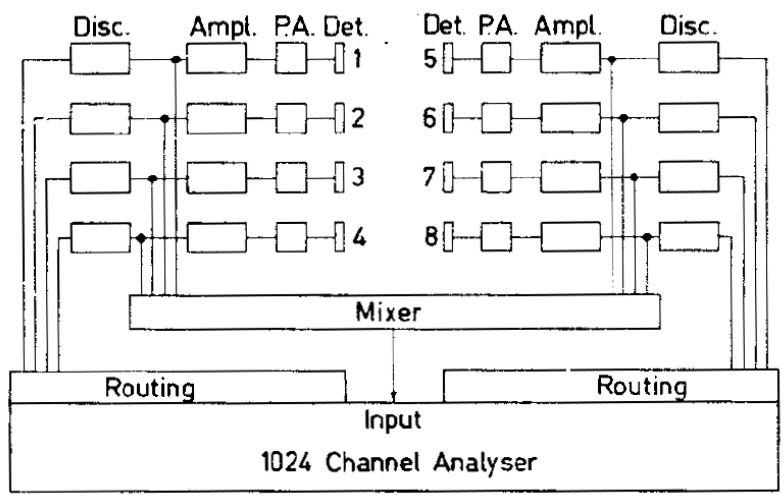

Figure 17.2. Block diagram of the electronic system used in measurements of deuteron polarization for Se isotopes. PA – preamplifier; Ampl. – amplifier; Disc. – discriminator.

Selenium targets were prepared using the method developed at Australian National University (see Chapter 14). In order to prevent sublimation of selenium under beam bombardment, targets were coated on both sides with carbon films of approximately 20 $\mu$g/cm$^2$. The target thickness was around 600 $\mu$g/cm$^2$. However, in order to reduce the data collection time, two targets of the same isotope were stacked to produce a combined thickness of approximately 1200 $\mu$g/cm$^2$. The isotopic enrichment of the target material was 86%, 98%, 94%, and 97% for $^{76}$Se, $^{78}$Se, $^{80}$Se, and $^{82}$Se, respectively.

Separation of the atomic substates was achieved in the field of a tapered sextupole magnet. The source was equipped with a weak-field and two strong-field RF transitions to produce vector and tensor polarization states of the deuteron beam (see the Appendix G). The spin direction was adjusted using a Wienfilter and switching the sign of the beam polarization was done every few seconds. This method eliminates first-order errors arising from geometrical effects and from inaccuracies of the required spin direction (see Chapter 15).

Typical beam current from the source was around 120 nA and on the target around 50 nA. The polarization of the deuteron beam was about 87% of the theoretical value (i.e. $p_z$ = ±0.58 or $\tau_{10}$ = ±0.71).

Experimental results are displayed in Figures 17.4 and 17.5. Preliminary data reduction was carried out at the time of measurements, and I well remember the excitement we felt when, point by point, they were gradually revealing a clear new behaviour. Our results show that contrary to the normal behaviour, the amplitudes of the analyzing power *increase* with increasing atomic mass number $A$ of the target nucleus. The clearest effect is for the vector analyzing powers $iT_{11}(\theta)$. However, the effect is also present for $T_{20}(\theta)$ and $T_{22}(\theta)$ at backward angles.





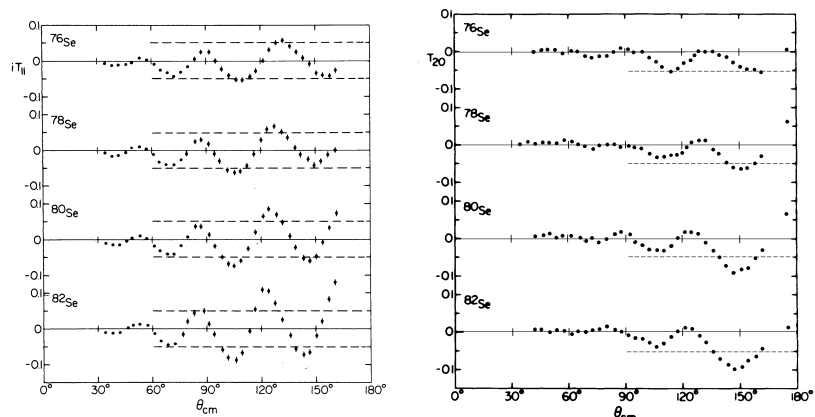

Figure 17.3. The unusual behaviour of the analyzing powers: the increasing amplitudes with the increasing mass of the target nuclei. Experimental angular distributions of the vector $iT_{11}(\theta)$ and tensor $T_{20}(\theta)$ analyzing powers for the elastic scattering of 12 MeV polarized deuterons from Se isotopes. The lines are to guide the eye. The horizontal lines are to help to see how the amplitudes of the analyzing powers increase with the increasing atomic mass number $A$ of the target nuclei.

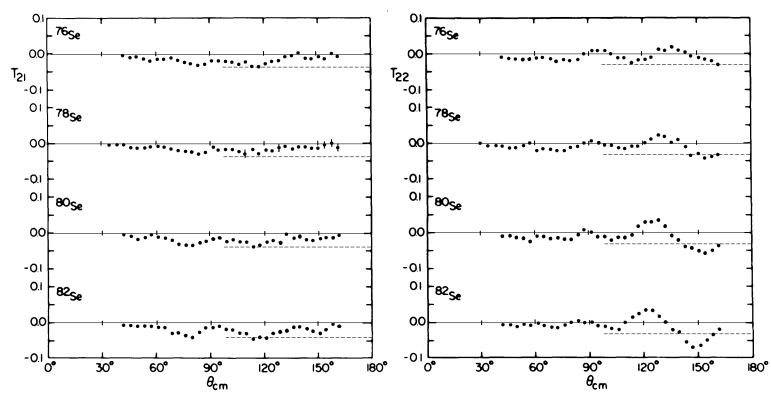

Figure 17.4. Experimental angular distributions of the $T_{21}(\theta)$ and $T_{22}(\theta)$ tensor analyzing powers for the elastic scattering of 12 MeV polarized deuterons from Se isotopes. The lines are to guide the eye. The values for the $T_{21}(\theta)$ component are too small to see the mass-dependence of its amplitudes but the $T_{22}(\theta)$ distributions show an increase in the absolute values of the analyzing power at backward angles with the increasing atomic mass number $A$.

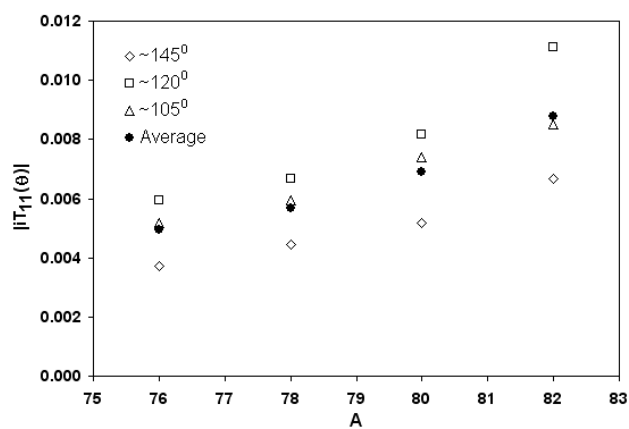

Figure 17.5. The unusual experimental results for Se isotopes: the absolute values $\left| iT_{11}(\theta) \right|$ of the vector analyzing power *increase* with the increasing atomic mass number $A$.





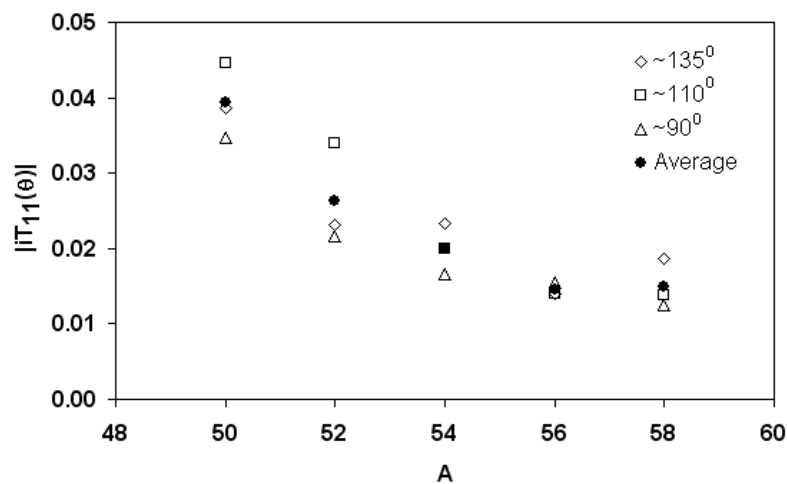

Figure 17.6. The normal behaviour of the absolute values $\left|iT_{11}(\theta)\right|$ of the vector analyzing power. The amplitudes *decrease* with the increasing atomic mass number $A$. This figure is based on our study for a wider range of target nuclei (Bürgi *et al.* 1980)

This interesting mass-dependence is also displayed in Figure 17.5, which shows the absolute amplitudes $\left|iT_{11}(\theta)\right|$. As can be seen, there is a clear and systematic increase of the polarization amplitudes with the increasing mass number $A$. For comparison, Figure 17.6 shows the normal and well-known behaviour of the absolute values of the amplitudes of the vector analyzing $\left|iT_{11}(\theta)\right|$ characterised by the *decreasing* amplitudes.

**Theoretical analysis**

The interpretation of the unusual behaviour of the analyzing powers for Se isotopes was not immediately obvious but it suggested a presence of a systematic reaction mechanism associated with changes in the internal structure of Se isotopes.

It so happened that around that time I visited the University of Colorado where I had a chance to talk to Peter Kunz who informed me about his calculations for the elastic scattering of polarized tritons. His calculations indicated an enhancement of the analyzing powers by second-order processes. This sounded interesting and I wondered whether our experimental results could be explained by the presence of such second-order processes.

Peter wrote a Coupled Channel Born Approximation program, CHUCK, which he kindly shared with me. I have brought his program to Canberra and I have adapted it to run on the Australian National University UNIVAC 1100/42 computer.

The program allows for an easy calculation of a variety of second-order processes. However, it can calculate only the differential cross sections and vector analyzing power $iT_{11}(\theta)$. It does not include calculations of higher order of the analyzing powers. This restriction did not seem to be important because our unusual experimental results were most prominent for the $iT_{11}(\theta)$ distributions.





I have used this program in the analysis of our experimental results for Se isotopes, but first I decided to carry out the conventional optical model calculations.

In all my calculations, I have used standard parameterisation of the optical model potential as described in Chapter 16 but I have used only three components: the central real component, the central surface-absorption component, and the real spin-orbit component. The potentials were described by 9 parameters: three potential depths parameters ($V$, $W_D$, and $V_{s.o.}$) and six geometrical parameters ($r_0$, $a_0$, $r_D$, $a_D$, $r_{s.o.}$, and $a_{s.o.}$).

Optical model analysis revealed that the differential cross sections and $iT_{11}(\theta)$ angular distributions could be fitted by adjusting only one parameter, the depth of the central imaginary component $W_D$. The final values of this parameter were: 17.25, 16.50, 15.50, and 14.50 MeV for $^{76}$Se, $^{78}$Se, $^{80}$Se, and $^{82}$Se isotopes, respectively. All parameters used in the optical model calculations and in the coupled channels analysis are listed in Table 17.1.

This was an interesting result, because it showed that the contribution of non-elastic processes in the deuteron-nucleus interaction, as represented by $W_D$, was decreasing gradually with the increasing atomic mass number $A$. Thus, by explicitly including at least some of these non-elastic channels in the calculations one might expect to describe the observed features of the elastic scattering by using a set of fixed parameters for all target isotopes.

Having done the standard optical model analysis, I have then carried out an extensive analysis of contributions from the second order processes (d,d')(d',d), (d,t)(t,d), and (d,p)(p,d). In this notation, the first process describes the elastic scattering via the inelastic scattering channels. The inelastic scattering (d,d') is quickly followed by a return of the target nucleus to its ground state and the outgoing deuteron is detected as if it were elastically scattered. This intermediate interaction cannot be detected by observing the energy of the outgoing deuteron but it might influence the observed angular distributions of the differential cross sections and analyzing powers.

In the second process, the neutron pickup reaction (d,t) is quickly followed by the neutron stripping (t,d) and the outgoing deuteron is also detected as if it were elastically scattered. In the third process the neutron stripping reaction (d,p) is followed by the neutron pickup (p,d).

In the intermediate inelastic channel, I have considered direct excitations of the first $2_1^+$ and $3_1^-$ levels. In the neutron pickup and stripping channels, I have considered neutron transfers via the $2p_{1/2}$ and $1g_{9/2}$ configurations. All these two-step processes are summarised in Figure 17.7. More complicated second-order processes could be considered but my calculations indicated that they were unnecessary. In fact, I have found (see Figure 17.9) that the only second-order process that has detectable influence on the elastic scattering is the inelastic scattering via the first excited state $2_1^+$.

Measurements of Coulomb excitation for selenium isotopes (Barrette *et al.* 1974) show that the deformation parameters derived from the $B(E\lambda)$ γ-transition probabilities for the $2_1^+$ and $3_1^-$ states decrease with the increasing number of neutrons in the target nuclei (see Figure 17.8). Spectra taken at a few angles for the inelastically scattered particles indicated relatively strong excitation of these





two states. The observed mass dependence of the $iT_{11}(\theta)$ distributions was therefore expected to be associated with the two-step scattering via these excited states.

Fig. 17.7. Two-step processes considered in my analysis of the elastic scattering of polarized deuterons from Se isotopes. Coupling strengths between the elastic and inelastic channels are proportional to the deformation parameters, which were derived from the $B(E\lambda)$ $\gamma$-transition probabilities as reported by Barrette *et al.* (1974). For the intermediate reaction channels, all calculations (except for a study of effects associated with the spectroscopic strength distributions) were carried out using total spectroscopic strengths for configurations corresponding to states located at the relevant centre-of-gravity energies.

Figure 17.8. Energy levels diagram for the $^{76,78,80,82}$Se isotopes. The pathways for gamma transitions and the relevant deformation parameters derived from the $B(E\lambda)$ $\gamma$-transition probabilities as reported by Barrette *et al.* (1974) are shown in the figure. It should be noted that the values of $\beta_2$ and $\beta_3$ parameters for the $2^+_1$ and $3^-_1$ states (used in my coupled channels calculations) decrease with the atomic number of the selenium isotope.





In the coupled channels calculations, the coupling strength for the two-step processes via the intermediate inelastic channels was defined by the deformation parameters $\beta_2$ and $\beta_3$ derived directly from the experimental values of the $B(E\lambda)$ $\gamma$-transition probabilities. In the case of the second-order processes involving transfer reactions I have assumed that the configurations 2p$_{1/2}$ and 1g$_{9/2}$ are located in just two states placed at their corresponding centre-or-gravity energies and that their respective coupling strengths are given by the sum of the spectroscopic factors. The normalization of the reaction amplitudes has been done assuming the coefficients $D_0$(d,p) = -122 MeV · fm$^{3/2}$ and $D_0$(d,t) = -182 MeV · fm$^{3/2}$ (Schneider, Burch, and Kunz 1976). In the more familiar form (see the Appendix E), where the square values are used, these coefficients correspond to $D_0^2 = 1.5 \times 10^4$ MeV$^2$ · fm$^3$ for the (d,p) reaction and $D_0^2 = 3.3 \times 10^4$ MeV$^2$ · fm$^3$ for the (d,t).

The reaction form factors have been calculated using standard parameters $r_0$ = 1.25 fm, $a$ = 0.65 fm and the potential depth adjusted to match the binding energies for the relevant states.

In order to see whether the assumption of a single state at the centre-of-gravity energy for a given configuration can produce valid results I have carried out also calculations using experimentally determined distributions of the single-particle configurations and their respective spectroscopic factors (Barbopoulos *et al.* 1979). Computation time increased considerably for these calculations but the final results were the same as produced by assuming that all the strength is concentrated in just one state for each of the two single-particle configurations.

In the preliminary calculations, I have used various sets of the optical-model potential parameters for protons, deuterons and tritons. They included both shallow and deep potentials with the surface or volume absorption, and they were taken from a variety of sources. I have found that, allowing for small adjustments of parameters, various combinations of parameter sets produced similar results. Furthermore, in the presence of the $0_1^+ \leftrightarrow 2_1^+$ coupling, calculated results were virtually insensitive to changes in either proton or triton parameters.

Thus, in my final analysis, I have used three selected sets of parameters. For protons, I used parameters derived by Becchetti and Greenlees (1969). For deuterons, I have used (as a starting values) the parameters I have just determined in my optical analysis (see above). For tritons, I used the parameters I have determined in my analysis of the (t,t) scattering (Nurzynski 1975; see also Chapter 11). However, I have added a spin-orbit component to the triton potential with parameters I received from Peter Kunz. All parameter sets are listed in the Table 17.1.

Having carried out a detailed analysis of the relative contributions of the two step processes (inelastic scattering via the $2_1^+$ and $3_1^-$ states and transfer reactions via the 2p$_{1/2}$ and 1g$_{9/2}$ configurations) I have found that only one intermediate channel (the scattering via the first $2_1^+$ states) is needed to explain the observed unusual behaviour of the amplitudes of the vector analyzing powers, $iT_{11}(\theta)$, i.e. their gradual increase with the increasing mass number $A$ of the target nuclei. This is illustrated in Figure 17.9.





Table 17.1

Optical model parameters used in the analysis of the $^{76,78,80,82}$Se($\vec{d},\vec{d}$) elastic scattering at 12 MeV

| Target Isotope | Particle | $V$ (MeV) | $r_0$ (fm) | $a_0$ (fm) | $W_D$ (MeV) | $r_D$ (fm) | $a_D$ (fm) | $V_{s.o.}$ (MeV) | $r_{s.o.}$ (fm) | $a_{s.o.}$ (fm) |
|---|---|---|---|---|---|---|---|---|---|---|
| $^A$Se | p | 44 | 1.25 | 0.650 | 13.50 | 1.25 | 0.470 | 7.5 | 1.25 | 0.47 |
| $^A$Se | t | 150 | 1.23 | 0.720 | 29.50 | 1.15 | 0.850 | 2.5 | 1.20 | 0.72 |
| $^{76}$Se [a] | d | 96 | 1.14 | 0.825 | 17.25 | 1.35 | 0.820 | 4.5 | 0.76 | 0.40 |
| $^{78}$Se [a] | d | 96 | 1.14 | 0.825 | 16.50 | 1.35 | 0.820 | 4.5 | 0.76 | 0.40 |
| $^{80}$Se [a] | d | 96 | 1.14 | 0.825 | 15.50 | 1.35 | 0.820 | 4.5 | 0.76 | 0.40 |
| $^{82}$Se [a] | d | 96 | 1.14 | 0.825 | 14.50 | 1.35 | 0.820 | 4.5 | 0.76 | 0.40 |
| $^A$Se [b] | d | 96 | 1.14 | 0.825 | 13.75 | 1.35 | 0.820 | 4.5 | 0.76 | 0.40 |

$^A$Se – means that the same set was used for all selenium isotopes.

[a] Parameters used in the conventional optical model analysis, i.e. without considering the second-order contributions.

[b] Parameters used in the coupled-channels calculations. They included the two-step the (d,d')2$^+$(d',d) intermediate scattering and explained the unusual experimental observations.

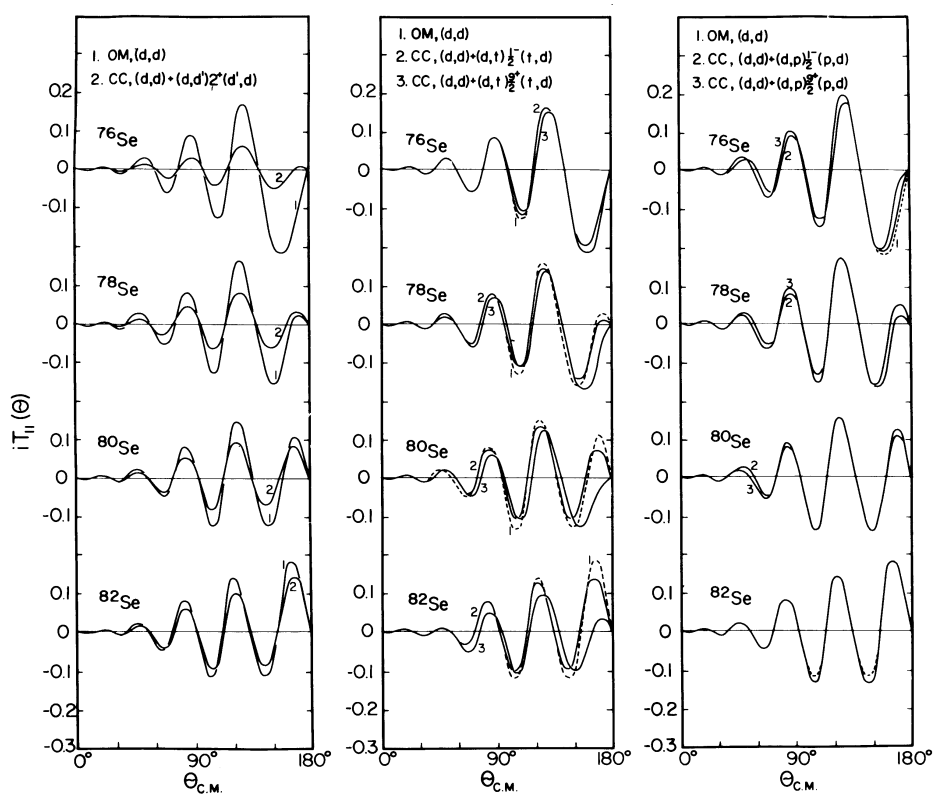

Figure 17.9. A study of the relative contributions of two-step processes in the elastic scattering of deuterons. All the calculations used the last set of parameters listed in Table 17.1 for deuterons. Proton and triton parameters are also listed in Table 17.1. The figure shows that only one intermediate process, i.e. the (d,d')2$^+$(d',d) two-step scattering, has a strong influence on the elastic scattering angular distributions of the $iT_{11}(\theta)$ vector analysis power.





The first panel of Figure 17.9 shows two sets of calculations. Group 1 is for the direct (d,d) elastic scattering as carried out using conventional optical model formalism. Group 2 is for the direct (d,d) elastic scattering combined with the two-step (d,d')2$^+$(d',d) scattering via the first excited states $2_1^+$ in the target nuclei. This panel shows clearly that the two sets of calculations differ considerably. It also shows that when the two-step process is added to the direct elastic scattering the amplitudes of the vector analyzing power $iT_{11}(\theta)$ *increase* with the increasing mass number of the target nuclei, which is in agreement with the observed behaviour.

The second panel shows the direct (d,d) scattering and compares it with the distributions resulting from adding the two-step contributions via the intermediate pickup (d,t) reaction involving the $2p_{3/2}$ or $1g_{1/2}$ configurations. Unlike the results shown in the first panel, these results show that there is hardly any difference between calculations for the direct scattering and for direct plus indirect scattering. It is clear that the two-step process via the neutron pickup reactions can be neglected in the calculations.

The last panel show similar results as displayed in the second panel but this time for the contributions from the two-step processes via the intermediate (d,p) stripping reactions. Again, as for the pickup contributions, the two-step elastic scattering via neutron stripping channel can be also neglected.

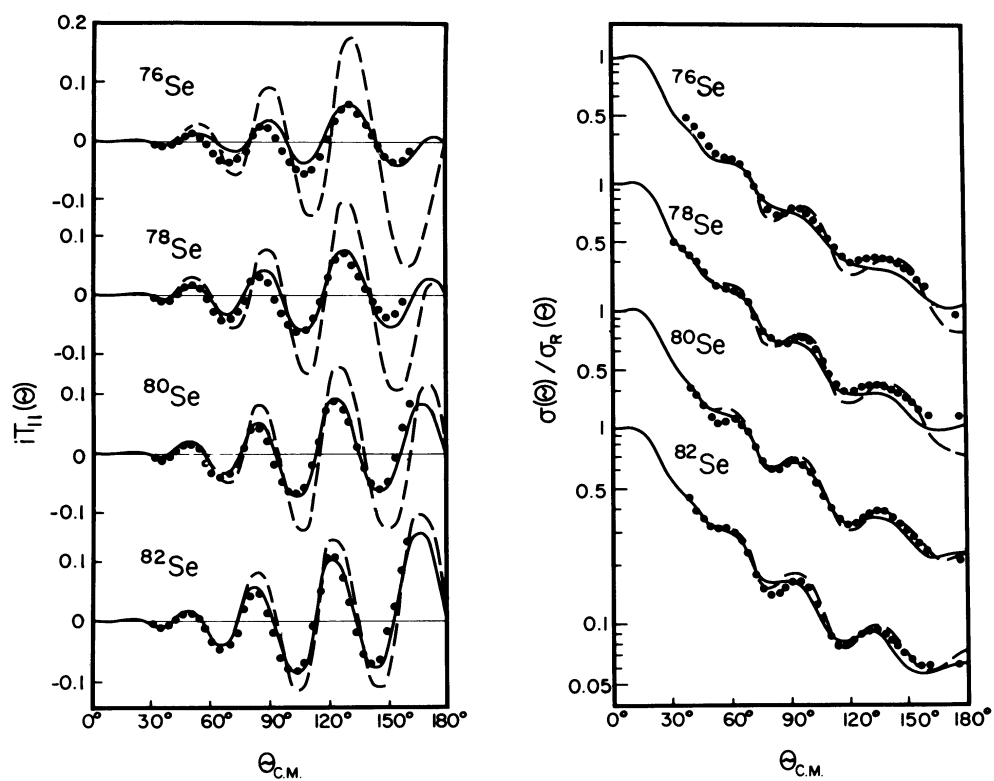

Figures 17.10. The direct and two-step elastic scattering. The calculations for the single step (d,d) elastic scattering (dashed lines) and the single-step plus two-step scattering, (d,d) + (d,d')2$^+$(d',d), via the first excited states $2_1^+$ in the target nuclei (continuous lines) are compared with the experimental data (dots). The figure shows that the unusual enhancement of the amplitudes of the vector analyzing powers $iT_{11}(\theta)$ can be explained as being due to the contributions from the two-step elastic scattering via the $2_1^+$ excited states.





In conclusion, therefore, only one type of the intermediate channel needs to be taken into account to explain the unusual behaviour of the experimentally observed vector analyzing powers $iT_{11}(\theta)$. This channel is the inelastic scattering via the first $2_1^+$ states in the target nuclei.

Final results of my analysis are shown in Figure 17.10. This figure shows two sets of theoretical distributions compared with the experimental distributions. One set (dashed lines) shows the results for the single-step (d,d) direct elastic scattering. The second set (continuous lines) is for the direct elastic scattering (d,d) *plus* the two-step (d,d')2$^+$(d',d) scattering via the first $2_1^+$ states in the target nuclei. This set of curves fits perfectly the observed angular distributions and shows that if the two-step process is included, results for all four isotopes can be reproduced theoretically using a fixed set of the optical model parameters (see Table 17.1).

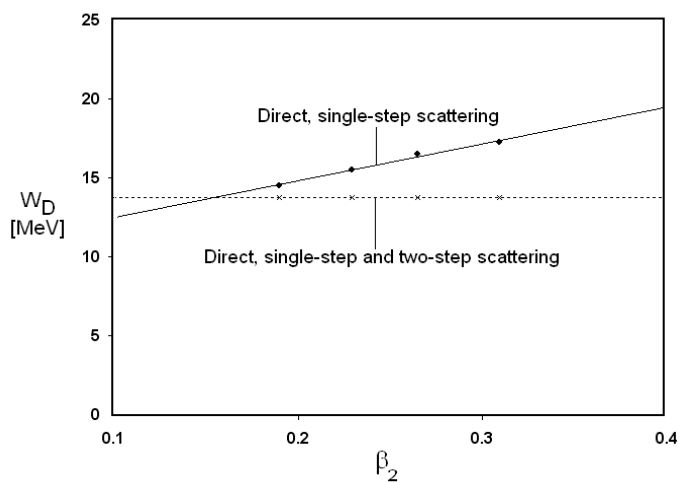

Figure 17.11. The effect of including or excluding the two-step scattering on the depth $W_D$ of the optical model potential.

The effect of including or excluding the two-step mechanism in the calculations is also illustrated in Figure 17.11. Without considering the two-step process explicitly in the calculations, $W_D$ depends linearly on $\beta_2$. The fit to the data gives the following dependence:

$$W_D = 10.12 + 23.44\beta_2 \quad \text{MeV}$$

By including the two-step process, the dependence on $\beta_2$ is removed, the description of the observed distributions is simplified, and only one set of optical potential parameters is required to fit the experimental results for all four selenium isotopes. The depth of the imaginary part of the central potential assumes then a fixed value of 13.75 MeV.

The imaginary component of the optical model accounts for contributions, which are not explicitly considered in the analysis of experimental data. Its constant value means that only one non-elastic scattering channel was important, the one which has now been included explicitly in the analysis. All other non-elastic channels have no effect on the observed distributions.





## Summary and conclusions

Measurements of the angular distributions of the differential cross sections and analyzing powers for the elastic scattering of 12 MeV polarized deuterons on $^{76,78,80,82}$Se isotopes revealed an unusual mass dependence. Contrary to the normal behaviour, where the amplitudes of the analyzing powers *decrease* with the increasing atomic mass number $A$, we have observed that for selenium isotopes the amplitudes *increase* with $A$. This is particularly clear for the angular distribution of the vector analyzing power $iT_{11}(\theta)$

I have carried out a theoretical analysis of the differential cross sections $\sigma_0(\theta)$ and vector analyzing powers $iT_{11}(\theta)$ assuming contributions from two-step processes via the intermediate inelastic, and single-neutron transfer channels. I have also carried out the conventional optical model analysis of the data. Parameters, which were used to fit the data are summarized in Table 17.1.

The coupled channels calculations included two-step scattering via the $2_1^+$ and $3_1^-$ states as well as the single-particle excitations of the $1g_{9/2}$ and $2p_{1/2}$ neutron configurations in the reactions (d,t)(t,d) and (d,p)(p,d). I have found that the two-step scattering (d,d')2$^+$(d',d) contributes most strongly to the elastic scattering. The influence of the $3_1^-$ state is negligible. The second-order scattering involving single-particle excitations has a noticeable but also relatively weak contribution, with the neutron pickup process being slightly stronger than the neutron stripping. However, in the presence of the two-step (d,d')2$^+$(d',d) scattering all other second-order processes studied here have a negligible influence on the calculated distributions.

By including the contributions from the two-step process (d,d')2$^+$(d',d), excellent fits to the $\sigma_0(\theta)$ and $iT_{11}(\theta)$ angular distributions have been obtained for all four selenium isotopes (see Figure 17.10) using a single, mass-independent set of the optical model parameters (the last set in Table 17.1). Results of the analysis show that the unusual mass-dependence of the $iT_{11}(\theta)$ amplitudes on the mass number $A$ of the selenium isotopes can be explained as being due to an interplay between the direct elastic scattering and the two-step scattering (d,d')2$^+$(d',d) via the first excited states in the target nuclei.

## References


Barbopoulos, L. O., Gebbie, D. W., Nurzynski, J., Borsaru, M. and Hollas, C. L. 1979, *Nucl. Phys.* **A331**:502.

Barrette, J., Barrette, M., Lamoureux, G. and Monaro, S. 1974, *Nucl. Phys.* **A235**:154.

Bürgi, H.R., Grüebler, W., Nurzynski, J., König, V., Schmelzbach, P. A., Risler, R., Jenny, B. and Hardekopf, R. A. 1980, *Nucl. Phys.* **A334**:413.

Nurzynski, J. 1975, *Nucl. Phys.* **A246**:333.

Schneider, M. J., Burch, J. D. and Kunz, P. O. 1976, *Phys. Lett.* **63B**:129.






<div align="center">**18**</div>

# Tensor Analyzing Power T$_{20}$(0$^0$) for the $^3$He(d,p)$^4$He Reaction at Deuteron Energies of 0.3 – 36 MeV

**Key feature:**

1. The $^3$He$(\vec{d}, p)$ $^4$He reaction is a useful medium for monitoring tensor polarization of deuteron beams produced by polarized ion sources.

2. The energy dependence of the $T_{20}(0^0)$ tensor analyzing power for this reaction, which was measured earlier at lower energies, has now been extended to higher energies of 11 - 36 MeV. Results can be fitted well using either second- or third-order polynomials. However, third-order polynomials give a better fit.

3. Previously (Schmelzbach, *et al.* 1976b) it has been shown that a set of *three* second-order polynomials has to be used to reproduce the experimental data for $T_{20}(0^0)$ at lower energies. In this new study, I have shown that to describe the data over the *whole* region of energies (0.3 – 36 MeV) a set of *five* second-order polynomials have to be used. However, a better description can be obtained by using just *three* third-order polynomials.

**Abstract:** Measurements of the $T_{20}(0^0)$ tensor analyzing power for the $^3$He$(\vec{d}, p)$ $^4$He reaction induced by the vector-polarized deuterons have been carried out at the incident deuteron energies 11-36 MeV. The analyzing power has been found to change smoothly from -0.92 at 11 MeV to -0.34 at 36 MeV. Excellent description of the experimental data over the whole range of the incident deuteron energies of 0.3 – 36 MeV has been obtained using a set of three third-order polynomials.

## Introduction

The reaction $^3$He$(\vec{d}, p)$ $^4$He induced by polarized deuterons is a useful medium for monitoring tensor polarization of deuterons produced by polarized ion sources. A known relationship between the tensor and vector polarizations of polarized ion sources can be employed in using this reaction also as a polarimeter in measurements of vector analyzing powers. This reaction has been used successfully over years in the Laboratorium für Kerphysik, ETH, Zürich, Switzerland.

At the reaction angle $\theta = 0^0$, all tensor analyzing powers except $T_{20}(0^0)$ are equal zero (Grüebler *et al.* 1971). At this angle, the general expression for the differential cross section (see Chapter 15) can be expressed by a simple form:

$$\sigma(0^0) = \sigma_0(0^0)\left[1 + \frac{1}{2}t_{20}T_{20}(0^0)(2\cos^2\beta - 1)\right]$$

where $t_{20}$ is the beam polarization, $T_{20}(0^0)$ is the analyzing power for the reaction $^3$He$(\vec{d}, p)$ $^4$He at $\theta = 0^0$, and $\beta$ the angle between the quantization axis and the direction of the incident beam (see Chapter 15).

Assuming that the $T_{20}(0^0)$ values are known, this reaction can be used as a convenient way to monitor tensor polarization $t_{20}$ of the incident deuteron beam.





However, one should notice that $3\cos^2 \beta - 1 = 0$ if $\cos^2 \beta = 1/3$, which corresponds to $\beta \approx 54.74^0$. Consequently, care should be taken to set up the quantization axis in such a way as to avid coming close to this angle.

In the polarimeter used at ETHZ, the $^3$He target was located at the end of the Faraday cup. The $^3$He cell had a diameter of 12 mm and was filled with $^3$He at a pressure of about 5 atm. The entrance window was made of a 6 $\mu$m Harvar foil. The $Q$-value of the $^3\text{He}(\vec{d}, p)^4\text{He}$ reaction is 18.35 MeV so the protons emerging at $0^0$ have a high energy. The end of cell $^3$He was in the form of a 0.4 mm stainless steel wall, which was thick enough to stop the deuteron beam but it absorbed only a small fraction of the energy of the emitted protons.

A surface-barrier detector was places on the beam axis at a distance of 12 cm from the $^3$He target centre to measure the emerging protons. The angular resolution of the detector was ±2.5$^0$. The detector was thick enough to stop protons with the energy of up to 12 MeV. Depending on the energy of accelerated deuterons, aluminium absorbers of 0.2 – 1 mm thick were placed in front of the proton detectors to lower the energy of detected protons. The tensor polarization of the beam $t_{20}$ was calculated from the ratio of counting rates for the positive and negative polarization using the known values of $T_{20}(0^0)$ for this reaction.

The advantage of using this reaction as a tensor polarization polarimeter are:

- The construction of the polarimeter is simple.
- Only one detector at $0^0$ is needed.
- Because of the high $Q$-value of the reaction, the protons spectrum is clean and there is no problem with the background.
- The absolute values of the analyzing power $T_{20}(0^0)$ are high at low energies and they vary smoothly with the incident deuteron energy.

Measurements of the $T_{20}(0^0)$ analyzing powers have been carried out earlier at low deuteron energies by Trainor, Clegg and Lisowski (1974) and Schmelzbach *et al.* (1976a). The analyzing power has been calibrated using the reactions $^{16}$O(d,$\alpha_1$)$^{14}$N* and $^4$He(d,d)$^4$He.

### Extension to higher energies

Due to the successful use of this reaction at tandem energies it was interesting to extend the measurements of $T_{20}(0^0)$ to higher energies available from the SIN[17] injector cyclotron. Some changes in the experimental setup have been necessary at these higher energies. A thicker entrance foil was used, which allowed for the pressure in the gas cell to be increased to 15 atm, which helped to compensate for lower differential cross sections for the reaction $^3\text{He}(\vec{d}, p)^4\text{He}$ at the higher energies available from the cyclotron. In order to stop the primary beam, different thicknesses of absorbers were used between the gas cell and proton detector, depending on the energy of accelerated deuterons. Two 5 mm thick detectors were used in coincidence to get clean spectra for the high-energy protons.

---

[17] Schweizerische Institut für Nuklearforschung





The tensor polarization of the SIN deuteron beam was determined to be $p_{zz} = 0.80$ by comparison with the Los Alamos measurements of d-$\alpha$ scattering at 17 MeV (Ohlsen *et al.* 1973). The results of measurements of the $T_{20}(0^0)$ tensor analyzing powers at these higher energies are shown in Figure 18.1. They show smooth energy dependence with the absolute value of $T_{20}(0^0)$ decreasing with the increasing deuteron energy indicating that this reaction might have limited application as a monitor of the deuteron polarization $p_{zz}$ at higher energies.

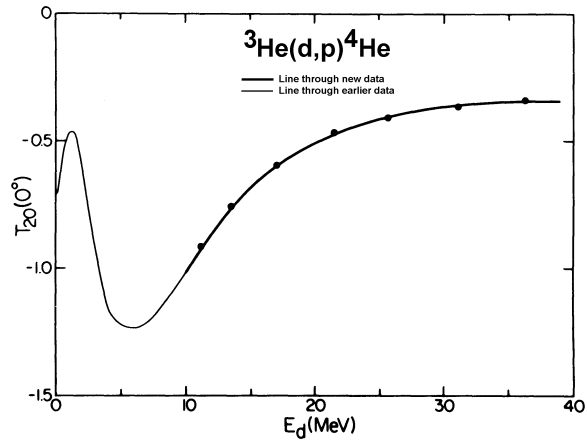

Figure 18.1. Energy dependence of the $T_{20}(0^0)$ tensor analyzing power for the reaction $^3$He$(\vec{d}, p)$$^4$He. The points represent our new measurements, which extend the earlier measurements to higher energies of 11-36 MeV. The combined statistical and systematic errors are smaller than the size of the points. The thick line is drawn as a guide for the eye. The thin line is drawn through the earlier data at lower incident deuteron energies (Schmelzbach *et al.* 1976a; Trainor, Clegg, and Lisowski 1974).

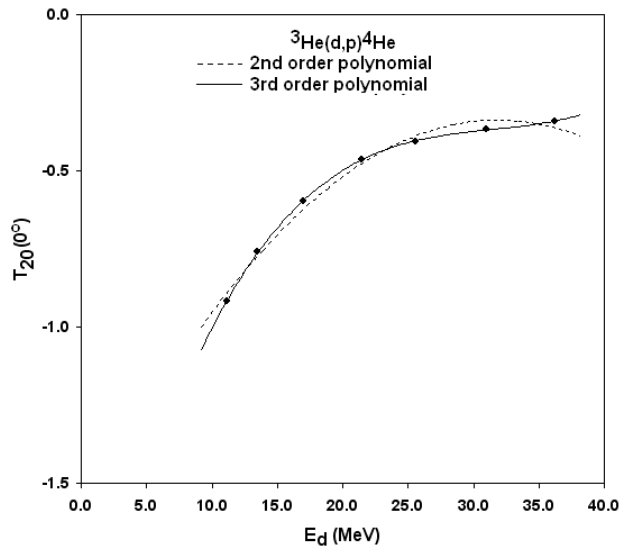

Figure 18.2. Results of our measurements of the $T_{20}(0^0)$ tensor analyzing power for the reaction $^3$He$(\vec{d}, p)$$^4$He in the energy range of 11-36 MeV are compared with the calculations using the second-order (dotted line) and third-order (continuous line) polynomials as listed in Table 18.1. The third order polynomial reproduces the data remarkably well.





For practical reasons, it is useful to parameterise the experimental data using simple mathematical formulae. Results at lower energies have been fitted using a set of three second-order polynomials (Schmelzbach *et al.* 1976b). It was therefore interesting to see whether the new set of data obtained now at higher deuteron energies could be also fitted using the same procedure. Figure 18.2 shows the fits obtained using the second- or third-order polynomial. Clearly, the third-order polynomial gives a better fit. The polynomials are listed in Table 18.1. The fit to the data is strongly sensitive to the highest order component in each polynomial.

Table 18.1

The third- and second-order polynomials used to fit the $T_{20}(0^0)$ data at the incident deuteron energies $E_d$ = 11-36 MeV (see Figure 18.1)

| Polynomials | $R^2$ |
|---|---|
| $T_{20}(0^0, E_d) = -1.6477340 + 0.0822609 E_d - 0.0012911 E_d^2$ | 0.9882675 |
| $T_{20}(0^0, E_d) = -2.2320672 + 0.1700278 E - 0.0052947 E^2 + 0.0000564 E^3$ | 0.9998055 |

## The $T_{20}(0^0)$ analyzing power for the energy range of 0.3-36 MeV

Having calculated the polynomials in this higher energy range I decided to join the new data with the previous measurements at lower energy and to see whether the formulae derived earlier using the *second-order* polynomials (Schmeltzbach *et al.* 1976b) could be still used. I had found that in order to represent both the old and new data using smooth transitions between various sets of polynomials, data had to be divided into different sections then used before and a new analysis had to be carried out.

Table 18.2

A set of the second-order polynomials describing the energy dependence of the tensor analyzing power $T_{20}(0^0)$ for the reaction $^3\text{He}(\vec{d}, p)^4\text{He}$ in the energy range of 0.3-36.2 MeV

| Energy Range (MeV) | Polynomials |
|---|---|
| $0.3 \leq E_d \leq 2.5$ | $T_{20}(0^0, E_d) = -0.7685 + 0.4785 E_d - 0.1900 E_d^2$ |
| $2.0 \leq E_d \leq 6.0$ | $T_{20}(0^0, E_d) = +0.3582 - 0.5800 E_d + 0.0517 E_d^2$ |
| $6.0 \leq E_d \leq 11.0$ | $T_{20}(0^0, E_d) = -1.4908 + 0.0247 E_d + 0.0023 E_d^2$ |
| $11.0 \leq E_d \leq 17.0$ | $T_{20}(0^0, E_d) = -2.0517 + 0.1309 E_d - 0.002653 E_d^2$ |
| $17.0 \leq E_d \leq 36.0$ | $T_{20}(0^0, E_d) = -1.3068 + 0.0562 E_d - 0.000821 E_d^2$ |

In my analysis, I have also included the data at very low energies, i.e. below 2.2 MeV, which were not described mathematically by Schmeltzbach *et al.* (1976b). I have found a new set of the second-order polynomial formulae (see Table 18.2)





that reproduce the observed energy dependence of $T_{20}(0^0)$ over the whole range of energies of 0.3-36 MeV (see Figure 18.3).

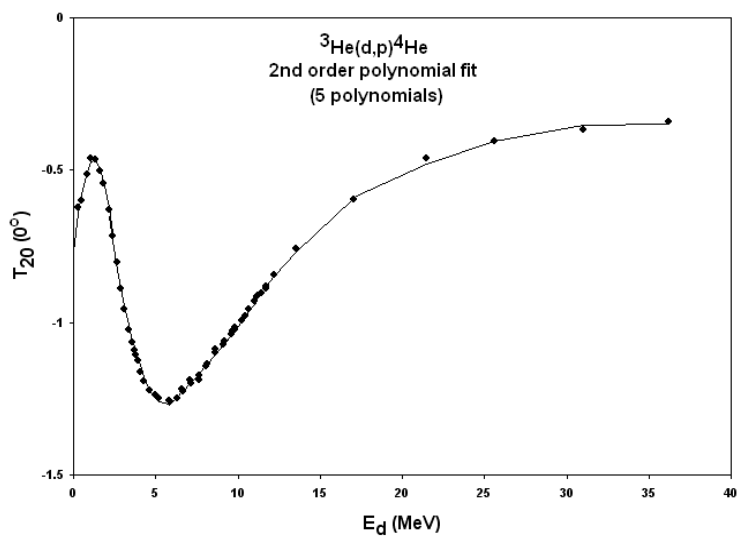

Figure 18.3. Results of measurements of the $T_{20}(0^0)$ tensor analyzing power for the reaction $^3$He$(\vec{d}, p)$ $^4$He in the energy range of 0.3-36 MeV are compared with a new set of the *second-order* polynomials. The continuous line is made of *five* polynomials listed in Table 18.2.

It is interesting to compare these new calculations with the calculations of Schmelzbach *et al.* (1976b) at low energies. This is done in Figure 18.4. Even though the two sets of calculations describe the experimental data sufficiently well, the new calculations appear to reproduce the experimental results a little better.

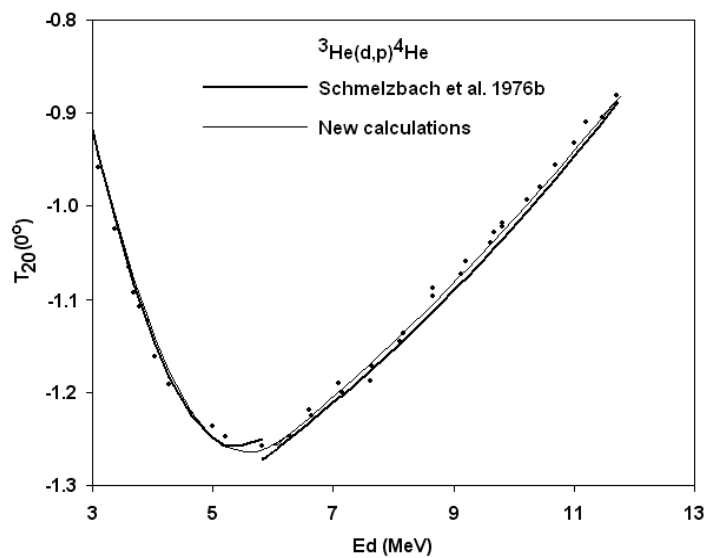

Figure 18.4. The calculations of Schmeltzbach *et al.* (1976b) are compared with the new calculations using the formulae listed in Table 18.2.

Results presented in Figure 18.2 suggest that third-order polynomials might give better fits to the data. Calculations over the whole range of the incident deuteron





energies are shown in Figure 18.5. The experimental data can now be fitted using just *three* functions (as compared to *five* if the second-order polynomials are used). Furthermore, the resulting theoretical line is now smoother than the line corresponding to a set of the second-order polynomials (cf Figures 18.3 and 18.5). The set of the *three* third-order polynomials describing the experimental data over the whole range of energies, 0.3 − 36 MeV, is given in Table 18.3.

It should be noted that the parameters for the last two equations in Table 18.2 and for the last equation in Table 18.3 are different than the relevant parameters in Table 18.1. This is because the energy range for the formulae in Tables 18.2 and 18.3 are different than the range for the formulae in Table 18.1.

Table 18.3

A set of the third-order polynomials describing the energy dependence of the tensor analyzing power $T_{20}(0^0)$ for the reaction $^3$He$(\vec{d},p)^4$He in the energy range of 0.3-36 MeV

| Energy Range (MeV) | Polynomials |
|---|---|
| $0.3 \leq E_d \leq 3.5$ | $T_{20}(0^0, E_d) = -0.8056 + 0.6166E_d - 0.3221E_d^2 + 0.0344E_d^3$ |
| $3.0 \leq E_d \leq 10.0$ | $T_{20}(0^0, E_d) = +0.5939 - 0.7875E_d + 0.1057E_d^2 - 0.0043E_d^3$ |
| $9.5 \leq E_d \leq 36.0$ | $T_{20}(0^0, E_d) = -2.2706 + 0.1755E_d - 0.0055E_d^2 + 0.0000586E_d^3$ |

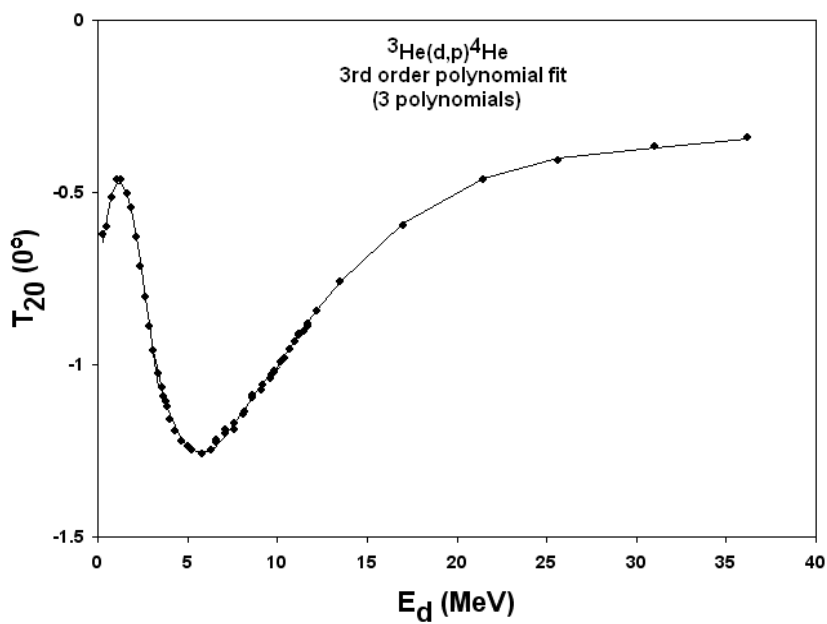

Figure 18.5. Results of the measurements of the $T_{20}(0^0)$ tensor analyzing power for the reaction $^3$He$(\vec{d},p)^4$He in the energy range of 0.3-36 MeV are compared with the *third-order* polynomial calculations. The continuous line is made of only *three* polynomials listed in Table 18.3.





**Summary and conclusions**

Earlier low-energy measurements of the $T_{20}(0^0)$ tensor analyzing power for the reaction $^3\text{He}(\vec{d}, p)$ $^4\text{He}$ have been extended to higher incident deuteron energies of up to 36 MeV. The energy dependence of the $T_{20}(0^0)$ analyzing power has been analysed using the second- and third-order polynomials. Experimental results over the whole range of energies of 0.3 − 36 MeV can be reproduced remarkably well by a set of only *three* third-order polynomials. The information supplied by these new set of data and the associated analysis can assist in using the $^3\text{He}(\vec{d}, p)$ $^4\text{He}$ as the polarization analyser.

**References**


Grüebler, W., König, V., Ruth, A., Schmelzbach, P. A., White, R. E. and Marmier, P. 1971, Nucl. Phys. **A176**:631.

Schmelzbach, P. A., Grüebler, W., König, V., Risler, R., Boerma, D. O. and Jenny, B.1976a, *Nucl. Phys.* **A264**:45.

Schmelzbach, P. A., Grüebler, W., König, V., Risler, R., Boerma, D. O. and Jenny, B. 1976b, *Proceedings of the Fourth International Symposium on Polarization Phenomena in Nuclear reactions*, 1975, eds W. Grüebler and V. Köning, Birkhauser Verlag, Basel, p. 899 and references therein.

Ohlsen, G. G., Lovoi, P. A., Salzmann, G. C., Meyer-Berkhout, U., Mitchell, C. K., and Grüebler, W. 1973, *Phys. Rev.* **C8**:1262.

Trainor, T. A., Clegg, T. B. and Lisowski, P. W. 1974, Nucl. Phys. **A220**:533.






<div align="center">**19**</div>

# The Maximum Tensor Analyzing Power $A_{yy} = 1$ for the ³He(d,p)⁴He Reaction

***Key feature:***

1. Points at which the analyzing powers reach their extreme values ( $A_y = 1$ and $A_{yy} = 1$ ) are of significant interest both experimentally and theoretically. Experimentally, they can serve as the reference points for the calibration of polarimeters. Theoretically, they can be used to test special conditions imposed by such extreme values.

2. We have found that a $A_{yy} = 1$ value for the ³He(d,p)⁴He reaction is located at $E_d$ = 9.28 MeV (lab) and $\theta$ = 23.6⁰ (lab).

3. We have found no $A_y = \pm 1$ point in the investigated range of energies and angles. However, our measurements have resulted in high-precision data, which can be used as reliable reference points in the calibration of polarimeters or in theoretical analyses.

4. We suggest that the extreme value of $A_{yy} = 1$ is associated with a resonance in ⁵Li.


**Abstract:** Very precise measurements of the vector and tensor analyzing powers, $A_y(E_d, \theta)$ and $A_{yy}(E_d, \theta)$ for the ³He($\vec{d}$, $p$)⁴He reaction have been carried out at the incident deuteron energies $E_d$ = 8.5-10.5 MeV and at angles $\theta$ = 12⁰ – 32⁰ (lab). The aim of this study was to search for the extreme values $A_y = 1$ and $A_{yy} = 1$. A maximum $A_{yy} = 1$ has been located at $(E_d, \theta)$ = (9.28 MeV, 23.6⁰) in the laboratory system of reference. No extreme value of $A_y = \pm 1$ was detected for this reaction.


## Introduction

Points where the analyzing powers reach their extreme values are of significant interest both experimentally and theoretically. For experimentalists, these points can be used in the absolute calibration of the experimental equipment used to measure beam polarization. For theorists, they allow to study special conditions for the $M$-matrix amplitudes (see the Appendix H).

The location of the extreme values of the analyzing powers have been investigated earlier for the elastic scattering of spin-1/2 and spin-1 projectiles from spin-0 target nuclei. For spin-1/2 projectiles, Plattner and Bacher (1971) showed analytically[18] that $A_y = \pm 1$ values must occur at three sets of energy-angle coordinates in the nucleon-$\alpha$ elastic scattering. For spin-1 particles, conditions for the maximum values of the analyzing power were investigated by Grüebler *et al.* (1975) for the d-$\alpha$ scattering.

---

[18] For an outline of the analytical determination of the extreme values of the analyzing powers see the Appendix I.





These authors have shown analytically that three $A_{yy} = 1$ points should exist in the deuteron energy range between 3 and 12 MeV. They have also located them experimentally.

In all these cases, there is a simple linear relation between just *two* elements of the transition matrix $M$, which connects the spin states of the incoming and outgoing channels.

However, the question is whether such points can also occur in nuclear transfer reactions. Seiler (1976) investigated the conditions imposed on the $M$-matrix elements for the spin configuration $1 + 1/2 \to 1/2 + 0$.[19] He found that the $A_{yy} = 1$ occurs if and only if two sums of the spin-non-flip amplitudes and the spin-flip amplitudes of $M$-matrix elements are equal zero.

$$M_{1,1/2;1/2} + M_{-1,1/2;1/2} = 0; \; M_{1,-1/2;1/2} + M_{-1,-1/2;1/2} = 0 \qquad (1)$$

In these equations, the indices denote the spin projections in the incoming and outgoing channels. This situation is different than the condition for the elastic scattering because two linear relations involving *four* $M$-matrix elements have to be satisfied at the same angle and energy. The existence of such a point cannot be proven analytically in the same way as the maxima for the elastic scattering.

Seiler (1976) suggested that the extreme values of $A_{yy} = 1$ should occur for the $^3\text{He}\,(\vec{d}, p)\,^4\text{He}$ reaction at the $(E_d, \theta)$ coordinates of around (9.0 MeV, 27⁰), and for the $^6\text{Li}\,(\vec{d}, p)\,^4\text{He}$ at around (5.5 MeV, 30⁰) and (9.0 MeV, 90⁰). To locate such maxima experimentally, precise mapping of the analyzing powers in the $(E_d, \theta)$ coordinates is required.

In our study, we have used the $^3\text{He}\,(\vec{d}, p)\,^4\text{He}$ reaction, which was investigated earlier by Grüebler *et al.* (1971). In our measurements, we have included the energy-angle mapping not only for the $A_{yy}$ component but also for $A_y$ because the previous measurements suggested that a $A_y = -1$ point might be present in the same region.

### Experimental procedure

The experimental method employed in this study was as described in Chapters 15, 17 and 18. The spin direction was perpendicular to the scattering plane. The exact spin position was adjusted by the Wienfilter to a precision better than 1° before starting the measurements. A polarimeter based on detecting protons from the $^3\text{He}\,(\vec{d}, p)\,^4\text{He}$ reaction at 0° was used to measure the beam polarization continuously (see Chapter 18). This reaction has been calibrated in the absolute sense with the precision better than 1%. A polarized deuteron beam with $p_y = 0.3$ and $p_{yy} = 0.9$ was delivered by the ETHZ atomic beam polarized source and EN tandem accelerator. The defining diaphragms in front of the detectors were 4 mm wide and 30 mm high and they were located at a distance of 256 mm from the middle of the gas target. The procedure of determining the analyzing powers is described extensively in a Chapter 15. This method, which uses detectors on the left- and right-hand side of the target and the frequent reversal of the sign

---

[19] The polarization formalism for the $1 + 1/2 \to 1/2 + 0$ structure is outlined in the Appendix F.





of the beam polarization, cancels instrumental asymmetries. The measurements are not affected by small deviations of the spin direction from the required position.

## Experimental results

### *Experimental data*

The analyzing powers $A_y$ and $A_{yy}$ have been measured between 8.5 and 10.5 MeV deuteron energy in the angular range $\theta_{lab} = 12° - 32°$. Results are shown in Figure 19.1. Extra care has been taken to carry out precise measurements, which is essential in the determination of the extreme values of the analyzing powers. The statistical errors are smaller than the size of the displayed data points. The curves represent the fits obtained by using polynomial functions. As discussed below, the polynomial interpolation of the experimental data was necessary to construct a two-dimensional map of the experimental data and to locate the $A_{yy}$ maxima. Since the measurements show clearly that for these energies there is no maximum for $A_y$ this component was not fitted with polynomial functions. However, our results represent very accurate data and consequently they can be used as reference points in calibrations of the vector polarization of polarized deuteron beams.

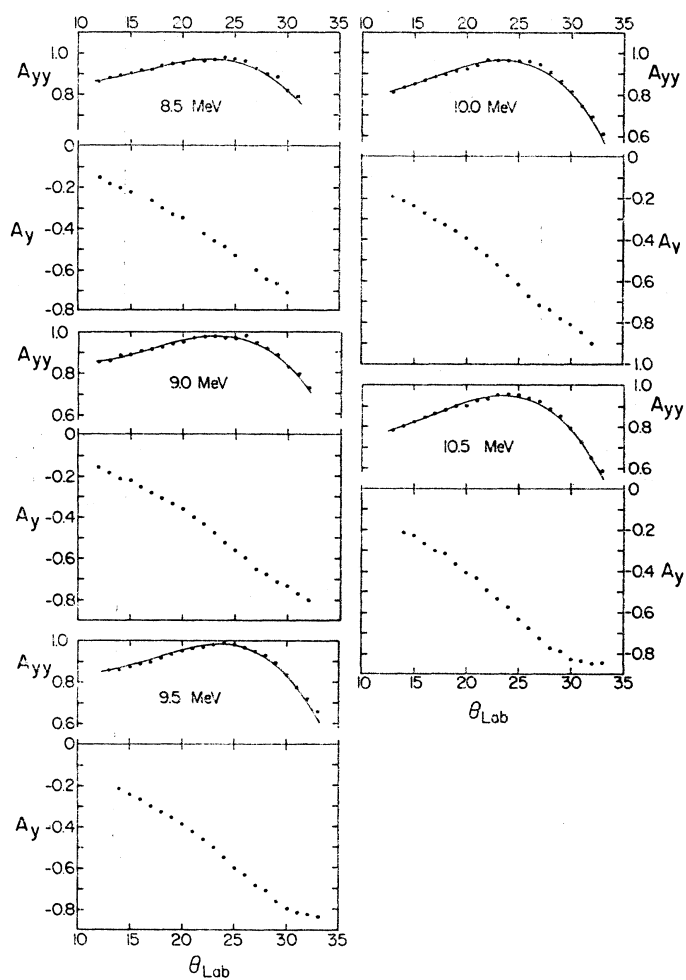

Fig. 19.1. Results of measurements of the analyzing powers $A_y$ and $A_{yy}$. The statistical errors are smaller than the displayed data points. The curves are the polynomial fits.





### Interpolation of the experimental data

The mapping of the region of measurements was carried out by interpolation calculations. In this procedure, each angular distribution was fitted by a polynomial the order of which was determined by the best fit to the data. We have found that in all cases the best fit was obtained by the third-order polynomials. The polynomial fits, together with the experimental points, were then used to construct a two-dimensional map for the $A_{yy}$ analyzing power. The map is shown in Figure 19.2.

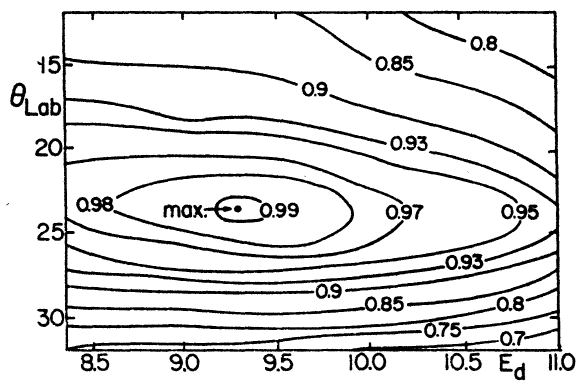

Figure 19.2. Contour plot of the tensor analyzing power $A_{yy}$ constructed using our experimental results and the polynomial interpolation method as described in the text.

### Corrections and uncertainties

The uncorrected interpolation gives a maximum value of $A_{yy} = 0.981 \pm 0.003$. This value has to be corrected for the finite solid angle used in the experiment. The quoted error does not include the uncertainties in the determination of the beam polarization. The necessary corrections and uncertainties, which have to be included in the final calculations of the $A_{yy}$ maximum, are shown in Table 19.1.

Table 19.1

Corrections and uncertainties for the measurements of the $A_{yy}$ analyzing power

| Corrections | |
|---|---|
| Correction due to the extended azimuthal angle $\phi$ | $+0.007$ |
| Correction due to the finite scattering angle $\theta$ | $+0.002$ |
| Total correction for $A_{yy}$ | $+0.009$ |

| Uncertainties | |
|---|---|
| Statistical uncertainty of $A_{yy}$ | 0.3 % |
| Statistical uncertainty in the determination of the beam polarization | 0.5 % |
| Uncertainty in the absolute value of the analysing power of the polarimeter | 0.9 % |
| Total uncertainty of $A_{yy}$ | $\pm 1.1$ % |

### Final results

Taking into account all possible errors of measurements, the final results are given in Table 19.2. They show the position of the $A_{yy} = 1$ point.





Table 19.2

Experimentally determined maximum value of $A_{yy}$

| | |
|---|---|
| $A_{yy}$ | 0.990 ± 0.011 |
| $E_{d,\,lab}$ | 9.28 ± 0.030 MeV |
| $\theta_{lab}$ | 23.6° ± 0.2° |
| $\theta_{c.m.}$ | 28.1° ± 0.2° |

## Discussion

Our study shows that within experimental errors a maximum tensor analyzing power $A_{yy} = 1$ for the reaction $^3\mathrm{He}(\vec{d}, p)\,^4\mathrm{He}$ has been located. This result can be tested by measurements of other observables, which must satisfy the expected theoretical conditions (Seiler 1976) imposed by the relations between the matrix elements, as given by equation (1) in the Introduction. Besides the vanishing of several polarization transfer coefficients and polarization correlation coefficients the proton polarization $P_y$ for unpolarized beam and target must be equal to the negative value of the analyzing power $A_{0,y}$ of a polarized $^3\mathrm{He}$ target. Although the relevant experimental data at the required angle and energy are not available, results in close proximity to the required condition appear to agree with the theoretical predictions (Brown and Haeberli 1963; Grüebler, König, and Schmelzbach, 1973; Hardekopf *et al.* 1973).

An analysis of the polarization transfer coefficients $K_x^{x'}$ and $K_z^{x'}$ measured by Hardekopf *et al.* (1973) at 8 MeV shows vanishing values of these quantities near the critical angle, as required by the eqn (1). All these data appear to support the result found in our study.

A further, experimental test of our result, would be to investigate the inverse reaction $^4\mathrm{He}(p,d)\,^3\mathrm{He}$, for which the emitted deuterons should have a maximum tensor polarization $p_{yy} = 1$ for the corresponding incident proton energy.

Having found the extreme value of $A_{yy} = 1$, it is interesting to consider the physical interpretation of the equations (1). The probability that the sums of two pairs of complex amplitudes vanish simultaneously accidentally is very low. A possible reason for the situation described by equations (1) may lie in the particular symmetric structure of these relations. A study of the d-α scattering (Grüebler *et al.* 1975) suggests that this feature might be explained by a resonance in the vicinity of this energy.

The analysis of the $^3\mathrm{He}(d,p)\,^4\mathrm{He}$ reaction in the energy range of between 2.8 to 11.5 MeV using Legendre polynomials seems to suggest a resonance behaviour (Grüebler *at al.* 1971). In this analysis, the tensor analyzing powers $T_{20}$, $T_{21}$, and $T_{22}$ show a resonance-like behaviour around 9 MeV deuteron energy corresponding to an orbital angular momentum $l = 2$. Examination of the energy levels of $^5\mathrm{Li}$ (Figure 19.3) suggests also that a resonance around the required excitation energy is possible.

The energy $E_d = 9.28$ MeV corresponds to the excitation energy of 22.2 MeV. There is an energy level at 22.06 MeV in $^5\mathrm{Li}$ with spin $^3/_2^-$. The spins in the entrance and exit channels are $S_i = S_d + S_{^3He} = 1/2$ or $3/2$, and $S_f = S_p + S_{^4He} = 1/2$. If we use the





notation $^{2S+1}L_J$ for the initial and final channels, then we can have three possible resonance transitions to this level (see Table 19.3). Thus, the expected reaction mechanism could be associated with the *P* and/or *F*-wave capture in the entrance channel.

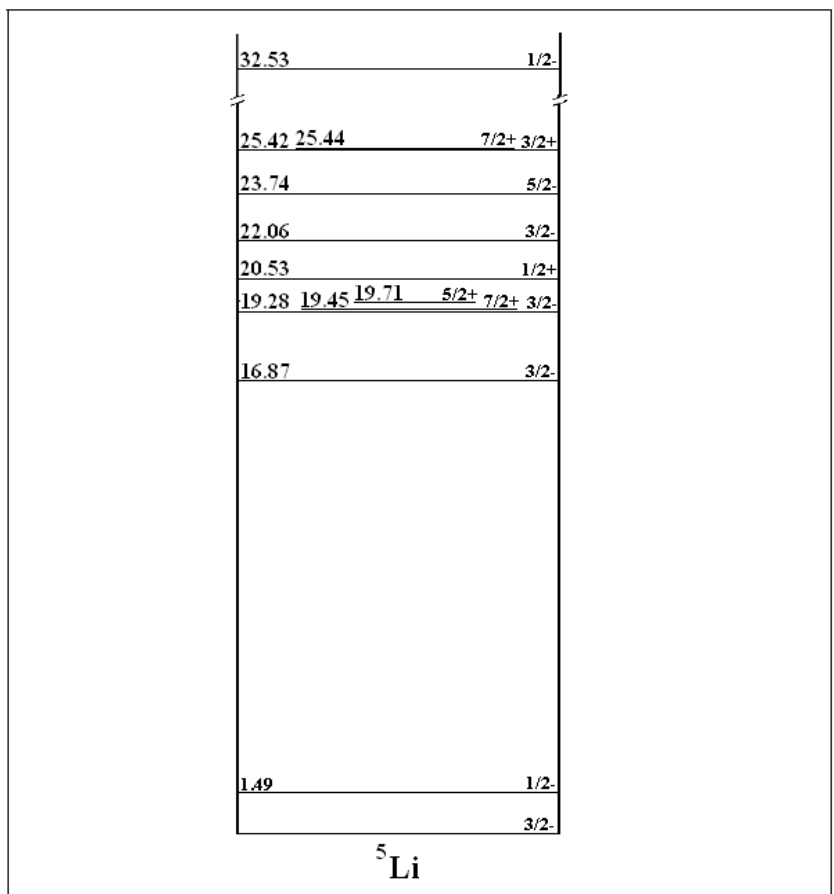

Figure 19.3. The energy level diagram for $^5$Li

Table 19.3
The possible resonant transitions at $E_d = 9.28$ MeV

| Transition | Resonant matrix element (Welton 1963) |
|---|---|
| | $(L_f\ S_f\ J^\pi | R | L_i\ S_i\ J^\pi)$ |
| $^2P_{3/2} \rightarrow {}^2P_{3/2}$ | $(1\ 1/2\ 3/2^- | R | 1\ 1/2\ 3/2^-)$ |
| $^4P_{3/2} \rightarrow {}^2P_{3/2}$ | $(1\ 1/2\ 3/2^- | R | 1\ 3/2\ 3/2^-)$ |
| $^4F_{3/2} \rightarrow {}^2P_{3/2}$ | $(1\ 1/2\ 3/2^- | R | 3\ 3/2\ 3/2^-)$ |

Transition: $^{2S_i+1}(L_i)_J \rightarrow {}^{2S_f+1}(L_f)_J$






**References**

Brown, R. I. and Haeberli, W. 1963, *Phys. Rev.* **130**:1163.

Goldfarb, L. J. B. 1958, *Nucl. Phys.* **7**:622.

Grüebler, W., König, V., Ruh, A., Schmelzbach, P. A., White, R. W. and Marmier, P. 1971, *Nucl. Phys.* **A176**:631.

Grüebler, W., König, V. and Schmelzbach, P. A. 1973, 'Results of measurements and analyses of nuclear reactions induced by polarized and unpolarized deuterons', Internal Report, ETH Zurich, May 1973.

Grüebler, W., Schmelzbach, P. A., König, V., Risler, R. Jenny, B. and Boerma, D. O. 1975, *Nucl. Phys.* **A242**:285.

Hardekopf, R. A., Armstrong, D. D., Grüebler, W., Keaton, P. W. and Meyer-Berkhout, U. 1973, *Phys. Rev.* **C8**:1629.

Plattner, G. R. and Bacher, A. D. 1971, *Phys. Lett.* **36B**:211.

Seiler, F. 1976, *Phys. Lett.* **61B**:144.

Welton, T. A. 1963, in *Fast Neutron Physics*, Vol. II, eds J. B. Marion and J. L. Fowler, Interscience, New York, p. 1317.






---



# Search for the $A_y = 1$ and $A_{yy} = 1$ Points in the $^6$Li(d,$\alpha$)$^4$He Reaction

***Key features:***

1. We have located *two* $A_{yy} = 1$ points for the reaction $^6$Li$(\vec{d}, \alpha)$ $^4$He, one at $E_d$ = 5.55 MeV (lab) and $\theta$ = 24.2$^0$ (lab) and one at $E_d$ = 8.80 MeV (lab) and $\theta$ = 76.8$^0$ (lab)

2. The $A_y = 1$ points, if present, should occur at the same energies and angles as $A_{yy} = 1$. No such points were observed for the $^6$Li$(\vec{d}, \alpha)$ $^4$He reaction indicating that two additional conditions for the *M*-matrix elements as required for $A_y = 1$ are not satisfied.

3. The suggested reaction mechanism responsible for the extreme values of the tensor analyzing power, $A_{yy} = 1$, is a resonance in $^8$Be.

**Abstract:** A search for the analyzing powers $A_y = 1$ and $A_{yy} = 1$ was carried out experimentally for the $^6$Li$(\vec{d}, \alpha)$ $^4$He reaction in the energy range of 5.0-6.5 MeV and 8.0-10.0 MeV. Two $A_{yy} = 1$ maxima had been found but no $A_y = 1$ points, which should occur at the same energies and angles, were detected. The precise values for the energies and angles of the $A_{yy} = 1$ maxima were determined by the polynomial interpolation of the experimental data.

## Introduction

Having successfully located the extreme value $A_{yy} = 1$ of the tensor analyzing power for the reaction $^3$He$(\vec{d}, p)$ $^4$He, we have decided to extend our search to the $^6$Li$(\vec{d}, \alpha)$ $^4$He reaction. In our previous study, we have established that the extreme values exist not only in the elastic scattering but also in nuclear reactions induced by polarized deuterons. It seemed therefore interesting to search for such extreme values in another transfer reaction. The importance of such points has been discussed in Chapter 19.

Inspection of the analyzing powers for the $^6$Li$(\vec{d}, \alpha)$ $^4$He reaction below 12 MeV of the incident deuteron energy (Seiler et al. 1976; Seiler, Rad, and Conzett 1976) suggested a possibility of the existence of the $A_{yy} = 1$ maxima near $E_d$ = 6 MeV ($\theta_{c.m.} \approx 35^0$) and $E_d$ = 9 MeV ($\theta_{c.m.} \approx 75^0$). A compilation of previous measurements is presented in Figure 20.1. The experimental data are not sufficiently precise but they seem to suggest that the analyzing power $A_{yy}$ reaches its extreme values in two places. In our study, we have included a search for $A_y = 1$ points, which if present should occur at the same energies and angles as the $A_{yy} = 1$ maxima.





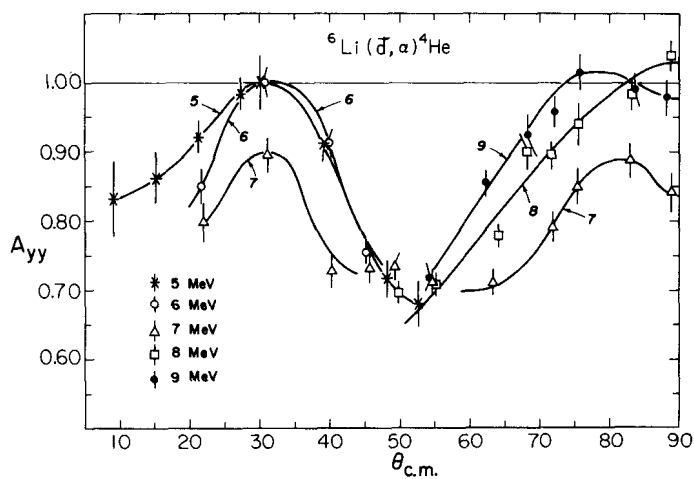

Figure 20.1. A compilation (Seiler *at al.* 1976) of the data for the $A_{yy}$ tensor analyzing power. The curves drown through the experimental points are labelled according to the incident deuteron energy. This figure shows that the tensor analyzing power $A_{yy}$ is likely to reach its extreme values of 1 at around 6 and 9 MeV and at the angles of around $30^0$ and $75^0$, respectively. Accurate measurements were required to confirm or refute this expectation.

## Experimental procedure

The general experimental arrangement and procedure are described in the previous Chapter. A self-supporting $^6$Li foil, enriched to 96%, with a thickness of about 300 μg/cm$^2$ was used as the target.

Due to the high $Q$ value for the $^6$Li$(\vec{d},\alpha)$ $^4$He reaction ($Q = 22.374$ MeV) the emitted $\alpha$ particles have high energy. Thus, by proper adjusting of the sensitive depth of the surface barrier detectors the background under the $\alpha$ peaks could be reduced to only a few percent. However, the cross section for this reaction is small (only about 0.3 mb/sr). This caused a significant problem at forward angles where the elastic scattering cross section is about three orders of magnitudes larger than the transfer reaction cross section. For this reason, electronic pile-up effects could not be prevented in all cases and had to be considered carefully in the analysis of the data.

The components $A_y$ and $A_{yy}$ were measured between 5.00 MeV and 6.50 MeV in the angular range of 20° to 50° in the c.m. system. Data were obtained in the energy steps of between 100 and 500 keV.

Measurements were also carried out for the incident deuteron energies of between 8.00 and 10.00 MeV in steps of 0.5 MeV. The c.m. angular range in this case was 60° to 105°.

## Results of measurements

Our experimental results for the lower energy range are presented in Figure 20.2. The statistical errors are smaller than the displayed data points. The displayed data have not yet been corrected for the finite geometry and electronics effects. The solid lines are polynomial curves fitted to the data in order to obtain the local maxima. The $A_y = 1$ point should occur at the same angle and energy as $A_{yy} = 1$ (Seiler *et al.* 1976).

It is clear that the $A_y$ component does not reach an extreme value of 1 for the $^6$Li$(\vec{d},\alpha)$ $^4$He





reaction. However, the uncorrected distributions for the $A_{yy}$ analyzing power come close to the extreme value.

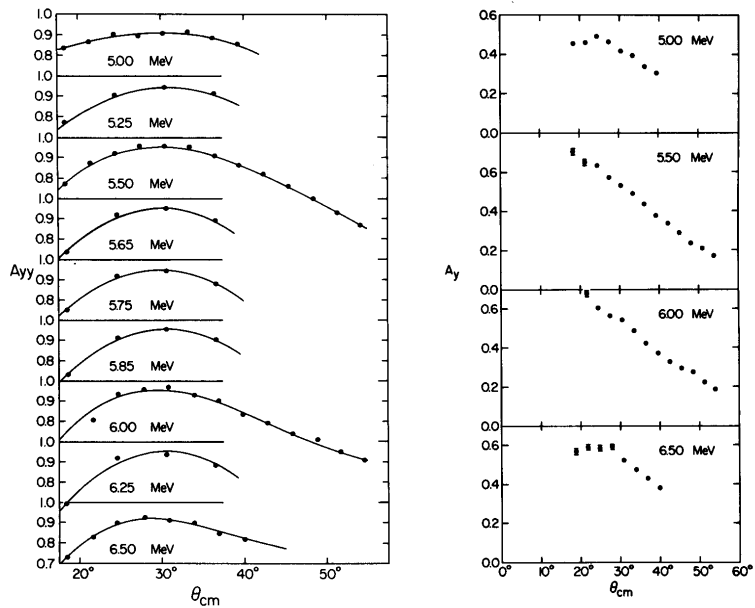

Figure 20.2. Tensor and vector analyzing powers measured in the energy range of 5.5-6.5 MeV. The distributions for the tensor analyzing power, $A_{yy}$, are shown on the left-hand side and for $A_y$ on the right-hand side of the figure. The statistical errors are smaller than the size of the points. The curves for the $A_{yy}$ component are the polynomial fits.

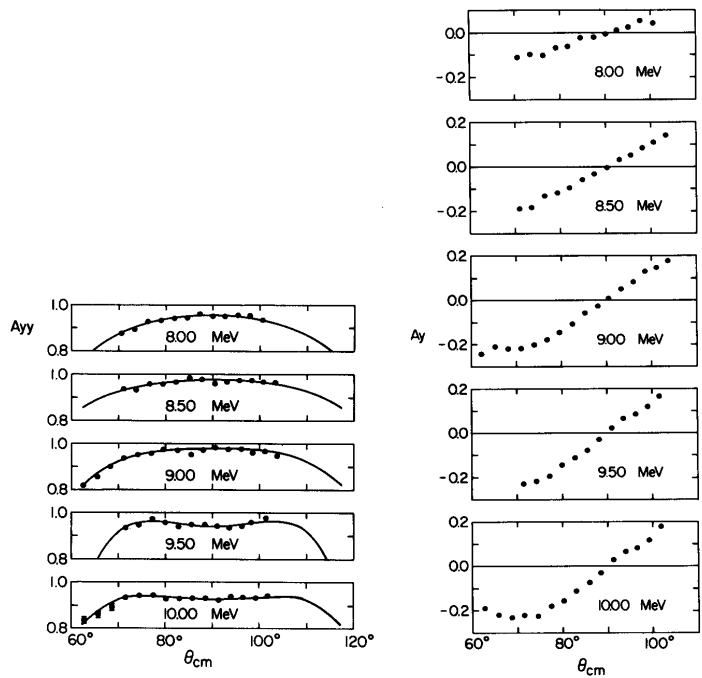

Figure 20.3. Vector and tensor analyzing powers measured in the energy range of 8.0-10.0 MeV. The distributions of the tensor analyzing power, $A_{yy}$, are shown on the left-hand side and for $A_y$ on the right-hand side of the figure. The statistical errors are smaller than the size of the data points. The curves for the $A_{yy}$ component are the polynomial fits.





Results obtained in the higher energy range are shown in Figure 20.3. The $A_y$ changes sign at 90°as expected for a reaction with two identical particles in the exit channel. The curves are again polynomial fits to the data.

Our data in this higher energy range demonstrate that while the vector component does not reach its extreme values $A_y = 1$ in the expected region of energies and angles, the tensor component $A_{yy}$ approaches the values of 1 at two energies.

### Finding the maxima of the $A_{yy}$ component.

The experimental mapping of the two energy regions was carried out using the interpolation procedure described in Chapter 19. At first the raw data were fitted with polynomial curves and the results are displayed in Figures 20.2 and 20.3. Next various necessary corrections were included.

The local maximum located between 5.0 and 6.5 MeV and at a small angle required a solid-angle correction of +1.2%. An additional correction was necessary because of the pile-up effects caused by the much higher counting rate of the elastically scattered deuterons. An investigation based on simulating the electronic pile-up effects by using two random pulse generators with pulse heights and pulse rates corresponding to the scattered deuterons and the alphas from the reaction, indicated a 3.5% counting loss in the simulated $\alpha$-peak in the spectrum and created an additional background on both sides of the peak. In spite of this large loss in the number of counts the calculated correction to the analyzing powers was only 0.7%.

At higher energies where a broad maximum is around 90° the calculated correction from the solid angle geometry is only +0.5%. Because of the much smaller elastic scattering cross section no pile-up loss correction was necessary.

The final absolute calibration at the maxima found at lower and higher energies was made with a beam calibrated using the analytically proved $A_{yy} = 1$ points in d-$\alpha$ scattering (Grüebler *at al.* 1975). This final calibration included also the measurement of the quantities $A_{xx}$ and $A_{zz}$. The relation $A_{xx} + A_{yy} + A_{zz} = 0$ can then be used as a consistency check of the data. Using our results, we get for the sums (0.0030 ± 0.0150) and (-0.0018 ± 0.0094) at the lower and higher energies, respectively.

The fits around the interpolated maxima are shown in Figure 20.4. The dots with error bars are the corrected and absolutely calibrated results of our measurements. The open circles represent the determined maximum values. The thick solid lines represent the values extracted from the polynomial mapping of the experimental results. The shades band represents the statistical errors of the measurements. An additional uncertainty of 0.007 caused mainly by the background subtraction and the uncertainty in the correction for pile-up losses must be applied to the results at 5.55 MeV. This is shown by the dashed curves.

Finally, the uncertainty of the beam polarization had to be also included. The energy of the accelerator had to be changed for the d-$\alpha$ calibration points and hence a slight change in the beam optics and a corresponding deviation of the beam polarization had to be taken into account. The total uncertainty of the calibration procedure was estimated at no more than 1 %. This uncertainty is shown as dashed horizontal lines in Figure 20.4. These lines show that due the uncertainly in the absolute value of the beam polarization the displayed value of 1 on the vertical scale of each figure can be located





anywhere between the two points indicated by the horizontal dashed line and the top frame of the figure. In numerical terms, the vertical scale in each figure can be shifted down by a maximum of 1%. The figure shows that within experimental errors we have located two $A_{yy} = 1$ points for the reaction $^6$Li(d,$\alpha$)$^4$He in the investigated ranges of energy and angles.

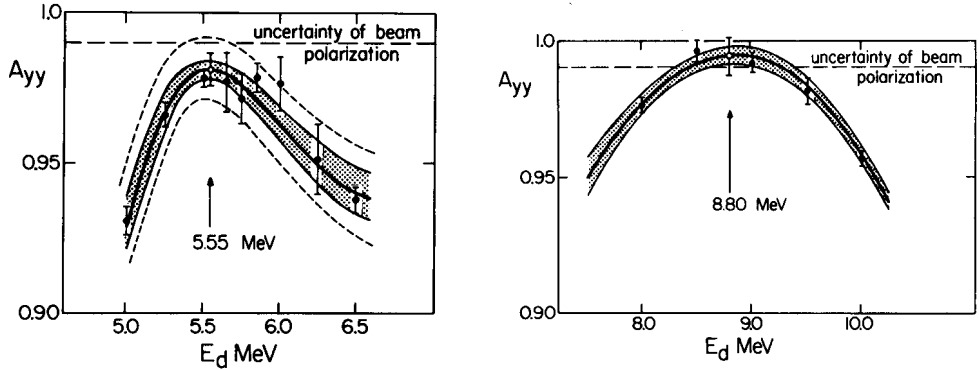

Figure 20.4. Experimental data and the interpolation curves based on the polynomial mapping. The shaded areas show statistical uncertainties. The dashed lines show the boundaries of uncertainties that include background substruction and pilup-up errors. They apply only to the lower energy data. The dashed horizontal lines show the maximum value of the beam polarization uncertainty caused by changing the energy between the $^6$Li($\vec{d},\alpha$) $^4$He measurements and d-$\alpha$ calibration. The interpretation of these lines is that the vertical scale can be shifted by the distance between the horizontal lines and the top of the frames.

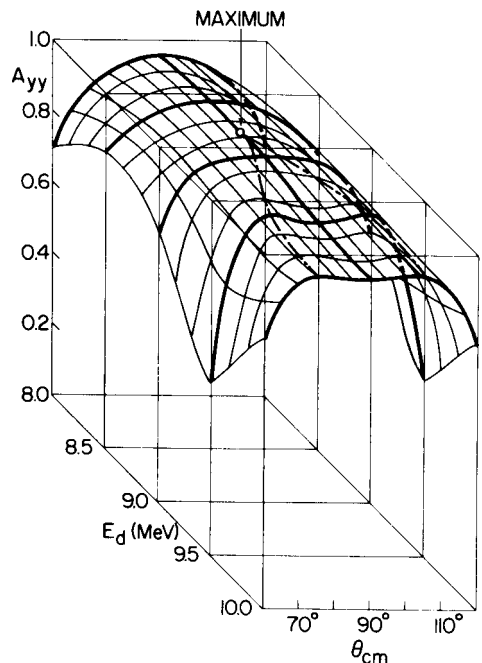

Figure 20.5. The three-dimensional representation of the experimentally determined tensor analyzing power $A_{yy}$. The maximum $A_{yy} = 1$ is indicated by an open circle.

As a sample of the polynomial mapping of the experimental data a three-dimensional representation of the experimentally determined $A_{yy}$ component is show in Figure





20.5. The maximum value of $A_{yy} = 1$ is shown as an open circle. The experimentally determined $A_{yy} = 1$ points are shown in Table 20.1.

Table 20.1

Experimentally determined location of the $A_{yy} = 1$ points for the reaction $^6$Li $(\vec{d}, \alpha)$ $^4$He

| $E_{d, \text{ lab}}$ (MeV) | $5.55 \pm 0.12$ | $8.80 \pm 0.25$ |
|---|---|---|
| $\theta_{\text{lab}}$ (deg) | $24.2 \pm 1.0$ | $76.8 \pm 1.0$ |
| $\theta_{\text{c.m.}}$ (deg) | $29.7 \pm 1.0$ | $90.0 \pm 1.0$ |

## Summary and discussion

The ground state of $^6$Li has spin 1, so the spin structure for the $^6$Li $(\vec{d}, \alpha)$ $^4$He reaction is $\vec{1} + 1 \rightarrow 0 + 0$. The necessary and sufficient condition for the existence of the $A_{yy} = 1$ maximum for this reaction is that

$$M_{11;00} + M_{1-1;00} = 0$$

where $M_{JM;J'M'}$ are the $M$-matrix elements. The physical meaning of this equation is that the absolute values of the amplitudes with parallel and antiparallel spins in the entrance channel should be equal but they should have opposite signs.

The $M$-matrix components depend on the incident energy $E_d$ and the reaction angle $\theta$. The same, but only necessary, condition applies also to the $A_y = 1$ maxima. Consequently, if $A_y = 1$ does exist it should occur at the same energy and angle as $A_{yy} = 1$. In our experimental survey of the $^6$Li $(\vec{d}, \alpha)$ $^4$He reaction we have located two $A_{yy} = 1$ maxima but no $A_y = 1$ maxima. Clearly, two other conditions (Seiler *et al.* 1976), which are also necessary for the presence of $A_y = 1$ points, are not satisfied.

Our results show that while the shape of the $A_{yy}(E_d, \theta)$ function is smooth and simple in the vicinity of 5.55 MeV, the behaviour of the same component around the 8.80 MeV is more complex. Near this energy, the angular distributions change from displaying only one maximum at 90° to two maxima located symmetrically on each side of 90° (see Figure 20.5). As the $A_{yy} = 1$ surface is nearly flat over a relatively large energy and angular range the location of this maximum is not well determined. One may speculate that the absolute maximum occurs at the forking point of the local maxima.

In our previous study, we have argued (see Chapter 19) that extreme maxima are probably associated with resonance interaction. Possible candidates for the resonance excitation in $^8$Be, which might be responsible for the observed maxima of the tensor analyzing powers are given in Table 20.2.





Table 20.2

Possible resonance excitations of the compound nucleus $^8$Be by the $^6$Li$(\vec{d}, \alpha)$ $^4$He reaction, which might be responsible for the observed $A_{yy} = 1$ maxima

| $E_d$ | $E_x$ | Resonant matrix element (Welton 1963) | Transition |
|---|---|---|---|
| | | $(L_f\, S_f\, J^\pi | R | L_i\, S_i\, J^\pi)$ | |
| 5.55 | 26.4 | $(4\ 0\ 4^+ | R | 2\ 2\ 4^+)$ | $^5D_4 \rightarrow {}^1G_4$ |
| 8.80 | 28.9 | $(6\ 0\ 6^+ | R | 4\ 2\ 6^+)$ | $^5G_6 \rightarrow {}^1I_6$ |

$E_d$ – incident deuteron energy.
$E_x$ – excitation energy of $^8$Be.
Transition – $^{2S_i+1}(L_i)_J \rightarrow {}^{2S_f+1}(L_f)_J$ .

$\hat{S}_i = \hat{S}_d + \hat{S}_t$ (channel spin) is a sum of the deuteron and $^6$Li spins.
$L_i$ – orbital angular momentum in the incident channel.
$J^\pi$ – spin-parity of the resonant state in the compound nucleus $^8$Be ($\hat{J} = \hat{S}_i + \hat{L}_i = \hat{S}_f + \hat{L}_f$).

$S_f$ – spin of $\alpha$ particles.
$L_f$ – orbital angular momentum in the exit channel.

The notation used in this Table is the same as used in Chapter 19. The resonant excitation of $^8$Be at these energies was suggested by the analysis of $\alpha$-$\alpha$ scattering (Bacher *at al.* 1972).

## References


Bacher, A. D., Resmini, F. G., Conzett, H. E., de Swiniarski, R., Meiner, H. and Ernst, J. 1972, *Phys. Rev. Lett.* **29**:1331.

Grüebler, W., Schmelzbach, P. A., König, V., Risler, R., Jenny, B. and Boerma, D. 1975, *Nucl. Phys.* **A242**:285.

Seiler, F., Rad, F. N., Conzett, H. E. and Roy, R., 1976, *Proc. Fourth Int. Symp. on Polarization Phenomena in Nuclear Reactions*, Zürich, eds W. Grüebler and V. König, Birkhauser Verlag, Basel, p.587

Seiler, F., Rad, F. N., and Conzett, H. E., 1976, *Proc. Fourth Int. Symp. on Polarization Phenomena in Nuclear Reactions*, Zürich, eds W. Grüebler and V. König, Birkhauser Verlag, Basel, p.897

Welton, T. A. 1963, in *Fast Neutron Physics*, Vol. II, eds J. B. Marion and J. L. Fowler, Interscience, New York, p. 1317.






---



# A Study of a Highly Excited Six-nucleon System with Polarized Deuterons

**Key feature:**

1. This work represents the first precise measurements of the angular distributions and excitation functions of the vector $A_y$ and tensor $A_{yy}$ and $A_{xx}$ analyzing powers in the energy range of 17-43 MeV. A total of 21 angular distributions have been measured and 14 excitation functions.

2. The data revealed large values of the analyzing powers at backward angles.

3. The extreme value of $A_y = 1$ at deuteron energy of around 28 MeV and at the reaction angle of around $135^0$ (lab) reported earlier by Conzett *at al.* (1976b) has not been confirmed. Even though the values for this component are large in this region they are well below the extreme value of 1.

4. Using our data and the earlier data at lower energy, we have constructed a contour map of the $A_{yy}$ analyzing power. The map suggests the existence of *five* $A_{yy} = 1$ points for energies of up to 50 MeV, one of which is the point at 35 MeV and $150^0$ (c.m.) revealed by our measurements. The remaining four points are below 15 MeV.

5. Angular distributions were analysed using the resonating-group theory and the three-body bake-up Faddeev formalism. The Faddeev formalism gives significantly better description of our data.

6. We have also carried out the phase-shift analysis of our data including spins $J$ = 1,2, 3 and 4 and $l$ = 0, 1, 2, 3 and 4. No clear resonance behaviour has been observed. Smaller energy steps would be required to study possible resonance excitations in this region of energies.

**Abstract:** Angular distributions of the vector analyzing power $A_y$ and of the tensor components $A_{yy}$ and $A_{xx}$ were measured in the energy range of between 17.0 and 42.8 MeV in steps of 4 to 5 MeV. A possible $A_{yy} = 1$ point was found near 35 MeV and 150° (in the centre-of-mass system). Our results are compared with predictions of the resonating-group formalism and the three-body Faddeev calculations.

## Introduction

Information on the structure of nuclei composed of a few nucleons and on the reaction mechanism of deuteron induced reactions with such light nuclei have been greatly enriched by measurements of polarization phenomena and by comparing experimental results with model calculations. Such microscopic calculations have been particularly successful for the six-nucleon system using a three-body model (Gammel, Hill and Thaler, 1960; Shanley, 1969; Chun, Han and Lin 1973); Charnomordic, Fayard and Lamot 1977; Elbaz, Fayard and Lamot 1978), refined cluster model (Hackenbroich, Heiss, and Le-Chi-Niem 1974; Hackenbroich 1975) or the resonating group formalism (Thompson and Tang 1973; Jacobs, Wildermuth and Wurster 1969; Lemere, Tang, and Thompson 1976). Most of these calculations are, however,





restricted to excitation energies in $^6$Li, which are below the threshold of the $^3$He-t break-up.

An extraordinary agreement with experimental polarization data for the d-$\alpha$ elastic scattering was obtained by Hackenbroich, Heiss, and Le-Chi-Niem (1974) who used the refined cluster model This calculation considered not only central and spin-orbit terms for the nucleon-nucleon interaction but also tensor interactions, and included coupling between the elastic channel and the $^5$He-p, $^5$Li-n and $^3$He-t channels.

In general, in the resonating-group formalism, these reaction channels are taken into account to a first approximation by phenomenological imaginary potentials. In most of these calculations, the spin-orbit and tensor terms in the nucleon-nucleon potential are neglected and thus no polarization observables are predicted.

A substantial improvement in theoretical study of the d-$\alpha$ scattering was made by Lemere, Tang, and Thompson (1976) who did include spin-orbit interaction. Their calculations yielded sets of phase-shifts for incident deuteron energies of up to 80 MeV. A study published by Charnomordic, Fayard and Lamot (1977) solved the three-body Faddeev equations for two nucleons and a structureless $\alpha$-particle. Their calculations of the d-$\alpha$ elastic scattering predict angular distributions of the differential cross sections, vector analyzing power and the three tensor analyzing powers for deuterons with energies of up to 43 MeV. Unfortunately, the validity of these predictions at higher energies could not be tested because of the lack of suitable data. At the time of our study, precise experimental results existed only for incident deuteron energies between 1 and 17 MeV (Meiner *et al.* 1967; Keller and Haeberli 1970; Grüebler *et al.* 1969; König *et al.* 1970; Grüebler *et al.* 1970; Ohlsen *et al.* 1973; Chang *et al.* 1973; Hardekopf *et al.* 1977; Grüebler *et al.* 1979)

At higher energies, measurements of only vector analyzing power $A_y$ have been reported by Conzett *et al.* (1976a) in the energy range of 15-45 MeV. Unfortunately, we shall see later that their reported values were grossly overestimated. Their data indicated a strong possibility of the existence of a $A_y = 1$ point at 28 MeV and at a backward angle (Conzett *et al.* 1976b), which implies that a $A_{yy} = 1$ point should also be present at the same energy and angle.

Using the procedure outlined in the Appendix H, and the $M$-matrix for the d-$\alpha$ scattering expressed in terms of the expansion coefficients (Ohlsen, Gammel, and Keaton 1972):

$$M = \frac{1}{2} \begin{bmatrix} A-B & -\sqrt{2}E & -(A+B) \\ -\sqrt{2}D & 2C & \sqrt{2}D \\ -(A+B) & \sqrt{2}E & A-B \end{bmatrix}$$

we can find that

$$A_{yy} = \frac{\left|A^2\right| - 2|B|^2 + |C|^2 + |D|^2 + |E|^2}{3\sigma_0}$$

and





$$A_y = \frac{2\,\mathrm{Im}(CD^* - AE^*)}{3\sigma_0}$$

where

$$3\sigma_0 = |A|^2 + |B|^2 + |C|^2 + |D|^2 + |E|^2$$

We can see that $A_{yy} = 1$ requires $B = 0$. However, $A_y = \pm 1$ requires not only $B = 0$ but also $A = \mp iE$ and $C = \pm iD$. It is clear therefore that if $A_y = 1$ then $A_{yy} = 1$ value should be also present at the same energy and angle because the two components share the common condition of $B = 0$. However, if $A_{yy} = 1$ it does not necessarily mean that $A_y = 1$ because for such a point to occur two additional conditions have to be satisfied for the vector analyzing power.

The $M$-matrix can be also expressed in terms of $M_{ij}$ components, where $ij$ are the deuteron spin projections in the incident and outgoing channels (see the Appendix I). Using this form of the $M$-matrix, we can find that (Grüebler *at al.* 1975b) $A_{yy} = 1$ if $M_{11} = -M_{1-1}$ and $A_y = \pm 1$ if not only $M_{11} = -M_{1-1}$ but also $M_{00} = \mp i\sqrt{2}M_{01}$ and $M_{10} = \mp i\sqrt{2}M_{11}$.

Unfortunately, the data of Conzett *et al.* (1976a and 1976b) are not precise enough to be sure of the existence of the $A_y = 1$ point.

Extensive phase-shift analysis of the d-$\alpha$ data was carried out for energies of 3-17 MeV (Grüebler *et al.* 1975a). These results could be compared with the theoretical phase shifts of Lemere, Tang, and Thompson (1976). However, from the discussion presented here it is clear that there was a need to extend the study of this few-nucleon system to higher deuteron energies.

**Experimental arrangement and method**

*Method of measurements*

The polarized deuterons were produced by a polarized ion source based on the atomic beam method (see the Appendix G). Deuterons were accelerated by the SIN[20] injector cyclotron. The spin direction of the polarized beam extracted from the cyclotron is fixed in the vertical direction by the magnetic field of the accelerator. The method of measurements of the analyzing powers is fully described in Chapter 15. However, I will summarise here some points that are relevant to this experiment.

The differential cross section for a scattering or reaction induced by polarized deuterons can then be written as:

$$\sigma(\theta, \varphi) = \sigma_0(\theta)\left[1 + \frac{3}{2}(\cos\varphi)\,p_z A_y(\theta) + \frac{1}{2}(\sin^2\varphi)\,p_{zz}A_{xx}(\theta) + \frac{1}{2}(\cos^2\varphi)\,p_{zz}A_{yy}(\theta)\right]$$

where $\varphi$ is the angle between the direction of the spin and the normal to the scattering plane, and $\sigma_0(\theta)$ is the cross section for an unpolarized beam. The quantities

---

[20] Schweizerische Institut für Nuklearforschung





$p_z$ and $p_{zz}$ are the source parameters describing the vector and tensor polarization of the deuteron beam, and $A_y$, $A_{xx}$ and $A_{yy}$ are the vector and tensor analyzing powers. In the case discussed here, these are the analyzing powers for the d-α reaction.

As can be seen, the cross section $\sigma(\theta, \varphi)$ is independent of the tensor analyzing power $A_{xz}$. This quantity is related to the spherical tensor $T_{21}$.

$$A_{xz} = \sqrt{3} T_{21}$$

To measure this component, the angle $\beta$, which is between the spin and the beam directions, should be $45^0$ (see Chapter 15). For cyclotrons, $\beta$ is either $0^0$ (for the horizontal reaction plane) or $90^0$ for the detectors in the vertical plane. Consequently, this component could not be measured in the present experiment.

The polarized ion source of the SIN injector cyclotron was equipped with three successive RF transitions (*cf* Appendix G). Table 21.1 shows the available configurations of the RF transitions and the associated theoretical maximum values of the beam polarization. As can be seen, the system can produce pure vector polarization with opposite signs and mixed vector and tensor polarization, also with opposite signs. Thus, the method described in Chapter 15 can be used to measure the analyzing powers. This method consists not only in using detectors located on two sides of the beam direction but also in changing the direction of the beam polarization.

Table 21.1

Configurations of the RF transitions and the corresponding theoretical values of the beam polarization

| Mode | WF | 2↔6 | 3↔5 | $p_z$ | $p_{zz}$ |
|------|-----|------|------|-------|----------|
| *a* | × | × | × | 0 | 0 |
| *b* | ✓ | × | × | -2/3 | 0 |
| *c* | × | ✓ | ✓ | +2/3 | 0 |
| *d* | ✓ | ✓ | × | -1/3 | +1 |
| *e* | × | × | ✓ | +1/3 | -1 |
| *f* | × | ✓ | × | +1/3 | +1 |

WF: weak field 1↔4 transition; 2↔6 and 3↔5 are the strong-field transitions; × means the RF is off; ✓ means the RF is on.

However, it should be pointed out that this method requires carefully tuned RF transitions in order to get the same absolute value of the polarization for both signs. So, for instance, the measurement of $A_y$ can be done with a purely vector polarized beam switching between modes *b* and *c* in Table 21.1. The advantage of this option is that the value of $p_z$ is larger than in any other case. On the other hand, using the mixed vector and tensor polarized beam in which $\left| p_z \right| = \frac{1}{3} \left| p_{zz} \right|$ one obtains the same





statistical accuracy for the $A_y$ and $A_{yy}$ components because the factor 3 in the vector term in the cross-section formula cancels the factor 1/3.

In principle one could measure all three analyzing powers simultaneously by placing detectors in the horizontal plane ($\beta = 0°$) for measuring the $A_y$ and $A_{yy}$ components, and in the vertical plane ($\beta = 90°$) for the $A_{zz}$ component. This method would require a complicated system of detectors and a more complex design of the gas target.

The same aim can be achieved by rotating a single-plane scattering chamber around the beam axis and measuring the three analyzing powers in two runs. This has the advantage that it is easier to place the detectors at extreme forward and backward angles and to measure more scattering angles simultaneously. In our study, measurements of the three analyzing powers $A_y$, $A_{xx}$ and $A_{yy}$ were taken in two runs:

(1) *Scattering chamber horizontal, $\beta = 0°$*. With the detectors placed on the left- and right-hand side of the beam direction, and with the positive and negative beam polarizations one has four counting rates $N_L^+$, $N_L^-$, $N_R^+$ and $N_R^-$ from which one can calculate the ratios $L$ and $R$ which are independent of the values of the solid angles of the detectors (*cf* Chapter 15):

$$L = \frac{N_L^+ - N_L^-}{N_L^+ + N_L^-} = +\frac{3}{2} p_z A_y(\theta) + \frac{1}{2} p_{zz} A_{yy}(\theta)$$

$$R = \frac{N_R^+ - N_R^-}{N_R^+ + N_R^-} = -\frac{3}{2} p_z A_y(\theta) + \frac{1}{2} p_{zz} A_{yy}(\theta)$$

From these equations one gets

$$A_y(\theta) = \frac{1}{3 p_z}(L - R)$$

$$A_{yy}(\theta) = \frac{1}{p_{zz}}(L + R)$$

(2) *Scattering chamber vertical, $\beta = 90°$*. With the detectors located up and down and with different signs of the beam polarization one has four counting rates $N_U^+$, $N_U^-$, $N_D^+$, and $N_D^-$, which can be used to calculated the ratios $U$ and $D$:

$$U = \frac{N_U^+ - N_U^-}{N_U^+ + N_U^-} = \frac{1}{2} p_{zz} A_{xx}(\theta)$$

$$D = \frac{N_D^+ - N_D^-}{N_D^+ + N_D^-} = \frac{1}{2} p_{zz} A_{xx}(\theta)$$





These ratios give

$$A_{xx} = \frac{1}{p_{zz}}(U + D)$$

To check the consistency of the data collection one can also change between modes $a$ and $b$ or $a$ and $c$ for the measurement of the vector analyzing power.

### *Detector arrangement*

The 70-cm diameter reaction chamber used in these measurements was virtually the same as described in Chapter 17. The polarized beam entered a 32-mm diameter gas $^4$He target through a collimation system with a final aperture of 4 mm in diameter. The beam was aligned by a four-way slit system at the entrance of the collimator. A cylindrical gas target had a 270° exit foil window. The entrance and exit widows were made of 6 $\mu$m Havar foils. The pressure in the gas cell was 1200 Torr. The beam was collected in a Faraday cup, which was equipped with an electrostatic suppressor electrode.

The detectors were mounted on two movable segments, which could be rotated independently by remote control. Each segment contained *four* detectors mounted 15° apart and 25 cm from the centre of the chamber. For particles emitted at small angles ($\theta_{lab}$ < 40°) additional segments fixed to the existing turntables were used with detectors positioned $5^0$ apart. These smaller segments also contained four detectors on each side of the beam line. The whole scattering chamber could be rotated around the beam axis, which allowed for the measurements to be taken in either horizontal or vertical plane.

The collimators in front of the silicon surface-barrier detectors were 9 mm wide by 38 mm high for the normal turntables and 4.5 mm wide for the small-angle segments. The heights of the collimators on smaller segments were suitably chosen to maintain the similar spread in the azimuthal angles as for detectors at larger angles.

For the detection of scattered deuterons, the detector thickness was between 1 and 5 mm and suitable aluminium absorbers were placed in front of them. The recoil $\alpha$-particles were stopped in the 0.2 to 1 mm thick detectors. In all cases the bias voltage of the detectors was adjusted just to stop the detected deuterons or $\alpha$-particles. In this way, clean spectra were obtained with a background level smaller than 3%.

The beam polarization was monitored continuously for the tensor component using a $^3$He(d,p)$^4$He polarimeter at 0° (see Chapter 18). The $^3$He gas cell was integrated with the Faraday-cup system, and the high-energy protons emitted at 0° were detected either by a 5-mm thick silicon surface-barrier detector combined with suitably chosen aluminium absorber or by a CsI scintillator. The absolute calibration of the polarization analyser was checked using measurements of the d-$\alpha$ scattering at Los Alamos at 15.0 and 17.0 MeV (Grüebler *at al.* 1979).

The detector outputs were combined in a multiplexer connected to an analog-to-digital converter (see Chapter 17). The resulting signals were routed to a PDP-11 computer, which performed an on-line data processing. The computer controlled also





the procedure for switching the sign of the beam polarization. This took place every few seconds, whenever the beam current integrator reached a preset charge. Although the deadtime correction was incorporated in the experimental procedure, the beam intensity (which was up to 80 nA) was adjusted in such a way as to maintain the deadtime corrections at the level of less than 10%.

## Results of measurements

### *Excitation functions*

Absolutely calibrated excitation functions for $A_y$, $A_{xx}$ and $A_{yy}$ analyzing powers are shown in Figures 21.1-21.3. The functions display smooth energy dependence for all three analyzing powers. It is interesting to notice that in the vicinity of $\theta = 150^0$ and for the incident deuteron energies of 33-38 MeV the component $A_{yy}$ appears to be reaching its extreme value of 1. However, contrary to the claim of Conzett at al. (1976b), the $A_y$ component remains well below the value of 1 in the whole energy range.

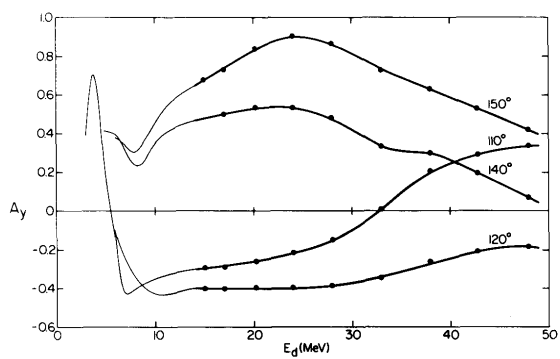

Figure 21.1. Excitations functions of the vector analyzing power $A_y$ for the d-α elastic scattering. The dots are the experimental results. The combined statistical and systematic errors are smaller than the size of the displayed points. The thick lines are to guide the eye. The thin curves were drawn through the earlier data taken at low energies (Grüebler *at al.* 1969; König at al. 1970; Grüebler *at al.* 1970; Ohlsen *at al.* 1973; Chang *at al.* 1973; Hardekopf *at al.* 1977; Grüebler *at al.* 1979). Contrary to the claim of Conzett *at al.* (1976b), the maximum value of $A_y$ at $150^0$ is well below the extreme value of 1.

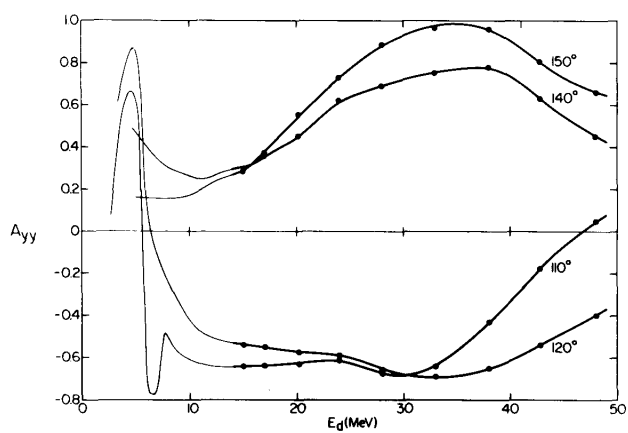

Figure 21.2. Excitation functions of the tensor analyzing power $A_{yy}$ for the d-α elastic scattering. See also the caption to Figure 21.1. These data show that the $A_{yy}$ component is probably reaching its extreme value of 1 at around 35 MeV and at the reaction angle of around $150^0$ (c.m.).





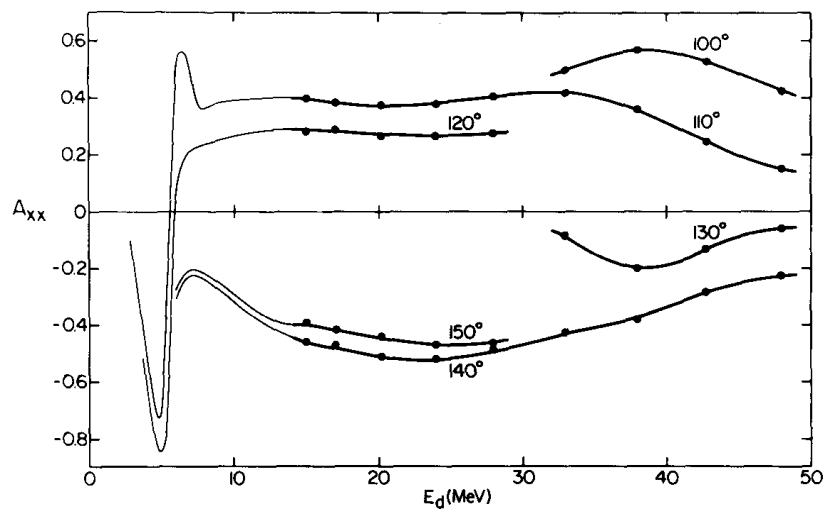

Figure 21.3. Excitation functions of the tensor analyzing power $A_{xx}$ for the d-α elastic scattering. See also the caption to Figure 21.1.

### *Angular distributions*

Angular distributions of the analyzing powers $A_y$, $A_{yy}$ and $A_{xx}$ for the elastic d-α scattering were measured at the incident deuteron energies of 17.0, 20.2, 24.0, 28.0, 33.0, 38.2 and 42.8 MeV. The scattered deuterons were detected in the angular range of between 20° and 80° (lab system) and the recoil α-particles between 10° and 40°. In this way two overlapping angular ranges of 30°-110° and 100°-160° (c.m. system) were obtained.

The range of statistical errors was between 0.001 to 0.010 with an average of about 0.005. In addition to statistical errors, a combined random error had to be also included. This error was caused by such effects as background subtraction, instrumental asymmetries, angle uncertainty, and by fluctuations of the beam polarization. By referring to the reproducibility of the data points and by using the overlapping angular range for the detection of deuterons and α-particles, a value of 0.005 was estimated for this random error. The lab angles for all the data are accurate to ±0.1°. The beam energy was determined to about ±100 keV. The energy spread of the polarized deuteron beam was about 100 keV.

The uncertainty of the absolute values of the measured analyzing powers are given by uncertainties of the beam polarization. This uncertainty ranges from about 2.0% at 17 MeV to about 4% at 42.8 MeV.

The distributions are shown in Figures 21.4 – 21.6. The shape of the angular distributions for all three analyzing powers changes slowly with the increasing energy. The $A_y$ and $A_{yy}$ components show large values at backward angles with a marked maximum around 150°. Between 33.0 MeV and 38.2 MeV, $A_{yy}$ reaches such high value that a maximum $A_{yy} = 1$ is possible in this region. Even though the values for $A_{xx}$ are not so large, the deep minimum at around 100⁰ and the large maximum near 150° could serve as good reference points for beam polarization analysers.





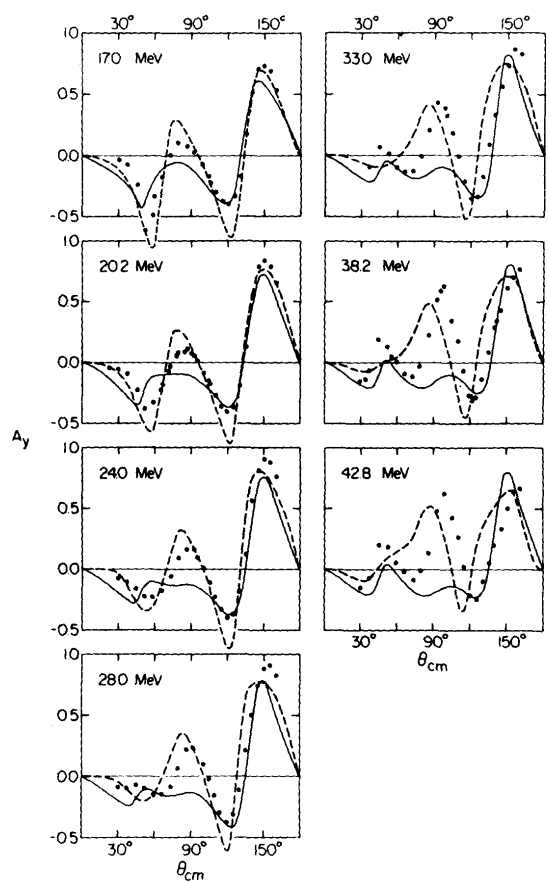

Figure 21.4. Angular distributions of the vector analyzing power $A_y$ for the d-α elastic scattering between 17.0 and 42.8 MeV. The statistical errors are smaller than the size of the data points. The dashed curves are the predictions based on the resonating-group theory (see Lemere, Tang and Thompson 1976). The solid curves are the predictions based on the Faddeev three-body brake up formalism (see Elbaz, Fayard and Lamot 1978).

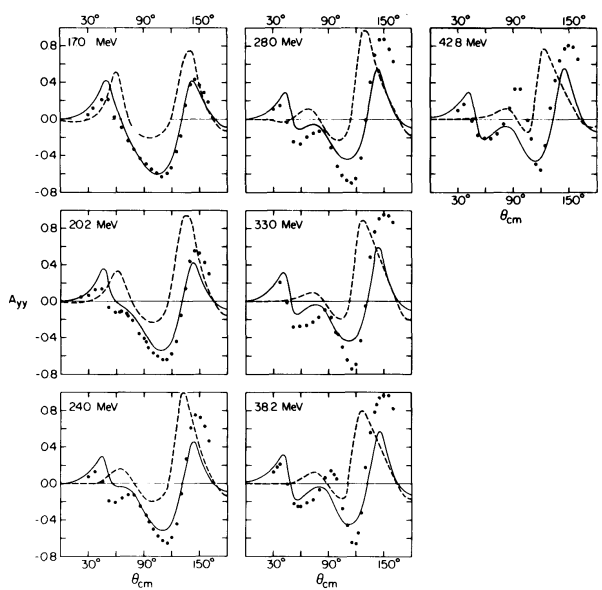

Figure 21.5. Angular distributions of the tensor analyzing power $A_{yy}$ for the d-α elastic scattering at deuteron energies of 17.0 - 42.8 MeV. The statistical errors are smaller than the displayed data points. The dashed curves are the predictions based on the resonating-group theory. The solid curves are predicted using on the Faddeev formalism.





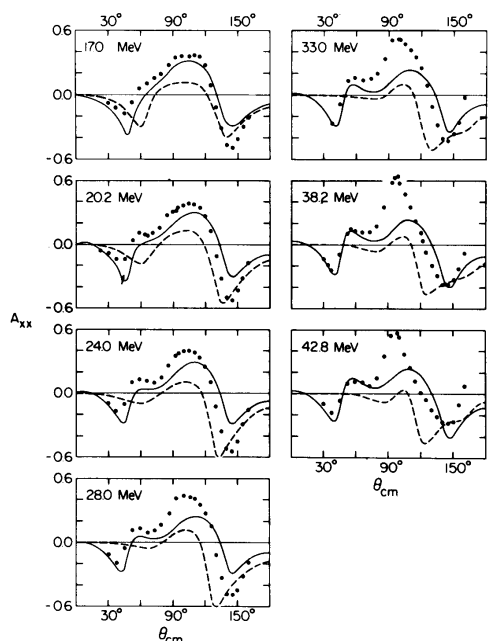

Figure 21.6. Angular distributions of the tensor analyzing power $A_{xx}$ for the d-α elastic scattering at deuteron energies of 17.0 - 42.8 MeV. The statistical errors are smaller than the displayed data points. The dashed curves are the predictions based on the resonating-group theory. The solid curves are predicted using the Faddeev formalism.

## Extreme values of the analyzing powers

Our study presents, for the first time, absolutely calibrated vector and tensor analyzing powers for the d-α scattering in the energy range of between 17 and 43 MeV. A total of 21 angular distributions and 14 excitation functions have been measured for the $A_y$, $A_{yy}$ and $A_{xx}$ components of the analyzing powers.

Earlier measurements (Conzett *et al.* 1976a and 1976b) suggested the existence of a $A_y = 1$ point at 28.6 MeV. These measurements are shown in Figure 21.7.

In the same figure, I have also presented our data. It is clear that the measurements of Conzett *et al.* produced significantly overestimated values of the $A_y$ component. Our values are much lower and they do not come close to the expected maximum of $A_y = 1$. The displayed curves are the polynomial fits to the data. The best fits were obtained using a second-order polynomial for our data and a third-order for the data of Conzett *et al.*

The figure contains also our data for the $A_{yy}$ component. Here, our data show a possible maximum $A_{yy} = 1$ at around 35 MeV as indicated by the polynomial fit. The best fit was obtained using a third-order polynomial.

Figure 21.7 has been constructed by taking measured values at the relevant maximum in the angular distributions. The figure should be compared with the measured excitation functions at 150[0] as shown in Figures 21.1 and 21.2.

Finally, using our data and the earlier data at lower energy, we have constructed a contour map of the tensor analyzing power $A_{yy}$. This map is shown if Figure 21.8.





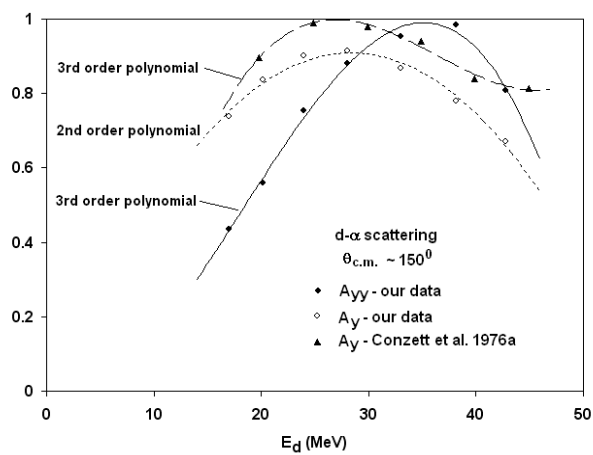

Figure 21.7. Energy dependence of the $A_y$ and $A_{yy}$ analyzing powers at the reaction angle of approximately $150^0$ (c.m.) extracted from the measured angular distributions. Our results were obtained using the distributions presented in Figures 21.4 and 21.5. The $A_y$ values of Conzett *at al.* (1976a) were extracted in a similar fashion, i.e. by using the last maximum of their angular distributions. The displayed lines are the best polynomial fits.

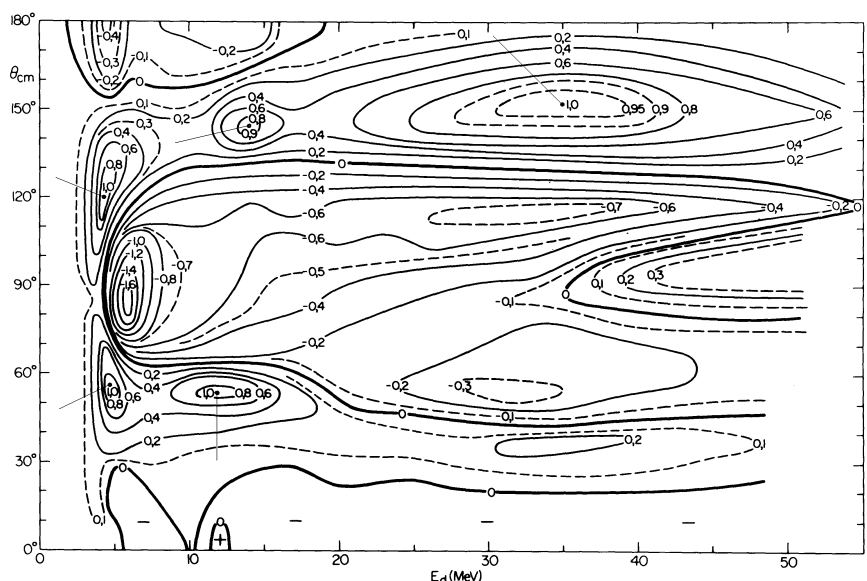

Figure 21.8. The contour map of the tensor analyzing power $A_{yy}$ constructed using our experimental data for the d-α elastic scattering and the earlier data obtained at low energies by Grüebler *at al.* (1975b). The short straight lines are to guide the eye to the expected $A_{yy} = 1$ points. In particular, the map shows a $A_{yy} = 1$ point near 35 MeV and 150° as suggested by the polynomial fit to our data presented in Figure 21.7.

## Theoretical calculations

We have carried out theoretical calculations for our results using the resonating group method (Lemere, Tang and Thompson 1976) and the three-body Faddeev formalism (Elbaz, Fayard and Lamot 1978). Results of our calculations are presented in Figures





21.4-21.6. The dashed curves are for the resonating group method and the solid curves are for the Faddeev formalism.

In the resonating group study, a single-channel approach was chosen. The effects of the open reaction channels were taken into account by using phenomenological imaginary potentials. The nucleon-nucleon potential contained a spin-orbit component, which was necessary to obtain non-zero values for the vector and tensor polarizations. The exchange Coulomb potential was also taken explicitly into consideration. However, no tensor interaction was included in the nucleon-nucleon force.

Even though the resonating-group formalism resulted in a fairly satisfactory agreement with experimental results at energies below 10 MeV (Lemere, Tang and Thompson 1976), at higher energies, as used in our measurements, only the component $A_y$, which depends mostly on the spin-orbit interaction, shows a similar quality of an agreement. For the components $A_{yy}$ and $A_{xx}$ the shape of the angular distributions is reproduced only approximately. This might be due to the absence of tensor forces in the theoretical calculations.

The figures show also calculations based on the Faddeev formalism. Here, the agreement with the experimental data is significantly better. In our calculations, we have used two-body N-N and N-$\alpha$ interactions. In addition, for practical reasons, a limited number of the two-body partial waves was used. In the N-$\alpha$ system, only $S_{1/2}$, $P_{1/2}$ and $P_{3/2}$ partial waves were used and in the N-N system only partial waves with isospin zero and $l \leq 1$ were included. For the latter system, the coupled $^3S_1$-$^3D$, partial waves with several different parameterisations were taken into account.

The agreement between experimental data and theoretical calculations deteriorates with the increasing deuteron energy. It is not clear whether it is because of the neglected higher partial waves or because of the break-up of the $\alpha$-particle. (The threshold for the $d + \alpha \rightarrow {}^3He + t$ break-up is 21.4 MeV.) Our study suggests that more detailed calculations and perhaps even more refined theories are needed in order to understand the structure of this highly excited few-nucleon system.

**Phase-shift analysis**

A new computer search code for the phase shift analysis of spin-one scattering from spinless targets was used to analyse the data. The calculation sections of this code were adapted from the older SPINONE program (McIntyre 1965) but we have added a general-purpose search routine MINUIT (James and Roos 1975) to allow for an automatic search. In addition, we have also altered the input data section to simplify the use of the program and to make the input clearer and more flexible.

Since phase shifts were available at 17 MeV (Grüebler *et al.* 1975a), we have used them as starting values for the 17.0 and 20.2 MeV searches in our preliminary analysis. At 17 MeV, the present data and the older Los Alamos data (Ohlsen *et. al.* 1973) were analysed separately in order to see if the results were compatible. Our data do not include the $A_{xz}$ component but they cover a wider range of angles. The earlier data were used as the basis for the normalization of our new measurements, and consequently no major differences in the phase shifts were expected. This expectation was confirmed by our calculations.





The remaining analysis was carried out using the phase shift results from lower energies as the starting values. Gradient searches were performed to minimize the $\chi^2$ function with all phases and mixing parameters varied simultaneously to achieve the best fit to the data. The searches were carried out in interrupted intervals so that intermediate results could be studied and the direction of the search monitored. The $\chi^2$ values are listed in Table 21.2. The energy dependence of the real parts of the phase shift parameters and the total reaction cross sections are presented in Figure 21.9. The corresponding fits to the data are shown in Figures 21.10-21.13.

Table 21.2
The $\chi^2$ values for the phase shift analysis

| $E_d$ (MeV) | $N$ | $\chi^2$ |
|---|---|---|
| 17.0 | 77 | 1.1 |
| 20.2 | 91 | 1.8 |
| 24.0 | 92 | 2.5 |
| 28.0 | 88 | 3.0 |
| 33.0 | 92 | 7.0 |
| 38.2 | 112 | 13.7 |
| 42.8 | 92 | 10.4 |

$N$ – The number of data points

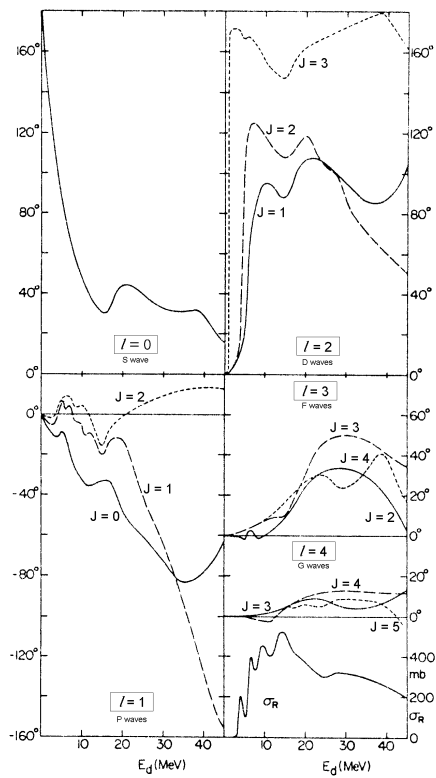

Figure 21.9. The energy dependence of the phase-shift parameters for the d-α elastic scattering in the energy range of 0-43 MeV. The values below 17 MeV are from Grüebler *et al.* (1975a). The higher-energy values are from our analysis. The energy dependence of the total reaction cross-section $\sigma_R$ calculated using the imaginary parts of phase shift parameters are also shown in the figure.





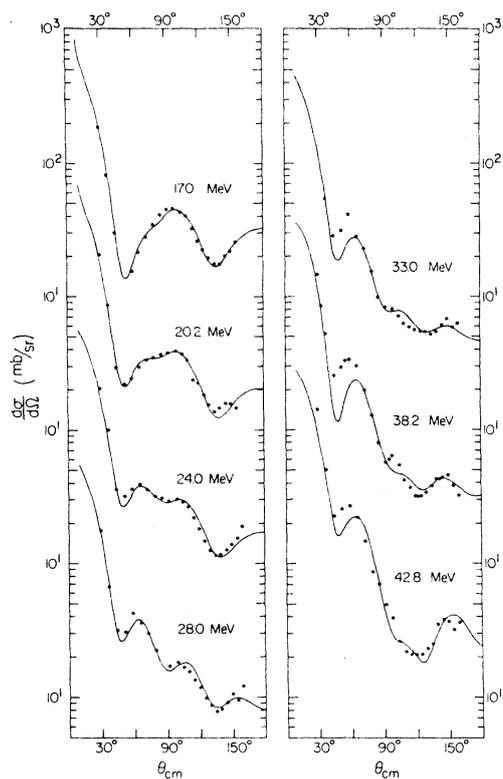

Figure 21.10. Differential cross-sections for the d-α scattering. The solid lines are the results of our phase-shift analysis.

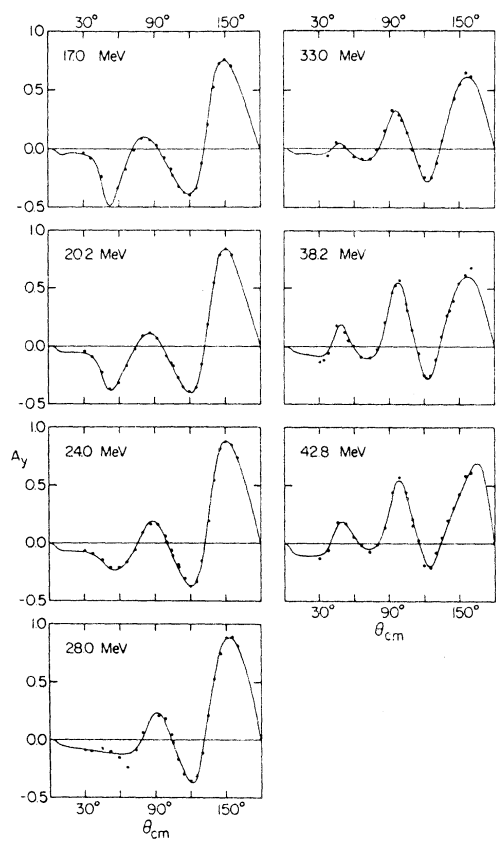

Figure 21.11. Angular distributions of the vector analyzing power $A_y$ (points) for the d-α elastic scattering are compared with the results of our phase-shift analysis (solid lines).





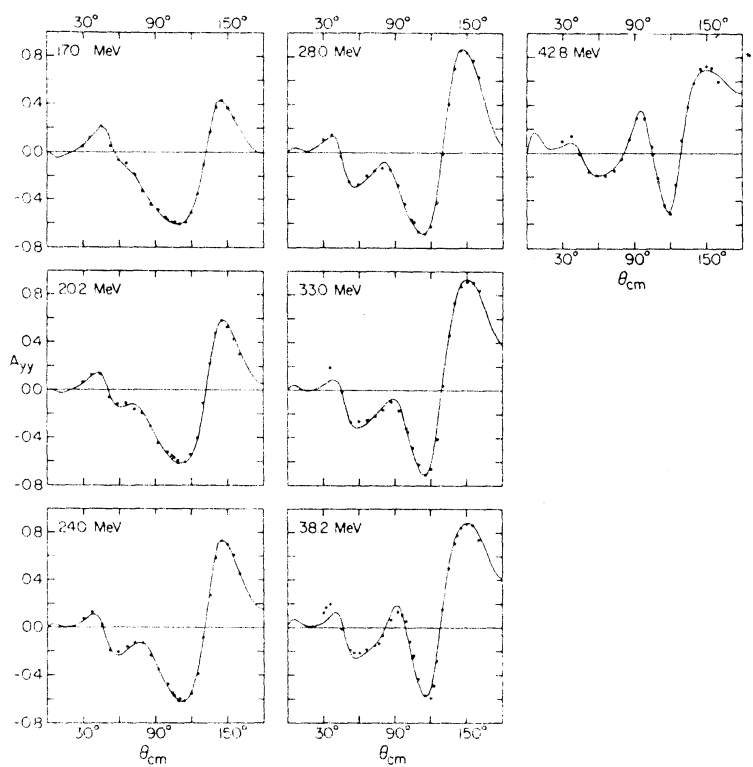

Figure 21.12. Angular distributions of the tensor analyzing power $A_{yy}$ (points) for d-α elastic scattering are compared with the results of our phase-shift analysis (solid lines).

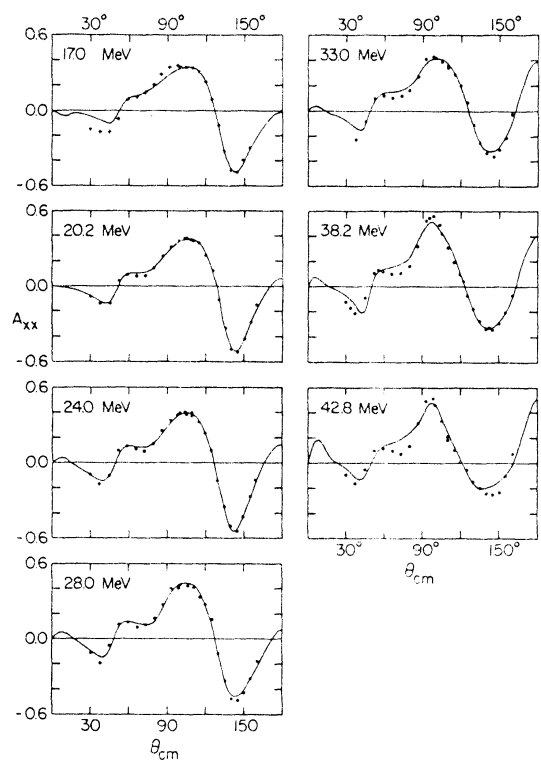

Figure 21.13. Angular distributions of the tensor analyzing power $A_{xx}$ (points) for the d-α elastic scattering are compared with the results of our phase-shift analysis (solid lines).





The energy range of our new data covers the excitation energy of 12.8-30.0 MeV in the compound $^6$Li nucleus. No resonances decaying to the d+$\alpha$ system are known in this range of the excitation energies. Some of the phase shift parameters show broad structures above 17 MeV, but it is not clear whether they can be associated with resonance interactions. The most striking change is in the $P_1$ ($J$ =1, $l$ = 1), which decreases from near zero at 17 MeV to around $-150^0$ at around 43 MeV (see Figure 21.9). It can be also noticed from Figures 21.10-21.13 that the fits to the data deteriorate at higher energies. This might indicate that higher $l$ - values are needed. However, our analysis shows that $G$ waves remain small over the whole range of energies.

The lack of indication for the presence of resonances in this analysis does not necessarily prove that there is no resonant behaviour in this region of the excitation energies. To uncover such interactions one would have to carry out measurements at significantly smaller energy steps. In particular, it would be necessary to study more closely the energy region around 35 MeV where the expected $A_{yy} = 1$ maximum is located.

## Summary and conclusions

We have carried out precise measurements of the angular distributions of the differential cross sections and analyzing powers ($A_y$, $A_{yy}$ and $A_{xx}$) for the d-$\alpha$ scattering in the energy range of 17-43 MeV. We have also measured excitation energies at selected reaction angles.

Our results show a possible $A_{yy} = 1$ maximum at around 35 MeV at a backward angle but do not confirm the previously claimed $A_y = 1$ maximum (Conzett *at al.* 1976b) at around 28 MeV. In fact, even though our results show a maximum around this energy, its value is well below the expected value of 1.

We have analysed our results using the resonating group theory and the Faddeev formalism. The calculations using the Faddeev formalism follow more closely our experimental data than the calculations based on the resonating group theory.

We have also carried out a phase shift analysis of our data using a wide range of spin values, $J$ = 1, 2, 3, and 4 and $l$ = 0, 1, 2, 3, and 4. The analysis does not reveal any clear resonance behaviour in this higher energy region. Measurements in smaller energy steps would be necessary to study the possible resonance behaviour. A more detailed study around 35 MeV would be desirable to investigate the demonstrated here possibility of the existence of the $A_{yy} = 1$ maximum.

## References


Charnomordic, B., Fayard, C. and Lamot, G. H., 1977, *Phys. Rev.* **C15**:864

Chun, D. S., Han, C. S. and Lin, D. L. 1973, *Phys. Rev.* **C7**:1329

Chang, C. C., Glavish, H. F., Avida, R. and Boyd, R. N. 1973, *Nucl. Phys.* **A212**:189

Conzett, H. E., W., Leemann, Ch., MacDonald, J. A. and Meulders, J. P. 1976a, Proc. Fourth Int. Symp. Phen. in Nuclear Reactions, Zürich, eds W. Grüebler and V. König (Birkhauser, Basel) p. 566







Conzett, H. E., Seiler, F., Rad, F. N., Roy, R. and Larimer, R. M. 1976b, *Proc. Fourth Int. Symp. Phen. in Nuclear Reactions*, Zürich, eds W. Grüebler and V. König (Birkhauser, Basel) p. 568.

Elbaz, E., Fayard, C. and Lamot, G. H. 1978, "Proc. Int. Conf. Nucl. Structure, Tokyo 1977", *J. Phys. Soc. Japan* **44**:304

Gammel, J. L., Hill, B. J. and Thaler, R. M. 1960, Phys. Rev. 119:267

Grüebler, W., König, V., Schmelzbach, P. A. and Marmier, P. 1969, *Nucl. Phys.* **A134**:686

Grüebler, W., König, V., Schmelzbach, P. A. and Marmier, P. 1970, *Nucl. Phys.* **A148**:391

Grüebler, W., Schmelzbach, P. A., König, V., Risler, R., Jenny, B. and Boerma, D. 1975a, *Nucl. Phys.* **A242**:265

Grüebler, W., Schmelzbach, P. A., König, V., Risler, R., Jenny, B. and Boerma, D. 1975b, *Nucl. Phys.* **A242**:285

Grüebler, W., Brown, R. E., Corell, F. D., Hardekopf, R. A., Jarmie, N. and Ohlsen, G. G. 1979, *Nucl. Phys.* **A331**:61

Hackenbroich, H. H. 1975, *Proc. 2nd Int. Conf. on Clustering Phenomena in Nuclei II*, Maryland, eds D. A. Goldberg, J. B. Marion, S. J. Wallace, ORO-4856-26, p. 107

Hackenbroich, H. H., Heiss, P. and Le-Chi-Niem, 1974, *Nucl. Phys.* **A221**:461

Hardekopf, R. A., Grüebler, W., Jenny, B., König, V., Risler, R., Bürgi, H. R. and Nurzynski, J. 1977, *Nucl. Phys.* **A287**:237

Jacobs, H., Wildermuth, K. and Wurster, E. 1969, *Phys. Lett.* **29B**:455

James, F. and Roos, M. 1975, *Comp. Comm.* **10**:343.

Keller, L. G. and Haeberli, W. 1970, *Nucl. Phys.* **A156**:465

König, V., Grüebler, W., Schmelzbach, P. A. and Marmier, P. 1970, *Nucl. Phys.* **A148**:380

Lemere, M., Tang, J. C. and Thompson, D. R. 1976, *Nucl. Phys.* **A266**:1

McIntyre, L. C. 1965, PhD Thesis, University of Wisconsin.

Meiner, H., Baumgartner, E., Darden, S. E., Huber, P. and Plattner, G. R. 1967, *Helv. Phys. Acta* **40**:483

Ohlsen, G. G., Gammel, J. L., and Keaton, 1972, *Phys. Rev.* **C5**:1205.

Ohlsen, G. G., Lovoi, P. A., Salzman, G. C., Meyer-Berkhout, U., Mitchell, C. K. and Grüebler, W. 1973, *Phys. Rev.* **C8**:262

Shanley, R. E. 1969, *Phys. Rev.* **187**:1328

Thompson, D. R. and Tang, Y. C. 1973, *Phys. Rev.* **C8**:1649






___



# A Study of the 5.65 MeV 1+ Resonance in 6Li

**Key features:**

1. The aim of this work was to investigate the 1+ resonance at the excitation energy of 5.65 MeV in 6Li.

2. The distinctive feature of our study is the high precision of experimental data.

3. We have measured the differential cross sections and analyzing powers ($iT_{11}$ and tensor $T_{20}$, $T_{21}$ and $T_{22}$) in the energy range of 6.04 to 7.05 MeV in small energy steps. The angular range of the data was 11° and 165° (c.m.). We have collected a total of 1126 high-precision data points

4. Using our data, we have carried out phase-shift analysis. Our data allowed us to obtain reliable sets of not only real but also imaginary components of phase-shift and mixing parameters.

5. Using the results of our phase-shift analysis as input data, we have carried out the *R*-matrix analysis. In these calculations, we have used a computer code of McIntyre (1965). However, we have modified it to allow for (for the first time) an automatic search of *R*-matrix parameters. We have obtained a close agreement between the *R*-matrix results and the phase shift and mixing parameters even for the imaginary components.

6. Our study yields precise values of the resonance parameters. It also demonstrates the sensitivity of the data not only to tensor forces but also to the number and the nature of assumed open channels.

7. Our results show that by carrying our precise measurements in the vicinity of a resonance one can obtain detailed information not only about the physical parameters of the resonance but also about the mechanism of nuclear interactions (the strength and nature of nuclear forces and the contributions of reaction channels).

*Abstract*: Previous analyses of the d - 4He elastic scattering have established the presence of a 1+ resonance near 5.7 MeV excitation energy in 6Li. The energy dependence of the *s*-wave to *d*-wave mixing parameters through this resonance gives an indication of the tensor force contributing to the interaction. In this work, we have made a detailed study of this parameter by measuring and then analyzing angular distributions of the differential cross section and all four analyzing powers for the 4He(d,d)4He elastic scattering at seven energies between 6 and 7 MeV. The phase-shift analysis of these data provides a detailed parameterization, which can be compared with theoretical calculations and shows for the first time the necessity for a complex mixing parameter. The *R*-matrix fits to the phase shifts establish more precisely the location and the width of this resonance.

## Introduction

The deuteron-α elastic scattering is a popular way to investigate the structure of 6Li. Measurements of the differential cross sections and analyzing powers for the d-α elastic scattering have been carried out for up to around 50 MeV (Grüebler *et al.* 1975; Grüebler *et al.* 1980; McIntyre and Haeberli 1967; Ohlsen *et al.* 1973; Schmelzbach *et al.* 1972). Phase-shift analyses of these data show several well-separated resonances, particularly for energies below 10 MeV. Of a particular interest for the investigation of tensor





forces is the $1^+$ resonance, which occurs near the excitation energy of 5.7 MeV, and which corresponds to the deuteron energy of 6 to 7 MeV (lab).[21]

Phase-shift analyses of experimental results in this region show a mixing between the *s*- and *d*-waves. Such a coupling between orbital angular momenta differing by two units is usually interpreted as a result of tensor interaction. Unfortunately, in earlier measurements, the energy steps in this region have been too large to determine the details of the mixing parameter.

In order to extract more accurate information about this resonance, we have carried out precise measurements of the differential cross section $\sigma_0$ and vector and tensor analyzing powers $iT_{11}$, $T_{20}$, $T_{21}$ and $T_{22}$ in the energy range of 6.04 to 7.05 MeV and in small steps of 100 to 200 keV. These independent observables were measured in the angular range of between 11° and 165° (c.m.). At each energy, about 160 data points were collected to support the intended detailed phase shift and *R*-matrix analyses.

### Experimental arrangement and method

The measurements were carried out using the ETH[22] atomic beam polarized ion source (see the Appendix G) and the EN tandem Van de Graaff accelerator. The average beam current on the target was 30 nA.

The polarized deuteron beam entered the gas target, 16 mm in diameter, through a collimation system with a final aperture of 2 mm in diameter. A gas cell had a 270° exit foil window. The entrance and exit windows were made of 1.25 μm Ni foils. A 60 x 20 mm oval-shaped gas cell with 6.5 μm Mylar foil was used for the extreme forward angles. The pressure in the cell was set at 600 Torr for the Ni foil and 200 Torr for the Mylar. Pressure variations during the measurements were less than 1 %. The energy of the polarized deuteron beam was determined in the middle of the gas cell to better than 20 keV.

A general description of the scattering chamber is given in Chapter 17. The arrangement of detectors was the same as described in Chapter 21 for the measurements at higher deuteron energies at SIN[23].

A $^3$He polarimeter set up at 0° was used to monitor the tensor polarization of the beam. Details of this polarimeter are described in Chapter 18. The method of measurements is given in Chapter 15. The essential parts of the procedure are the use of the left and right detectors at the same scattering angles, adjusting the spin direction of polarized deuterons by a Wien filter to an optimal position for the measurement of each specific analyzing power component, and switching the sign of the beam polarization every few seconds. This method eliminates first-order errors from geometrical effects and from deviations from the required spin direction.

### Experimental results

Results of measurements together with results of our phase-shift analysis are presented in Figures 22.1-22.5. The errors are smaller than the displayed data points.

---

[21] The difference between the binding energies of $^6$Li, deuterons, and alpha particles is 1.4743 MeV.

$E_{lab} = 1.5 E_{c.m.}$ for the d-α scattering. Therefore, the excitation of the 5.65 MeV state in $^6$Li corresponds to $E_{lab} = 6.3$ MeV.

[22] Laboratorium für Kernphysik, Eidg. Techische Hochschule, Zürich, Switzerland.

[23] Schweizerische Institut für Nuklearforschung





A total of 1126 data points has been collected. It is interesting to notice that all observables measured in the energy range of the 1⁺ resonance do not show any dramatic energy dependence. Consequently, on the basis of the experimental data alone, i.e. without carrying out a phase-shift analysis, one would not expect a resonance. Such behaviour is typical for broad resonances in light nuclei.

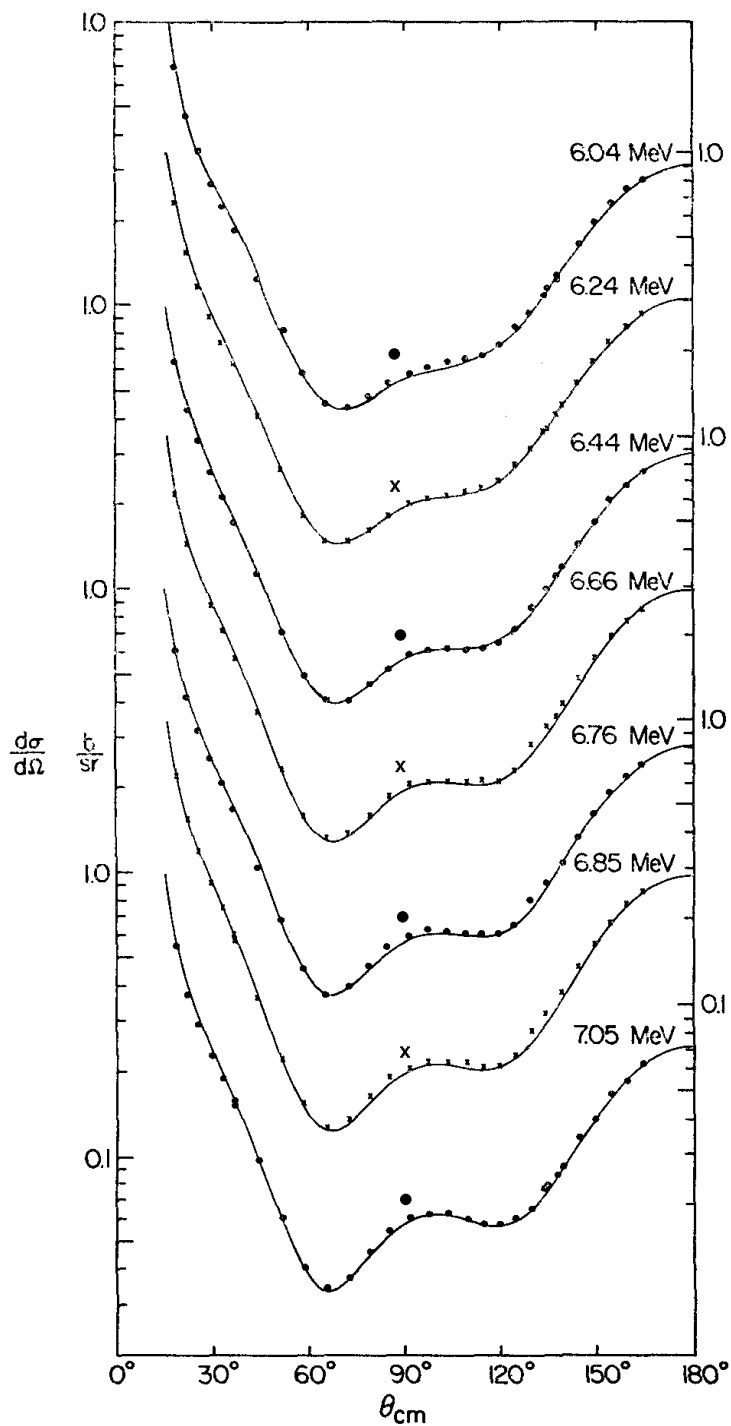

Figure 22.1. Angular distributions of the differential cross-sections for the d-α scattering between 6 and 7 MeV. The dots and crosses are larger than the statistical errors. The distributions identified by large filled-in circles refer to the left-hand side scale; crosses refer to the right-hand side scale. The curves are the phase-shift fits.





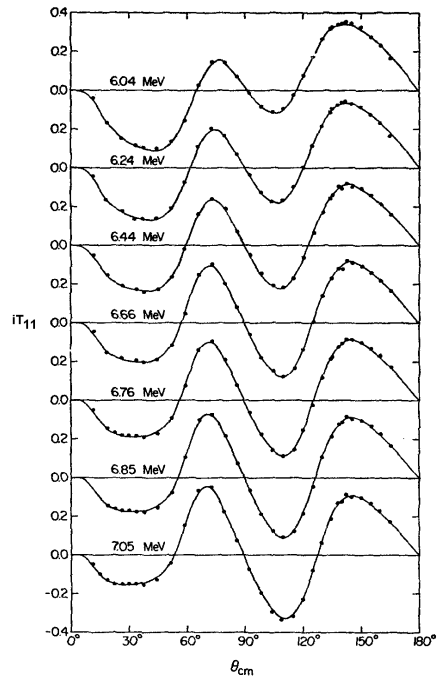

Figure 22.2. The vector analyzing powers $iT_{11}$ for the d-α scattering between 6 and 7 MeV. The dots are larger than the statistical errors. The curves are the phase-shift analysis fits.

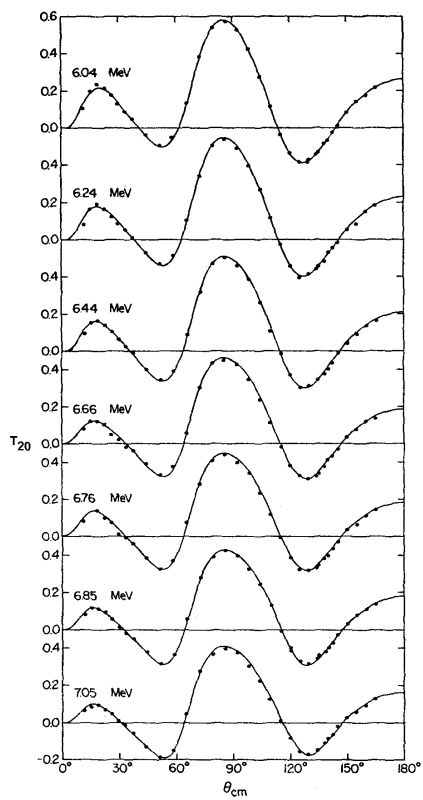

Figure 22.3. The tensor analyzing powers $T_{20}$ for the d-α scattering between 6 and 7 MeV. The dots are larger than the statistical errors. The curves are the phase-shift analysis fits.





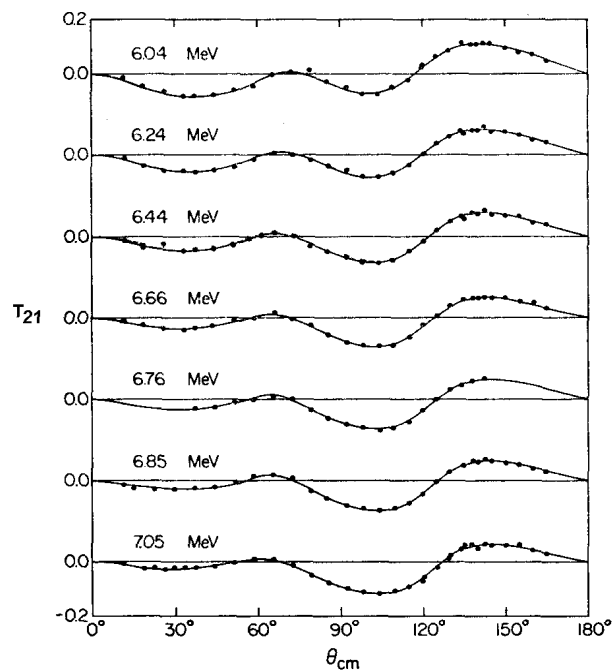

Figure 22.4. The tensor analyzing powers $T_{21}$ for the d-α scattering between 6 and 7 MeV. The dots are larger than the statistical errors. The curves are the phase-shift analysis fits.

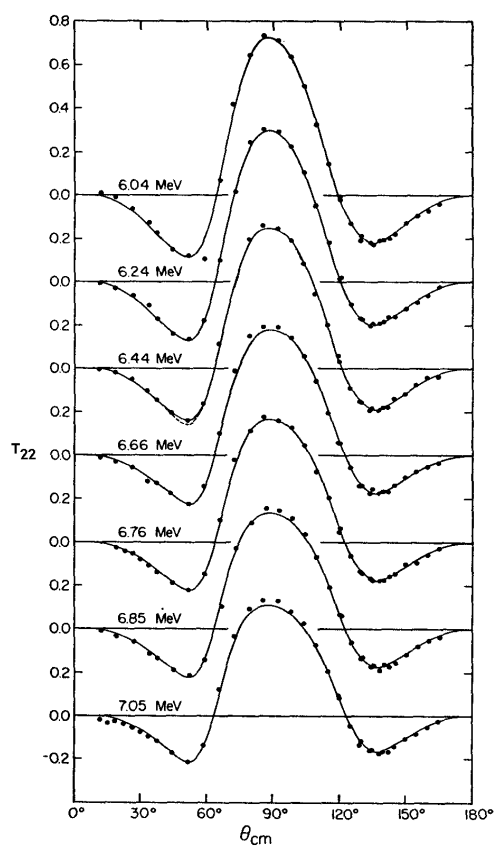

Figure 22.5 The tensor analyzing powers $T_{22}$ for the d-α scattering between 6 and 7 MeV. The dots are larger than the statistical errors. The curves are the phase-shift analysis fits.





For the purpose of the phase shift analysis, it is useful to separate the errors of the measurements into relative and absolute errors. The relative errors describe uncertainties in the relative positions of data points. The absolute errors (the scale errors) give information about the uncertainty of the overall normalization of the data.

For the differential cross-sections, relative errors arise from such effects as statistical uncertainty in the number of counts, variation in the gas target pressure, errors in background subtraction, beam-current integration, detector position uncertainty, relative solid angles. The statistical uncertainties from the number of counts were very small. All relative errors are not larger than 2%.

The scale error of the differential cross-section data is also estimated at 2.0%. This includes such effects as target gas purity, calibration of the current integrator and target pressure gauge.

For the measurements of the analyzing-powers, the statistical uncertainties from the number of counts are also very small, generally less than 0.005. Although most large systematic errors cancel for the analyzing-power data obtained with symmetric detector systems and beam polarization reversal, some small false asymmetries remain. The scale error i.e. the normalization of the analyzing-power measurements arises only from the calibration of the polarimeter that monitored the beam polarization. This uncertainty is estimated at 1 %.

## The phase-shift analysis

### *General discussion and procedure*

As the experimental data have become more complete, the phase shifts used to parameterise d-$\alpha$ scattering have become more precise. Early analyses of differential cross sections were superseded by McIntyre and Haeberli (1967) who included tensor polarizations measured by double scattering. With the development of polarized ion sources, vector analyzing powers and more accurate tensor analyzing powers were added to the data. The analysis of Schmelzbach *et al.* (1972) used for the first time a fairly complete set of measurements of differential cross sections, vector and all three tensor analyzing powers to determine the d-$\alpha$ phase shifts between 3 and 11.5 MeV. Grüebler *et al.* (1975) extended the analysis to 17 MeV and improved its accuracy by adding new experimental data from Los Alamos (Ohlsen *et al.* 1973). The data and analysis have been later extended to higher energies of up to around 40 MeV (Grüebler *et al.* 1980).

Descriptions of the phase-shift parameterization and the search procedure may be found in publications of McIntyre and Haeberli (1967) and Schmeltzbach *et al.* (1972). Of particular relevance to our analysis is the use of the Blatt-Biedenharn prescription (1952) to account for the off-diagonal terms in the scattering matrix, which depend not only on the eigen phase shifts but also on mixing parameters (see the Appendix J). These off-diagonal terms represent coupling between angular momenta $l$ differing by two units. They are therefore a measure of the tensor force in the interaction.

For the 1$^+$ resonance studied here, McIntyre and Haeberli (1967) showed how the mixing parameter is affected by the ratio of the $s$- and $d$-wave reduced widths. Their calculations are shown in Figure 22.6.





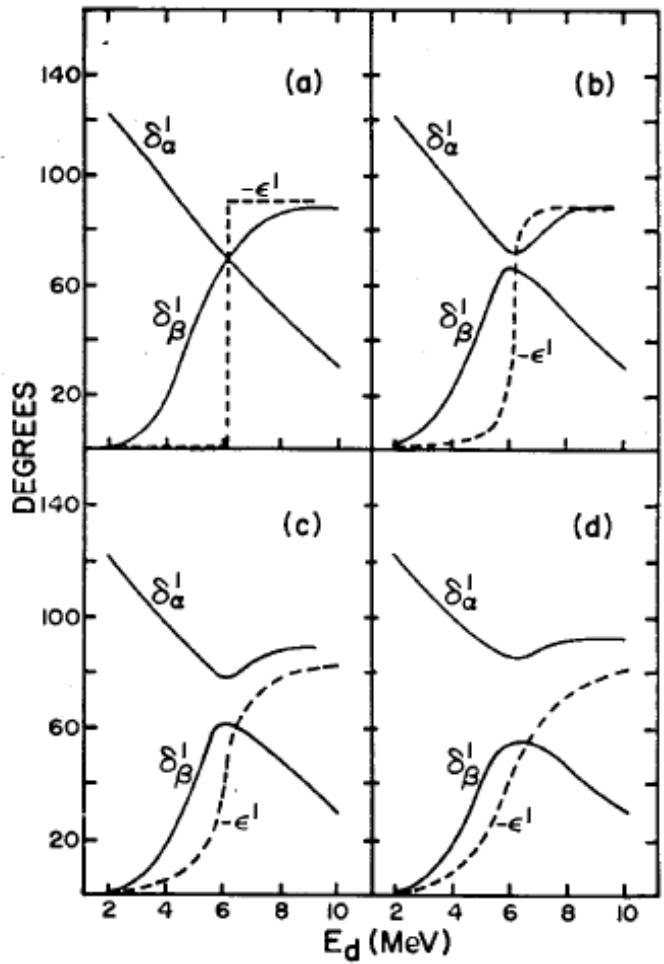

Figure 22.6. The dependence of the calculated phase shifts $\delta_\alpha^1$ and $\delta_\beta^1$, and of the mixing parameter $\varepsilon^1$ on the ratio of the $s$ - wave to $d$ - wave reduced widths. The rations are: (a) 0, (b) 0.001, (c) 0.010 and (d) 0.035. The other parameters are: the bombarding energy at resonance – 6.2 MeV; the total laboratory energy at resonance – 7 MeV; and the interaction radius – 4.5 fm.

For the pure $d$-wave resonance, i.e. with no tensor force, $\varepsilon^1$ is either zero or $\pi/2$ (which just exchanges the roles of the two eigen phase shifts $\delta_\alpha^1$ and $\delta_\beta^1$).[24] In this case, there is a sudden transition from $\varepsilon^1 = 0$ to $\varepsilon^1 = \pi/2$ as one crosses the resonance from the lower to higher incident energies. Increasing the ratio of the $s$-wave to $d$-wave reduced widths smoothed out the transition between $\varepsilon^1 = 0$ and $\varepsilon^1 = \pi/2$. With an increased accuracy of the experimental data, greater angular range, and closer energy steps in our measurements, we expected to determine the shape of this mixing parameter with a greater precision than it was previously possible and thus to obtain more precise information on the ratio of the wave functions and on the tensor coupling.

---

[24] The notation used here is $\varepsilon^J$, where $J$ is the spin of the resonant state. The same applies to $\delta^J$. The distinction is also made between the phase shifts with or without mixing. Phase shifts with mixing are denoted as $\delta_k^J$ where $k = \alpha, \beta, \ldots$, whereas phase shifts without mixing are denoted as $\delta_l^J$ where $l$ is the orbital angular momentum of the incoming wave.





### Description of the phase-shift formalism

For an outline of the phase-shift formalism see the Appendix J. Here, I am presenting only selected points relevant to our analysis.

With no mixing between partial waves, each diagonal element of the scattering or collision matrix $U_{l'l}^J$ can be described by the corresponding phase shift

$$U_{l'l}^J = e^{2i\delta_l}$$

If one considers only elastic scattering, the phase shifts are real, automatically giving a unitary collision matrix. Allowing the phase shifts to be complex accounts for absorption into inelastic channels. This is often represented by the inelastic parameter

$$\eta_l^J = e^{-2\operatorname{Im}\delta_l^J}$$

For $J = 1^+$, the scattering matrix with non-zero off-diagonal elements has the following form (Blatt-Biedenharn 1952):[25]

$$U = \begin{bmatrix} U_{00}^1 & U_{02}^1 \\ U_{20}^1 & U_{22}^1 \end{bmatrix} = \begin{bmatrix} (\cos^2 \varepsilon^1)e^{2i\delta_\alpha^1} + (\sin^2 \varepsilon^1)e^{2i\delta_\beta^1} & \dfrac{1}{2}(\sin 2\varepsilon^1)(e^{2i\delta_\alpha^1} - e^{2i\delta_\beta^1}) \\ \dfrac{1}{2}\sin(2\varepsilon^1)(e^{2i\delta_\alpha^1} - e^{2i\delta_\beta^1}) & (\sin^2 \varepsilon^1)e^{2i\delta_\alpha^1} + (\cos^2 \varepsilon^1)e^{2i\delta_\beta^1} \end{bmatrix}$$

The unitary condition of the collision matrix containing complex phase-shift parameters has to be modified. The new condition is satisfied if the imaginary parts of the phase shifts are positive ($0 \le \eta_l^J \le 1$). For complex mixing parameters, additional inequalities that must be satisfied were given by Arvieux (1967). However, Seyler (1969) pointed out that only one of these relations is sufficient:

$$\left| e^{2i\delta_\beta} - e^{2i\delta_\alpha} \right| \sinh^2(\operatorname{Im} 2\varepsilon) \le (1 - \eta_\beta^2)(1 - \eta_\alpha^2)$$

Examination of this inequality leads to the conclusion that the imaginary part of the mixing parameter can be non-zero only when there is absorption in both of the mixed channels, i.e. when $\operatorname{Im}\delta_\alpha$ and $\operatorname{Im}\delta_\beta > 0$. The inequality proposed by Seyler was included in our phase-shift analysis as a required constraint.

Precision data are needed to determine not only the real but also the imaginary component of the mixing parameter. As will be shown later, our data yielded sufficiently high-quality values for this component to include it in our *R*-matrix calculations.

### Phase-shift results and discussion

The starting phase-shift parameters were taken from the results of Grüebler *et al.* (1975). We have also included some earlier data (Schmelzbach *et al.* 1972) at both ends of our energy range. The phase-shift analysis was carried out using a modified code of McIntyre (1965). The program used in the calculations minimises the $\chi^2$ function for the data consisting of the differential cross sections and of spin-one analyzing powers $iT_{11}$, $T_{20}$, $T_{21}$ and $T_{22}$.

---

[25] Blatt and Biedenharn use letter *S* for this matrix (see also the Appendix J).

A summary of the quality of fits generated by the phase-shift analysis is shown in Table 22.1. As can be seen by looking at the $\chi^2$ and weighted variance values, the resulting fits to the old data (Grüebler *et al.* 1975) were generally poor. In contrast, the fits to the new data were significantly better. The reason is that the new data are of much higher quality.

Table 22.1
Summary of the quality of the phase-shift analysis

| $E_d$ | Number of data points | $\chi^2$ | Weighted variance |
|---|---|---|---|
| 4.81[a]) | 61 | 2.2 | 3.4 |
| 5.32[a]) | 59 | 1.5 | 2.4 |
| 5.83[a]) | 65 | 1.5 | 2.3 |
| 6.04 | 162 | 0.67 | 0.82 |
| 6.24 | 163 | 0.75 | 0.91 |
| 6.44 | 161 | 0.71 | 0.86 |
| 6.66 | 160 | 0.80 | 0.98 |
| 6.76 | 151 | 0.66 | 0.82 |
| 5.85 | 161 | 0.67 | 0.81 |
| 7.05 | 168 | 1.18 | 1.42 |
| 7.86[a]) | 83 | 1.1 | 1.7 |

[a]) Old data (Grüebler *et al.* 1975)

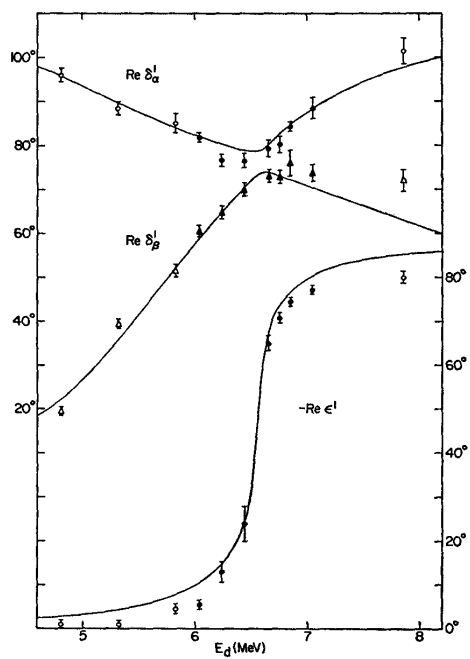

Figure 22.7. The real parts of the eigen phase shifts $\delta_\alpha^1$, $\delta_\beta^1$ and of the mixing parameter $\varepsilon^1$ for the d - α elastic scattering. The dots and solid triangles are the result of the analysis of the present data; open circles and triangles are for the calculations based on the data of Schmelzbach *et al.* (1972). The scale for $\varepsilon^1$ is on the right-hand side. The curves are the single-level *R*-matrix fits.

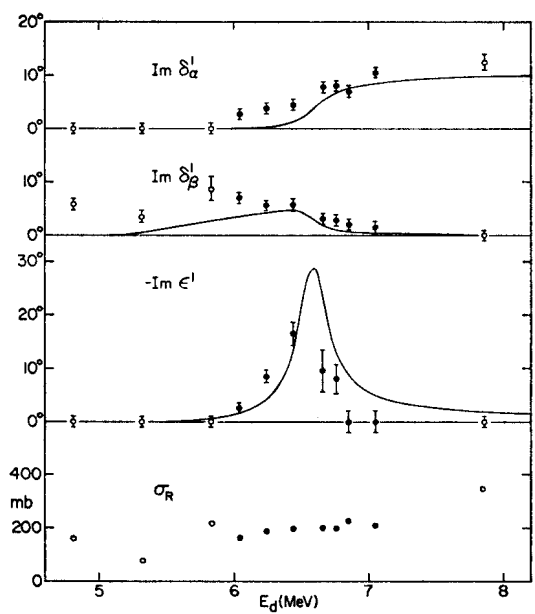

Figure 22.8. The imaginary parts of the eigen phase shifts $\delta_\alpha^1$, $\delta_\beta^1$ and of the mixing parameter $\varepsilon^1$ for the d-$\alpha$ elastic scattering. The dots and solid triangles are the result of analysis of the present data; open circles and triangles for the calculations based on the data of Schmelzbach *et al.* (1972). The curves are the single-level *R*-matrix fits. The total reaction cross sections calculated from the complete set of phase shifts are shown in the lowest section of the figure.

The good quality of fits to the data points can be also seen in Figures 22.1-22.5. The only systematic deviation in the fits occurs near the maximum in $T_{22}$. All other observables are well reproduced by the phase shift calculations. The overall $\chi^2$ value for the new data was 0.78 and the weighted variance 0.95.

The phase parameters that are affected by the 1+ resonance under investigation are shown in Figures 22.7 and 22.8. The curves in these figures are results of the single-level *R*-matrix fit to the phases, as discussed later.

The deuteron break-up $^5$He+p channel is open in the entire range of energies used in the present experiment. The $^5$Li+n channel opens at 6.3 MeV and the clear change in $\mathrm{Im}\,\delta_\alpha^1$ might be associated with the opening of this channel. The imaginary part of $\varepsilon^1$ is negligible except when the real part passes through 45°. It might be of interest to mentions that the imaginary component produces a noticeable improvement in the fits to the data in this region. Above and below the resonance the unitary condition forces $\mathrm{Im}\,\varepsilon^1$ to be zero because one of the inelastic phases approaches zero at these energies.

**The R-matrix analysis**

Level parameters of the 1+ resonance in $^6$Li can be extracted using the single-level approximation of the *R*-matrix theory. As mentioned earlier, the shape of the mixing parameter depends on the ratio of two partial widths contributing to the resonance, and using our more precise data we expected to determine this ratio with a higher precision than before. In addition, we expected a more precise description of the 1+ resonance.

To achieve these goals, we have carried out $R$-matrix calculations using our phase shift parameters. All the calculations were done using the single-level $R$-matrix code LEV21 (McIntyre 1965), which we have modified by combining it with the general search routine MINUIT (James and Roos 1975). The modified program performed an automatic search of $R$-matrix parameters while fitting the generated earlier phase shifts.

The resulting level parameters are listed in Table 22.2. They are compared with the parameters obtained in previous analyses (McIntyre and Haeberli 1967; Schmeltzbach *et al.* 1972). The calculated values for the phase parameters are shown as solid lines in Figures 22.7 and 22.8. The parameter errors from our fits were of the order of one part in the last significant digit for values listed in the Table 22.2. As can be seen from this table, the ratio of the $s$-wave to $d$-wave reduced widths is 0.002, confirming the very small effect of the tensor interaction in this resonance.

Table 22.2
Single-level parameters of the $J = 1^+$ resonance

|  |  | a) | b) | This work |
|---|---|---|---|---|
| Deuteron energy at resonance | $E_d$ | 6.4 | 6.4 | 6.26 |
| Energy eigenvalue | $E_\lambda$ | 8.4 | 9.8 | 8.75 |
| Excitation energy in $^6$Li | $E_x(^6\text{Li})$ | 5.7 | 5.7 | 5.65 |
| Reduced widths: |  |  |  |  |
| elastic s-wave | $\gamma^2_{\lambda, l=0}$ | 0.01 | 0.003 | 0.005 |
| elastic d-wave | $\gamma^2_{\lambda, l=2}$ | 2.1 | 3.1 | 2.5 |
| inelastic proton emission | $\gamma^2_{\lambda, l=1}$ | 0.5 | 0.8 | 0.8 |
| inelastic neutron emission | $\gamma^2_{\lambda, l=1}$ | 0.5 | 0.8 | 0.8 |
| Interaction radius (fm) | $a$ | 4.3 | 4.1 | 4.12 |

All parameters, except for the interaction radius, are in MeV.
a) – McIntyre and Haeberli (1967); b) – Schmeltzbach *et al.* (1972)

Even though the imaginary component of the mixing parameter is not well determined in this search, it is encouraging that the $R$-matrix calculations give its correct sign and shape. Our ability to include this component in our study shows the importance of precise experimental results in this type of analyses.

## Comparison with microscopic calculations

The earliest resonating group calculations for the d-$\alpha$ scattering used no spin-orbit or tensor force in the nucleon-nucleon potential and thus could not predict polarizations. Lemere, Tang and Thompson (1976) added a spin-orbit component to the potential in their calculations that extended to around 54 MeV (c.m.). The most complete calculation below $E_{c.m.}$ = 8 MeV was made by Hackenbroich *et al.* (1974) who not only used a nucleon-nucleon potential containing spin-orbit and tensor terms, but also coupled several reaction channels in their cluster-model scheme. These authors were able to predict the mixing between the $l = 0$, and $l = 2$ partial waves. In their calculations, the break-up of the deuteron was assumed to proceed through the $^5$He-p and $^5$Li-n fragmentations. They have also considered the excited states of $^5$He and $^5$Li in such break-up channels.

Some results of Hackenbroich *et al.* (1974) are shown in Figure 22.9 together with our values for the $\varepsilon^1$ mixing parameter. The dashed curve presents an interesting case. It shows that in principle, if one includes all inelastic channels, strong coupling between the *l* = 0 and *l = 2* partial waves is possible without a tensor forces. Referring to the calculations presented in Figure 22.6, this strong coupling is demonstrated by the strong smoothing out of the shape of the energy dependence of the $\varepsilon^1$ mixing parameter. However, these results appear to be physically meaningless for the studied resonance because the calculated curve does not even come close to the experimentally determined values of the mixing parameter.

The fit to the experimentally determined mixing parameters is greatly improved if tensor force is included in the calculations. However, the resulting fit is still far from perfect (see the full line in Figure 22.9). The best fit is obtained if tensor force is retained but if inelastic break-up fragmentations ($^5$He*-p and $^5$Li*-n) are excluded from the calculations. As the energies used in our measurements are below the thresholds for these inelastic fragmentations their exclusion is not surprising. What is surprising, however, is that their inclusion produces such a large effect.

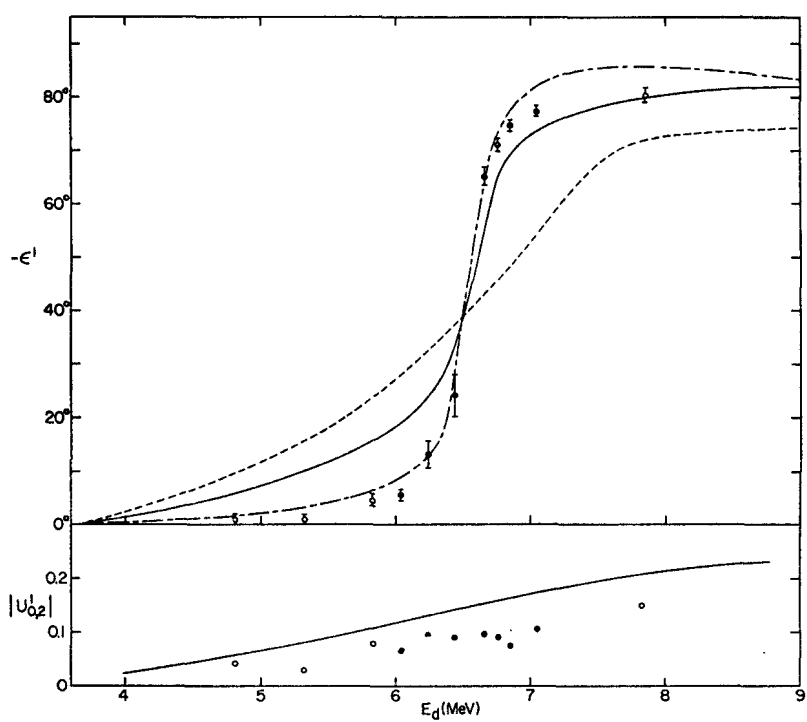

Figure22.9. Comparison of the results of our phase shift analysis (dots) for the *J* = 1 mixing parameter with the microscopic calculations of Hackenbroich *et al.* (1974) (curves). Solid line: calculations *with* tensor potential and with *all* fragmentations. Dashed line: calculations *without* tensor potential but with *all* fragmentations. Dashed-dotted line: calculations *with* tensor potential but *without* $^5$He*-p and $^5$Li*-n fragmentations. The lower section of the figure shows our results (dots) for the absolute values of the off-diagonal element $\left| U_{02}^1 \right|^2$ of the scattering matrix compared with the calculation of Schütte *et al.* (1976) who used tensor potential and all fragmentations.

It is also useful to look at the magnitude of the off-diagonal matrix element $\left| U_{02}^1 \right|^2$. This element gives a direct measure of the mixing between the two partial waves and is not complicated by the rapid variation of the $\varepsilon^1$ mixing parameter near the phase crossing region. In the lower part of Figure 22.9 we have plotted this quantity as generated by our analysis (dots) together with the calculations of Schütte *et al.* (1976) who included tensor force and all fragmentations. Again it is clear that the calculated mixing is too high when coupling to all channels is included.

In conclusion, we have demonstrated that tensor interaction in the d-$\alpha$ scattering at energies corresponding to the 1$^+$ resonance in $^6$Li is small but that it still has a decisive influence on the coupling between the $s$- and d-waves. We have also demonstrated that in order to study the details of resonance interactions it is essential to have precise data for the distributions of the differential cross sections and analyzing powers. It is by having a full complement of such measurements that we were able to gain detailed insights into the 1$^+$ resonance in $^6$Li and into the accompanying reaction mechanism. Our data can also serve as a reliable basis for more refined analyses.

## References

Arvieux, J. 1967, *Nucl. Phys.* **A102**:513.

Blatt, J. M. and Biedenharn, L. C. 1952, *Rev. Mod. Phys.* **24**:258.

Grüebler, W., Schmelzbach, P. A., König, V., and Boerma, D. 1975, *Nucl. Phys.* **A242**:265.

Grüebler, W., König, V., Schmelzbach, Jenny, B., Bürgi, H. R., Hardekopf, R. A., Nurzynski, J. and Risler, R. 1980, *Nucl. Phys.* **A334**:365.

Hackenbroich, H. H., Heiss, P. and Le-Chi-Niem, 1974, *Nucl. Phys.* **A221**:461.

James, F. and Roos, M. 1975, *Comp. Comm.* **10**:343.

Lemere, M., Tang, Y. C. and Thompson, D. R. 1976, *Nucl. Phys.* **A266**:1.

McIntyre L. C. 1965, PhD thesis, University of Wisconsin

McIntyre, L. C. and Haeberli, W. 1967, *Nucl. Phys.* **A91**:382.

Ohlsen, G. G., Lovoi, P. A., Salzman, G. C., Meyer-Berkhout, U., Mitchell, C. K. and Grüebler, W. 1973, *Phys. Rev.* **C8**:1262.

Schmelzbach, P. A., Grüebler, W., König, V. and Marmier, P. 1972, *Nucl. Phys.* **A184**:193.

Schütte, W., Hackenbroich, H. H., Stöwe, H. Heiss, P. and Aulenkamp, H. 1976, *Phys. Lett.* **65B**:214.

Seyler, R. G. 1969, *Nucl. Phys.* **A124**:253.